\documentclass[12pt,a4paper]{article}
\usepackage[numbers]{natbib}
\usepackage{authblk}
\usepackage[margin=1in]{geometry}

\RequirePackage{amsthm,amsmath,amsfonts,amssymb}
\RequirePackage{graphicx}

\usepackage[noend]{algpseudocode}
\usepackage{algorithm}
\usepackage{algorithmicx}
\usepackage{color}
\usepackage{enumerate}
\usepackage{graphics}
\usepackage{pagecolor}
\usepackage{soul}
\usepackage{xcolor}
\usepackage{xspace}
\usepackage[normalem]{ulem}
\usepackage{setspace} 
\usepackage{hypbmsec}
\usepackage[all]{xy}
\usepackage{bm}
\usepackage[colorlinks=true]{hyperref}
\usepackage{bbm}
\usepackage{amsmath}
\usepackage{mathtools}
\DeclareMathOperator{\Tr}{Tr}

\RequirePackage{epsfig}
\RequirePackage{amssymb}
\RequirePackage{natbib}
\RequirePackage{graphicx}
\RequirePackage{verbatim}
\RequirePackage{algorithm}
\RequirePackage[noend]{algpseudocode}
\RequirePackage{bbm}
\RequirePackage{esdiff}
\RequirePackage{bm}
\RequirePackage{physics}
\RequirePackage{titlesec}
\RequirePackage{xpatch}
\usepackage{mathtools}
\usepackage{caption}
\usepackage{subcaption}

\RequirePackage{epsfig}
\RequirePackage{amssymb}
\RequirePackage{natbib}
\RequirePackage{graphicx}

\RequirePackage{amsmath}
\RequirePackage{mathtools}
\RequirePackage{subcaption}
\RequirePackage{xspace}
\RequirePackage[all]{xy}
\RequirePackage{bm}
\RequirePackage{physics}
\RequirePackage{algorithm}
\RequirePackage{enumerate}
\RequirePackage[noend]{algpseudocode}
\RequirePackage{bbm}
\RequirePackage{esdiff}
\RequirePackage{titlesec}
\RequirePackage{xpatch}

\newcommand{\tX}{\ensuremath{\mathfrak{X}}\xspace} 
\newcommand{\x}{{\boldsymbol x}} 
\newcommand{\vecX}[2]{{ \vec {\x}}_{#1}^{#2}} 
\newcommand{\extended}[1]{g_{#1}}
\newcommand{\proposal}[1]{h_{#1}}
\newcommand{\ddspace}[1]{\text{ }\dd#1}
\newcommand{\children}[1]{\ensuremath{\textrm{Ch}(#1)}\xspace}
\newcommand{\leaf}[1]{\ensuremath{\textrm{Leaf}(#1)}\xspace}
\newcommand{\troot}[1]{\ensuremath{\textrm{Root}(#1)}\xspace}
\newcommand{\tree}{\ensuremath{\mathbb{T}}\xspace}
\newcommand{\vertices}{\ensuremath{\mathcal{V}}\xspace}
\newcommand{\edges}{\ensuremath{\mathcal{E}}\xspace}
\newcommand{\Rlang}{\textbf{\textsf{R}}}
\newcommand{\Rcpplang}{\textbf{\textsf{Rcpp}}}
\newcommand{\Cpplang}{C\nolinebreak\hspace{-.05em}\raisebox{.4ex}{\tiny\bf +}\nolinebreak\hspace{-.10em}\raisebox{.4ex}{\tiny\bf +}}

\newcommand{\plim}{\text{\emph{plim}}}

\newcommand{\vecXX}[2]{{ \vec {\bm{X}}}_{#1}^{#2}} 
\newcommand{\vecM}[2]{{ \vec {\bm{M}}}_{#1}^{#2}} 
\newcommand{\CESS}[1]{\text{CESS}_{#1}}
\newcommand{\ESS}{\text{ESS}}
\newcommand{\IAD}{\text{IAD}}
\newcommand{\SH}[1]{\allowbreak\text{SH}(#1)}
\newcommand{\SSH}[1]{\allowbreak\text{SSH}(#1)}

\newcommand{\gmcf}{Generalised Monte Carlo Fusion\xspace}
\newcommand{\gmcfa}{GMCF\xspace}
\newcommand{\gmcfl}{Generalised Monte Carlo Fusion (GMCF)\xspace}
\newcommand{\hmcf}{Divide-and-Conquer Generalised Monte Carlo Fusion\xspace}
\newcommand{\hmcfa}{D\&C-GMCF\xspace}

\newcommand{\generalisedbf}[1]{\allowbreak\hyperref[alg:GBF]{\ensuremath{\texttt{gbf}(#1)}}\xspace}
\newcommand{\dcgbf}[1]{\allowbreak\ensuremath{\hyperref[alg:dc_gbf]{\texttt{D\&C-Fusion}(#1)}}\xspace}
\newcommand{\gbf}{Generalised Bayesian Fusion\xspace}
\newcommand{\gbfa}{GBF\xspace}
\newcommand{\gbfl}{Generalised Bayesian Fusion (GBF)\xspace}
\newcommand{\hgbf}{Divide-and-Conquer Fusion\xspace}
\newcommand{\hgbfa}{D\&C-Fusion\xspace}
\newcommand{\hgbfl}{Divide-and-Conquer Fusion (D\&C-Fusion)\xspace}

\newcommand{\radonniko}{Radon-Nikod\'{y}m }

\newcommand{\iid}{\allowbreak\text{iid}}

\newcommand{\algoref}[1]{\hyperref[#1]{Algorithm \ref{#1}}}
\newcommand{\stepref}[1]{\hyperref[#1]{Step \ref{#1}}}
\newcommand{\algstepref}[2]{\hyperref[#1]{Algorithm \ref{#1} Step \ref{#2}}}
\newcommand{\figref}[1]{\hyperref[#1]{Figure \ref{#1}}}
\newcommand{\apxref}[1]{\hyperref[#1]{Appendix \ref{#1}}}
\newcommand{\condref}[1]{\hyperref[#1]{Condition \ref{#1}}}
\newcommand{\defnref}[1]{\hyperref[#1]{Definition \ref{#1}}}
\newcommand{\thmref}[1]{\hyperref[#1]{Theorem \ref{#1}}}
\newcommand{\cororef}[1]{\hyperref[#1]{Corollary \ref{#1}}}
\newcommand{\remref}[1]{\hyperref[#1]{Remark \ref{#1}}}
\newcommand{\propositionref}[1]{\hyperref[#1]{Proposition \ref{#1}}}
\newcommand{\secref}[1]{\hyperref[#1]{Section \ref{#1}}}
    
\newtheorem{theorem}{\bf Theorem}[section]

\newtheorem{proposition}{\bf Proposition}[section]
\newtheorem{definition}{\bf Definition}[section]
\newtheorem{condition}{\bf Condition}[section]

\newtheorem{remark}{\bf Remark}[section]
\newtheorem{corollary}{\bf Corollary}[section]

\providecommand{\keywords}[1]
{
  \small	
  \textbf{\textit{Keywords---}} #1
}

\title{Divide-and-Conquer Fusion}

\author[1,3]{Ryan S.Y. Chan\footnote{Corresponding author. Email: rchan@turing.ac.uk}}
\author[2,3]{Murray Pollock}
\author[1,3]{Adam M. Johansen}
\author[1,3]{Gareth O. Roberts}
\affil[1]{Department of Statistics, University of Warwick, Coventry, CV4 7AL.}
\affil[2]{School of Mathematics, Statistics and Physics, Newcastle University, Newcastle-upon-Tyne, United Kingdom, NE1 7RU.}
\affil[3]{The Alan Turing Institute, British Library, London, United Kingdom, NW1 2DB.}

\begin{document}

\maketitle


\begin{abstract}
Combining several (sample approximations of) distributions, which we term \emph{sub-posteriors}, into a single distribution proportional to their product, is a common challenge. Occurring, for instance, in distributed `big data' problems, or when working under multi-party privacy constraints. Many existing approaches resort to approximating the individual sub-posteriors for practical necessity, then find either an analytical approximation or sample approximation of the resulting (product-pooled) posterior. The quality of the posterior approximation for these approaches is poor when the sub-posteriors fall out-with a narrow range of distributional form, such as being approximately Gaussian. Recently, a \emph{Fusion} approach has been proposed which finds an exact Monte Carlo approximation of the posterior, circumventing the drawbacks of approximate approaches. Unfortunately, existing Fusion approaches have a number of computational limitations, particularly when unifying a large number of sub-posteriors. In this paper, we generalise the theory underpinning existing Fusion approaches, and embed the resulting methodology within a recursive  divide-and-conquer sequential Monte Carlo paradigm. This ultimately leads to a competitive Fusion approach, which is robust to increasing numbers of sub-posteriors.
\end{abstract}




\section{Introduction} \label{sec:introduction}

In this paper, we are interested in the following $d$-dimensional (\emph{product-pooled}) target density (which we term the \emph{fusion density}),
\begin{align}
    f(\bm{x}) \propto f_{1}(\bm{x}) \cdots f_{C}(\bm{x}) = \prod_{c=1}^{C} f_{c}(\bm{x}), \label{eq:fusion_density}
\end{align}
where $\bm{x} \in \mathbb{R}^{d}$, $f_{c}(\bm{x})$ for $c\in\{1,\dots,C\}$ represent the individual densities which we wish to unify (termed \emph{sub-posteriors} in deference to the fact that a major application of this technique will be the setting in which the posterior is proportional to the product of these factors), and $C$ represents the total \emph{number} of sub-posteriors. We assume that we have access to independent realisations from each sub-posterior, and that it is possible to evaluate each sub-posterior pointwise up to its normalising constant. Although typically, one would only have approximate samples from each sub-posterior, we will discuss later that neither of these assumptions form limiting factors for our methodology.

The need to unify several (sample approximations of) distributions, over a common parameter space, into a single sample approximation of the distribution in the manner of \eqref{eq:fusion_density} is surprisingly common. For instance, it arises classically in expert elicitation \citep{Albert_et_al_2012, Berger_1980, Genest_Zidek_1986} and meta-analysis \citep{Fleiss_1993}. However, it has proven to be challenging methodologically in a number of modern settings due to problem specific constraints. These include when dealing with the \emph{privacy constraints} of the individual sources \citep{Yildrim_et_al_2019}, in cases where the sheer \emph{number of sources} is overwhelming, or if the \emph{networking constraints} of the sources are truly \emph{distributed} \citep{Scott_et_al_2016}. This in turn has motivated a range of problem specific and pragmatic \emph{approximations}. These approximations are invariably distributional, and imposed at the level of the individual source (for instance, the sub-posteriors being approximately Gaussian). Such approximations limit the applicability of methodological approaches to particular settings, and outside those settings the unified results can be poorly understood, and even misleading. We instead focus on developing methodology for an \emph{exact} Monte Carlo approximation of the unified distribution \eqref{eq:fusion_density}---one which provides robust inference in a wide range of practical problems, and yet is amenable to use alongside any problem specific constraints.

The majority of the recent methodological developments for representing or sampling from \eqref{eq:fusion_density} have been focused on tackling distributed \emph{`big data'} problems \citep[see for instance][]{Scott_et_al_2016, Neiswanger_et_al_2014, Wang_and_Dunson_2013, Minsker_et_al_2014, Srivastava_et_al_2015, Nemeth_and_Sherlock_2018}. In this setting, due to its sheer size, the data is split across a number of cores (say $C$ cores), inference is separately conducted on each core (often using MCMC), and then the respective methodologies attempt to unify the sample approximations of the distribution (as per \eqref{eq:fusion_density}, and typically using a convenient approximation). In this paper, we will compare our methodology with a number of the most popular approaches, and so will briefly describe these here. The \emph{Consensus Monte Carlo (CMC)} approach of \citet{Scott_et_al_2016} produces approximate samples from \eqref{eq:fusion_density} by means of a weighted average of sub-posterior samples. It can be shown that CMC is \emph{exact} when each sub-posterior is Gaussian, and can be useful in settings where each sub-posterior is approximately Gaussian, which is often the case in big data settings \citep{Walker_1969, Johnson_1970, Le_Cam_1986, Van_der_Vaart_1998, Le_Cam_Lo_2000}. However, it has been shown to exhibit large bias in other settings \citep{Wang_and_Dunson_2013}. \citet{Neiswanger_et_al_2014} suggest a strategy (which we term the \emph{Kernel Density Estimate Monte Carlo (KDEMC)} approach) based on using a kernel density estimate to approximate the sub-posterior densities, and in effect approximating \eqref{eq:fusion_density} by implicitly sampling from the product of non-parametric density estimates. Finally, the \emph{Weierstrass sampler} of \citet{Wang_and_Dunson_2013} provides an alternative method for approximating \eqref{eq:fusion_density} by means of using the product of Weierstrass transforms for each sub-posterior. Interestingly, we find empirically that for a cheap and crude approximation of \eqref{eq:fusion_density} then the (simplest) CMC approach outperforms all other methodologies, but in cases where accuracy is a concern then our (more computationally expensive) \emph{Fusion} approach should be used. 

The \emph{Fusion approach} \citep{Dai_et_al_2019, dai_et_al_2023} constructs a direct sample approximation of \eqref{eq:fusion_density} itself, rather than seeking to obtain an adhoc approximation of $f$ by combining approximations of the sub-posteriors. Underpinning the Fusion approach is the simple observation that if we sampled (independently) $\bm{X}^{(c)} \sim f_c$ for $c\in\{1,\dots,C\}$ then conditional on the event that $\bm{X}^{(1)} = \dots = \bm{X}^{(C)}$, we have that $\bm{X}^{(1)}$ has density $f$ given in \eqref{eq:fusion_density}.

Clearly the difficulty with exploiting this observation is that we are conditioning on an event of probability $0$. The \emph{Monte Carlo Fusion (MCF)} approach of \citet{Dai_et_al_2019} provides a framework for practically enforcing this conditioning. This is achieved by initialising $C$ stochastic processes (independently from one another) using a single realisation from each sub-posterior (i.e.\ $\bm{X}^{(c)}_0 \sim f_c$ for $c\in\{1,\dots,C\}$ where the subscript is a temporal index, noting that $\bm{X}^{(1)}_0 \neq \dots \neq \bm{X}^{(C)}_0$), evolving the processes in such a manner that (i) these processes \emph{coalesce} at some fixed future time (i.e.\ $\bm{X}^{(1)}_T = \dots = \bm{X}^{(C)}_T$); and (ii), the common marginal distribution at the coalescence time, $T$, is $f$. By repeating this approach multiple times, MCF provides multiple i.i.d. draws from $f$. 

The \emph{Bayesian Fusion (BF)} approach of \citet{dai_et_al_2023} re-examined the theoretical underpinnings of MCF by introducing a \emph{stochastic differential equation (SDE)} describing the coalescence of the $C$ stochastic processes, and exploited this theory together with methodology for \emph{sequential Monte Carlo (SMC)} to \emph{gradually} coalesce the stochastic processes. The resulting output of the BF approach is a number of \emph{correlated} and \emph{weighted} draws from $f$. BF is a far more practical and robust algorithm than MCF. A key advantage of BF over MCF is that it is possible to give considerable user guidance in its implementation.

Although BF provides significant improvements over MCF, the applicability of the methodology is still limited by factors including: (i) the numbers of sub-posteriors being combined; (ii) the level of sub-posterior correlation; (iii) the dimensionality of the sub-posteriors; (iv) the degree to which the sub-posteriors \emph{conflict}; and (v) the computational cost of the approach even when the user-specified tuning parameters are optimally chosen. 
In this paper, we make two key contributions to address the limitations of MCF and BF: (i) we significantly improve upon the computational efficiency of BF by allowing the user to incorporate global information about each sub-posterior within the SDE formulation, and unify \emph{subsets} of the sub-posteriors at any one time---we term this approach \emph{\gbfl}, and present it in \secref{sec:GBF} and \algoref{alg:GBF}; (ii) using the flexibility given by (i) in which sub-posteriors can be partially unified, we embed our \gbfa methodology within the \emph{divide-and-conquer} paradigm of \citet{Lindsten_et_al_2017}, allowing the user to combine sub-posteriors in \emph{stages} to recover the fusion density $f$. We term this \emph{\hgbfl}, and present it in \secref{sec:dc_gbf} and \algoref{alg:dc_gbf}. 

The remainder of the paper is organised as follows: In \secref{sec:GBF_guidance} we present detailed guidance on implementing our \gbfa and \hgbfa approaches, and in particular choosing any tuning parameters. In \secref{sec:dc_gbf_examples} we present applications of our methodology for a variety of models, comparing them to competing approximate methodologies. We conclude by outlining a variety of ways or Fusion approach could be extended, and used in other application settings. All technical proofs and detailed calculations are collated in the appendices.

Statistical computations for this paper were written in \Rlang\, \citep{R_lang}, \Cpplang\, and \Rcpplang\, \citep{Eddelbuetell_2013}. The code for this paper can be found on GitHub at \href{https://github.com/rchan26/DCFusion}{\texttt{https://github.com/rchan26/DCFusion}}.


\section{A generalisation of the Fusion approach} \label{sec:GBF}

In this section we develop theory and methodology to generalise and improve upon the BF approach of \citet{dai_et_al_2023}, by incorporating information about the covariance of the sub-posteriors within the SDE formulation. For completeness in Appendix \ref{app:connections} we more fully outline the connections of our methodology to the earlier MCF and BF works, highlighting explicitly the advantages of our approach, but for ease of presentation here we instead present our approach directly. In this section we also consider the more abstract problem of sampling from the density $f^{(\mathcal{C})} \propto \prod_{c\in\mathcal{C}} f_{c}$, where $\mathcal{C}$ is an index set representing the sub-posteriors we want to unify, and we assume we can sample (independently) $\bm{X}^{(c)} \sim f_c$ for $c \in{\mathcal{C}}$. This abstraction is useful for the methodology we develop in \secref{sec:dc_gbf}. 

For the purposes of simplifying the subsequent notation, we denote by $\vecX{t}{(\mathcal{C})}\in\mathbb{R}^{|\mathcal{C}| \times d}$ a vector composed of $\bm{x}_{t}^{(c)}\in\mathbb{R}^{d}$ for $c\in\mathcal{C}$ (in particular, we have $\vecX{t}{(\mathcal{C})}:=(\bm{x}_{t}^{(c_{1})}, \dots, \bm{x}_{t}^{(c_{|\mathcal{C}|})})$, with $c_{i}$ denoting the $i^\text{th}$ element of the index set $\mathcal{C}$). We further assume that for $c\in\{1,\dots,C\}$, $f_c$ is nowhere zero and everywhere differentiable, and that we can compute $A_{c}(\bm{x}):=\log f_{c}(\bm{x})$, $\nabla A_{c}(\bm{x})$, and $\nabla^{2} A_{c}(\bm{x})$ pointwise (where $\nabla$ is the gradient operator and $\nabla^{2}$ is the Hessian). A fuller discussion of these assumptions is given in Appendix \ref{app:connections}, but note that they match those of the earlier works of \citet{Dai_et_al_2019,dai_et_al_2023}.

We begin by describing the joint distribution of $|\mathcal{C}|$ coalescing stochastic processes on $[0,T]$ that at time $T$ have the common marginal $f^{(\mathcal{C})} \propto \prod_{c\in\mathcal{C}} f_{c}$. We term this the \emph{fusion measure}, $\mathbb{F}$. To aid in the development of the subsequent methodology, we require that the stochastic processes can be simulated, and so this is done by considering a \radonniko correction of the so-called \emph{proposal measure} ($\mathbb{P}$), which is defined to be the probability law induced by $|\mathcal{C}|$ interacting $d$-dimensional parallel continuous-time Markov processes in $[0,T]$ where each process is given by the SDE,
\begin{equation}
    \label{eq:proposal_SDE}
    \dd \bm{X}_{t}^{(c)} = \frac{\tilde{\bm{X}}_{t}-\bm{X}_{t}^{(c)}}{T-t} \dd t + \mathbf{\Lambda}_{c}^{\frac{1}{2}} \dd \bm{W}_{t}^{(c)}, \qquad \bm{X}_{0}^{(c)}:=\bm{x}_{0}^{(c)} \sim f_{c}, \quad t \in [0,T],
\end{equation}
where $\mathbf{\Lambda}_{c}$ are (positive semi-definite) user-specified matrices associated to sub-posterior $f_{c}$ for $c\in\mathcal{C}$ with $\mathbf{\Lambda}_{c}^{1/2}$ being the (positive semi-definite) square root of $\mathbf{\Lambda}_c$ where $\mathbf{\Lambda}_{c}^{1/2} \mathbf{\Lambda}_{c}^{1/2} = \mathbf{\Lambda}_{c}$. Note that for the purposes of our numerical simulations later we use the Schur decomposition. Furthermore, $\{\bm{W}_{t}^{(c)}\}_{c\in\mathcal{C}}$ denotes independent Brownian motions, and
\begin{equation*}
    \tilde{\bm{X}}_{t}^{(c)}:=\left(\sum_{c\in\mathcal{C}}\mathbf{\Lambda}_{c}^{-1}\right)^{-1}\left(\sum_{c\in\mathcal{C}}\mathbf{\Lambda}_{c}^{-1}\bm{X}_{t}^{(c)}\right),
\end{equation*}
denoting the weighted average of the processes at time $t$. In practice we typically take $\mathbf{\Lambda}_{c}$ to be a user estimate of the covariance matrix of the sub-posterior, $\hat{\mathbf{\Sigma}}_{c}$ which can be computed using the available  sub-posterior samples for $f_{c}$ thereby incorporating problem-specific information about covariance structure. We will see that the choice for these matrices influences the efficiency of the algorithm but not the target distribution itself and thus incurs no bias. Realisations of the proposal measure are denoted as $\tX:=\{\vecX{t}{(\mathcal{C})}, t\in[0,T]\}$. For the purposes of exposition, we defer discussion on the practical simulation of $\mathbb{P}$ to \secref{subsec:GBF_proposal_simulation}.

Now, we let the \emph{Fusion measure} $\mathbb{F}$ be simply the measure induced by the following \radonniko derivative:
\begin{equation}
    \label{eq:fusion_measure}
    \frac{\dd \mathbb{F}}{\dd \mathbb{P}}(\tX) \propto \rho_{0}\left(\vecX{0}{(\mathcal{C})}\right) \cdot \prod_{c\in\mathcal{C}} \left[ \exp \left\{ - \int_{0}^{T} \phi_{c} \left(\bm{X}_{t}^{(c)}\right) \dd t \right\} \right],
\end{equation}
where $\{\bm{X}_{t}^{(c)}, t\in[0,T]\}$ is a Brownian bridge from $\bm{X}_{0}^{(c)}:=\bm{x}_{0}^{(c)} \sim f_{c}$ to $\bm{X}_{T}^{(c)}:=\bm{x}_{T}^{(c)}$ with covariance matrix $\mathbf{\Lambda}_{c}$ and
\begin{equation}
    \label{eq:rho_0_gbf}
    \rho_{0}\left(\vecX{0}{(\mathcal{C})}\right) := \exp \left\{ - \sum_{c\in\mathcal{C}} \frac{(\tilde{\bm{x}}_{0}^{(\mathcal{C})}-\bm{x}_{0}^{(c)})^{\intercal} \mathbf{\Lambda}_{c}^{-1} (\tilde{\bm{x}}_{0}^{(\mathcal{C})}-\bm{x}_{0}^{(c)})}{2T} \right\},
\end{equation}
where 
\begin{equation}
    \label{eq:x_mean}
    \tilde{\bm{x}}_{t}^{(\mathcal{C})}:=\left(\sum_{c\in\mathcal{C}}\mathbf{\Lambda}_{c}^{-1}\right)^{-1}\left(\sum_{c\in\mathcal{C}}\mathbf{\Lambda}_{c}^{-1}\bm{x}_{t}^{(c)}\right),
\end{equation}
and
\begin{equation}
    \label{eq:phi}
    \phi_{c}(\bm{x}) := \frac{1}{2} \left( \nabla \log f_{c}(\bm{x})^{\intercal} \mathbf{\Lambda}_{c} \nabla \log f_{c}(\bm{x}) + \Tr(\mathbf{\Lambda}_{c} \nabla^{2} \log f_{c}(\bm{x})) \right).
\end{equation}

Now, considering the time $T$ marginal of $\tX\sim\mathbb{F}$ we (almost surely) have:

\begin{theorem}
    \label{theorem:fusion_measure_GBF}
    Under the fusion measure $\mathbb{F}$, the ending points of the $|\mathcal{C}|$ interacting, parallel processes have a common value at time $T$, $\bm{y}^{(\mathcal{C})}$ which has density $f^{(\mathcal{C})}$ and $\bm{y}^{(\mathcal{C})} = \bm{x}_{T}^{(c_{1})}=\dots=\bm{x}_{T}^{(c_{|\mathcal{C}|})}$ almost surely.
\proof See \apxref{app:fusion_measure_GBF}. \hfill $\blacksquare$
\end{theorem}

\thmref{theorem:fusion_measure_GBF} suggests that we can simulate from the fusion target density $f^{(\mathcal{C})}$ by simulating $\tX \sim \mathbb{F}$ and retaining the $T$ time marginal, $\bm{y}^{(\mathcal{C})}$. As suggested by the theory, we do so by means of simulating a number of proposals $\tX \sim \mathbb{P}$ and accepting (or importance weighting) the terminal time marginal $\bm{y}^{(\mathcal{C})}$ with probability proportional to the \radonniko derivative in \eqref{eq:fusion_measure}. As such, we need to consider: (i), how to simulate proposals from $\tX \sim \mathbb{P}$ (outlined in \secref{subsec:GBF_proposal_simulation}); and (ii), how to compute the \radonniko correction \eqref{eq:fusion_measure} (outlined in \secref{subsec:GBF_radonniko}). We then present our proposed complete methodology in \secref{subsec:GBF_methodology}. We discuss possible extensions of our approach in \secref{subsec:GBF_practicalities}.

\subsection{Simulating from the Proposal Measure} \label{subsec:GBF_proposal_simulation}

First, we consider how to simulate proposals from $\tX \sim \mathbb{P}$. We begin by noting that the initialisation of the proposal measure given by \eqref{eq:proposal_SDE} at time $t=0$ only requires independent draws from the $|\mathcal{C}|$ sub-posteriors that we wish to unify, which in this paper we assume we have access to. If independent sampling is not feasible, it is possible to obtain approximate sub-posterior samples using MCMC (see \citet[Section 3.6]{dai_et_al_2023} for a discussion on the impacts of using approximate sub-posterior samples for Fusion). Further, although paths $\tX \sim \mathbb{P}$ are infinite dimensional random variables (and so we cannot draw entire sample paths from $\mathbb{P}$), it is sufficient for our needs to simulate (exactly) the paths at a finite collection of times provided we can ensure that we are able to simulate the path (\emph{exactly}) at time $T$. For clarity, we only consider simulating $\tX$ at times given by the following auxiliary temporal partition,
\begin{equation}
    \label{eq:partition}
    \mathcal{P} = \{t_{0}, t_{1}, \dots, t_{n}: 0 =: t_{0} < t_{1} < \cdots < t_{n} := T\}.
\end{equation}

We let $\Delta_{j}:=t_{j}-t_{j-1}$ and for notational simplicity, subscripts are suppressed when considering the processes at times given in the temporal partition. In particular, let $\bm{x}_{j}^{(c)}$ denote $\bm{x}_{t_{j}}^{(c)}$, and let $\vecX{j}{(\mathcal{C})}$ denote $\vecX{t_{j}}{(\mathcal{C})}$. We will see from the following proposition, that algorithmically, to simulate from $\mathbb{P}$ at the time points in $\mathcal{P}$, we can simply initialise the $|\mathcal{C}|$ paths with $\bm{x}_{0}^{(c)} \sim f_{c}$ for $c\in\mathcal{C}$ and sequentially simulate from the Normal distributions given in \propositionref{prop:proposal_simulation_GBF}\ref{prop:proposal_simulation_GBF:a} (for $j\in\{1,\dots,n-1\}$) and \ref{prop:proposal_simulation_GBF:b} (for $j=n$). The following proposition tells us how to simulate from the transition density of $\mathbb{P}$:

\begin{proposition}
    \label{prop:proposal_simulation_GBF}
    Let $\mathcal{C}:=(c_{1},\dots,c_{|\mathcal{C}|})$ denote the index set representing the sub-posteriors we wish to unify, then if $\tX$ satisfies \eqref{eq:proposal_SDE}, then under the proposal measure, $\mathbb{P}$, we have
    \begin{enumerate}[(a)]
        \item \label{prop:proposal_simulation_GBF:a} For $s<t<T$,
        \begin{equation}
            \left. \vecXX{t}{(\mathcal{C})} \middle| \left(\vecXX{s}{(\mathcal{C})}=\vecX{s}{(\mathcal{C})}\right) \right. \sim \mathcal{N}_{|\mathcal{C}|d} \left(\vecM{s,t}{(\mathcal{C})}, \bm{V}_{s,t} \right),
        \end{equation}
        where $\vecM{s,t}{(\mathcal{C})}\in\mathbb{R}^{|\mathcal{C}|\times d}:=\left(\bm{M}_{s,t}^{(c_{1})},\dots,\bm{M}_{s,t}^{(c_{|\mathcal{C}|})}\right)$ with
        \begin{equation}
            \label{eq:M_GBF}
            \bm{M}_{s,t}^{(c)} = \frac{T-t}{T-s} \bm{x}_{s}^{(c)} + \frac{t-s}{T-s} \tilde{\bm{x}}_{s},
        \end{equation}
        and
        \begin{equation}
            \label{eq:V_GBF}
            \bm{V}_{s,t} =
            \begin{pmatrix}
            \bm{\Gamma}_{11} & \bm{\Gamma}_{12} & \dots & \bm{\Gamma}_{1|\mathcal{C}|} \\
            \bm{\Gamma}_{21} & \bm{\Gamma}_{22} & \dots & \bm{\Gamma}_{2|\mathcal{C}|} \\
            \vdots & \vdots & \ddots & \vdots \\
            \bm{\Gamma}_{|\mathcal{C}|1} & \bm{\Gamma}_{|\mathcal{C}|2} & \dots & \bm{\Gamma}_{|\mathcal{C}||\mathcal{C}|}
            \end{pmatrix} \in \mathbb{R}^{|\mathcal{C}|d \times |\mathcal{C}|d},
        \end{equation}
        where for $i,j=1,\dots,|\mathcal{C}|$,
        \begin{align}
            \bm{\Gamma}_{ii} & = \frac{(t-s)(T-t)}{T-s} \mathbf{\Lambda}_{c_{i}} + \frac{(t-s)^{2}}{T-s}\mathbf{\Lambda}_{\mathcal{C}} \in \mathbb{R}^{d \times d}, \\
            \mathbf{\Gamma}_{ij} & = \frac{(t-s)^{2}}{T-s}\mathbf{\Lambda}_{\mathcal{C}} \in \mathbb{R}^{d \times d}.
        \end{align}
        \item \label{prop:proposal_simulation_GBF:b} For $s<t=T$, $\bm{y}^{(\mathcal{C})}:=\bm{x}_{T}^{(c_{1})}=\dots=\bm{x}_{T}^{(c_{|\mathcal{C}|})} \sim \mathcal{N}_{d}(\tilde{\bm{x}}_{s}, \mathbf{\Lambda}_{\mathcal{C}})$.
        \item \label{prop:proposal_simulation_GBF:c} For each $c\in\mathcal{C}$, the distribution of $\{\bm{X}_{q}^{(c)}, s \leq q \leq t\}$ given endpoints $\bm{X}_{s}^{(c)}=\bm{x}_{s}^{(c)}$ and $\bm{X}_{t}^{(c)}=\bm{x}_{t}^{(c)}$ is a Brownian bridge with covariance matrix $\mathbf{\Lambda}_{c}$, so
        \begin{equation}
            \left. \bm{X}_{u}^{(c)} \middle| \left(\bm{x}_{s}^{(c)}, \bm{x}_{t}^{(c)} \right) \right. \sim \mathcal{N}_{d} \left( \frac{(t-q)\bm{x}_{s}^{(c)}+(q-s)\bm{x}_{t}^{(c)}}{t-s}, \frac{(t-q)(q-s)}{t-s}\mathbf{\Lambda}_{c}\right).
        \end{equation}
    \end{enumerate}
    \proof See \apxref{app:proposal_simulation_GBF}. \hfill $\blacksquare$
\end{proposition}

As we can initialise a draw from $\mathbb{P}$, and from \propositionref{prop:proposal_simulation_GBF} we can simulate from its transition density, we can now explicitly express the $d(n|\mathcal{C}|+1)$-dimensional density of the $|\mathcal{C}|d$-dimensional Markov process at the $(n+1)$ time marginals given by the temporal partition under $\mathbb{P}$, by iterative simulation from the transition density:
\begin{equation}
    \label{eq:h_GBF}
    \proposal{\mathcal{C}} \left( \vecX{0}{(\mathcal{C})}, \dots, \vecX{n-1}{(\mathcal{C})}, \bm{y}^{(\mathcal{C})} \right) \propto f\left(\vecX{0}{(\mathcal{C})}\right) \cdot \prod_{j=1}^{n-1} \mathcal{N}_{|\mathcal{C}|d}\! \left( \vecX{j}{(\mathcal{C})} \middle| \vecM{j}{(\mathcal{C})}, \bm{V}_{j} \right) \cdot \mathcal{N}_{d} \left( \bm{y}^{(\mathcal{C})} \middle| \tilde{\bm{x}}_{n-1}^{(\mathcal{C})}, \mathbf{\Lambda}_{\mathcal{C}} \right),
\end{equation}
where $f\left(\vecX{0}{(\mathcal{C})} \right) \propto  \prod_{c\in\mathcal{C}} f_{c} \left( \bm{x}_{0}^{(c)} \right)$, and $\mathcal{N}_{d}(\bm{x}|\bm{\mu}, \mathbf{\Sigma})$ denotes the density of a $d$-dimensional Normal distribution (evaluated at $\bm{x}$) with mean $\bm{\mu}$ and covariance $\mathbf{\Sigma}$. For notational convenience we let $\vecM{j}{(\mathcal{C})}=\vecM{t_{j-1},t_{j}}{(\mathcal{C})}$ and $\bm{V}_{j}=\bm{V}_{t_{j-1},t_{j}}$.

\subsection{\radonniko correction of the Proposal} \label{subsec:GBF_radonniko} 

Now, we direct our consideration to the second step: computing the \radonniko correction of \eqref{eq:fusion_measure}, given we have drawn our proposal from $\mathbb{P}$ restricted to the times given by the partition $\mathcal{P}$. Factorising the \radonniko derivative in \eqref{eq:fusion_measure} according to the temporal partition $\mathcal{P}$, the $d(n|\mathcal{C}|+1)$-dimensional density under $\mathbb{F}$ is
\begin{equation}
    \label{eq:g_GBF}
    \extended{\mathcal{C}} \left( \vecX{0}{(\mathcal{C})}, \dots, \vecX{n-1}{(\mathcal{C})}, \bm{y}^{(\mathcal{C})} \right) \propto \proposal{\mathcal{C}} \left( \vecX{0}{(\mathcal{C})}, \dots, \vecX{n-1}{(\mathcal{C})}, \bm{y}^{(\mathcal{C})} \right) \cdot \prod_{j=0}^{n} \rho_{j},
\end{equation}
where $\rho_{0}$ is given in \eqref{eq:rho_0_gbf} and for $j\in\{1,\dots,n\}$,
\begin{equation}
    \label{eq:rho_j_gbf}
    \rho_{j} \left( \vecX{j-1}{(\mathcal{C})}, \vecX{j}{(\mathcal{C})} \right) = \prod_{c\in\mathcal{C}} \mathbb{E}_{\mathbb{W}_{\mathbf{\Lambda}_{c},j}} \left[ \exp \left\{ -\int_{t_{j-1}}^{t_{j}} \left( \phi_{c} \left( \bm{X}_{t}^{(c)} \right) - \mathbf{\Phi}_{c} \right) \right\} \right] \in (0,1],
\end{equation}
and where $\mathbb{W}_{\mathbf{\Lambda}_{c},j}$ is the law of a Brownian bridge $\{\bm{X}_{t}^{(c)}, t \in (t_{j-1},t_{j})\}$ from $\bm{X}_{t_{j-1}}:=\bm{x}_{j-1}^{(c)}$ to $\bm{X}_{t_{j}}:=\bm{x}_{j}^{(c)}$ with covariance $\mathbf{\Lambda}_{c}$, and $\mathbf{\Phi}_{c}<\infty$ is a constant such that $\phi_{c}(\bm{x}) \geq \mathbf{\Phi}_{c}$ for all $\bm{x}$ and each $c\in\mathcal{C}$. We note that the terms $\mathbf{\Phi}_{c}$ for $c\in\mathcal{C}$ (the global lower bounds of the respective $\phi_{c}$ for $c\in\mathcal{C}$) in \eqref{eq:rho_j_gbf} can be absorbed into normalising constants and, hence, as we apply sequential Monte Carlo methodology, they need not be evaluated (as shown more explicitly in \secref{subsec:GBF_methodology}). 

Whilst $\rho_{0}$ (given by \eqref{eq:rho_0_gbf}) can be computed easily, direct computation of $\rho_{j}$ in \eqref{eq:rho_j_gbf} for $j\in\{1,\dots,n\}$ is not possible as it requires evaluation of path integrals of Brownian motion. However, it is possible to construct non-negative unbiased estimators for \eqref{eq:rho_j_gbf} (with finite variance and computable in finite cost) in a similar fashion to \citet{Beskos_et_al_2008, Fearnhead_2008, Dai_et_al_2019, dai_et_al_2023}. To do so, we require for a given sample path $\bm{X}_{[t_{j-1},t_{j}]}^{(c)} \sim \mathbb{W}_{\mathbf{\Lambda}_{c},j}$ that we have upper and lower bounds for $\phi_{c}(\bm{X}_{t}^{(c)})$ for each $c\in\mathcal{C}$. In general, it is not possible to find global bounds for $\phi_{c}$, so we follow the approach of \citet{Beskos_et_al_2008} and \citet{Pollock_et_al_2016} who noted that if we can bound a sample path $\bm{X}_{[t_{j-1},t_{j}]}^{(c)}\sim\mathbb{W}_{\mathbf{\Lambda}_{c},j}$, then conditional on these \emph{layers} (or bounds) of the sample path, then we will be able to find \emph{local} upper and lower bounds of $\phi_{c}$ denoted $U_{j}^{(c)}$ and $L_{j}^{(c)}$, respectively, such that $\phi_{c}(\bm{X}_{t}^{(c)})\in[L_{j}^{(c)}, U_{j}^{(c)}]$ for $t\in[t_{j-1},t_{j}]$. In order to practically implement this, we need to simulate Brownian bridges jointly with a compact region which almost surely constrains their path (a mechanism for doing this is described in \citet[Sections 7 and 8]{Pollock_et_al_2016}). We now describe one approach for doing this. 

To achieve this, let $R_{c}:=R_{c}(\bm{X}_{[t_{j-1},t_{j}]})$ denote the \emph{layer information} (i.e the compact region in which $\bm{X}_{t}^{(c)}$ is constrained in time $[t_{j-1},t_{j}]$). We note that it is possible to partition the sample space into disjoint sets and simulate from associated distribution function (without having to sample the underlying path), $R_{c} \sim \mathcal{R}_{c}$. If $\mathbf{\Lambda}_{c}=\mathbb{I}_{d}$, we can simulate a layer to which $\bm{X}_{t}^{(c)} \in R_{c}$ for $t\in[t_{j-1},t_{j}]$ by using algorithms outlined in \citet[Sections 7 and 8]{Pollock_et_al_2016} (for instance \citet[Algorithm 14]{Pollock_et_al_2016}). In the case where $\mathbf{\Lambda}_{c}\neq\mathbb{I}_{d}$, we can still simulate $R_{c}$ by appealing to a suitable transformation (which we detail fully in \apxref{app:unbiased_estimators} and \algoref{alg:unbiased_estimator_rho_j}). Furthermore, once we have simulated layer information for $\bm{X}_{t}^{(c)}$ for $t\in[t_{j-1},t_{j}]$, we can simulate the path at any required time marginals conditional on the simulated layer, $\bm{X}_{t}^{(c)} \sim \mathbb{W}_{\mathbf{\Lambda}_{c},j}|R_{c}$ (via a transformation and applying for instance \citet[Algorithm 15]{Pollock_et_al_2016}).

Although it is possible to find tight local bounds for $\phi_{c}$ in a problem specific manner by exploiting specific structure, there are some generic strategies that can be followed. In sufficiently regular settings one might construct the partition necessary by first partitioning the domain of $\phi_{c}$ and then looking at the pre-image of that partition under $\phi_{c}$, thereby reducing the problem to a univariate one. Alternatively, it is helpful in practice to note that it is possible to find generic (less tight) bounds given by the following proposition:
\begin{proposition} \label{prop:bounds}
    For all $c\in\mathcal{C}$ and $\bm{x} \in R_{c}$, we have $\phi_{c} \left( \bm{x} \right) \in \left[ L_{j}^{(c)}, U_{j}^{(c)} \right]$, where
    \begin{equation}
        \label{eq:lower_bound}
        L_{j}^{(c)} := - \frac{1}{2} \left( d \cdot P^{\mathbf{\mathbf{\Lambda}}_{c}} \right),
    \end{equation}
    \begin{equation}
        \label{eq:upper_bound}
        U_{j}^{(c)} := \frac{1}{2} \left[ \left( \norm{\mathbf{\Lambda}_{c}^{\frac{1}{2}} \nabla \log f_{c} \left( \hat{\bm{x}}^{(c)} \right)} + \max_{\bm{x} \in R_{c}} \norm{\mathbf{\Lambda}_{c}^{-\frac{1}{2}} \left( \bm{x} - \hat{\bm{x}}^{(c)} \right)} \cdot P^{\mathbf{\Lambda}_{c}} \right)^{2} + d \cdot P^{\mathbf{\Lambda}_{c}} \right],
    \end{equation}
    where $d$ denotes the dimension of $\bm{x}$, $\norm{\cdot}$ is the Euclidean norm, $\hat{\bm{x}}^{(c)}$ is a user-specified point central to $R_{c}$, and where $P^{\mathbf{\Lambda}_{c}}$ is a quantity such that
    \begin{equation}
        \label{eq:P}
        P^{\mathbf{\Lambda}_{c}} \geq \max_{\bm{x} \in R_{c}} \gamma \left( \mathbf{\Lambda}_{c} \nabla^{2} \log f_{c} \left( \bm{x} \right) \right),
    \end{equation}
    with $\gamma$ denoting the matrix norm, defined as
    \begin{equation}
        \gamma(A):=\max_{\norm{\bm{x}} \neq 0} \frac{\norm{A\bm{x}}}{\norm{\bm{x}}}.
    \end{equation}
    \proof See \apxref{app:bounds}. \hfill $\blacksquare$
\end{proposition}

Once local bounds for $\phi_{c}$ are obtained, we can unbiasedly estimate $\rho_{j}$ \eqref{eq:rho_j_gbf} for $j\in\{1,\dots,n\}$ by letting $\Delta_{j}:=t_{j}-t_{j-1}$ and computing $a_{j}\tilde{\rho}_{j}$, where $a_{j}:=\exp\left(\sum_{c\in\mathcal{C}}\mathbf{\Phi}_{c}\Delta_{j}\right)$ and 
\begin{equation}
    \label{eq:rho_tilde_j_gbf}
    \tilde{\rho}_{j} \left( \vecX{j-1}{(\mathcal{C})}, \vecX{j}{(\mathcal{C})} \right) := \prod_{c\in\mathcal{C}} \left( \frac{\Delta_{j}^{\kappa_{c}} \cdot e^{-U_{j}^{(c)}\Delta_{j}}}{\kappa_{c}! \cdot p \left( \kappa_{c} | R_{c} \right)} \cdot \prod_{k_{c}=1}^{\kappa_{c}} \left[U_{j}^{(c)} - \phi_{c} \left( \bm{X}_{\xi_{c, k_{c}}}^{(c)} \right) \right] \right),
\end{equation}
where $R_{c}$ is the simulated layer information for the Brownian bridge sample path $\bm{X}_{t}^{(c)} \sim \mathbb{W}_{\mathbf{\Lambda}_{c},j}$ from $\bm{x}_{j-1}^{(c)}$ to $\bm{x}_{j}^{(c)}$, $L_{j}^{(c)}$ and $U_{j}^{(c)}$ are constants such that $L_{j}^{(c)} \leq \phi \left( \bm{X}_{t}^{(c)} \right) \leq U_{j}^{(c)}$ for all $\bm{X}_{t}^{(c)} \sim \mathbb{W}_{\mathbf{\Lambda}_{c},j}|R_{c}$, $\kappa_{c}$ is a discrete random variable with conditional probabilities $\mathbb{P}[\kappa_{c} = k_{c} | R_{c}] := p(\kappa_{c} | R_{c})$ (which at this stage we allow to be arbitrary) and $\xi_{c,1},\dots,\xi_{c,\kappa_{c}} \overset{\text{iid}}{\sim} \mathcal{U}[t_{j-1},t_{j}]$ for all $c\in\mathcal{C}$. 
\begin{theorem}
    \label{theorem:unbiased_estimator_GBF}
    Let $a_{j}:=\exp\left(\sum_{c=1}^{C}\mathbf{\Phi}_{c}\Delta_{j}\right)$, then for every $j=1,\dots,n$, $a_{j}\tilde{\rho}_{j}$ is an unbiased estimator of $\rho_{j}$. In particular, we have
    \begin{align}
        \rho_{j}
        & = \mathbb{E}\left[ \mathbb{E}\left[\mathbb{E}\left[ \mathbb{E}\left[ a_{j}\tilde{\rho}_{j} \middle\vert \{\mathcal{R}_{c},\bm{X}_{[t_{j-1},t_{j}]}^{(c)},\kappa_{c}\}_{c\in\cal{C}}\right]\middle\vert \{\mathcal{R}_{c},\bm{X}_{[t_{j-1},t_{j}]}^{(c)}\}_{c\in\cal{C}}, \right] \middle\vert \{\mathcal{R}_{c}\}_{c\in\cal{C}}\right]\right] \nonumber \\
        & = \mathbb{E}_{\bar{\mathcal{R}}}\mathbb{E}_{\bar{\mathbb{W}}|\bar{\mathcal{R}}}\mathbb{E}_{\mathbb{\bar{\mathbb{K}}}}\mathbb{E}_{\bar{\mathbb{U}}}\left[a_{j}\tilde{\rho}_{j}\right],
    \end{align}
    where (for readability) the expectation subscript denotes the law with which they are taken. Here, $\mathcal{R}$ denotes the law of $\{R_{c}\sim\mathcal{R}_{c}:c=1,\dots,C\}$, $\bar{\mathbb{W}}$ denotes the law of the $C$ Brownian bridges $\{\mathbb{W}_{\mathbf{\Lambda}_{c},j}:c=1,\dots,C\}$, $\bar{\mathbb{K}}$ denotes the law of $\{\kappa_{c}:c=1,\dots,C\}$ and $\bar{\mathbb{U}}$ denotes the law of $\{\xi_{c,1},\dots,\xi_{c,\kappa_{c}}:c=1,\dots,C\}\overset{\iid}{\sim}\mathcal{U}[t_{j-1},t_{j}]$.
    \proof See \apxref{app:unbiased_estimator_GBF_proof}. \hfill $\blacksquare$
\end{theorem}

We note that this unbiased estimator for $\rho_{j}$ allows for significant flexibility in choosing the law $\mathbb{K}$. Following the discussion in \citet[Appendix B]{dai_et_al_2023}, there are two natural choices of unbiased estimators that could be used by making particular choices for the distribution of the discrete random variable used to simulate $\kappa_{c}$ for $c\in\mathcal{C}$. We denote these $\tilde{\rho}_{j}^{(a)}$ and $\tilde{\rho}_{j}^{(b)}$ and are based, respectively, upon the GPE-1 and GPE-2 estimators of \citet{Fearnhead_2008}:

\begin{definition}
    \label{cond:GPE1}
    (GPE-1 for $\rho_{j}$ \eqref{eq:rho_j_gbf}): Choosing the law of $\kappa_{c} \sim \textrm{{\upshape Poi}} \big( (U_{j}^{(c)}-L_{j}^{(c)}) \Delta_{j} \big)$ for $c\in\mathcal{C}$ leads to the following estimator:
    \begin{equation}
        \tilde{\rho}_{j}^{(a)}\left( \vecX{j-1}{(\mathcal{C})}, \vecX{j}{(\mathcal{C})} \right) := \prod_{c\in\mathcal{C}} \left( e^{-L_{j}^{(c)} \Delta_{j}} \cdot \prod_{k_{c}=1}^{\kappa_{c}} \left[ \frac{ U_{j}^{(c)} - \phi_{c} \left( \bm{X}_{\xi_{c,k_{c}}}^{(c)} \right)}{U_{j}^{(c)}-L_{j}^{(c)}} \right] \right),
    \end{equation}
    where $\exp\{\sum_{c=1}^{C}\mathbf{\Phi}_{c}\Delta_{j}\}\cdot\tilde{\rho}_{j}^{(a)}$ is an unbiased estimator for $\rho_{j}$.
\end{definition}

\begin{definition}
    \label{cond:GPE2}
    (GPE-2 for $\rho_{j}$ \eqref{eq:rho_j_gbf}): Choosing the law of $\kappa_{c} \sim \textrm{{\upshape NB}}(\gamma_{c}, \beta_{c})$ for $c\in\mathcal{C}$ with
    \begin{equation}
        \label{eq:NB_mean}
        \gamma_{c} := U_{j}^{(c)}\Delta_{j} - \int_{t_{j-1}}^{t_{j}} \phi_{c} \left(\bm{x}_{j-1}^{(c)} \cdot \frac{t_{j}-s}{\Delta_{j}} + \bm{x}_{j}^{(c)} \cdot \frac{s-t_{j-1}}{\Delta_{j}} \right) \dd s,
    \end{equation}
    leads to the following estimator:
    \begin{equation}
        \tilde{\rho}_{j}^{(b)}\left( \vecX{j-1}{(\mathcal{C})}, \vecX{j}{(\mathcal{C})} \right) := \prod_{c\in\mathcal{C}} \left( e^{-U_{j}^{(c)}\Delta_{j}} \cdot \frac{\Delta_{j}^{\kappa_{c}} \cdot \Gamma(\beta_{c}) \cdot (\beta_{c}+\gamma_{c})^{\beta_{c}+\kappa_{c}}}{\Gamma(\beta_{c}+\kappa_{c}) \beta_{c}^{\beta_{c}} \gamma_{c}^{\kappa_{c}}} \cdot \prod_{k_{c}=1}^{\kappa_{c}} \left[ U_{j}^{(c)} - \phi_{c} \left( \bm{X}_{\xi_{c,k_{c}}}^{(c)} \right) \right] \right),
    \end{equation}
    where $\exp\{\sum_{c=1}^{C}\mathbf{\Phi}_{c}\Delta_{j}\}\cdot\tilde{\rho}_{j}^{(b)}$ is an unbiased estimator for $\rho_{j}$.
\end{definition}

The estimators $\tilde{\rho}_{j}^{(a)}$ and $\tilde{\rho}_{j}^{(b)}$ can be computed as detailed in \apxref{app:unbiased_estimators}, and by means of Algorithm \ref{alg:unbiased_estimator_rho_j}, by appealing to \citet[Algorithm 4 and Appendix. B]{dai_et_al_2023}. $\tilde{\rho}_{j}^{(a)}$ and $\tilde{\rho}_{j}^{(b)}$ have particularly desirable properties (by choosing $L_{j}^{(c)}$ and $U_{j}^{(c)}$ as in \propositionref{prop:bounds}):
\begin{proposition}
    \label{prop:unbiased_est}
    Let $a_{j}:=\exp\{\sum_{c=1}^{C}\mathbf{\Phi}_{c}\Delta_{j}\}$, then $a_{j}\tilde{\rho}_{j}^{(a)}(\vecX{}{(\mathcal{C})}, \bm{y}^{(\mathcal{C})})$ and $a_{j}\tilde{\rho}_{j}^{(b)}(\vecX{}{(\mathcal{C})}, \bm{y}^{(\mathcal{C})})$ are unbiased estimators of ${\rho}_{j}(\vecX{}{(\mathcal{C})}, \bm{y}^{(\mathcal{C})})$ which are positive with finite variance. In addition, $\tilde{\rho}_{j}^{(a)}(\vecX{}{(\mathcal{C})}, \bm{y}^{(\mathcal{C})})\in[0,1]$.
    \proof See \citet{Fearnhead_2008}. \hfill $\blacksquare$
\end{proposition}

As we discuss in \secref{subsec:GBF_methodology}, the critical consideration when choosing the law $\mathbb{K}$ is to minimise the variance of the estimator. In our subsequent simulations, we will typically choose the GPE-2 estimator in \condref{cond:GPE2} as it has been empirically shown to have superior performance in \citet[Section 5]{Fearnhead_2008} and \citet[Section 3.5]{dai_et_al_2023}. Note that the mean run time for both the estimators $\tilde{\rho}_{j}^{(a)}$ and $\tilde{\rho}_{j}^{(b)}$ will be random, but will be finite and proportional to $\kappa_c$ for a given layer $R_{c}\sim\mathcal{R}_c$.

\subsection{Methodology} \label{subsec:GBF_methodology}

As we outlined earlier in this section, \thmref{theorem:fusion_measure_GBF} suggests that we can simulate from the fusion target density $f^{(\mathcal{C})}$ by simulating $\tX \sim \mathbb{F}$ and retaining the $T$ time marginal, $\bm{y}^{(\mathcal{C})}$. This can be achieved by simulating a number of proposals $\tX \sim \mathbb{P}$ and accepting (or importance weighting) the terminal time marginal $\bm{y}^{(\mathcal{C})}$ with probability proportional to the \radonniko derivative in \eqref{eq:fusion_measure}. We are now able to implement each of these steps (as discussed in Sections \ref{subsec:GBF_proposal_simulation} and \ref{subsec:GBF_radonniko} respectively), but we have considerable freedom over the details of the methodological approach.

The simplest approach is a \emph{rejection sampler}: simulate a proposal from $\proposal{\mathcal{C}}$ \eqref{eq:h_GBF} (by utilising \propositionref{prop:proposal_simulation_GBF}), accept this proposal with probability $\prod_{j=0}^{n}\rho_{j}$, and conditional on acceptance return $\bm{y}^{(\mathcal{C})}$. As more fully discussed in \apxref{app:connections}, this coincides methodologically with MCF if we set $\mathbf{\Lambda}_{c} = \mathbb{I}_d$ for $c\in\{1,\dots,C\}$ (although the formulation is different). The benefit of such a rejection sampler is it returns i.i.d. draws from $f^{(\mathcal{C})}$. However, it suffers from several inefficiencies. In particular, we would expect the acceptance probabilities $\rho_{j}$ given in \eqref{eq:rho_0_gbf} to decay geometrically with increasing number of sub-posteriors, $|\mathcal{C}|$, as each term in this product is bounded by $1$. Furthermore, the acceptance probability $\prod_{j=0}^{n}\rho_{j}$ will typically decay exponentially with increasing $T$. Consequently, a rejection sampling approach for this problem will ultimately be impractical in many practical settings as it will have very small acceptance probabilities. Similarly, the naive importance sampling adaptation of this approach (in which the proposal of the rejection sampler are all retained with a un-normalised importance weight of $\prod_{j=0}^{n}\tilde{\rho}_{j}$) will ultimately suffer from the same issues of robustness in practice. 

Inspired by the importance sampling approach, the BF approach of \citet{dai_et_al_2023} introduced the  auxiliary temporal partition $\mathcal{P}$ in order to simulate from $\extended{\mathcal{C}}$ using SMC: allowing for the \emph{gradual} coalescence of the $C$ stochastic processes. In particular, we can initialise an SMC algorithm by simulating $N$ particles from the time $0$ marginal in $\proposal{\mathcal{C}}$ (which consists of composing $|\mathcal{C}|$ samples from each of the sub-posterior densities to obtain $\vecX{0}{(\mathcal{C})}$), and assigning them an initial un-normalised importance weight given by $w_{0,i}^{\prime}:=\rho_{0}(\vecX{0,i}{(\mathcal{C})})$ for $i\in\{1,\dots,N\}$. This initial particle set constitutes an approximation to the time $0$ marginal of $\extended{\mathcal{C}}$, and can be sequentially propagated $n$ times (i.e.\ $|\mathcal{P}|-1$ times) through the temporal time mesh $\mathcal{P}$ by simulating $\vecX{j,i}{(\mathcal{C})} | \vecX{j-1,i}{(\mathcal{C})} \sim \mathcal{N}_{d} \left( \vecM{j,i}{(\mathcal{C})}, \bm{V}_{j} \right)$ as per \eqref{eq:M_GBF} and \eqref{eq:V_GBF} in \propositionref{prop:proposal_simulation_GBF}. In our SMC formulation at each iteration ($j\in\{1,\dots,n\}$) the un-normalised importance weight of every particle is updated by a factor of $\tilde{\rho}_{j}(\vecX{j-1,i}{(\mathcal{C})},\vecX{j,i}{(\mathcal{C})})$ as per \eqref{eq:rho_tilde_j_gbf}. Upon normalisation, the resulting weighted particle set after the $n$th iteration is an approximation of both the time $T$ marginal of $\extended{\mathcal{C}}$ and our fusion target $f^{(\mathcal{C})}$. In particular,
\begin{equation}
    f^{(\mathcal{C})}(\bm{y}) \dd\bm{y} \approx \sum_{i=1}^{N} w_{n,i}^{(\mathcal{C})} \cdot\delta_{\bm{y}_{i}^{(\mathcal{C})}} (\dd \bm{y}).
\end{equation}

As we remarked upon in \secref{subsec:GBF_radonniko}, due to this normalisation of the particle set weights we can avoid the need to explicitly compute the constants $\mathbf{\Phi}_{c}$ in \eqref{eq:rho_j_gbf}, as they are simply constants which cancel.

As is common in SMC, to avoid \emph{weight degeneracy} in which the variance of the importance weights degrades rapidly in $n$, we employ a \emph{resampling} strategy (see for instance \citet{Gerber_et_al_2019} for a recent investigation of the properties of many resampling schemes). In particular, we monitor the particle set for weight degeneracy by estimating its \emph{effective sample size ($\ESS$)} \citet{Kong_1994}. If the ESS falls below some user-specified threshold then at the beginning of the next iteration we resample the particle set to get $N$ equally weighted particles. In all of our simulations in the subsequent sections, we used residual resampling \citep{Higuchi_1997, Liu_Chen_1998, Whitley_1994}.

We term our resulting Fusion approach \emph{\gbfl} and summarise it in \algoref{alg:GBF}. 

\begin{algorithm}
    \caption{\generalisedbf{\mathcal{C},\{\{\bm{x}^{(c)}_{0,i}, w^{(c)}_i\}_{i=1}^{M},\mathbf{\Lambda}_{c}\}_{c\in\mathcal{C}},N,\mathcal{P}}: \gbfl.}
    \label{alg:GBF}
    \begin{enumerate}
        \item \textbf{Initialisation} ($j=0$):
        \begin{enumerate}
            \item \textbf{Input:} Importance weighted realisations $\{\bm{x}^{(c)}_{0,i}, w^{(c)}_i\}_{i=1}^{M}$ for $c\in\mathcal{C}:=(c_{1},\dots,c_{|\mathcal{C}|})$, the user-specified matrices, $\{\mathbf{\Lambda}_{c}: c\in\mathcal{C}\}$, the number of particles required, $N$, and temporal partition $\mathcal{P}:=\{t_{0},t_{1},\dots,t_{n}:0=:t_{0}<t_{1}<\cdots<t_{n}:=T\}$.
            \item Compose the importance weighted realisations $\{\vecX{0,k}{(\mathcal{C})}, w^{(\mathcal{C})\prime}_{0,k}\}_{k=1}^{M}$ where $w^{(\mathcal{C})\prime}_{0,k} := \big(\prod_{c\in\mathcal{C}} w_{k}^{(c)}\big) \cdot \rho_{0}(\vecX{0,k}{(\mathcal{C})})$ for $k \in \{1,\dots,M\}$ as per \eqref{eq:rho_0_gbf}. \label{alg:GBF:initial_weights}
            \item \textbf{$w_{0,k}^{(\mathcal{C})}$:} For $k$ in $1$ to $M$, compute normalised weight $w_{0,k} = w^{(\mathcal{C})\prime}_{0,k} / \sum_{k^{\prime}=1}^{M} w^{(\mathcal{C})\prime}_{0,k^{\prime}}$.
            \item \textbf{$g_{0}^{M}$}: Set $g_{0}^{M}(\dd\vecX{0}{(\mathcal{C})}):=\sum_{k=1}^{M} w_{0,k}^{(\mathcal{C})}\cdot\delta_{\vecX{0,k}{(\mathcal{C})}}(\dd\vecX{0}{(\mathcal{C})})$
            \item \textbf{$\vecX{0,i}{(\mathcal{C})}$:} If $M \neq N$, for $i=1,\dots,N$, resample $\vecX{0,i}{(\mathcal{C})} \sim g_{0}^{M}$ and reset $w_{0,i}^{(\mathcal{C})}=\frac{1}{N}$. \label{alg:GBF:resample_0}
        \end{enumerate}
        \item \textbf{Iterative updates}. For $j\in\{1,\dots,n\}$: \label{alg:GBF:iterations}
        \begin{enumerate}
            \item \textbf{Resample:} If the $\ESS:=\left(\sum_{i=1}^{N} {w_{j-1,i}^{(\mathcal{C})}}^{2}\right)^{-1}$ breaches the lower user-specified threshold, then for $i=1,\dots,N$, resample $\vecX{j-1,i}{(\mathcal{C})} \sim g_{j-1}^{N}$ and reset $w_{j-1,i}^{(\mathcal{C})}=\frac{1}{N}$. \label{alg:GBF:resample_j}
            \item For $i$ in $1$ to $N$, \label{alg:GBF:propose_and_assign}
            \begin{enumerate}
                \item \textbf{$\vecX{j,i}{(\mathcal{C})}$:} Simulate $\vecX{j,i}{(\mathcal{C})} \sim \mathcal{N}_{d} \left( \vecM{j,i}{(\mathcal{C})}, \bm{V}_{j} \right)$ as per \propositionref{prop:proposal_simulation_GBF}. \label{alg:GBF:propagate}
                \item \textbf{$w_{j,i}^{(\mathcal{C})\prime}$:} Compute un-normalised weight $w_{j,i}^{(\mathcal{C})\prime} = w_{j-1,i}^{(\mathcal{C})} \cdot \tilde{\rho}_{j}(\vecX{j-1,i}{(\mathcal{C})}, \vecX{j,i}{(\mathcal{C})})$ as per \eqref{eq:rho_tilde_j_gbf} (using \algoref{alg:unbiased_estimator_rho_j}).
            \end{enumerate}
            \item \textbf{$w_{j,i}^{(\mathcal{C})}$:} For $i$ in $1$ to $N$, compute normalised weight $w_{j,i}^{(\mathcal{C})} = w^{(\mathcal{C})\prime}_{j,i} / \sum_{k^{\prime}=1}^{N} w^{(\mathcal{C})\prime}_{j,k^{\prime}}$. \label{alg:GBF:update_weight}
            \item \textbf{$g_{j}^{N}$}: Set $g_{j}^{N}(\dd\vecX{j}{(\mathcal{C})}):=\sum_{i=1}^{N} w_{j,i}^{(\mathcal{C})}\cdot\delta_{\vecX{j,i}{(\mathcal{C})}}(\dd\vecX{j}{(\mathcal{C})})$.
        \end{enumerate} 
        \item \textbf{Output:} $\left\{\vecX{0,i}{(\mathcal{C})}, \dots, \vecX{n-1,i}{(\mathcal{C})}, \bm{y}_{i}^{(\mathcal{C})}, w_{n,i}^{(\mathcal{C})} \right\}_{i=1}^{N}$, where $\hat{f}^{(\mathcal{C})}(\dd \bm{y}) := g_{n}^{N}(\dd\bm{y}) \approx f^{(\mathcal{C})}(\bm{y})\dd\bm{y}$.
    \end{enumerate}
\end{algorithm}

\subsection{Practical extensions of Generalised Bayesian Fusion} \label{subsec:GBF_practicalities}

We now consider the practicalities of Algorithm \ref{alg:GBF}. To generalise the algorithm further (and make it amenable to a recursive divide-and-conquer approach as in \hgbf), we assume we have access to $M$ importance weighted realisations of each sub-posterior, $\{\bm{x}_{0,k}^{(c)}, w_{k}^{(c)}\}_{k=1}^{M}$ for $c\in\mathcal{C}$. To initialise the algorithm, we start by composing $M$ initial weighted particles by pairing the draws from each sub-posterior $\{\vecX{0,k}{(\mathcal{C})}\}_{k=1}^{M}$, and compute the associated (un-normalised) partial weights $\{w^{(\mathcal{C})\prime}_{0,k}\}_{k=1}^{M} $ where $w^{(\mathcal{C})\prime}_{0,k} := \big(\prod_{c\in\mathcal{C}} w_{k}^{(c)}\big) \cdot \rho_{0}(\vecX{0,k}{(\mathcal{C})})$ for $k=1,\dots,M$. If we have $M \neq N$, we resample to obtain $N$ samples from each sub-posterior, otherwise, we choose to only resample if the ESS is below some user-specified threshold. Note that in the \textbf{Input} step of Algorithm \ref{alg:GBF}, we may have access to different numbers of samples from each sub-posterior: say $M_c$ importance weighted samples for sub-posterior $f_{c}$ (for $c \in \mathcal{C}$). In order to compose our $M$ partial proposals in \stepref{alg:GBF:initial_weights}, there are a number of approaches we could take. As presented above, if $M_c=M$ for $c\in\mathcal{C}$, we simply pair the sub-posterior draws index-wise. This is a basic merging strategy of the sub-posterior realisations and has the advantage that it can be implemented in $O(M)$ cost (and if $M_c \neq M$ for every $c\in\mathcal{C}$ one could simply sub-sample to obtain a common number of samples from each sub-posterior). However, as noted in \citet{Lindsten_et_al_2017}, while this approach has a low computational cost, it can lead to high variance when the product $\prod_{c \in \mathcal{C}} f_{c}(\bm{x}^{(c)})$ differs substantially from the corresponding marginal of $f^{(\mathcal{C})}$ --- which one might expect to be the case in our setting when the sub-posteriors disagree.

We found this simple approach more than adequate in our simulations, but there are more sophisticated options available should they be required in still more challenging settings. In particular, as described in \citet[Section 4.1]{Lindsten_et_al_2017}, at the expense of a computational cost $O(\prod_{c\in\mathcal{C}} M_c)$, one could instead compose all possible permutations of the samples from each sub-posterior before weighting and then resampling to reduce the number of points in the approximation back to a pre-specified number, arriving at a better approximation at a greater cost. They termed this approach ``mixture resampling'' and also detailed a ``lightweight mixture resampling'' approach in which more than one permutation, but not all possible permutations, are used and found it to work well; as noted by \citet{Kuntz_et_al_2021} such a strategy can be connected directly with the theory of incomplete $U$-statistics and consequently one might hope to realise much of the benefit of mixture resampling at a much reduced cost \citep{Kong_Zheng_2021}.


\section{A divide-and-conquer approach to Fusion} \label{sec:dc_gbf}

A key drawback of the Monte Carlo Fusion and Bayesian Fusion approaches of \citet{Dai_et_al_2019, dai_et_al_2023}, and the \gbf approach we introduced and outlined in \secref{sec:GBF}, is that it lacks robustness with increasing number of sub-posteriors, $|\mathcal{C}|$. This is unsurprising as the extended target and proposal densities ($\extended{\mathcal{C}}$ and $\proposal{\mathcal{C}}$) are $d(n|\mathcal{C}|+1)$-dimensional, and these become increasingly mismatched with increasing dimension. In particular, as a consequence of the definition of $\rho_{j}$ in \eqref{eq:rho_j_gbf}, the acceptance probability of any rejection-based scheme will decrease geometrically with increasing $|\mathcal{C}|$. Fundamentally, importance sampling variants of this will not address this bottleneck.

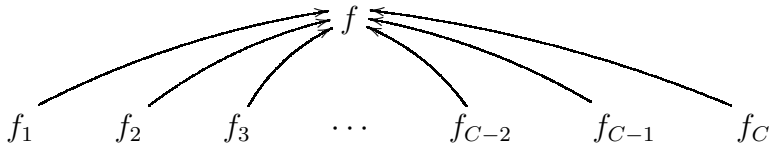
\begin{figure}[ht]
	\makebox[\textwidth]{\xymatrix{
& & & {f} & & & \\
{f_1} \ar@<0.9ex>@/^/[urrr] & {f_2} \ar@<0.35ex>@/^/[urr] & f_3 \ar@/^/[ur] & \cdots & {f_{C-2}} \ar@/_/[ul] & {f_{C-1}} \ar@<-0.35ex>@/_/[ull] & {f_C} \ar@<-0.9ex>@/_/[ulll]
}}
\caption{A tree representation of the fork-and-join approach of Monte Carlo Fusion.} \label{fig:fork_and_join}
\end{figure}

As presented both in \citet{Dai_et_al_2019, dai_et_al_2023}, Fusion is an example of a \emph{fork-and-join} approach---\emph{all} of the sub-posteriors are unified in a single step. In particular, within the GBF framework of \secref{sec:GBF} we set $\mathcal{C}:=\{1,\dots,C\}$. This is illustrated in the \emph{tree} diagram of \figref{fig:fork_and_join}, where the \emph{leaves} of the tree represent the available sub-posterior densities, the directed edges are used to illustrate the computational flow of MCF, and the \emph{root} vertex of the tree is the desired fusion density, $f$ (as given in \eqref{eq:fusion_density}).

\begin{figure}[ht]
     \centering
     \begin{subfigure}[b]{0.475\textwidth}
	    {\makebox[\textwidth]{\xymatrix@C=1.5em@R=2.25em{
	         &  & f &  &   \\
	         & f_1 f_2 \ar@/^/[ur]|{\phantom{f}\cdots\phantom{f}} & \cdots & f_{C-1}f_C \ar@/_/[ul]|{\phantom{f}\cdots\phantom{f}} & \\
            f_1 \ar@/^/[ur] & f_2 \ar[u] & \cdots & f_{C-1} \ar[u] & f_C \ar@/_/[ul]
        }}}
         \caption{A balanced-binary tree.}
    \label{fig:balanced_binary}
    \end{subfigure}
    \hfill
     \begin{subfigure}[b]{0.475\textwidth}
	    {\makebox[\textwidth]{\xymatrix@C=1em@R=1em{	
	        & & & & f\\
	        &  &  & \prod^{C-1}_{c=1} f_c \ar@/^/[ur] & \\
	        & f_1 f_2 \ar@/^/[urr]|{\phantom{f}\cdots\phantom{f}}  & & & \\
            {f_1} \ar@/^/[ur] & {f_2} \ar[u] & \cdots & {f_{C-1}} \ar[uu] & f_C \ar[uuu]
        }}}
         \caption{A progressive tree.}
        \label{fig:progressive}
    \end{subfigure}
        \caption{Illustrative hierarchies for the Fusion problem of \eqref{eq:fusion_density}.}
    \label{fig:hierarchies}
\end{figure}
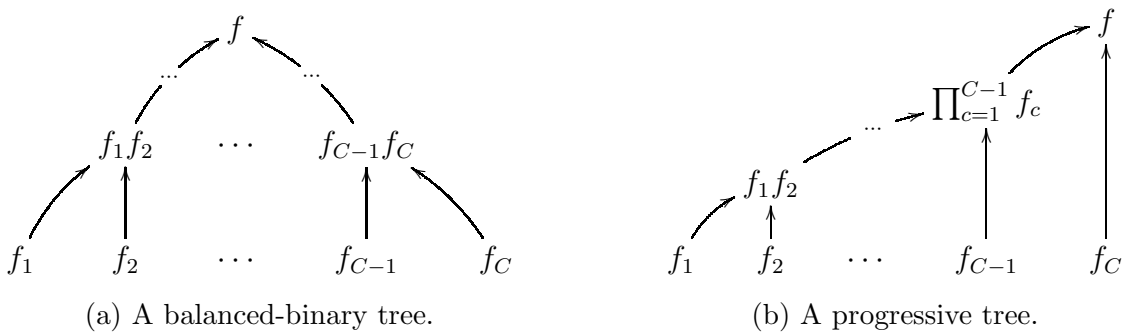

As the goal of the methodology is to approximate $f$ in \eqref{eq:fusion_density}, one could envision a recursive \emph{divide-and-conquer} approach in which the sub-posteriors are combined in stages to recover $f$. There are a number of possible orderings in which we could combine sub-posteriors, and so we represent these orderings in \emph{tree diagrams}, and term these \emph{hierarchies} (see \figref{fig:hierarchies}). For instance as illustrated in \figref{fig:balanced_binary}, one approach would be to combine two sub-posteriors at a time (we term this a \emph{balanced-binary tree} approach). In \figref{fig:balanced_binary}, the intermediate vertices represent intermediate (\emph{auxiliary}) densities up to proportionality. The approximation of the distribution associated with any non-leaf vertex is obtained by an application of Fusion methodology to the densities of the children of that vertex. A balanced-binary tree approach is perhaps the most natural way to combine sub-posteriors in a \emph{truly distributed setting} (where the simulation of each sub-posterior has been conducted separately, and so the inferences we wish to combine are distributed). Another approach is given in \figref{fig:progressive}, whereby sub-posteriors are fused one at a time (which we term a \emph{progressive tree} approach). This is perhaps the most natural approach for an \emph{online setting}. We focus on applying GBF to these two natural hierarchies for the remainder of this paper, although other hierarchies are certainly possible within our framework, and there is no limitation in unifying more than two vertices at any level of a tree (as suggested by both \secref{sec:GBF} and \figref{fig:fork_and_join}).

From this recursive perspective, sample approximations of auxiliary densities obtained at one level of any tree are themselves treated as sub-posteriors at the next level up. As such, one can iteratively apply the Fusion methodology of \secref{sec:GBF}, working through the levels of the tree from the leaves to the root, using at each stage the output of one step as the input for the subsequent step. An advantage of our divide-and-conquer approach is that as fewer sub-posteriors are combined at each stage, we avoid (at each stage) the rapidly diminishing and variable importance weights. 

A divide-and-conquer variant of Sequential Monte Carlo (D\&C-SMC) was recently introduced in \citet{Lindsten_et_al_2017}. D\&C-SMC generalises the classical SMC framework from sequences (or chains) to trees, such as those in Figures \ref{fig:fork_and_join} and \ref{fig:hierarchies}. The theoretical properties of D\&C-SMC are increasingly-well characterized and include a strong law of large numbers, finite sample $L_p$ errors bounds as well as a $\sqrt{N}$-central limit theorem under mild conditions (see \citet{Kuntz_et_al_2021b}). We thus embed our \gbfa approach within a D\&C-SMC algorithm to address the robustness of Fusion with increasing $|\mathcal{C}|$, albeit this being a trade-off with the cost of the repeated application of the methodology. In our recursive setting, we unify distributed sample approximations by operating on a tree of \emph{auxiliary} Fusion densities. Let $\tree = (\vertices,\edges)$ denote a tree with vertices $\vertices$ and (directed) edge set $\edges$. Let $\leaf{\tree}$ denote the leaves of the tree (which represent the sub-posteriors $f_1,\dots,f_C$), $\troot{\tree}$ denote the root of the tree (which represents $f$) and $\children{v}$ denote the children of vertex $v \in \vertices$ where $\children{t} = \emptyset$ if $t$ is a leaf. Let $\vertices = \{ v_0, v_1,\ldots,v_C,\ldots\}$ be the set of vertices, with $v_0 = \troot{\tree}$, $\{v_1,\ldots,v_C\} = \leaf{\tree}$ and as many intermediate vertices as are required to specify the tree.

For the purposes of utilising the methodology developed in \secref{sec:GBF}, we define the following notation for non-leaf vertices (i.e.\ $v \notin \leaf{\tree}$): let $\mathcal{C}_v := \cup_{u\in \children{v}} \mathcal{C}_u$ denote the index set representing the sub-posteriors that we want to unify for vertex $v \notin \leaf{\tree}$. In addition, to simplify the notation and avoid an unnecessary level of subscripts, we index densities and other quantities by $v$ rather than $\mathcal{C}_v$ when it is clear what is intended. In particular, let $\mathbf{\Lambda}_{v} := \mathbf{\Lambda}_{\mathcal{C}_{v}}$, $\vecX{t}{(v)} := \vecX{t}{(\mathcal{C}_{v})}$, $\tilde{\bm{x}}_{t}^{(v)} := \tilde{\bm{x}}_{t}^{(\mathcal{C}_{v})}$, $\bm{y}^{(v)} := \bm{y}^{(\mathcal{C}_{v})}$ where $\bm{y}^{(v)} \sim f_{v} := f^{(\mathcal{C}_{v})}$ for $v \notin \leaf{\tree}$. Let $\mathbb{W}_{\mathbf{\Lambda}_{v},j}$ denote the law of a Brownian bridge $\{\bm{X}_{t}^{(v)}, t \in [t_{j-1},t_{j}]\}$ with $\bm{X}_{t_{j-1}}^{(v)}:=\bm{x}_{j-1}^{(v)}$ and $\bm{X}_{t_{j}}^{(v)}:=\bm{x}_{j}^{(v)}$ with covariance $\mathbf{\Lambda}_{v}$ for $j\in\{1,\dots,n\}$. The extended target and proposal densities for vertex $v \notin \leaf{\tree}$ are denoted $\extended{v} := \extended{\mathcal{C}_{v}}$ and $\proposal{v} := \proposal{\mathcal{C}_{v}}$, respectively. Lastly, the importance sampling weights for $v \notin \leaf{\tree}$ are given by $\rho_{0}^{(v)}(\vecX{}{(v)}) := \rho_{0}(\vecX{}{(\mathcal{C}_{v})})$ and $\rho_{j}^{(v)}(\vecX{}{(v)}, \bm{y}^{(v)}) := \rho_{j}(\vecX{}{(\mathcal{C}_{v})}, \bm{y}^{(\mathcal{C}_{v})})$ for all $j$.

To describe our \emph{\hgbfl} approach, we specify an algorithm that is carried out at each vertex $v \in \vertices$ which leads to a recursive procedure; an initial call to $\dcgbf{\troot{\mathcal{V}},\ldots}$ carries out the overall approach. For $v\in\vertices$, we define a procedure (as given in \algoref{alg:dc_gbf}), which returns a weighted particle set $\{\vecX{0,i}{(v)}, \dots, \vecX{n-1,i}{(v)}, \bm{y}_{i}^{(v)}, w_{n,i}^{(v)}\}_{i=1}^{N}$ where $w_{n,i}^{(v)}$ denotes the normalised importance weight of particle $i$ for vertex $v\in\vertices$. From this particle set, we can take the marginal weighted samples for $\bm{y}^{(v)}$ to approximate the fusion density $f_{v} \propto \prod_{u\in\children{v}} f_{u}$ for vertex $v \in \vertices$. Recall that the leaf vertices, $v_{c}$ for $c\in\{1,\dots,C\}$, represent each of the sub-posteriors. It is possible to additionally incorporate importance sampling for the leaf vertices, but for simplicity we assume that we have access to unweighted samples for the sub-posteriors. Therefore, at these leaf vertices, we simply sample from the sub-posteriors. If independent sampling is not feasible, one could use MCMC to obtain unweighted sample approximations at the leaves. Formal arguments (under appropriate regularity conditions) could in principle follow an approach analogous to that in \cite{Finke_2020}. If $v$ is a non-leaf vertex, we simply call \algoref{alg:GBF} by inputting the importance weighted samples $\{\bm{y}_{i}^{(u)}, w_{i}^{(u)}\}_{i=1}^{N}$ for $u \in \children{v}$. As in standard SMC, although the auxiliary distributions are defined on larger spaces we do not need to retain sampled values which are not subsequently used; to obtain a more computationally manageable algorithm, we can choose to retain only the final parameter space marginal at each vertex (i.e.\ only returning $\{\bm{y}^{(v)}_{i}, w_{i}^{(v)}\}_{i=1}^{N}$) since we only require this to compute the importance weights in \algoref{alg:GBF} at each vertex $v \notin \leaf{\tree}$.

\begin{algorithm}
    \caption{\dcgbf{v,N,\mathcal{P}}: \hgbfl.}
    \label{alg:dc_gbf}
    {\bf Given:} Sub-posteriors, $\{f_u\}_{u \in \leaf{\tree}}$, and preconditioning matrices $\{\mathbf{\Lambda}_u\}_{u\in\tree}$.\\
    {\bf Input:} Node in tree, $v$, the number of particles $N$, and (optionally) the temporal mesh partitions $\{\mathcal{P}_{u}\}_{u\in\children{v}}$, $\mathcal{P}_{v}$.
    \begin{enumerate}
        \item For $u \in \children{v}$,
        \begin{enumerate}
            \item 
            $\left\{\vecX{0,i}{(u)}, \dots, \vecX{n-1,i}{(u)}, \bm{y}^{(u)}_{i}, w_{n,i}^{(u)}\right\}_{i=1}^{N} \leftarrow \dcgbf{u, N, \mathcal{P}_{u}}$. \label{alg:hgbf:recursion}
        \end{enumerate}
        \item If $v \in \leaf{\tree}$, \label{alg:dc_gbf:leaf_vertices}
        \begin{enumerate}
            \item For $i=1,\dots,N$, sample $\bm{y}^{(v)}_{i} \sim f_{v}(\bm{y})$.
            \item {\bf Output:} $\{\emptyset, \bm{y}^{(v)}_{i}, \frac{1}{N}\}_{i=1}^{N}$.
        \end{enumerate}
        \item If $v \notin \leaf{\tree}$, \label{alg:dc_gbf:non_leaf_vertices}
        \begin{enumerate}
        \item If $\mathcal{P}_{v}$ is not inputted, apply guidance from \secref{subsec:GBF_T_guidance} and \secref{subsec:GBF_mesh_guidance}.
        \item {\bf Output:} Call \generalisedbf{\children{v},\{\{\bm{y}_{i}^{(u)}, w^{(u)}_i\}_{i=1}^{N},\mathbf{\Lambda}_u\}_{u\in\children{v}},N,\mathcal{P}_{v}}.
        \end{enumerate}
    \end{enumerate}
\end{algorithm}

Note that in \algoref{alg:dc_gbf}, we allow the user to specify different temporal partitions at each node and level (i.e.\ $\{\mathcal{P}_{u}\}_{u\in\children{v}}$, $\mathcal{P}_{v}$). As we explore fully in  \secref{sec:GBF_guidance}, when we develop guidance for user chosen tuning parameters, having this flexibility on the temporal partition can lead to a far more robust and efficient implementation of \algoref{alg:dc_gbf}.


\section{Implementational guidance for Generalised Bayesian Fusion} \label{sec:GBF_guidance}

In this section we develop guidance for choosing the parameter $T$ and the temporal partition $\mathcal{P}$ (and so $n$ implicitly) for our \gbfl approach (\algoref{alg:GBF}), the guidance for which can be used directly \emph{at each node} within our \hgbf approach (\algoref{alg:dc_gbf}). As \gbfa is fundamentally a sequential Monte Carlo (SMC) algorithm, we want to choose these hyperparameters in such a way to ensure that the discrepancy between subsequent proposal and target distributions are not degenerate. For this reason, and in common with \citet[Section 3]{dai_et_al_2023}, we look at the incremental weight changes and study the \emph{current effective sample size (CESS)} associated with these weights:
\begin{equation}
    \label{eq:CESS_j_def}
    \CESS{j} := \frac{\left(\sum_{i=1}^{N} \tilde{\rho}_{j,i}\right)^{2}}{\sum_{i=1}^{N} \tilde{\rho}_{j,i}^{2}} \text{ for } j=1,\dots,n; \qquad \CESS{0} := \frac{\left(\sum_{i=1}^{N} \rho_{0,i}\right)^{2}}{\sum_{i=1}^{N} \rho_{0,i}^{2}},
\end{equation}
where $\rho_{0,i}$ and $\tilde{\rho}_{j,i}$ are given in \eqref{eq:rho_0_gbf} and \eqref{eq:rho_tilde_j_gbf} respectively.

In order to develop heuristics to choose hyper-parameters, we consider the idealised setting of combining multivariate Gaussian sub-posteriors with mean vector $\bm{a}_{c}$ and covariance matrix ${b|\mathcal{C}|}\mathbf{\Lambda}_{c} / m$, for some $b>0$, for $c\in\mathcal{C}$. The target is $f \sim \mathcal{N}_{d}(\tilde{\bm{a}}, {b|\mathcal{C}|}\mathbf{\Lambda}_{\mathcal{C}}/m)$, where $\tilde{\bm{a}}:=\left(\sum_{c\in\mathcal{C}}\mathbf{\Lambda}_{c}^{-1}\right)^{-1}\left(\sum_{c\in\mathcal{C}}\mathbf{\Lambda}_{c}^{-1}\bm{a}_{c}\right)$ and $\mathbf{\Lambda}_{\mathcal{C}}:=\left(\sum_{c\in\mathcal{C}}\mathbf{\Lambda}_{c}^{-1}\right)^{-1}$.

In BF, this idealised setting was used to help select $T$ and $n$ and, by imposing an additional assumption that the partition was a \emph{regular} mesh, in turn $\mathcal{P}$. In this section we instead develop guidance for $T$ (see \secref{subsec:GBF_T_guidance}) in the more sophisticated \gbfa setting, and then in \secref{subsec:GBF_mesh_guidance} investigate the more challenging selection of $\mathcal{P}$ without assumption on its regularity (i.e.\ permitting an \emph{irregular} choice of mesh)---and so we instead \emph{implicitly find $n$}. These ideas can also be directly applied to improve BF itself, which we show later in our numerical results. 

In our idealised setting, the key consideration is  the degree to which the sub-posteriors disagree with one another. To measure how significant the \emph{sub-posterior conflict} is we define
\begin{equation}
    \label{eq:sigma_a}
    \sigma^{2}_{\bm{a}}:=\frac{1}{|\mathcal{C}|}\sum_{c\in\mathcal{C}} (\bm{a}_{c}-\tilde{\bm{a}})^{\intercal}\mathbf{\Lambda}_{c}^{-1}(\bm{a}_{c}-\tilde{\bm{a}}).
\end{equation}

We further consider the two following conditions in order to explore how the algorithm hyperparameters should change according to sub-posterior heterogeneity:
\begin{condition} \label{cond:SH} $\SH{\lambda}$. The sub-posteriors obey the $\SH{\lambda}$ condition (for some constant $\lambda>0$) if
\begin{equation}
    \sigma_{\bm{a}}^{2} = \frac{b(|\mathcal{C}|-1)\lambda}{m}.
\end{equation}
\end{condition}

\begin{remark} 
    \label{remark:lambda_choice_GBF}
    \emph{Interpretation of $\lambda$.} Of course $\SH{\lambda}$ will always hold for some $\lambda $, and this condition can alternatively be interpreted as a definition of $\lambda$. We will be particularly interested in moderate values of $\lambda$ close to $1$ which will indicate only weak or no sub-posterior discrepancy.
    $\SH{\lambda}$ is a natural condition, arising for instance if $\frac{m}{|\mathcal{C}|}$ of the data is randomly allocated to each sub-posteriors then $\sigma_{\bm{a}}^{2} \sim \frac{b}{m} \chi_{|\mathcal{C}|-1}^{2}$ and have mean $\frac{b(|\mathcal{C}|-1)}{m}$. The $\lambda $ for which 
    $\SH{\lambda}$ holds is therefore $\chi_{|\mathcal{C}|-1}^{2}/(|\mathcal{C}|-1)$ and therefore has mean $1$ and variance $2/(|\mathcal{C}|-1)$.
    Consequently for large $|\mathcal{C}|$, we would expect $\lambda$ to be close to $1$. In this idealised i.i.d. case, these arguments duplicate classical ANOVA calculations.
    
    However the $\SH{\lambda}$ condition for moderate $\lambda >1$ is also of interest indicating weak discrepancy between sub-posteriors. This would occur (for instance) if the data consisted of disjoint segments of a long ergodic stationary sequence with no long-range dependence where, in this case, $\lambda$ is an estimate of the integrated auto-correlation time of the sequence. For this reason, the scenario
    $\lambda <1$ would not normally occur (particularly for large $|\mathcal{C}|$).

    In the examples later on, we will set $\lambda=1$ as default, since this is the natural iid scenario. However, as noted above, if we suspect that there is weak discrepancy between the sub-posteriors, or there is some dependency between the subsets of data, we may also choose $\lambda$ to be slightly greater than $1$ or alternatively estimate it from the data.
\end{remark}

The defining characteristic of $\SH{\lambda}$ is that $\lambda$ is stable for large data sizes (large $m$). However for stronger sub-posterior discrepancy, just as the power of ANOVA tests become larger for larger data sets, $\lambda$ will become much larger with $m$ where there is a systematic difference in the data distributions between sub-posteriors. Now $\SH{\lambda}$ will not adequately describe this dependence, and so we consider the following scenario instead:

\begin{condition} \label{cond:SSH} $\SSH{\gamma}$. The sub-posteriors obey the super sub-posterior heterogeneity $\SSH{\gamma}$ condition (for some constant $\gamma>0$) if
\begin{equation}
    \sigma_{\bm{a}}^{2} = b\gamma.
\end{equation}
\end{condition}
As with $\SH{\lambda}$, this can alternatively be seen as a definition of $\gamma$.
This setting can arise if the sub-posterior heterogeneity does not decay with data size $m$.

\begin{remark}
    \label{remark:b_choice_GBF}
    \emph{Choice of $b$:} In the case that the user-specified matrices $\{\mathbf{\Lambda}_{c}\}_{c\in\mathcal{C}}$ are chosen to be the estimated covariance matrices for each sub-posterior, then we would set $b=\frac{m}{|\mathcal{C}|}$. Therefore, the sub-posteriors $f_{c}\sim\mathcal{N}_{d}(\bm{a}_{c}, \frac{b|\mathcal{C}|}{m}\mathbf{\Lambda}_{c})$ have variance which closely matches the sub-posterior variance. In general, we want to choose $b$ such that $\frac{b|\mathcal{C}|}{m}\mathbf{\Lambda}_{c}$ is close to the variance of sub-posterior $f_{c}$ for $c\in\mathcal{C}$.
\end{remark}

We study empirically our choices of tuning parameter ($T$, $n$ and $\mathcal{P}$) in the idealised settings described by the $\SH{\lambda}$ condition (of \condref{cond:SH}) and $\SSH{\gamma}$ condition (of \condref{cond:SSH}) in Sections  \ref{subsec:GBF_similar_means}--\ref{subsec:GBF_dissimilar_means} respectively. 

Note that the implementational guidance we provide in this section is for the general application and tuning of \gbfa methodology. In many practical settings there will be additional constraints which require further modification to \gbfa. This includes settings where latency between cores is problematic, or in scenarios where functional evaluations of the sub-posterior densities $f_{c}$ are not available. In \apxref{app:practical_considerations} we provide further direction on some of what we envisage to be the most common modifications.

\subsection{Guidance for choosing \texorpdfstring{$T$}{T}} \label{subsec:GBF_T_guidance}

In this section, we develop guidance on selecting $T$ for the two idealised settings, $\SH{\lambda}$ and $\SSH{\gamma}$, defined in Conditions \ref{cond:SH} and \ref{cond:SSH}, respectively. In each setting, by first specifying the lower bound on the initial effective sample size that we desire, we can compute a minimum value of $T$ which should be used in \algoref{alg:GBF}. As choosing a larger value of $T$ typically results in more iterations in GBF, we suggest using the minimum value of $T$ which is suggested. The time horizon $T$ only directly affects the initial weighting given to each of the $N$ particles through $\rho_{0}$ in \eqref{eq:rho_0_gbf}. Thus, to develop guidance for $T$ we study $\CESS{0}$ in \eqref{eq:CESS_j_def}:
\begin{theorem}
    \label{theorem:T_guidance}
    Let $f_{c}\sim\mathcal{N}_{d}(\bm{a}_{c}, \frac{b|\mathcal{C}|}{m}\mathbf{\Lambda}_{c})$ for $c\in\mathcal{C}$, then considering the initial conditional effective sample size $\CESS{0}$ we have that as $N\rightarrow\infty$, the following convergence in probability holds
    \begin{equation}
        N^{-1}\CESS{0} \overset{p}{\rightarrow} \exp\left\{-\frac{\sigma_{\bm{a}}^{2}\left(\frac{b}{m}\right)}{\left(\frac{T}{|\mathcal{C}|}+\frac{b}{m}\right)\left(\frac{T}{|\mathcal{C}|}+\frac{2b}{m}\right)}\right\} \cdot \left[ 1 + \frac{\left(\frac{|\mathcal{C}|b}{Tm}\right)^{2}}{1+\frac{2|\mathcal{C}|b}{Tm}} \right]^{-\frac{(|\mathcal{C}|-1)d}{2}}. \label{eq:CESS_0}
    \end{equation}
    \proof See \apxref{app:guidance_proofs}. \hfill $\blacksquare$
\end{theorem}

The following corollary considers the effect of $T$ on $\CESS{0}$ in the $\SH{\lambda}$ and $\SSH{\gamma}$ settings:

\begin{corollary}
    \label{corollary:T_guidance}
    If for some constant $k_{1}>0$, $T$ is chosen such that $T \geq \frac{b|\mathcal{C}|^{3/2}k_{1}}{m}$ for some constant $k_{1}$, then the following lower bounds on $\CESS{0}$ hold:
    \begin{enumerate}[(a)]
        \item If $\SH{\lambda}$ holds for some $\lambda>0$, then
        \begin{equation}
            \label{eq:CESS_0_SH}
            \lim_{N\rightarrow\infty} N^{-1}\CESS{0} \geq \exp \left\{ - \frac{\lambda}{k_{1}^{2}} - \frac{d}{2k_{1}^{2}} \right\}.
        \end{equation}
        \item If $\SSH{\gamma}$ holds for some $\gamma>0$, and $T \geq k_{2}|\mathcal{C}|^{\frac{1}{2}}$ for some constant $k_{2}$, then
        \begin{equation}
            \label{eq:CESS_0_SSH}
            \lim_{N\rightarrow\infty} N^{-1}\CESS{0} \geq \exp \left\{ - \frac{b\gamma}{k_{1}k_{2}} - \frac{d}{2k_{1}^{2}} \right\}.
        \end{equation}
    \end{enumerate}
    \proof See \apxref{app:guidance_proofs}. \hfill $\blacksquare$
\end{corollary}

We choose $k_{1}$ and $k_{2}$ by means of \remref{remark:T_guidance_k1_k2}, which in turn allows us to determine $T$. As required by \remref{remark:T_guidance_k1_k2} we first set $\lambda=1$ (see
\remref{remark:lambda_choice_GBF}), $b$ (using \remref{remark:b_choice_GBF}), and $\sigma_{\bm{a}}^{2}$ as per \eqref{eq:sigma_a}.
\begin{remark}
    \label{remark:T_guidance_k1_k2}
    Choice of $k_{1},k_{2}$: To choose $k_{1}$ and $k_{2}$, we first specify $\zeta\in(0,1)$ to be a lower bound on the initial relative effective sample size that we would desire. We then can consider which situation that we are likely to be in, and then:
    \begin{enumerate}
        \item Under $\SH{\lambda}$, suppose we want to ensure $N^{-1}\CESS{0}$ is above $\zeta\in(0,1)$, from \eqref{eq:CESS_0_SH}, we have $\exp\left\{-\frac{\lambda}{k_{1}^{2}}-\frac{d}{2k_{1}^{2}}\right\}=\zeta$, which implies we choose $k_{1}=\sqrt{-\frac{(\lambda+\frac{d}{2})}{\log(\zeta)}}$.
        \item Under $\SSH{\gamma}$, suppose we want to ensure $N^{-1}\CESS{0}$ is above $\zeta\in(0,1)$, then from \eqref{eq:CESS_0_SSH}, we have
        \begin{equation}
            \exp \left\{ - \frac{b\gamma}{k_{1}k_{2}} - \frac{d}{2k_{1}^{2}} \right\} = \zeta. \label{eq:CESS_0_SSH_zeta}
        \end{equation}
        Recall that for $\SSH{\gamma}$, we must have $T \geq\max\left\{\frac{b|\mathcal{C}|^{3/2}k_{1}}{m}, |\mathcal{C}|^{\frac{1}{2}}k_{2}\right\}$. Since we wish $T$ to be small, we would like $k_{1}$ and $k_{2}$ to be small, and thus we set these two terms equal to each other and find $k_{2}=\frac{b|\mathcal{C}|k_{1}}{m}$. Substituting into \eqref{eq:CESS_0_SSH_zeta}, we then choose $k_{1}=\sqrt{-\frac{\left(\frac{\gamma m}{C}+\frac{d}{2}\right)}{\log(\zeta)}}$.
    \end{enumerate}
\end{remark}

Given $k_{1}$ and $k_{2}$, $T$ can be chosen such that $T \geq \frac{b|\mathcal{C}|^{3/2}k_{1}}{m}$ if $\SH{\lambda}$ holds, and $T \geq \max\left\{\frac{b|\mathcal{C}|^{3/2}k_{1}}{m}, |\mathcal{C}|^{\frac{1}{2}}k_{2}\right\}$ if $\SSH{\gamma}$ holds. Typically we want to minimise iterations of \algstepref{alg:GBF}{alg:GBF:iterations}, and so we choose the smallest $T$ which satisfies the user-specified $\zeta\in(0,1)$.

\subsection{Guidance for choosing \texorpdfstring{$\mathcal{P}$}{the temporal mesh}} \label{subsec:GBF_mesh_guidance}

In order to choose the temporal mesh $\mathcal{P}$ we consider two approaches, each of which is considered and optimised by means of our CESS of \eqref{eq:CESS_j_def}: i) by first fixing $n$ and assuming a \emph{regular} mesh (as in \citet{dai_et_al_2023}), we then optimise for $n$ by reference to the maximally tolerable degradation of $\CESS{j}$ over any single iterate (see \secref{subsubsec:regular_mesh}); (ii) by starting at $t_0=0$ we decide on the placement of $t_1$ such that we do not violate the maximally tolerable degradation of $\CESS{1}$, and then iterate until we reach $T$, and so leading to a \emph{irregular} (\emph{adaptive}) mesh and implicitly choosing $n$ (see \secref{subsubsec:adaptive_mesh}). We summarise these two mesh constructions in Algorithms \ref{alg:regular_mesh} and \ref{alg:adaptive_mesh}

To simplify the analysis of \algoref{alg:GBF}, for which there is considerable flexibility in the choice of proposal distribution for our unbiased estimator of the importance weights (see \thmref{theorem:unbiased_estimator_GBF} of \secref{subsec:GBF_radonniko}), we assume that we have access to the \emph{optimal} unbiased estimator. \citet[Theorem 1]{Fearnhead_2008} (and \citet[Appendix B]{dai_et_al_2023}) show that the variance of the unbiased estimator $a_{j}\tilde{\rho}_{j}$ is minimised when $p(\kappa_{c}|R_{c}) \sim \textrm{{\upshape Poi}}(\lambda_{c})$, where
\begin{equation}
    \label{eq:optimal_lambda_c}
    \lambda_{c}:=\left[\Delta_{j}\int_{t_{j-1}}^{t_{j}}\left(U_{j}^{(c)}-\phi\left(\bm{X}_{t}^{(c)}\right)\right)^{2} \dd t \right]^{\frac{1}{2}},
\end{equation}
for $c\in\mathcal{C}$. With this choice the second moment is finite and $\mathbb{E}\left[\left(a_{j}\tilde{\rho}_{j}\right)^{2}\right] \leq 1 < \infty$. In practice choosing this optimal distribution for $\mathbb{K}$ is not possible since the integral in \eqref{eq:optimal_lambda_c} cannot be evaluated directly. This is why in \secref{subsec:GBF_radonniko} we choose alternative simulatable distributions (as described in Conditions \ref{cond:GPE1}--\ref{cond:GPE2}), which try to match this optimal distribution. 

With this optimal choice, we establish the following theorem: 
\begin{theorem}
    \label{theorem:mesh_guidance}
    Let $p(\kappa_{c}|R_{c})$ in \eqref{eq:rho_tilde_j_gbf} be a Poisson distribution with intensity given in \eqref{eq:optimal_lambda_c}, for $c\in\mathcal{C}$, and $k_{3}$, $k_{4}$ be positive constants. If $\lim_{\Delta_{j}\rightarrow0}$ is taken over sequences of $\Delta_{j}:=t_{j}-t_{j-1}\rightarrow0$ with
    \begin{equation}
        \label{eq:mesh_guidance}
        t_{j}-t_{j-1} \leq \tilde{\Delta}_{j} := \min \left\{ \frac{b^{2}|\mathcal{C}|k_{3}}{\mathbb{E}\left[\nu_{j}\right]m^{2}}, \left(\frac{b^{2}|\mathcal{C}|k_{4}}{2m^{2}d}\right)^{\frac{1}{2}} \right\},
    \end{equation}
    where
    \begin{equation}
        \label{eq:nu_j}
        \nu_{j} := \frac{1}{|\mathcal{C}|} \sum_{c\in\mathcal{C}} \left(\bm{x}_{j-1}^{(c)}-\bm{a}_{c}\right)^{\intercal}\mathbf{\Lambda}_{c}^{-1}\left(\bm{x}_{j-1}^{(c)}-\bm{a}_{c}\right),
    \end{equation}
    and the expectation $\mathbb{E}[\nu_{j}]$ is taken over $\vecX{j-1}{(\mathcal{C})}$, we have
    \begin{equation}
        \plim_{\Delta_{j}\rightarrow0} \plim_{N\rightarrow\infty} N^{-1}\CESS{j} \geq e^{-k_{3}-k_{4}},
    \end{equation}
    where $\plim$ denotes a limit in probability.
    \proof See \apxref{app:guidance_proofs}. \hfill $\blacksquare$
\end{theorem}

\begin{remark} \label{rem:nuj_est}
    In \thmref{theorem:mesh_guidance}, $\nu_{j}$ (as defined in \eqref{eq:nu_j}) describes the \emph{scaled/weighted} average variation of the $|\mathcal{C}|$ trajectories of the distribution of their proposed update locations with respect to their individual sub-posterior means (i.e.\ describing how far $\bm{x}_{j-1}^{(c)}$ is from $\bm{a}_{c})$. Since the \gbfa approach has $|\mathcal{C}|$ trajectories which are initialised from their respective sub-posterior distributions and coalesce to a common end point, this variation is mainly determined by a combination of: (i) how large the time horizon $T$ is; (ii) how large the interval we are simulating over for this iteration $(t_{j-1},t_{j}]$; and (iii) how much the sub-posteriors conflict which we determine by looking at the variation in their means as per \eqref{eq:sigma_a}. Given a weighted particle set from the $(j-1)$th iteration of the algorithm, $\{\vecX{j-1,i}{(\mathcal{C})}, w_{j-1,i}^{(\mathcal{C})}\}_{i=1}^{N}$, a natural estimator for $\mathbb{E}\left[\nu_{j}\right]$ is
    \begin{equation}
        \label{eq:E_nu_j_approximate}
        \widehat{\mathbb{E}\left[\nu_{j}\right]} = \sum_{i=1}^{N} w_{j-1,i}^{(\mathcal{C})} \left( \frac{1}{|\mathcal{C}|} \sum_{c\in\mathcal{C}} \left(\bm{x}_{j-1,i}^{(c)}-\bm{a}_{c}\right)^{\intercal}\mathbf{\Lambda}_{c}^{-1}\left(\bm{x}_{j-1,i}^{(c)}-\bm{a}_{c}\right) \right).
    \end{equation}
\end{remark}

Following \thmref{theorem:mesh_guidance} and \remref{rem:nuj_est} we now have the additional problem of specifying $k_{3}$ and $k_{4}$, and using the result to develop practical guidance. We do so by means of letting the user choose the meaning parameter $\zeta^{\prime}\in(0,1)$, which is we define to be \emph{a lower bound on the conditional effective sample size that they would tolerate}. We can then select $k_{3}$ and $k_{4}$ such that $e^{-k_{3}-k_{4}}=\zeta^{\prime}$ and compute
\begin{equation}
    t_{j}=\min\left\{T, t_{j-1}+\tilde{\Delta}_{j}\right\}
\end{equation}
recursively at each iteration until $j=n$ such that $t_{n}=T$.
\begin{remark}
    Note that we expect to have very different performance with different choices of $k_{3}$ and $k_{4}$. For instance, we can obtain a very high $\CESS{j}$ by simply choosing $k_{3}$ very small and set $k_{4}=-\log(\zeta^{\prime})-k_{3}$, which ultimately leads to having very small intervals sizes $\tilde{\Delta}_{j}$. Choosing small interval sizes may help computationally simulating $\tilde{\rho}_{j}$, but this comes at the cost of having more iterations of the algorithm, leading to an increased communication between the cores. Natural choices for jointly specifying $k_{3}$ and $k_{4}$ are ones which lead to the largest interval size which still satisfies $N^{-1}\CESS{j}\geq\zeta^{\prime}\in(0,1)$, as this minimises the number of iterations of \algstepref{alg:GBF}{alg:GBF:iterations}.
\end{remark}

We now consider the previously outlined \emph{regular} and \emph{irregular} (\emph{adaptive}) mesh selection of $\mathcal{P}$ in \secref{subsubsec:regular_mesh} and \secref{subsubsec:adaptive_mesh} respectively. 

\subsubsection{A regular mesh construction} \label{subsubsec:regular_mesh}

Imposing an additional assumption that the temporal partition $\mathcal{P}$ is \emph{regular} simplifies \algoref{alg:GBF} as it avoids us having to dynamically compute \eqref{eq:mesh_guidance} at each iteration of \stepref{alg:GBF:iterations}. In particular, $\tilde{\Delta}_{j}=\Delta$ for each $j\in\{1,\dots,n\}$ where $n=\lceil T/\Delta \rceil$ (where $\lceil x \rceil$ denotes the smallest integer greater than or equal to $x$). This simplification of regularity was suggested in \citet[Remark 6]{dai_et_al_2023}. They noted that for large datasets in which observations were randomly allocated to sub-posteriors, that one would expect sub-posterior heterogeneity to be small. Hence one would expect $\mathbb{E}[\nu_{j}]$ to be small (of $\mathcal{O}(m^{-1})$). In their simulations, \citet[Section 3 and 4]{dai_et_al_2023} set $k_{3}=k_{4}=1$ and $\Delta:=t_{j-1}-t_{j}=\sqrt{(b^{2}|\mathcal{C}|k_{4})/(2m^{2}d)}$ for all $j$. The rationale presented for these choices in \citet[Remark 6]{dai_et_al_2023} does not hold in full generality so in this section, we instead develop a more systematic way to construct a regular mesh. In particular, setting $k_3 = k_4$ as they suggest is sub-optimal. 

Given a user specified lower bound on $\CESS{j}$ that they would tolerate (i.e.\ some $\zeta^{\prime}\in(0,1)$), we want to minimise the number of iterations of \algstepref{alg:GBF}{alg:GBF:iterations}. This is achieved with reference to \thmref{theorem:mesh_guidance} (and in particular \eqref{eq:mesh_guidance}). In particular, we choose a combination of $k_{3}$ and $k_{4}$ such that: (i) $\exp\{-k_{3}-k_{4}\} \geq \zeta^{\prime}$ (i.e.\ $\CESS{j}$ for any $j$ does not violate the chosen $\zeta^{\prime})$; and (ii), $\frac{b^{2}|\mathcal{C}|k_{3}}{\mathbb{E}\left[\nu_{j}\right]m^{2}} \geq \sqrt{\frac{b^{2}|\mathcal{C}|k_{4}}{2m^{2}d}}$ for each $j$.

The difficulty here is that at each iteration, we need the average variation of the trajectories $\mathbb{E}[\nu_{j}]$. Of course, this is not possible directly and so an estimate $\widehat{\mathbb{E}[\nu_{j}]}$ is computed as per the guidance of \remref{rem:nuj_est}. To ensure the chosen $\zeta^{\prime}$ is not violated at \emph{any} iteration we follow the guidance of \eqref{eq:mesh_guidance} by using a supremum over all intervals of this estimator (i.e.\ $\sup_j \widehat{\mathbb{E}[\nu_{j}]}$). This choice allows us to specify $k_3$ and $k_4$, and so in turn $n$ and $\mathcal{P}$. 

\begin{remark}
    For ease of practically implementing \algoref{alg:GBF}, it is desirable to avoid any recursive definitions of $n$ and $\mathcal{P}$ (i.e.\ they are specified prior to calling \algstepref{alg:GBF}{alg:GBF:iterations} where they are required). In this setting we would need to estimate $\sup_j \widehat{\mathbb{E}[\nu_{j}]}$ based upon \emph{only} the \emph{initial} (weighted) sub-posterior realisations $\{\vecX{0,i}{(\mathcal{C})}, w_{0,i}^{(\mathcal{C})}\}_{i=1}^{M}$ obtained in \algstepref{alg:GBF}{alg:GBF:initial_weights}.
    
    Following \remref{rem:nuj_est}, we would expect $\mathbb{E}[\nu_{j}]$ to be maximised at $t=T$ (corresponding to \eqref{eq:E_nu_j_max_1}), but in some instance may also occur at $t=0$ (corresponding to \eqref{eq:E_nu_j_max_2}). In most practical applications of \gbfa it will be at $t=T$ as the proposal for the coalescence of the $|\mathcal{C}|$ stochastic processes has a Gaussian distribution with mean $\tilde{\bm{x}}_{0}^{(\mathcal{C})}$ with variance $T\mathbf{\Lambda}_{\mathcal{C}}$ (as a consequence of \propositionref{prop:proposal_simulation_GBF} and considering $s=0$ and $t=T$). On the other hand, if the sub-posterior means are very close together, the largest variation in the trajectories from their respective means could occur at the start of the bridge. As such, we propose taking the larger value of those two scenarios to arrive at the following approximation:
    \begin{equation}
        \label{eq:E_nu_j_max}
        \sup_{j}\widehat{\mathbb{E}\left[\nu_{j}\right]} \approx \max\{\Psi_1, \Psi_2\},
    \end{equation}
    where
    \begin{align}
        \Psi_1 & := \sum_{i=1}^{M} w_{0,i}^{(\mathcal{C})}  \frac{1}{|\mathcal{C}|} \sum_{c\in\mathcal{C}} \left(\tilde{\bm{x}}_{0,i}^{(\mathcal{C})}-\bm{a}_{c}\right)^{\intercal}\mathbf{\Lambda}_{c}^{-1}\left(\tilde{\bm{x}}_{0,i}^{(\mathcal{C})}-\bm{a}_{c}\right), \label{eq:E_nu_j_max_1} \\
        \Psi_2 & := \sum_{i=1}^{M} w_{0,i}^{(\mathcal{C})}  \frac{1}{|\mathcal{C}|} \sum_{c\in\mathcal{C}} \left(\bm{x}_{0,i}^{(c)}-\bm{a}_{c}\right)^{\intercal}\mathbf{\Lambda}_{c}^{-1}\left(\bm{x}_{0,i}^{(c)}-\bm{a}_{c}\right), \label{eq:E_nu_j_max_2}
    \end{align}
    and where $\tilde{\bm{x}}_{0,i}^{(\mathcal{C})}$ is defined in \eqref{eq:x_mean} and $w_{0,i}^{(\mathcal{C})}$ are the initial particle weights given in \algstepref{alg:GBF}{alg:GBF:initial_weights}.

    Our approximation of $\sup_{j}\widehat{\mathbb{E}\left[\nu_{j}\right]}$ has obvious limitations: it may not be conservative enough to ensure the user chosen $\zeta^\prime$ is not breached; it may be too conservative and lead to choosing $n$ too high. In practice we have found it to be a robust approximation. 
\end{remark}

Once we have a suitable estimate of $\sup_{j}\widehat{\mathbb{E}\left[\nu_{j}\right]}$, we need to find a suitable choice for $k_{3}$ and $k_{4}$ to ensure that we always choose the RHS side of \eqref{eq:mesh_guidance} (as that leads to a regular mesh) and satisfies $\zeta^{\prime}$. As there are many combinations of $k_{3}$ and $k_{4}$ which can return a regular mesh, we aim to find the combination which returns the largest interval size. We can do this by means of the following proposition which considers the $j$th interval of the partition:
\begin{proposition}
    \label{prop:k4_choice}
    Considering the $j$th interval of $\mathcal{P}$ (i.e.\ $[t_{j-1},t_j]$), given a user-specified threshold $\zeta^{\prime}\in(0,1)$ and estimate $\widehat{\mathbb{E}[\nu_{j}]}$ of $\mathbb{E}[\nu_{j}]$, then the largest interval size which satisfies $N^{-1}\CESS{j}\geq\zeta^{\prime}$ is given by
    \begin{equation*}
       \tilde{\Delta}_{j} = \sqrt{\frac{b^{2}|\mathcal{C}|k_{4,j}}{2m^{2}d}},
    \end{equation*}
    where,
    \begin{equation}
        \label{eq:k4_choice}
        k_{4,j} := \frac{\left(\frac{\widehat{\mathbb{E}[\nu_{j}]}^{2}m^{2}}{2b^{2}|\mathcal{C}|d}-2\log(\zeta^{\prime})\right) - \sqrt{\left(2\log(\zeta^{\prime})-\frac{\widehat{\mathbb{E}[\nu_{j}]}^{2}m^{2}}{2b^{2}|\mathcal{C}|d}\right)^{2}-4\log(\zeta^{\prime})^{2}}}{2}.
    \end{equation}
    \proof See \apxref{app:guidance_proofs}. \hfill $\blacksquare$
\end{proposition}

Using \propositionref{prop:k4_choice}, we can substitute our estimate of $\sup_{j}\widehat{\mathbb{E}[\nu_{j}]}$ into \eqref{eq:k4_choice}, and subsequently compute the regular interval size $\Delta:=\sqrt{\frac{b^{2}|\mathcal{C}|k_{4}}{2m^{2}d}}$ and hence $n=\lceil T/\Delta \rceil$. In effect, here we are setting $k_4 = \sup_j k_{4,j}$. This process is summarised in \algoref{alg:regular_mesh}.
\begin{algorithm}
    \caption{Computing regular mesh $\mathcal{P}$.}
    \label{alg:regular_mesh}
    \textbf{Input:} Time $T>0$ and importance weighted particles $\{\vecX{0,i}{(\mathcal{C})},w_{0,i}^{(\mathcal{C})}\}_{i=1}^{M}$.
    \begin{enumerate}
        \item Compute estimate of $\sup_{j}\widehat{\mathbb{E}[\nu_{j}]}$ as per \eqref{eq:E_nu_j_max}. \label{alg:regular_mesh:estimate}
        \item Compute $k_{4}$ using the estimate from \stepref{alg:regular_mesh:estimate} as per \eqref{eq:k4_choice}.
        \item Compute $\Delta:=\sqrt{\frac{b^{2}|\mathcal{C}|k_{4}}{2m^{2}d}}$ and let $n=\lceil T/\Delta \rceil$.
        \item For $j\in\{1,\dots,n\}$, let $t_{j}=\min\{T, t_{j-1}+\Delta\}$.
        \item \textbf{Output:} $\mathcal{P}:=\{t_{0},\dots,t_{n}\}$.
    \end{enumerate}
\end{algorithm}

\subsubsection{An adaptive mesh construction} \label{subsubsec:adaptive_mesh}

Our presentation of \secref{subsubsec:regular_mesh} (as opposed to that of \citet{dai_et_al_2023}), naturally suggests an \emph{adaptive} approach, leading to a partition $\mathcal{P}$ with an \emph{irregular} mesh. Since the construction of the regular mesh is based upon the \emph{worst} case scenario of the trajectory variation, this leads to an excessive resolution of $\mathcal{P}$. In this section we will address this.

Instead, suppose we are at the beginning of the $j$th iteration of \algstepref{alg:GBF}{alg:GBF:iterations}. At this point we have in effect simulated our $|\mathcal{C}|$ stochastic processes up to time $t_{j-1}<T$. We can now consider the placement of the next point in the partition (i.e.\ $\min(t_j,T)$) with reference to the user chosen $\zeta^{\prime}\in(0,1)$. In particular, we want the interval to be as large as possible while ensuring that the $\CESS{j}$ does not degrade by more than $\zeta^{\prime}$. To do this we can compute an estimate of $\mathbb{E}[\nu_{j}]$ as per \eqref{eq:E_nu_j_approximate} and appeal to \propositionref{prop:k4_choice} in order to choose $k_{4,j}$, and consequently the interval size $\Delta_{j}$ in order to set $t_j = \min(t_{j-1} + \Delta_j, T)$. Once we reach $T$ we simply halt iterating \algstepref{alg:GBF}{alg:GBF:iterations}. 

In contrast to the regular mesh construction in \secref{subsubsec:regular_mesh}, we cannot compute the temporal mesh prior to \algstepref{alg:GBF}{alg:GBF:iterations}. Therefore, the computation of the interval size for iteration $j$ must be done immediately after \stepref{alg:GBF:resample_j} and prior to \stepref{alg:GBF:propose_and_assign} of \algoref{alg:GBF}. In this setting, the number of steps in \algoref{alg:GBF}, $n$, is not known in advance. Given the construction of the regular mesh assumes the \emph{worst case} interval in selecting the mesh size, we would expect that $n$ would be lower in our adaptive approach. Indeed, we show this empirically in our later simulation studies. We summarise this approach in \algoref{alg:adaptive_mesh}.
\begin{algorithm}
    \caption{Computing adaptive mesh $\mathcal{P}$ (computing $\Delta_{j}$ at iteration $j$ immediately after \algstepref{alg:GBF}{alg:GBF:resample_j}).}
    \label{alg:adaptive_mesh}
    \textbf{Input:} Time $T>0$ and importance weighted particles $\{\vecX{j-1,i}{(\mathcal{C})}, w_{j-1,i}^{(\mathcal{C})}\}_{i=1}^{N}$.
    \begin{enumerate}
        \item Compute $\widehat{\mathbb{E}[\nu_{j}]}$ as per \eqref{eq:E_nu_j_approximate}. \label{alg:adaptive_mesh:estimate}
        \item Compute $k_{4}$ with the estimate from \stepref{alg:adaptive_mesh:estimate} as per \eqref{eq:k4_choice}.
        \item Compute $t_{j}=\min\left\{T, t_{j-1}+\sqrt{\frac{b^{2}|\mathcal{C}|k_{4}}{2m^{2}d}}\right\}$.
        \item \textbf{Output:} $\Delta_{j}:=t_{j}-t_{j-1}$.
    \end{enumerate}
\end{algorithm}



\section{Examples} \label{sec:dc_gbf_examples}

In this section we consider a number of models applied to a variety datasets, and suppose the dataset is randomly split into $C$ (disjoint) subsets. We compare the performance of our Fusion methodologies (\gbfa and \hgbfa) with other established (approximate) methodologies. To compare performance, we consider their computational run-times and \emph{Integrated Absolute Distance (IAD)}. To compute the IAD we average across each dimension the difference between the true target (\emph{fusion}) density ($f$), and a kernel density estimate of the draws realised using a given methodology ($\hat{f}$). In particular,
\begin{equation}
    \label{eq:IAD}
    \IAD = \frac{1}{2d} \sum_{j=1}^{d} \int \abs{\hat{f}(\theta_{j}) - f(\theta_{j})} \dd \theta_{j} \in [0,1].
\end{equation}
In the case where the true marginal density is not available analytically, we take as a proxy for the target $f$ a kernel density estimate of $f$ (for instance, obtained by sampling from $f$ or using the output of an MCMC run). As a benchmark for the target $f$ we use \texttt{Stan} \citep{Carpenter_et_al_2017} to implement an MCMC sampler for the target posterior distribution using the full dataset. In implementing Fusion methodologies we use the GPE-2 variants of \algoref{alg:GBF:update_weight} and \algoref{alg:unbiased_estimator_rho_j} as before. Our implementation is as presented in Sections \ref{sec:GBF} and \ref{sec:dc_gbf}, following the guidance presented in \secref{sec:GBF_guidance} (but without the inclusion of any adaptions such as those presented in \apxref{app:practical_considerations}). In implementing \hgbfa we use the balanced-binary tree hierarchy. For brevity of the main paper, all detailed derivations required for these specific examples have been put in \apxref{app:phi_calulations}.

The established methodologies we consider are \emph{Consensus Monte Carlo (CMC)} \citep{Scott_et_al_2016} (implemented using the \texttt{parallelMCMCcombine} package in \Rlang\, \citep{Miroshnikov_Conlon_2014}), the kernel density averaging approach of \citet{Neiswanger_et_al_2014} (which we term \emph{KDEMC}, and also implemented using the \texttt{parallelMCMCcombine} \Rlang\, package), and the \emph{Weierstrass Rejection Sampler (WRS)} \citep{Wang_and_Dunson_2013} (implemented using their \Rlang\, code available at \href{https://github.com/wwrechard/weierstrass}{\texttt{https://github.com/wwrechard/weierstrass}}).

\subsection{Simulation studies} \label{subsec:simulation_studies}

In \apxref{app:dc_gbf_simulation_studies}, we study empirically the robustness of our Fusion algorithms in our two idealised settings---the $\SH{\lambda}$ setting (\condref{cond:SH}) and $\SSH{\gamma}$ setting (\condref{cond:SSH}). We consider a range of different hyperparameter choices and illustrate in both settings that utilising both the guidance for $T$ and the mesh $\mathcal{P}$, developed in \secref{sec:GBF_guidance}, drastically improves the performance of both BF and \gbfa. By comparing the regular and adaptive meshes, we found that the adaptive mesh generally performed better as it provided similar performance to the regular mesh but at a much reduced computational cost. The full details of these experiments can be found in Sections \ref{subsec:GBF_similar_means} and \ref{subsec:GBF_dissimilar_means}. In \secref{subsec:GBF_dimension_study} we compare the performance of Fusion methodologies with increasing dimensionality and found that our \gbfa and \hgbfa approaches offer the best performance with regards to dimension.

\subsection{Robust regression} \label{subsec:robust_reg}

In this section we consider the \emph{`Combined Cycle Power Plant'} dataset available from the \emph{UCI Machine Learning Repository} \citep{Kaya_et_al_2012, Tufekci_2014}. The dataset comprises $m=9568$ records of the \textbf{net hourly electrical output} of a combined cycle power plant over 6 years between 2006 and 2011, together with four (hourly averaged) \emph{ambient} variables: \textbf{temperature}; \textbf{ambient-pressure}; \textbf{relative-humidity}; and, \textbf{exhaust-vacuum}.

To model electrical output using the ambient variables, we use a robust regression model: 
\begin{align*}
    y_{i} \sim \text{t}(\nu, X_{i} \bm{\beta}, \sigma), & \quad i=1,\dots,n, \\
    \beta_{j^{\prime}} \sim \mathcal{N}_{1}\left(\mu_{\beta_{j^{\prime}}}, \sigma_{\beta_{j^{\prime}}}^{2}\right), & \quad j^{\prime}=0,\dots,p,
\end{align*}
where $\bm{y}\in\mathbb{R}^{n}$ is the dependent variable (electrical output), $X\in\mathbb{R}^{n \times (p+1)}$ is the design matrix, $\bm{\beta}\in\mathbb{R}^{p+1}$ is the vector of predictor (ambient) variables which we want to perform inference on. For simplicity, we assume that $\nu$, $\sigma$, $\mu_{j^{\prime}}$ and $\sigma_{\beta_{j^{\prime}}}^{2}$ for $j^{\prime}=0,\dots,p$ are known.

\begin{figure}[ht]
    \centering
    \begin{subfigure}[b]{0.49\textwidth}
        \centering
        \includegraphics[width=\textwidth]{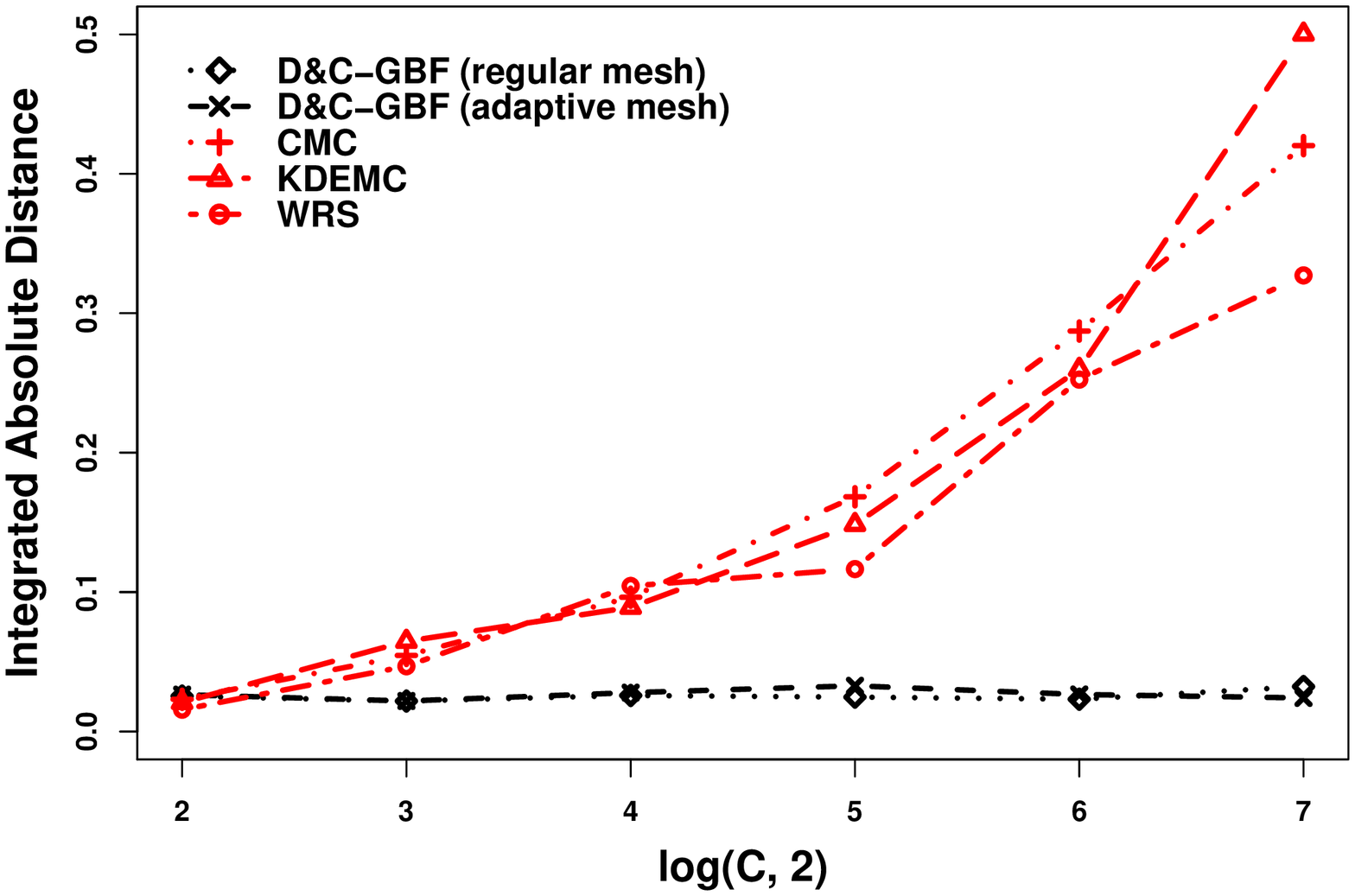}
        \caption{Integrated Absolute Distance.}
    \label{fig:robust_reg_pp_IAD}
    \end{subfigure}
    \hfill
    \begin{subfigure}[b]{0.49\textwidth}
        \centering
        \includegraphics[width=\textwidth]{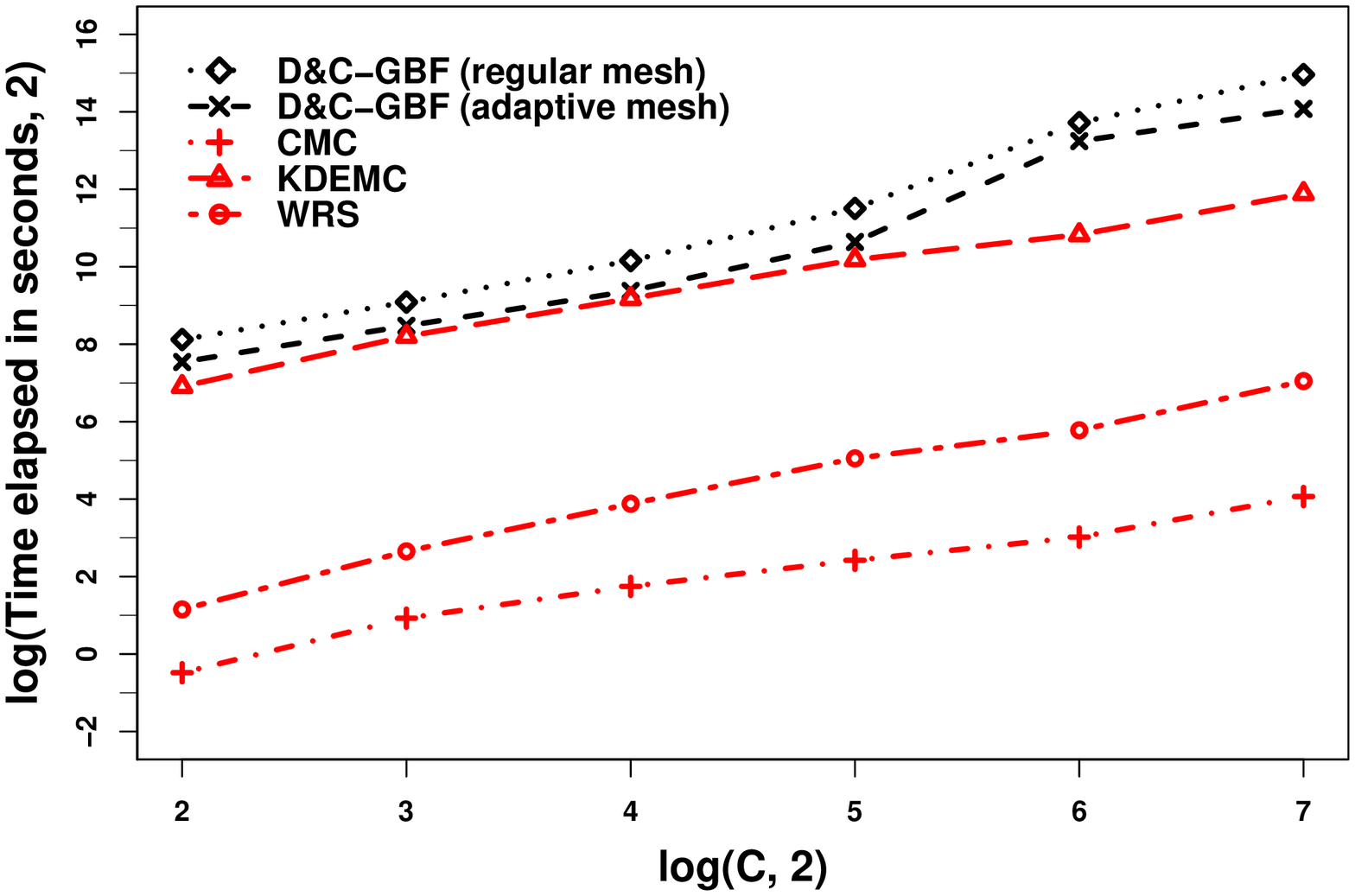}
        \caption{Computational cost.}
        \label{fig:robust_reg_pp_time}
    \end{subfigure}
    \caption{Comparison of competing methodologies to \hgbfa applied to a robust regression model using the power plant dataset (see \secref{subsec:robust_reg}).}
    \label{fig:robust_reg_pp}
\end{figure}

For our dataset $p=4$, and so $d=5$. We consider $C\in\{4,8,16,32,64,128\}$ cores, each of which is assigned a random split of the data. We use \texttt{Stan} to sample the sub-posteriors (with $\mu_{j^{\prime}}=0$ and $\sigma_{\beta_{j^{\prime}}}^{2}=10C$ for $j^{\prime}=0,\dots,p$), which we will attempt to unify as in \eqref{eq:fusion_density}. We use the approximate CMC, KDEMC and WRS approaches to do this, together with our \hgbfa approach. In implementing \hgbfa we set $N=10000$, $\zeta=0.5$, $\zeta^{\prime}=0.05$, and consider both the regular and adaptive mesh variants of the temporal partition, $\mathcal{P}$. We resample if the ESS falls below $0.5N$. The results are presented in \figref{fig:robust_reg_pp}.

\figref{fig:robust_reg_pp_IAD} clearly shows that of all the approaches considered, \hgbfa provides the highest quality and most reliable sample approximation for $f$, and is the most robust to increasing $C$. Although more expensive computationally, \hgbfa has a cost which grows at the same rate as the approximate methodologies considered. 

\subsection{Negative Binomial regression} \label{subsec:NB_reg}

Here we consider the \emph{`Bike Sharing'} dataset available on the \emph{UCI Machine Learning Repository} \citep{Fanaee_2014}. The dataset contains $m=17379$ records of the \textbf{total count of bikes on rental each hour}, together with seven variables: \textbf{seasonality} (a categorical variable with four levels: spring, summer, autumn, winter); \textbf{weekend} (binary, taking value $1$ if a weekend, and $0$ if not); \textbf{holiday}; (binary, taking value $1$ if a holiday, and $0$ if not); \textbf{rush-hour} (binary, taking value $1$ if recorded on a weekday between $7$AM-$9$AM or $4$PM-$7$PM, and $0$ if not); \textbf{weather} (binary, taking value $1$ if `clear', and $0$ if not); \textbf{temperature} (continuous); and, \textbf{wind-speed} (continuous). We use treatment contrast coding to encode the \textbf{seasonality} via three binary variables.

To model the total count of bikes on rental, we use the following Negative binomial (NB) regression model:
\begin{align*}
    y_{i} \sim \text{NB}(\mu_{i}, r), & \qquad \text{where } \log(\mu_{i}) = X_{i}\bm{\beta}, \qquad i\in\{1,\dots,n\}, \\
    \beta_{j^{\prime}}  \sim \mathcal{N}_{1}\left(\mu_{\beta_{j^{\prime}}}, \sigma_{\beta_{j^{\prime}}}^{2}\right), & \qquad j^{\prime}\in\{0,\dots,p\},
\end{align*}
where $\bm{y}\in\mathbb{R}^{n}$ is our total count of bikes on rental, $X\in\mathbb{R}^{n \times (p+1)}$ is the design matrix, $\bm{\beta}\in\mathbb{R}^{p+1}$ is the vector of predictor variables. For simplicity, $r$, $\mu_{\beta_{j^{\prime}}}$, $\sigma_{\beta_{j^{\prime}}}^{2}$ for $j^{\prime}=0,\dots,p$ are assumed known.

For this data set $p=9$, and so $d=10$. As in \secref{subsec:robust_reg}, we split the dataset amongst $C\in\{4,8,16,32,64,128\}$ cores, and use \texttt{Stan} with $\mu_{j^{\prime}}=0$ and $\sigma_{\beta_{j^{\prime}}}^{2}=10C$ for $j^{\prime}=0,\dots,p$, to recover the respective sub-posteriors. To implement \hgbfa we set $N=10000$, $\zeta=0.2$, $\zeta^{\prime}=0.05$, and consider both regular and adaptive mesh variants of $\mathcal{P}$, and resample if the ESS drops below $0.5N$. 

\begin{figure}[ht]
    \centering
    \begin{subfigure}[b]{0.49\textwidth}
        \centering
        \includegraphics[width=\textwidth]{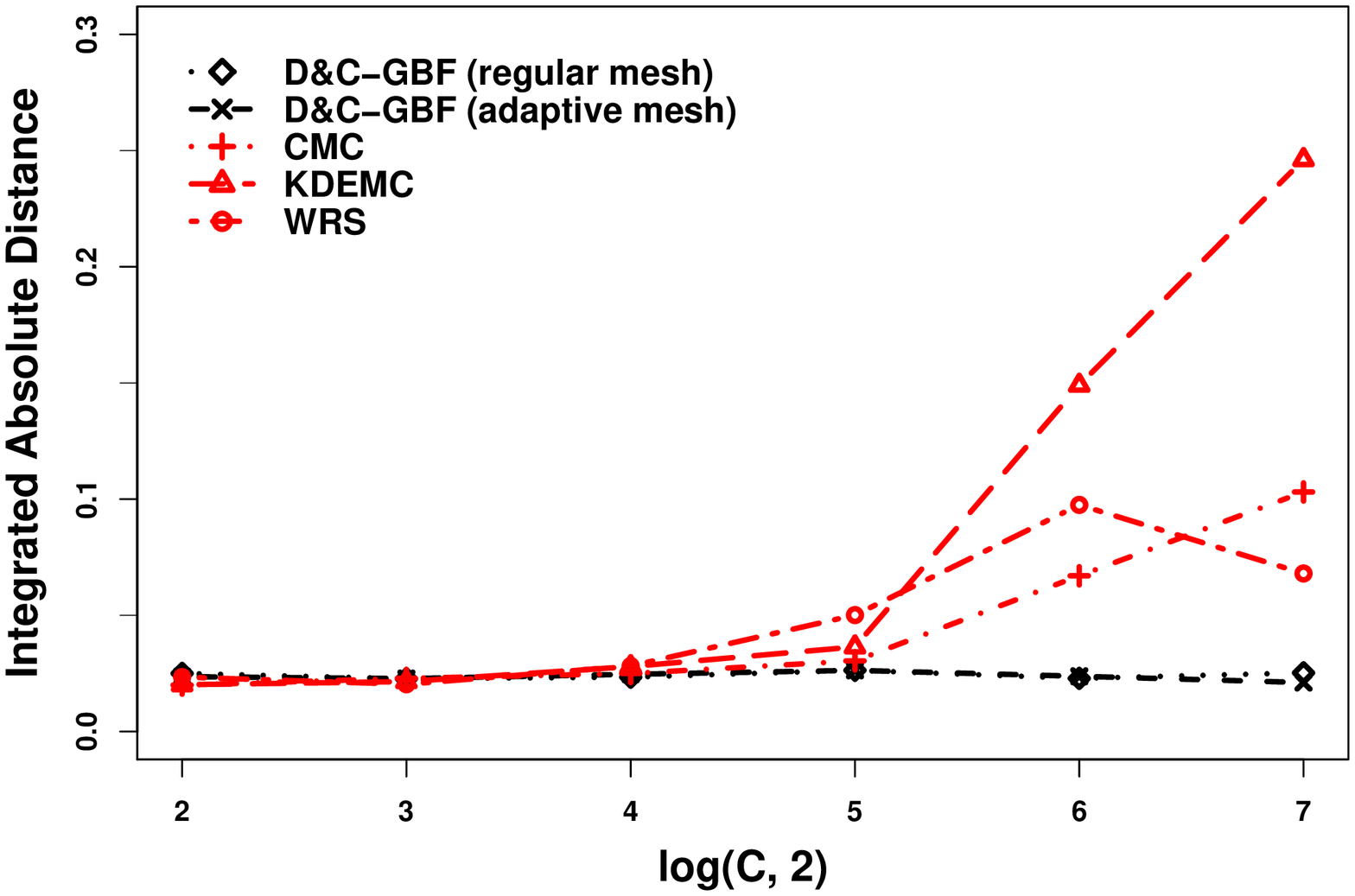}
        \caption{Integrated Absolute Distance.}
    \label{fig:NB_reg_bs_IAD}
    \end{subfigure}
    \hfill
    \begin{subfigure}[b]{0.49\textwidth}
        \centering
        \includegraphics[width=\textwidth]{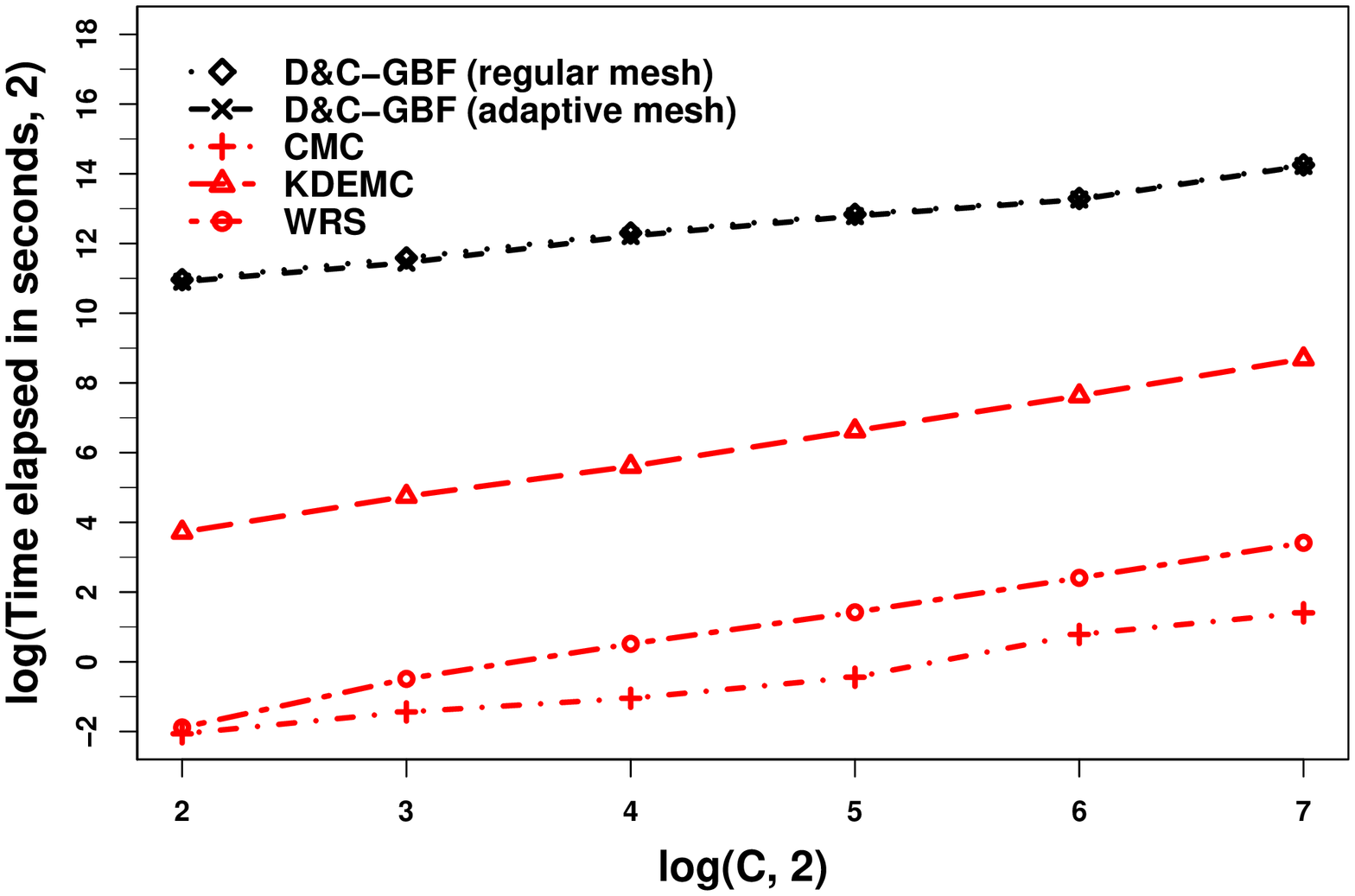}
        \caption{Computational cost.}
        \label{fig:NB_reg_bs_time}
    \end{subfigure}
    \caption{Comparison of competing methodologies to \hgbfa applied to a Negative Binomial regression model using the bike sharing dataset (see \secref{subsec:NB_reg}).}
    \label{fig:NB_reg_bs}
\end{figure}

The results in \figref{fig:NB_reg_bs} again show that, when contrasted with existing (approximate) approaches, \hgbfa provides the most accurate sample approximation, and is robust and consistent with increasing $C$. 

\subsection{Logistic regression} \label{subsec:log_reg}

In this section, we apply a logistic regression model to two different datasets (each of which highlight an aspect of Fusion):
\begin{equation}
    \label{eq:log_reg}
    y_{i} = 
    \begin{cases}
    1 \qquad \text{with probability } \frac{\exp \{\mathbf{x}_{i}^{\intercal} \bm{\beta}\}}{1+\exp \{\mathbf{x}_{i}^{\intercal} \bm{\beta}\}}, \\
    0 \qquad \text{otherwise}.
    \end{cases}
\end{equation}

\subsubsection{Small data} \label{subsec:log_reg:sd_dcgbf}

Here we consider a \emph{small data size} scenario ($m=1000$), in which the data is simulated from a logistic regression model \eqref{eq:log_reg}. This is a variant of \citet[Section 4.3]{Scott_et_al_2016}, and is of interest as when the data is (randomly) split among the available cores both \emph{exact} and \emph{approximate} Fusion approaches struggle. This is due to the resulting sub-posteriors being naturally conflicting and lacking fully overlapping support with one another.

Each record of the simulated design matrix contained four covariates in addition to an intercept. The $i$th entry of the design matrix is given by $\mathbf{x}_{i} = [1,\zeta_{i,1},\zeta_{i,2},\zeta_{i,3},\zeta_{i,4}]^{\intercal}$, where $\zeta_{i,1},\zeta_{i,2},\zeta_{i,3},\zeta_{i,4}$ are random variables generated from a mixture density with a point-mass at zero (and so are either \emph{activated} or not). In particular, we have for $j\in\{1,\dots,4\}$ that $\zeta_{i,j}\sim p_{j} \mathcal{N}_{1}(1,1) + (1-p_{j})\delta_0$. For this example we chose $p_{1}=0.2$, $p_{2}= 0.3$, $p_{3}=0.5$ and $p_{4}=0.01$ (corresponding to a rarely activated covariate). Upon simulating the design matrix, binary observations were obtained by simulation using the parameters $\mathbf{\beta}=[-3,1.2,-0.5,0.8,3]^{\intercal}$. In total there were a relatively small number of positive responses ($\sum_{i}y_{i}=129$). 

To conduct Fusion we first equally split the data between $C\in\{4,8,16,32,64\}$ cores. We again use \texttt{Stan} with Gaussian prior distributions with mean $0$ and variance $C$ on each parameter to find a sample approximation of each sub-posterior. 

\begin{figure}[ht]
    \centering
    \begin{subfigure}[b]{0.49\textwidth}
        \centering
        \includegraphics[width=\textwidth]{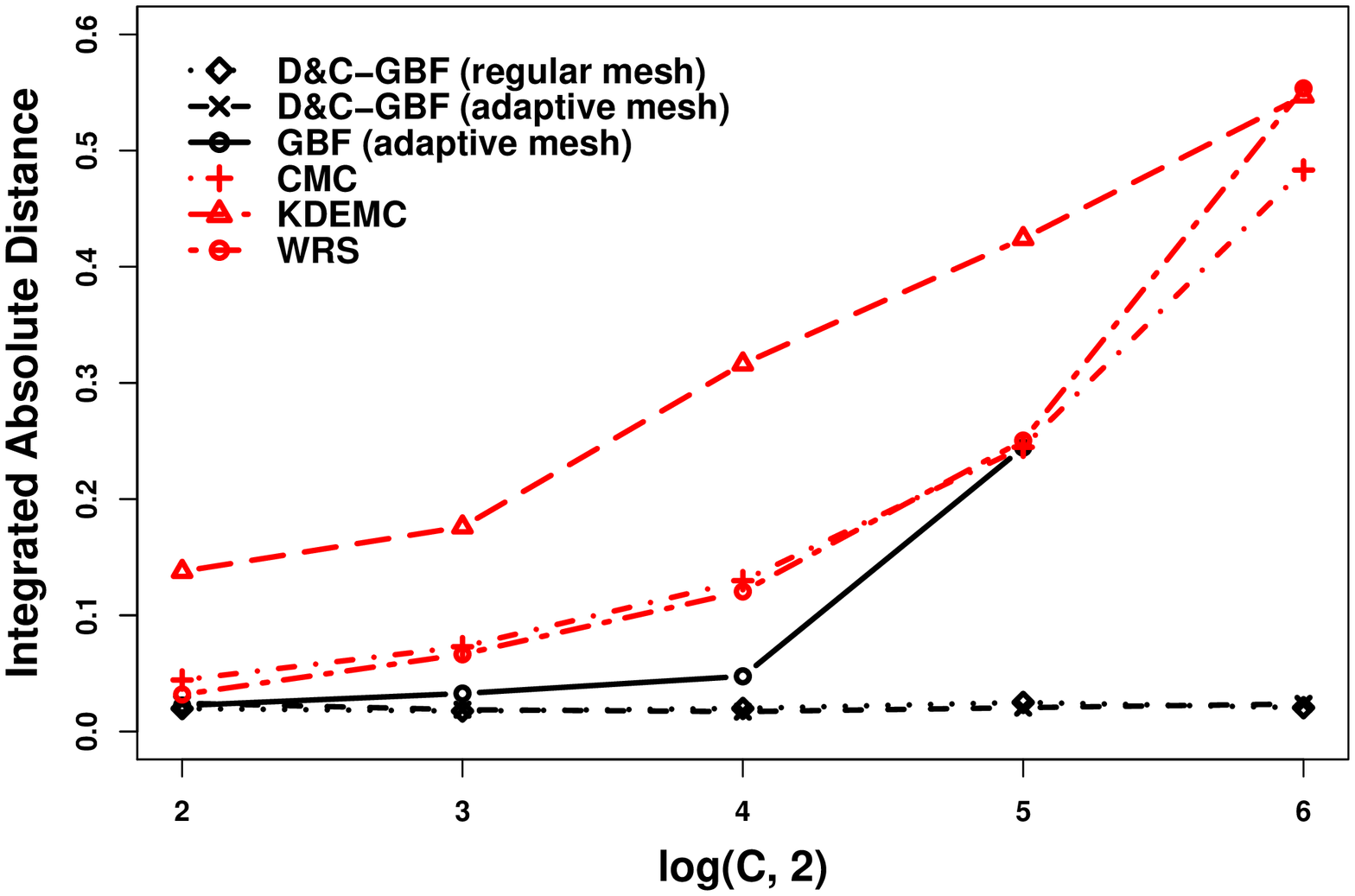}
        \caption{Integrated Absolute Distance.}
    \label{fig:log_reg_sd_dcgbf_IAD}
    \end{subfigure}
    \hfill
    \begin{subfigure}[b]{0.49\textwidth}
        \centering
        \includegraphics[width=\textwidth]{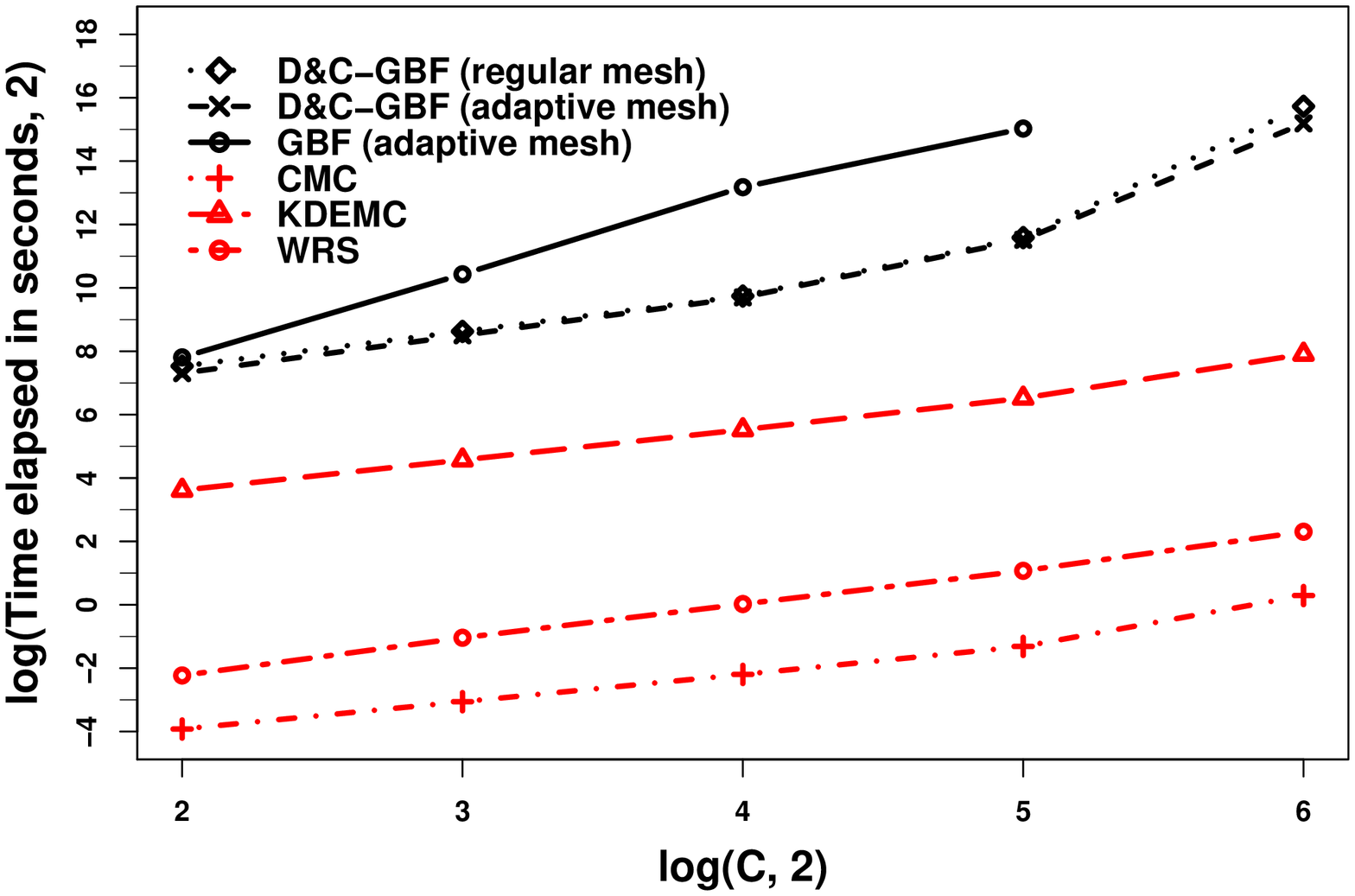}
        \caption{Computational cost.}
        \label{fig:log_reg_sd_dcgbf_time}
    \end{subfigure}
    \caption{Comparison of competing methodologies to \hgbfa applied to a logistic regression model using a simulated small data set (see \secref{subsec:log_reg:sd_dcgbf}).}
    \label{fig:log_reg_sd_dcgbf}
\end{figure}

Together with the approximate methodologies, we implemented our \hgbfa approach with $N=10000$, $\zeta=0.2$, $\zeta^{\prime}=0.05$, and both regular and adaptive temporal partition meshes. Here, we also consider applying \gbfl (i.e.\ directly applying \algoref{alg:GBF} with $\mathcal{C}:=\{1,\dots,C\}$ (which is equivalent to \hgbfa within a fork-and-join tree hierarchy, as per \figref{fig:fork_and_join})). We present the results in \figref{fig:log_reg_sd_dcgbf}.

Considering \figref{fig:log_reg_sd_dcgbf_IAD}, we see again that \hgbfa achieves the best sample approximation, and the quality of the sample approximation is robust to increasing $C$. Note that our divide-and-conquer framework offers significant gains, with \hgbfa outperforming \gbfa in terms of robustness with $C$ (even with the same tuning parameter guidance being followed). Note that CMC outperforms all other approximate methodologies, which leaves the practitioner with a clear decision: if a cheap but approximate methodology is needed use CMC, but if accuracy is the goal then \hgbfa should be used. 

\subsubsection{NYC Flights 2013 Data} \label{subsec:log_reg:nyc}

\begin{figure}[ht]
    \centering
    \begin{subfigure}[b]{0.49\textwidth}
        \centering
        \includegraphics[width=\textwidth]{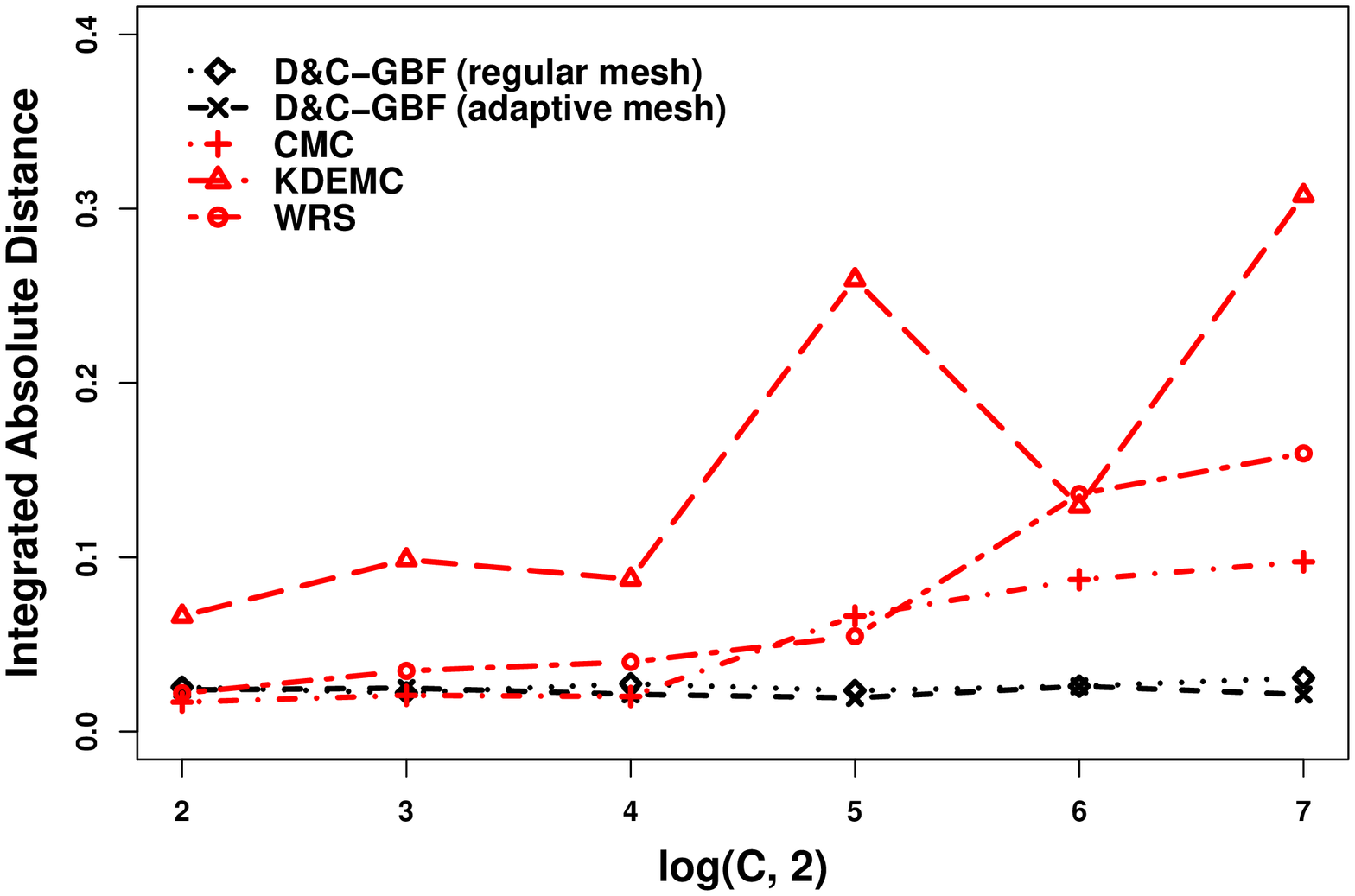}
        \caption{Integrated Absolute Distance.}
    \label{fig:log_reg_nyc_IAD}
    \end{subfigure}
    \hfill
    \begin{subfigure}[b]{0.49\textwidth}
        \centering
        \includegraphics[width=\textwidth]{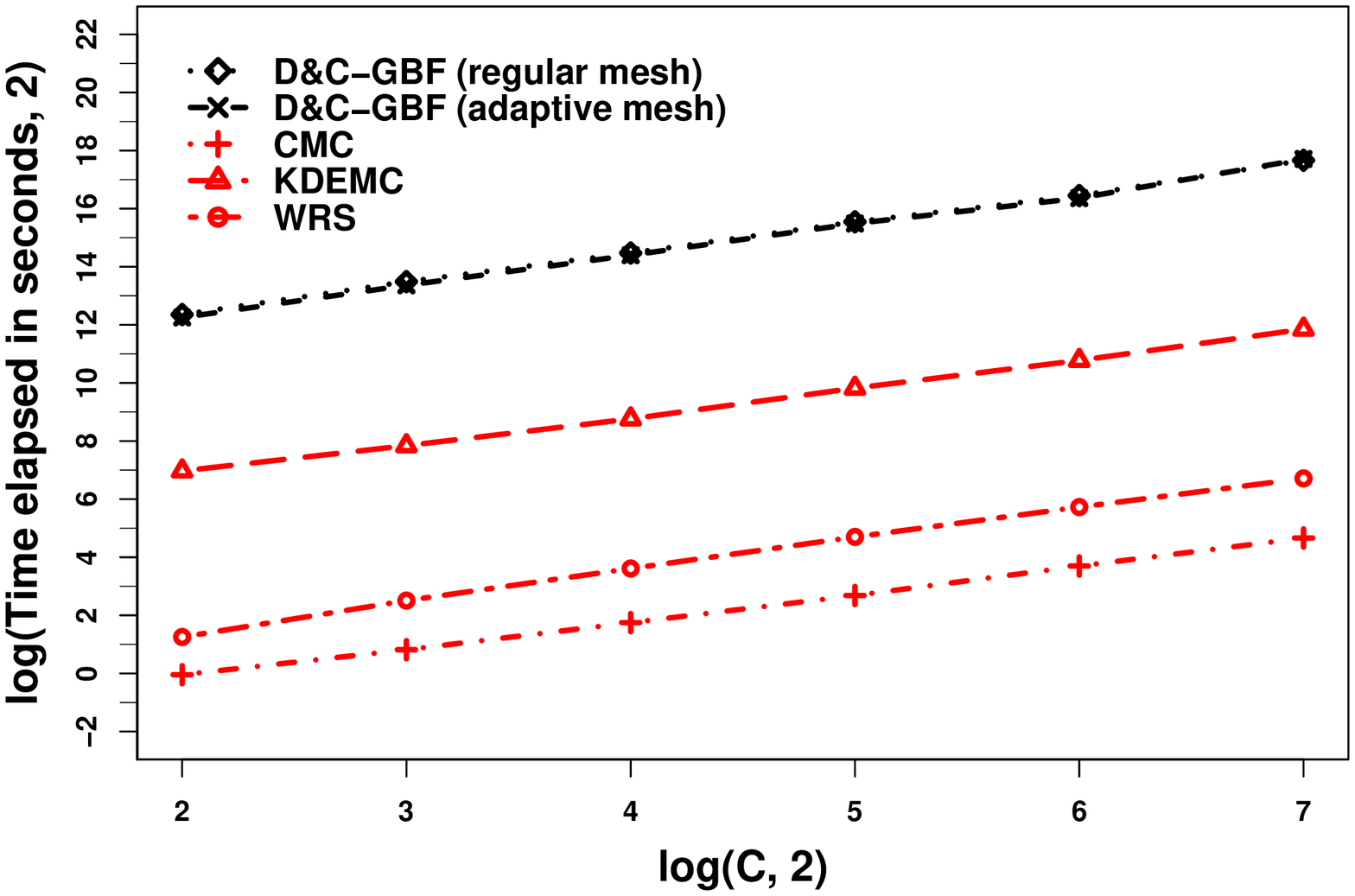}
        \caption{Computational cost.}
        \label{fig:log_reg_nyc_time}
    \end{subfigure}
    \caption{Comparison of competing methodologies to \hgbfa applied to a logistic regression model using the \texttt{nycflights13} dataset (see \secref{subsec:log_reg:nyc}).}
    \label{fig:log_reg_nyc}
\end{figure}

Finally, we study a logistic regression model \eqref{eq:log_reg} applied to the \texttt{nycflights13} dataset (obtained from the \texttt{nycflights13} \Rlang\, package available on CRAN \citep{Wickham_2021}). In this study we predict on-time arrival of airplanes, by creating binary observations for \textbf{arrival-delay} (taking the value $1$ if the flight arrived $1$ minute or more late, and $0$ otherwise). We model this using $p=20$ predictor variables (so $d=21$). After removing any entries with \texttt{NA} values, in total the dataset was of size $m=327346$. This dataset was split randomly across $C\in\{4,8,16,32,64,128\}$ cores, and we used \texttt{Stan} to find sample approximations of each sub-posterior (using Gaussian priors with mean $0$ and variance $C$ for each parameter). \hgbfa was implemented with $N=30000$, $\zeta=0.2$ and $\zeta^{\prime}=0.05$. The results are shown in \figref{fig:log_reg_nyc}.

As before, \hgbfa provides the best sample approximation and is robust to increasing $C$, but comes at the expense of increased computational cost. Although approximate methodologies have been specifically developed to tackle Bayesian big-data problems, here we see that they struggle to recover $f$ even in this idealised scenario. They additionally (and critically) lack robustness when scaled with $C$.

\begin{figure}
    \centering
    \includegraphics[width=0.5\textwidth]{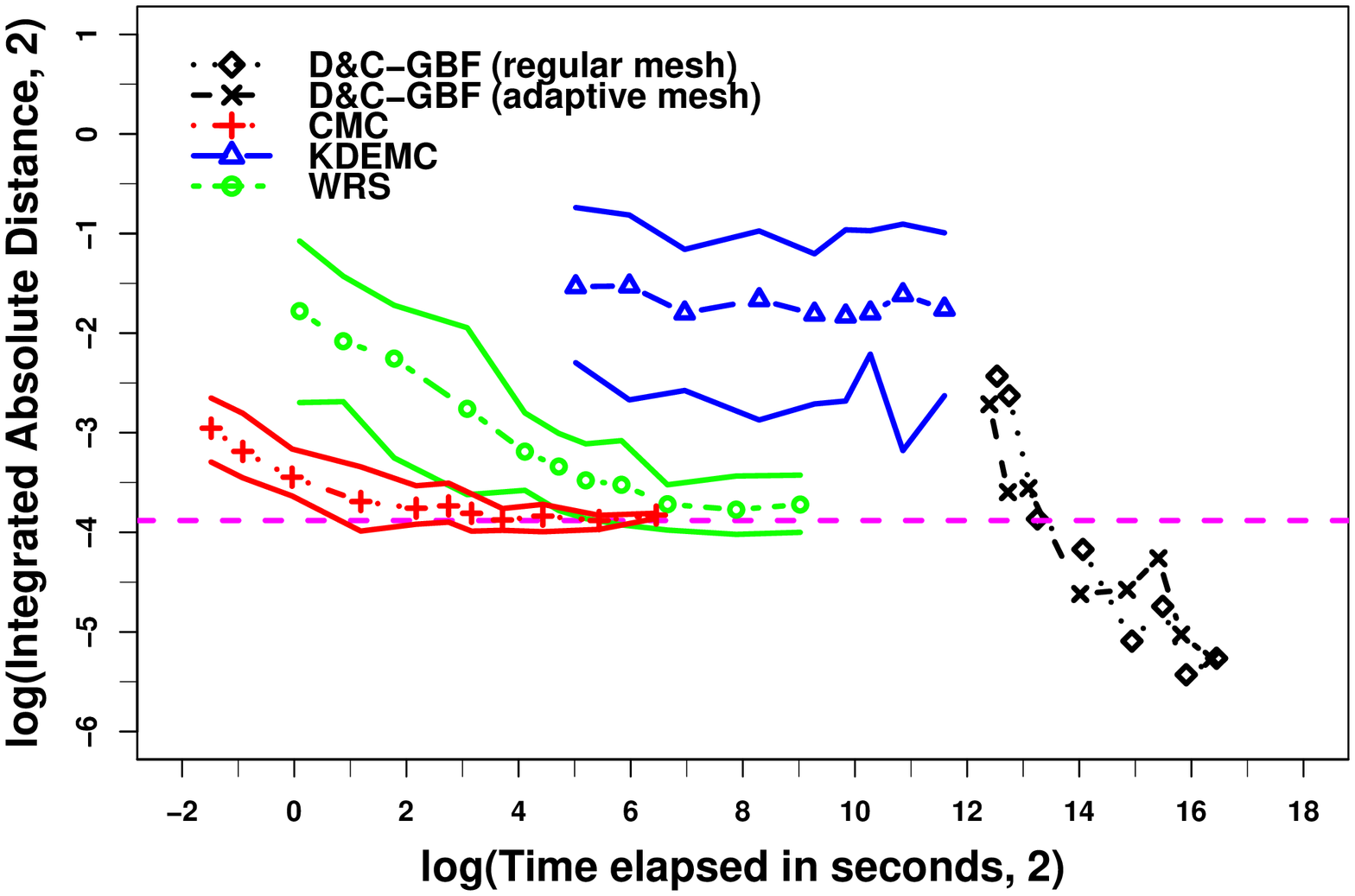}
    \caption{Integrated absolute distance against computational budget for competing methodologies to \hgbfa applied to a logistic regression model using the \texttt{nycflight} and fixing $C=64$ (see  \secref{subsec:log_reg:nyc}).}
    \label{fig:log_reg_nyc_varying_N}
\end{figure}

To further compare the methodologies, we consider fixing $C=64$ and varying the computational budget for each method by varying the sample size $N$ in order to study the effect of increased computation on IAD. We again compute the IAD against the same benchmark for the target $f$ that was used above (based upon $N=30000$ samples using \texttt{Stan}). Our results are shown in \figref{fig:log_reg_nyc_varying_N}. 

The IAD of the approximate methodologies considered in \figref{fig:log_reg_nyc_varying_N} had large variance, and so we run each of these methods 10 times and took an average of the IAD. To show the variability we also plot the minimum and maximum IAD achieved in the 10 runs. The longest run for each approximate methodology was one hour, or when it had become apparent that further computation was not improving IAD. As such, the CMC and KDEMC approaches were considered for a range of sample sizes from $N=500$ to $N=200000$, but KDEMC was only considered for $N=500$ to $N=50000$. For CMC and WRS, the average and variance of the IAD decreases with more computation, but both methods quickly reach a point where IAD no longer decreases. In  \figref{fig:log_reg_nyc_varying_N} we additionally plot a pink dashed line which is the minimum mean value IAD achieved for CMC, as this seems to the point which the IAD of CMC converges to. For KDEMC increased computation does not improve the average IAD or its variability. As such CMC is clearly the best of the approximate methods, with it achieving the lowest computational cost and lowest IAD.  For \hgbfa, we considered $N=500$ to $N=30000$. We did not perform replicate runs for \hgbfa due to the comparative lack of variability in results for this methodology. Of course, being an exact methodology Monte Carlo error can be further decreased by simply increasing $N$, but does achieve better IAD than CMC for its increased computational cost. 

As there is a reasonably large number of data points on each core ($m/64 \approx 5000$) the sub-posteriors are approximately Gaussian, and hence CMC performs unsurprisingly well. Taking into account accuracy and computational budget, then CMC performs the best out of all approximate methodologies here. We are however left with the same conclusion: if the practitioner values accuracy, or they have a poor understanding of the biases induced by an approximate approach, then our \hgbfa methodology should be used.


\section{Conclusion} \label{sec:conclusion}

The Fusion approach to unifying sub-posteriors into a coherent sample approximation of the posterior (as in \eqref{eq:fusion_density}), offers fundamental advantages over approximation based approaches. In particular, Fusion avoids imposing any distributional approximation on the sub-posteriors, and so is more robust to a wider range of models, and circumvents needing to understand the impact of imposed approximations on the unified posterior. To date, Fusion approaches have had impractical computational cost in realistic settings, lacking robustness when considering: the number of sub-posteriors being unified; when unifying highly correlated sub-posteriors; the dimensionality of the sub-posteriors; and when considering conflicting sub-posteriors. In this paper, we have substantially addressed the practical issues of Fusion approaches by means of several theoretical and methodological extensions.

In \secref{sec:GBF} we introduced \emph{\gbfl}, which is a sequential Monte Carlo algorithm that incorporates available global information for each sub-posterior in order to construct informative proposals. As shown in \secref{subsec:simulation_studies}, \gbfa addresses the lack of robustness when the sub-posteriors have strong correlation structure. By embedding \gbfa within the Divide-and-Conquer Sequential Monte Carlo (D\&C-SMC) framework \citep{Lindsten_et_al_2017,Kuntz_et_al_2021b} in \secref{sec:dc_gbf}, we introduced \emph{\hgbfl}, together with a number of \emph{tree hierarchies}, which allow the sub-posteriors to be combined in stages to recover $f$. By using the provided guidance for selecting the hyperparameters required for the \gbfa approach (and developed in \secref{sec:GBF_guidance}), we saw in \secref{subsec:GBF_dimension_study} that our \hgbfa approach was the most scalable Fusion approach to date with regards to dimension. In \secref{sec:dc_gbf_examples}, we applied our \hgbfa methodology to a variety of models with realistic data sets and compared its performance with competing approximate methodologies. In all of these settings, our implementation of \hgbfa offered the best performance in terms of Integrated Absolute Distance to an appropriate benchmark, at a modest computational cost. Furthermore, the examples in \secref{sec:dc_gbf_examples} showed that \hgbfa is a robust approach to unifying large numbers of sub-posteriors.

There are a number of interesting avenues for extending the work of this paper. Perhaps most interesting is to adapt the \hgbfa approach to constraints in practical settings. As discussed in the introduction, one particularly promising direction is when considering \eqref{eq:fusion_density} under \emph{privacy constraints} of the individual sources \citep{Yildrim_et_al_2019}. In this setting, we may have a number of parties that wish to combine their distributional analysis on a common parameter space and model but cannot reveal their distribution due to confidentiality. This of course requires careful modifications to our approach and is an active area of research of the authors, and motivates variant \emph{tree hierarchies} in \hgbfa. 

Another application is when considering a truly distributed \emph{`big data'} setting where we have much larger datasets than ones considered in \secref{sec:dc_gbf_examples}. In such settings, we may consider a large number of sub-posteriors $C$ since the computational benefit of parallelisation for a divide-and-conquer method is typically proportional to the number of available processors \citep{Nemeth_and_Sherlock_2018}. Although our divide-and-conquer approach is scalable with $C$, communication between different cores is expensive in a parallel setting \citep{Scott_et_al_2016}. We discussed several practical implementation considerations in \apxref{sec:GBF_guidance:communication}, which aim to limit the amount of communication between cores for \algoref{alg:GBF}, but a considered implementation of these techniques have yet to be explored. To make our Fusion methodology more applicable to large data settings, it would be particularly interesting to investigate embedding a sub-sampling approach within the Fusion algorithms (akin to the approaches of \citet{Pollock_et_al_2020, Bouchard_et_al_2018, Baker_et_al_2019, Bierkens_et_al_2019}). First steps in integrating sub-sampling into our \hgbfa are considered in \apxref{sec:GBF_guidance_unbiased_estimators}. We also note that there is a growing literature on implementing SMC approaches in parallel and distributed settings (see for instance \citet[Section 7.5.3]{Doucet_and_Lee_2018}) which may also be interesting to integrate within Fusion.

From a theoretical perspective, current Fusion methodologies only consider sub-posteriors on a common parameter space. One direction of interest is extending Fusion methodology to combine sub-posteriors with varying dimension. The \emph{Markov Melding} framework of \citet{Goudie_et_al_2019} where separate sub-models (potentially of differing dimension) are fitted to different data sources and then joined, is promising. In this setting, the \emph{tree hierarchies} could be defined by the model itself. To mitigate computational robustness of Fusion with increasing dimension in this setting, it may be possible to further utilise the methodology in \citet{Lindsten_et_al_2017}. 

\section*{Acknowledgements}

We would like to thank Louis Aslett, Hector McKimm, Krzysztof \L{}atuszy\'{n}ski, Nicolas Chopin and Hongsheng Dai for helpful discussions on aspects of the paper. This work was supported by the Engineering and Physical Sciences Research Council under grant numbers EP/K034154/1, EP/K014463/1, EP/N510129/1, EP/R034710/1, EP/R018561/1 and EP/T004134/1 and by The Alan Turing Institute Doctoral Studentship and two Alan Turing Institute programmes; the Lloyd’s Register Foundation programme on `Data-centric engineering' and the UK Government's `Defence and security' programme.

The data used in this paper are openly available: the code to reproduce the simulated data used for the simulation studies can be found at \href{https://github.com/rchan26/DCFusion}{\texttt{https://github.com/rchan26/DCFusion}}; the `Combined Cycle Power Plant' dataset (\secref{subsec:robust_reg}) can be accessed via the UCI Machine Learning Repository at \href{https://archive.ics.uci.edu/ml/datasets/Combined+Cycle+Power+Plant}{\texttt{https://archive.ics.uci.edu/ml/datasets/Combined+Cycle+Power+Plant}}; the `Bike Sharing' dataset (\secref{subsec:NB_reg}) was accessed via the UCI Machine Learning Repository at \href{https://archive.ics.uci.edu/ml/datasets/Combined+Cycle+Power+Plant}{\texttt{https://archive.ics.uci.edu/ml/datasets/bike+sharing+dataset}}; the \texttt{nycflights13} dataset (\secref{subsec:log_reg:nyc}) is available through the \texttt{nycflights13} \Rlang\, package on CRAN available at \href{https://github.com/tidyverse/nycflights13}{\texttt{https://github.com/tidyverse/nycflights13}}.

\appendix


\begin{appendix}

\section{Connections with Monte Carlo Fusion and Bayesian Fusion} \label{app:connections}

In this appendix, we more explicitly draw connections with the earlier \emph{Monte Carlo Fusion (MCF)} approach of \citet{Dai_et_al_2019}, and \emph{Bayesian Fusion (BF)} approach of \citet{dai_et_al_2023}. In particular, we outline how our \gbfl approach, which we develop in \secref{sec:GBF}, improves upon these approaches. We do so by considering several toy examples to illustrate the benefits of the algorithmic developments we have presented in this paper.

Firstly, the theory and methodology developed in \secref{sec:GBF} admits the Monte Carlo Fusion \citet{Dai_et_al_2019} and Bayesian Fusion \citep{dai_et_al_2023} approaches as a special case and is established in the following corollaries:

\begin{corollary}
    \label{corollary:MCF_equivalence}
    Setting $\mathcal{P}:=\{0,T\}$, $\mathbf{\Lambda}_{c} = \mathbb{I}_{d}$ for $c\in\mathcal{C}:=\{1,\dots,C\}$, where $\mathbb{I}_{d}$ is the identity matrix of dimension $d$ and accepting a proposal $\bm{y}^{(\mathcal{C})}$ as a sample from \eqref{eq:fusion_density} with probability $(\rho_{0}\cdot\tilde{\rho}_{1}^{(a)})(\vecX{}{(\mathcal{C})}, \bm{y}^{(\mathcal{C})})\cdot \exp\{\sum_{c\in\mathcal{C}} \mathbf{\Phi}_{c} T\}$, we recover the Monte Carlo Fusion approach of \citet[Algorithm 1]{Dai_et_al_2019}.
\end{corollary}

\begin{corollary}
    \label{corollary:BF_equivalence}
    Setting $\mathbf{\Lambda}_{c}=\mathbb{I}_{d}$ for $c\in\mathcal{C}:=\{1,\dots,C\}$, where $\mathbb{I}_{d}$ is the identity matrix of dimension $d$, and applying the approach outlined in \algoref{alg:GBF} recovers the Bayesian Fusion approach of \citet[Algorithm 1]{dai_et_al_2023}.
\end{corollary}

We note however that the MCF formulation to arrive at this algorithm is different and is based on the following proposition:
\begin{proposition} \label{prop:extended}
    Suppose that $p_{c}$ is the transition density of a Markov chain on $\mathbbm{R}^d$ with a stationary probability density proportional to $f_{c}^{2}$. Then
    the $(|\mathcal{C}|+1)d$-dimensional probability density proportional to the integrable function
    \begin{align}
        \extended{\mathcal{C}}^{MCF}\big(\vecX{}{(\mathcal{C})}, \bm{y}^{(\mathcal{C})}\big) 
        & := \prod_{c\in\mathcal{C}} \left[f_{c}^{2}\big(\bm{x}^{(c)}\big) \cdot p_{c}\big(\bm{y}^{(\mathcal{C})} \big| \bm{x}^{(c)}\big) \cdot \frac{1}{f_{c}(\bm{y}^{(\mathcal{C})})} \right], \label{eq:g}
    \end{align}
    admits marginal density $f^{(\mathcal{C})} \propto \prod_{c \in \mathcal{C}}f_c$ for $\bm{y}^{(\mathcal{C})}\in \mathbb{R}^{d}$.
\end{proposition}

\proof By integrating out $\vecX{}{(\mathcal{C})}$, we have
\begin{align}
    & \int_{\mathbb{R}_{d}} \cdots \int_{\mathbb{R}_{d}} \extended{\mathcal{C}}^{MCF}\left(\vecX{}{(\mathcal{C})}, \bm{y}^{(\mathcal{C})}\right) \ddspace{\bm{x}^{(c_{1})}} \cdots \ddspace{\bm{x}^{(c_{|\mathcal{C}|})}} \nonumber \\
    & \qquad = \prod_{c\in\mathcal{C}} \left[ \int_{\mathbb{R}_{d}} f_{c}^{2}\left(\bm{x}^{(1)}\right) \cdot p_{c}\left(\bm{y}^{(\mathcal{C})}\middle|\bm{x}^{(c)}\right) \cdot \frac{1}{f_{c}\left(\bm{y}^{(\mathcal{C})}\right)} \ddspace{\bm{x}^{(c)}} \right] \nonumber \\
    & \qquad = \prod_{c\in\mathcal{C}} \left[ \frac{f_{c}^{2}\left(\bm{y}^{(\mathcal{C})}\right)}{f_{c}\left(\bm{y}^{(\mathcal{C})}\right)} \right] \nonumber \\
    & \qquad = \prod_{c\in\mathcal{C}} f_{c}\left(\bm{y}^{(\mathcal{C})}\right) = f^{(\mathcal{C})}\left(\bm{y}^{(\mathcal{C})}\right).
\end{align} 
Hence, $\bm{y}^{(\mathcal{C})}$ has marginal density $f^{(\mathcal{C})}$. \hfill $\blacksquare$

\citet{Dai_et_al_2019} exploited \propositionref{prop:extended} by noting that if the index set $\mathcal{C}:=\{1,\dots,C\}$, then we recover the target \emph{fusion} density $f$ (as given in \eqref{eq:fusion_density}). Since $\extended{\mathcal{C}}^{MCF}$ will not typically be accessible directly, \citet{Dai_et_al_2019} proposed sampling from $\extended{\mathcal{C}}^{MCF}$ by constructing a suitable $(|\mathcal{C}|+1)d$-dimensional proposal density (say, $\proposal{\mathcal{C}}$) for use within a rejection sampling algorithm \citep[Algorithm 1]{Dai_et_al_2019}, and then simply retaining the $\bm{y}^{(\mathcal{C})}$ marginal of any accepted draw as a realisation of $f^{(\mathcal{C})}$. \citet{Dai_et_al_2019} showed that if $p_{c}$ in \propositionref{prop:extended} was chosen to be the transition density of a \emph{constant volatility Langevin diffusion} at time $T$ with invariant measure $f_{c}^2$ for each $c\in\mathcal{C}$ respectively, then for a (easily accessible) proposal $\proposal{\mathcal{C}}$ constructed by sampling a single draw from each sub-posterior ($\bm{x}^{(c)}\sim f_{c}$ for $c\in\mathcal{C}$), and then a single Gaussian random variable parameterised by the sub-posterior realisations (corresponding to the $\bm{y}^{(\mathcal{C})}$-marginal), the acceptance probability was readily computable. Although the formulation of the approach was different, this corresponds algorithmically to a rejection sampling variant of \gbfa and setting $\mathbf{\Lambda}_{c}=\mathbb{I}_{d}$ for all $c\in\mathcal{C}$ with $\mathcal{P}:=\{0,T\}$.

The advantage of the BF and \gbfa approaches is that our formulation allows for a general temporal mesh $\mathcal{P}$. We have seen in \secref{sec:GBF_guidance} and \secref{subsec:simulation_studies} that the choice of $T$ and $\mathcal{P}$ can drastically alter the performance of the algorithm and so a clear advantage of BF and \gbfa over MCF is having a greater flexibility in hyperparameter selection. By being restricted to choosing $\mathcal{P}:=\{0,T\}$, in many cases, it will not be possible to choose a $T$ which is large enough for initialisation of the algorithm (i.e.\ to ensure $\CESS{0}$ is large) \emph{and} to choose $T$ small enough for $\CESS{1}$ to be sufficently large. This trade-off in choice of $T$ in MCF is ultimately why it can fail in many practical settings. 

As discussed in the introduction, the existing MCF and BF approaches lack robustness in various key practical settings. For the remainder of this section, we re-visit these settings, and with the aid of illustrative examples, show that our new approach addresses these key bottlenecks. Since many of the key limitations are present with both approaches, we focus on comparing against \emph{MCF} approach by setting $\mathcal{P}:=\{0,T\}$ in \algoref{alg:GBF}. We call this variant of \gbfa with $\mathcal{P}:=\{0,T\}$ \emph{\gmcfl}. In particular, in \secref{subsec:varying_corr_example} we consider the effect of increasing sub-posterior correlation, in \secref{subsec:varying_C_example} we consider the robustness with increasing numbers of sub-posteriors, and in \secref{subsec:conflicting_subposteriors} we consider how to address conflicting sub-posteriors. Throughout this section we use the GPE-2 estimator of $\rho_{j}$ as given in \defnref{cond:GPE2}, and use the Trapezoidal rule to estimate the mean $\gamma_{c}$ in \eqref{eq:NB_mean} and fix $\beta_{c}=10$ for $c\in\mathcal{C}$. To compare the methodology, we compute both the computational run-times of each methodology and a metric which we term the \emph{Integrated Absolute Distance} (IAD) \eqref{eq:IAD}.
 
\subsection{Effect of correlation} \label{subsec:varying_corr_example}

\begin{figure}[ht]
    \centering
    \includegraphics[width=0.5\textwidth]{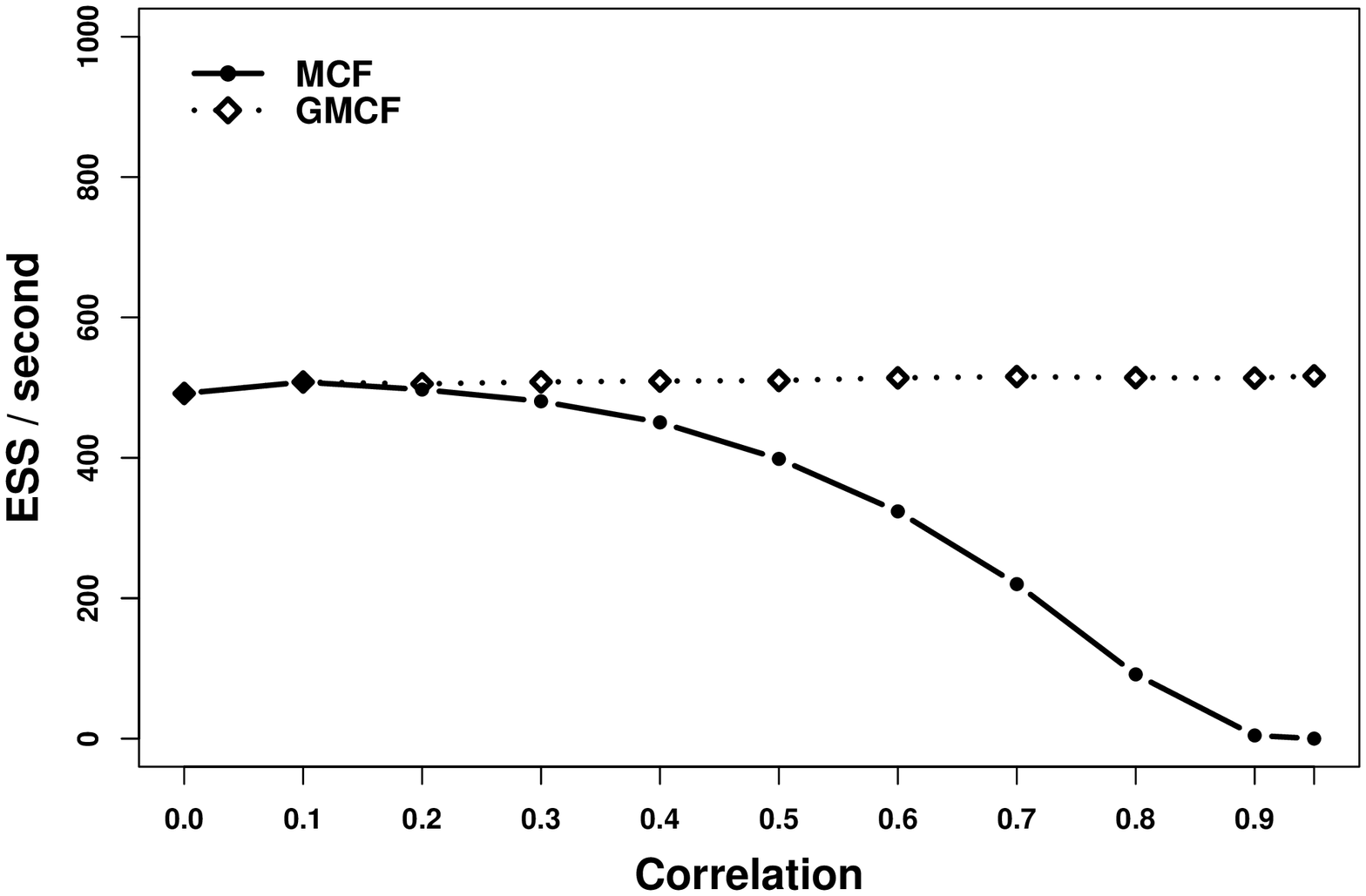}
    \caption{$\ESS$ per second (averaged over $50$ runs) when contrasting Monte Carlo Fusion and \gmcf, along with increasing sub-posterior correlation, as per the example in  \secref{subsec:varying_corr_example}.}
    \label{fig:bivariate_Gaussian_ESS_per_sec}
\end{figure}

One of our key contributions in this paper was the generalisation of BF which incorporated covariance information of the sub-posteriors within our algorithm. In this example, we focus on the illustrative case in which we wish to recover a bi-variate Gaussian target distribution, $f \propto f_{1}f_{2}$, where $f_{c} \sim \mathcal{N}_{2}(\bm{0}, \mathbf{\Sigma})$ with,
\begin{equation*}
    \mathbf{\Sigma} = 
    \begin{pmatrix} 
        1.0 & \rho_{\text{corr}} \\
        \rho_{\text{corr}} & 1.0
    \end{pmatrix}.
\end{equation*}
As we are only considering combining two sub-posteriors in this section, we in effect consider only the \gmcfa approach. To study the impact of sub-posterior correlation on the robustness of MCF and \gmcfa (Algorithm \ref{alg:GBF} with $\mathcal{P}:=\{0,T\}$) we can simply consider varying the single parameter $\rho_{\text{corr}}$, and compute the \emph{Effective Sample Size} (ESS) per second averaged across $50$ runs in order to compare the efficiency of each methodology. For simplicity, we assume we are able to sample directly from each sub-posterior, and for both methodologies we set $T=1$. For the purposes of implementing \gmcfa, we simply set $\mathbf{\Lambda}_{c} = \hat{\mathbf{\Sigma}}_{c}$, where $\hat{\mathbf{\Sigma}}_{c}$ is the \emph{estimated} covariance matrix from the sub-posterior samples for $c=1,2$ (and so in effect we have incorporated global information into our proposals), and use a particle set size of $N=10000$. The results are presented in \figref{fig:bivariate_Gaussian_ESS_per_sec}, which clearly show that \gmcfa is robust to increasing sub-posterior correlation, and offers a significant computational advantage over MCF (which in this case exhibits a strong degradation in efficiency and performance).

\subsection{Effect of hierarchy} \label{subsec:varying_C_example}

In our new formulation outlined in \secref{sec:GBF}, we consider the more abstract setting of sampling from $f^{(\mathcal{C})} \propto \prod_{c\in\mathcal{C}} f_{c}$, where $\mathcal{C}$ is the index set of sub-posteriors which we want to unify. This abstraction  of combining facilitates the recursive use of the algorithms to develop our \hgbfl approach in \secref{sec:dc_gbf} - benefits of which are highlighted clearly in \secref{subsec:log_reg:sd_dcgbf} (Here \hgbfa outperforms \gbfa as we increase $C$).

\begin{figure}[ht]
     \centering
     \begin{subfigure}[b]{0.48\textwidth}
         \centering
         \includegraphics[width=\textwidth]{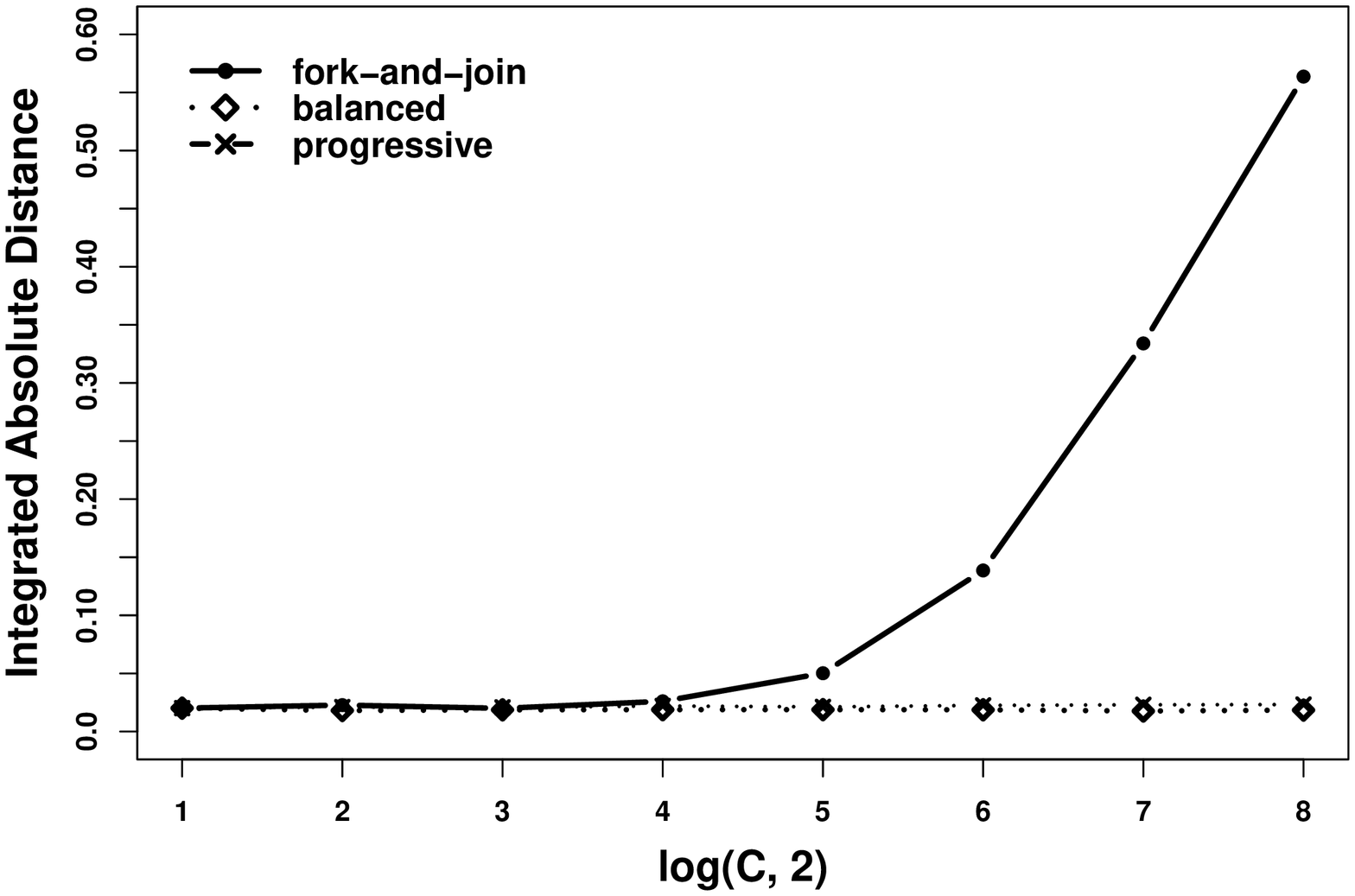}
         \caption{Integrated Absolute Distance.}
    \label{fig:varying_C_IAD}
     \end{subfigure}
     \hfill
     \begin{subfigure}[b]{0.48\textwidth}
         \centering
         \includegraphics[width=\textwidth]{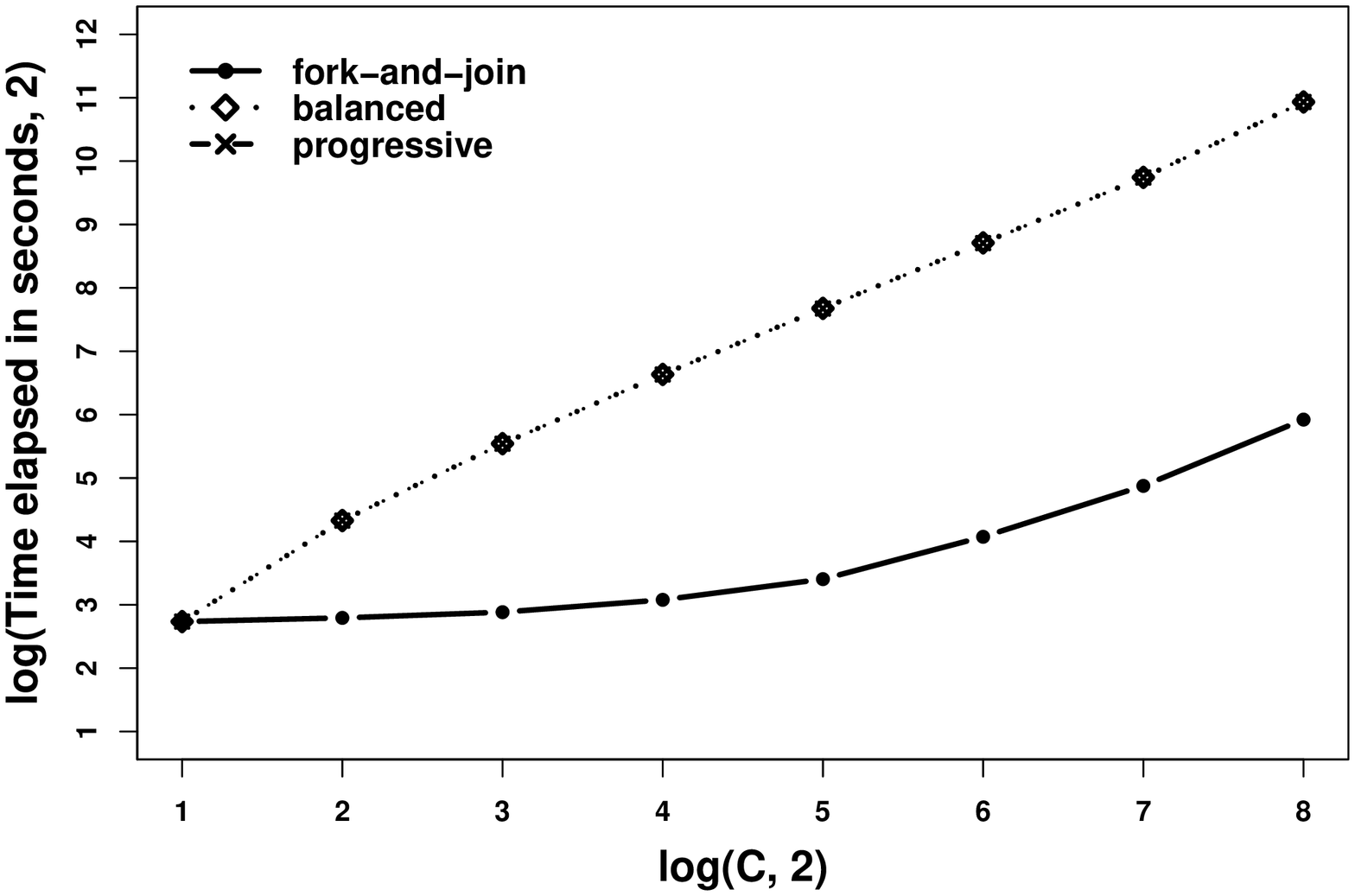}
         \caption{Computational cost.}
        \label{fig:varying_C_time}
     \end{subfigure}
    \caption{Illustrative comparison of the effect of using different hierarchies in \secref{subsec:varying_C_example} (averaged over $50$ runs).}
    \label{fig:varying_C}
\end{figure}

In this example, we consider the illustrative case of attempting to recover a univariate standard Gaussian target distribution. In particular, we have $f \propto \prod^C_{c=1} f_c$, where $f_{c} \sim \mathcal{N}(0, C)$ for $c=1,\dots,C$. By simply varying $C$, we can study the robustness with increasing numbers of sub-posteriors of MCF (in effect the fork-and-join approach illustrated in \figref{fig:fork_and_join}), and both our suggested versions of \hmcf (the balanced-binary tree approach illustrated in \figref{fig:balanced_binary}, and the progressive tree approach illustrated in \figref{fig:progressive}). Note that in our chosen idealised setting, there is no advantage conferred with our embedded \gmcf methodology of \secref{sec:GBF}, and so we are simply contrasting hierarchies. In all cases we use a particle set of size $N=10000$ with resampling if $\ESS < N/2$, set $T=1$, use an appropriately scaled identity as the preconditioning (scalar) matrix, and average across $50$ runs. The results are presented in \figref{fig:varying_C}, which clearly show that, in contrast to the fork-and-join tree approach, both the balanced-binary tree and progressive tree approaches are robust in recovering the correct posterior distribution in the case of increasing $C$ at the cost of modestly increased computational cost.

\subsection{Dealing with conflicting sub-posteriors} \label{subsec:conflicting_subposteriors}

Directly unifying $C$ \emph{conflicting} sub-posteriors (sub-posteriors which have little common support and have high total-variation distance) using a fork-and-join approach as in MCF and \figref{fig:fork_and_join} is impractical. This can be understood with reference to \eqref{eq:rho_0_gbf} and \eqref{eq:rho_j_gbf}, which indicates that importance weights will degrade rapidly in this setting. 

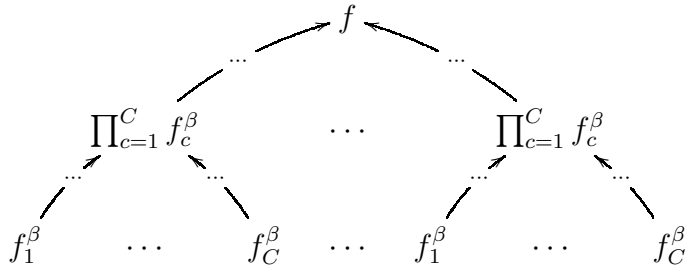
\begin{figure}[ht]
     \centering
	    {\makebox[\textwidth]{\xymatrix@C=1em@R=2em{
	        & & & f & & & \\
            & \prod^{C}_{c=1} f_c^{\beta} \ar@/^/[urr]|{\phantom{f}\cdots\phantom{f}}  & & \cdots & &   \prod^{C}_{c=1} f_c^{\beta} \ar@/_/[ull]|{\phantom{f}\cdots\phantom{f}}  & \\
            f_1^{\beta} \ar@/^/[ur]|{\phantom{f}\cdots\phantom{f}} & \cdots & f_C^{\beta} \ar@/_/[ul]|{\phantom{f}\cdots\phantom{f}} & \cdots & f_1^{\beta} \ar@/^/[ur]|{\phantom{f}\cdots\phantom{f}} & \cdots & f_C^{\beta} \ar@/_/[ul]|{\phantom{f}\cdots\phantom{f}} \\
        }}}
        \caption{Illustrative tree approach for the Fusion problem in the case of conflicting sub-posteriors as in \secref{subsec:conflicting_subposteriors}. $1/\beta$ copies of the $C$ tempered (and over-lapping) sub-posteriors represent the leaves of the tree, which are unified into $1/\beta$ tempered versions of $f$ (using a suitable tree and \hmcfa as in \secref{sec:dc_gbf}), and then unified again (using another tree, and \hmcfa) to recover $f$.} 
    \label{fig:tempertree}
\end{figure}

An approach to deal with conflicting sub-posteriors is to temper the sub-posteriors (to an inverse temperature $\beta \in (0,1]$ such that there is sufficient sub-posterior overlap), and then propose a suitable tree for which the recursive \hmcf approach we introduced in \secref{sec:dc_gbf} could then be applied to recover \eqref{eq:fusion_density}. In particular,
\begin{equation}
    f(\bm{x}) \propto  \prod_{i=1}^{1/\beta} \left[ \prod_{c=1}^{C} f_{c}^{\beta}(\bm{x}) \right], \qquad \text{for } \frac{1}{\beta} \in \mathbb{N}.
\end{equation}
One such generic tree is provided in \figref{fig:tempertree}, in which the tempered sub-posteriors are first unified into $1/\beta \in \mathbb{N}$ tempered posteriors, which are then again unified into $f$.

\begin{figure}[ht]
     \centering
     \begin{subfigure}[b]{0.48\textwidth}
         \centering
         \includegraphics[width=\textwidth]{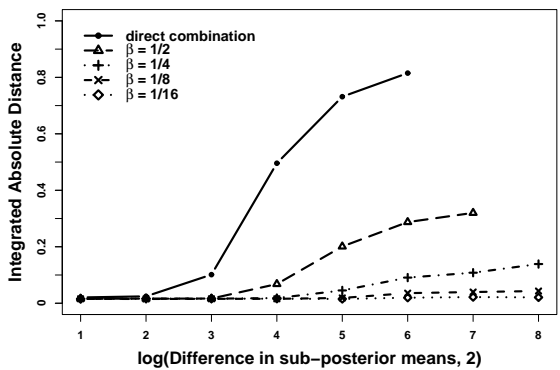}
         \caption{Integrated Absolute Distance.}
    \label{fig:separate_modes_IAD}
     \end{subfigure}
     \hfill
     \begin{subfigure}[b]{0.48\textwidth}
         \centering
         \includegraphics[width=\textwidth]{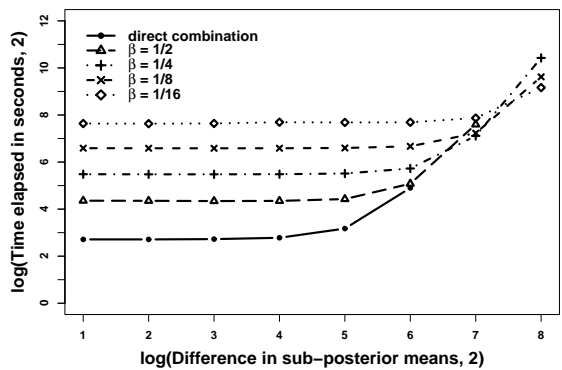}
         \caption{Computational cost.}
        \label{fig:separate_modes_time}
     \end{subfigure}
    \caption{Illustrative comparison of using no tempering (solid line), and tempering at $4$ different levels together with \hmcfa, to combat conflicting sub-posteriors as per \secref{subsec:conflicting_subposteriors} (averaged over $50$ runs).  }
    \label{fig:seprate_modes}
\end{figure}

To illustrate the advantage of our \hmcfa and tempering approach in the case of conflicting sub-posteriors, we consider the scenario of unifying two Gaussian sub-posteriors with the same variance ($1$), but with different mean ($\pm \mu$). In particular, we have $f \propto f_{1}f_{2}$ where $f_{1} \sim \mathcal{N}(-\mu, 1)$ and $f_{2} \sim \mathcal{N}(\mu, 1)$. By simply increasing $\mu$ we can emulate increasingly conflicting sub-posteriors and study how MCF (which is equivalent to the fork-and-join approach of \figref{fig:fork_and_join}), behaves in terms of the IAD metric and computational time. We contrast this with our tempering approach, considering a range of temperatures $1/\beta\in\{2,4,8,16\}$, and then following the guidance of \figref{fig:tempertree}. In particular, we use our \hmcfa approach to unify the tempered sub-posteriors with the balanced-binary approach of \figref{fig:balanced_binary} for both the first and second stage in \figref{fig:tempertree}. In all cases, we use a particle set size of $N=10000$ with resampling if $\text{ESS}<N/2$, set $T=1$, and average across $50$ runs. The results are presented in \figref{fig:seprate_modes}, and show clearly that our \hmcfa approach is significantly more robust to conflicting sub-posteriors than the MCF approach where no tempering is applied. A natural trade-off arises when applying the tempering approach suggested, in that decreasing $\beta$ results in tempered sub-posteriors which are less conflicting and are easier to combine, but there is an increased computational cost in recovering $f$ as an increased number of levels are added to the resulting tree.

\section{Proof of \texorpdfstring{\thmref{theorem:fusion_measure_GBF}}{Theorem 1}} \label{app:fusion_measure_GBF}

\proof Following the approach of \citet[Appendix A]{Dai_et_al_2019}, we begin by proving that the law of $|\mathcal{C}|$ independent Brownian motions initialised at $\bm{x}_{0}^{(c)} \sim f_{c}$ for $c\in\mathcal{C}$ and conditioned to coalesce at time $T$ satisfies \eqref{eq:proposal_SDE}. Here, we use Doob $h$-transforms \citep[Chapter IV, Section 6.39]{Rogers_and_Williams_2000} and define the following space-time harmonic function
\begin{equation}
    h\left(t, \vecX{t}{(\mathcal{C})}\right) = \int \prod_{c\in\mathcal{C}} \frac{1}{\sqrt{2\pi(T-t)\abs{\mathbf{\Lambda}_{c}}}} \exp\left\{-\frac{(\bm{y}-\bm{x}_{t}^{(c)})^{\intercal}\mathbf{\Lambda}_{c}^{-1}(\bm{y}-\bm{x}_{t}^{(c)})}{2(T-t)}\right\} \ddspace{\bm{y}},
\end{equation}
which represents the integrated density of coalescence at time $T$ given the current state $\vecX{t}{(\mathcal{C})}$. Then the $|\mathcal{C}|$ conditioned processes satisfy a SDE of the form,
\begin{equation}
    \dd \vecXX{t}{(\mathcal{C})} = \vec{\mathbf{\Lambda}}^{\frac{1}{2}} \ddspace{\vec{\bm{W}}_{t}^{(\mathcal{C})}} + \vec{\mathbf{\Lambda}} \nabla \log \left(h(t, \vecXX{t}{(\mathcal{C})})\right) \ddspace{t},
\end{equation}
where $\nabla \log(h(t, \vecX{t}{(\mathcal{C})})) =: \left(\bm{v}_{t}^{(c_{1})}, \dots, \bm{v}_{t}^{(c_{|\mathcal{C}|})}\right)$ is a collection of $|\mathcal{C}|d$-dimensional vectors and
\begin{equation*}
    \vec{\mathbf{\Lambda}}^{\frac{1}{2}} =
    \begin{pmatrix}
    \mathbf{\Lambda}_{c_{1}}^{\frac{1}{2}} & \bm{0}_{d \times d} & \dots & \bm{0}_{d \times d} \\
    \bm{0}_{d \times d} & \mathbf{\Lambda}_{c_{2}}^{\frac{1}{2}} & \dots & \bm{0}_{d \times d} \\
    \vdots & \ddots & \vdots & \vdots \\
    \bm{0}_{d \times d} & \bm{0}_{d \times d} & \dots & \mathbf{\Lambda}_{c_{|\mathcal{C}|}}^{\frac{1}{2}}
    \end{pmatrix}, \qquad
    \vec{\mathbf{\Lambda}} =
    \begin{pmatrix}
    \mathbf{\Lambda}_{c_{1}} & \bm{0}_{d \times d} & \dots & \bm{0}_{d \times d} \\
    \bm{0}_{d \times d} & \mathbf{\Lambda}_{c_{2}} & \dots & \bm{0}_{d \times d} \\
    \vdots & \ddots & \vdots & \vdots \\
    \bm{0}_{d \times d} & \bm{0}_{d \times d} & \dots & \mathbf{\Lambda}_{c_{|\mathcal{C}|}}
    \end{pmatrix},
\end{equation*}
where $\mathbf{\Lambda}_{c}^{\frac{1}{2}}$ is the (positive semi-definite) square root of $\mathbf{\Lambda}_{c}$ where $\mathbf{\Lambda}_{c}^{\frac{1}{2}}\mathbf{\Lambda}_{c}^{\frac{1}{2}}=\mathbf{\Lambda}_{c}$ for $c\in\mathcal{C}$, and $\bm{0}_{d \times d}$ denotes the $d \times d$ matrix with all elements equal to $0$.

Considering the $c$th term and letting $\mathbf{\Lambda}_{\mathcal{C}}^{-1}=\sum_{c\in\mathcal{C}}\mathbf{\Lambda}_{c}^{-1}$, then
\begin{align*}
    \bm{v}_{t}^{(c)} 
    & = \frac{\int \left(\frac{\mathbf{\Lambda}_{c}^{-1}(\bm{y}-\bm{x}_{t}^{(c)})}{T-t}\right) \prod_{c\in\mathcal{C}} \frac{1}{\sqrt{2\pi(T-t)\abs{\mathbf{\Lambda}_{c}}}} \exp\left\{-\frac{\left(\bm{y}-\bm{x}_{t}^{(c)}\right)^{\intercal}\mathbf{\Lambda}_{c}^{-1}\left(\bm{y}-\bm{x}_{t}^{(c)}\right)}{2(T-t)}\right\} \dd \bm{y}}{\int \prod_{c\in\mathcal{C}} \frac{1}{\sqrt{2\pi(T-t)\abs{\mathbf{\Lambda}_{c}}}} \exp\left\{-\frac{\left(\bm{y}-\bm{x}_{t}^{(c)}\right)^{\intercal}\mathbf{\Lambda}_{c}^{-1}\left(\bm{y}-\bm{x}_{t}^{(c)}\right)}{2(T-t)}\right\} \dd \bm{y}} \nonumber \\
    & = \frac{\int \left(\frac{\mathbf{\Lambda}_{c}^{-1}\bm{y}}{T-t}\right) \exp\left\{-\frac{(\bm{y}-\tilde{\bm{x}}_{t})^{\intercal}\mathbf{\Lambda}_{\mathcal{C}}^{-1}(\bm{y}-\tilde{\bm{x}}_{t})}{2(T-t)}\right\} \dd \bm{y}}{\int \exp\left\{-\frac{(\bm{y}-\tilde{\bm{x}}_{t})^{\intercal}\mathbf{\Lambda}_{\mathcal{C}}^{-1}(\bm{y}-\tilde{\bm{x}}_{t})}{2(T-t)}\right\} \dd \bm{y}} - \frac{\mathbf{\Lambda}_{c}^{-1}\bm{x}_{t}^{(c)}}{T-t} \nonumber \\
    & = \frac{\mathbf{\Lambda}_{c}^{-1}\left(\tilde{\bm{x}}_{t}-\bm{x}_{t}^{(c)}\right)}{T-t}.
\end{align*}
Consequently, we have
\begin{equation}
    \nabla \log \left(h(t, \vecX{t}{(\mathcal{C})})\right) = \left( \frac{\mathbf{\Lambda}_{c_{1}}^{-1}\left(\tilde{\bm{x}}_{t}-\bm{x}_{t}^{(c_{1})}\right)}{T-t}, \dots, \frac{\mathbf{\Lambda}_{c_{|\mathcal{C}|}}^{-1}\left(\tilde{\bm{x}}_{t}-\bm{x}_{t}^{(c_{|\mathcal{C}|})}\right)}{T-t} \right),
\end{equation}
and \eqref{eq:proposal_SDE} holds.

Next, we show that under $\mathbb{F}$ this common value has density $f$. Since $\mathbb{P}$ is the measure for $|\mathcal{C}|$ coalesced Brownian motions (shown above), from \eqref{eq:fusion_measure}, we can write $\mathbb{F}$ as
\begin{align}
    \dd\mathbb{F}(\tX)
    & \propto \dd\mathbb{P}(\tX) \cdot \rho_{0}\left(\vecX{0}{(\mathcal{C})}\right) \cdot \prod_{c\in\mathcal{C}} \left[ \exp \left\{ - \int_{0}^{T} \phi_{c} \left(\bm{X}_{t}^{(c)}\right) \dd t \right\} \right] \nonumber \\
    & \propto \left[ \prod_{c\in\mathcal{C}} f_{c}\left(\bm{x}_{0}^{(c)}\right) \right] \cdot \exp \left\{ - \frac{(\bm{y}^{(\mathcal{C})}-\tilde{\bm{x}}_{0}^{(\mathcal{C})})^{\intercal}\mathbf{\Lambda}_{\mathcal{C}}^{-1}(\bm{y}^{(\mathcal{C})}-\tilde{\bm{x}}_{0}^{(\mathcal{C})})}{2T} \right\} \cdot \dd\bar{\mathbb{W}}_{\mathbf{\Lambda}}(\tX) \nonumber \\
    & \qquad \cdot \exp \left\{ - \sum_{c\in\mathcal{C}} \frac{(\tilde{\bm{x}}_{0}^{(\mathcal{C})}-\bm{x}_{0}^{(c)})^{\intercal} \mathbf{\Lambda}_{c}^{-1} (\tilde{\bm{x}}_{0}^{(\mathcal{C})}-\bm{x}_{0}^{(c)})}{2T} \right\} \cdot \prod_{c\in\mathcal{C}} \left[ \exp \left\{ - \int_{0}^{T} \phi_{c} \left(\bm{X}_{t}^{(c)}\right) \dd t \right\} \right] \nonumber \\
    & = \left[ \prod_{c\in\mathcal{C}} f_{c}\left(\bm{x}_{0}^{(c)}\right) \right] \cdot \exp \left\{ - \sum_{c\in\mathcal{C}} \frac{(\bm{y}^{(\mathcal{C})}-\bm{x}_{0}^{(c)})^{\intercal} \mathbf{\Lambda}_{c}^{-1} (\bm{y}^{(\mathcal{C})}-\bm{x}_{0}^{(c)})}{2T} \right\} \cdot \dd\bar{\mathbb{W}}_{\mathbf{\Lambda}}(\tX) \nonumber \\
    & \qquad \cdot \prod_{c\in\mathcal{C}} \left[ \exp \left\{ - \int_{0}^{T} \phi_{c} \left(\bm{X}_{t}^{(c)}\right) \dd t \right\} \right],
\end{align}
where $\bar{\mathbb{W}}_{\mathbf{\Lambda}}$ denotes the law of $|\mathcal{C}|$ independent Brownian bridges $\{\bm{X}_{t}^{(c)},t\in[0,T]\}_{c\in\mathcal{C}}$ starting at $\bm{X}_{0}^{(c)}:=\bm{x}_{0}^{(c)}$ and ending at $\bm{X}_{T}^{(c)}:=\bm{y}^{(\mathcal{C})}$ (with covariance $\mathbf{\Lambda}_{c}$). Let $\extended{\mathcal{C}}(\vecX{0}{(\mathcal{C})},\bm{y}^{(\mathcal{C})})$ denote the marginal distribution of $\mathbb{F}$ at $\vecX{0}{(\mathcal{C})}$ and $\vecX{T}{(\mathcal{C})}=:\bm{y}^{(\mathcal{C})}$, then we have
\begin{align}
    \extended{\mathcal{C}}(\vecX{0}{(\mathcal{C})},\bm{y}^{(\mathcal{C})})
    & \propto \prod_{c\in\mathcal{C}} \left[ f_{c}\left(\bm{x}_{0}^{(c)}\right) \right] \cdot \exp \left\{ - \sum_{c\in\mathcal{C}} \frac{(\bm{y}^{(\mathcal{C})}-\bm{x}_{0}^{(c)})^{\intercal} \mathbf{\Lambda}_{c}^{-1} (\bm{y}^{(\mathcal{C})}-\bm{x}_{0}^{(c)})}{2T} \right\} \nonumber \\
    & \qquad \cdot \prod_{c\in\mathcal{C}} \left[ \exp \left\{ - \int_{0}^{T} \phi_{c} \left(\bm{X}_{t}^{(c)}\right) \dd t \right\} \right], \nonumber \\
    & = \prod_{c\in\mathcal{C}} \left[f_{c}^{2}\left(\bm{x}_{0}^{(c)}\right) \cdot p_{c}\left(\bm{y}^{(\mathcal{C})} \middle| \bm{x}_{0}^{(c)}\right) \cdot \frac{1}{f_{c}\left(\bm{y}^{(\mathcal{C})}\right)} \right], \label{eq:g_extended}
\end{align}
where
\begin{align}
    p_{c} \left( \bm{y}^{(\mathcal{C})} \middle| \bm{x}_{0}^{(c)} \right) & \propto \frac{f_{c}\left(\bm{y}^{(\mathcal{C})}\right)}{f_{c} \left(\bm{x}_{0}^{(c)}\right)} \cdot \exp \left\{ -\frac{(\bm{y}^{(\mathcal{C})}-\bm{x}_{0}^{(c)})^{\intercal} \mathbf{\Lambda}_{c}^{-1} (\bm{y}^{(\mathcal{C})}-\bm{x}_{0}^{(c)})}{2T} \right\} \nonumber \\
    & \qquad \cdot \mathbb{E}_{\mathbb{W}_{\mathbf{\Lambda}_{c}}} \left[ \exp \left\{ - \int_{0}^{T} \phi_{c} \left(\bm{X}_{t}^{(c)}\right) \dd t \right\} \right].
\end{align}
Using the Dacunha-Castelle representation \citep[Lemma 1]{Dacunha_et_al_1986}, this is the transition density density of a Langevin diffusion with covariance matrix $\mathbf{\Lambda}_{c}$ over time $t\in[0,T]$. Critically, this diffusion process has invariant density proportional to $f_{c}^{2}$, so
\begin{equation*}
    \int p\left(\bm{y}^{(\mathcal{C})}\middle|\bm{x}_{0}^{(c)}\right) f_{c}^{2}\left(\bm{x}_{0}^{(c)}\right) \ddspace{\bm{x}_{0}^{(c)}} = f_{c}^{2}\left(\bm{y}^{(\mathcal{C})}\right).
\end{equation*}
By integrating out $\vecX{0}{(\mathcal{C})}$ in \eqref{eq:g_extended}, we can see that $\extended{\mathcal{C}}(\vecX{0}{(\mathcal{C})},\bm{y}^{(\mathcal{C})})$ admits $f^{(\mathcal{C})}$ as a marginal. \hfill $\blacksquare$

\section{Proof of \texorpdfstring{\propositionref{prop:proposal_simulation_GBF}}{Theorem 2}} \label{app:proposal_simulation_GBF}

\proof For part \ref{prop:proposal_simulation_GBF:a}, we begin by deriving the joint density of $\vecXX{t}{(\mathcal{C})}$ conditional on the state at time $s$, $\vecX{s}{(\mathcal{C})}$. Firstly, consider the $d(|\mathcal{C}|+1)$ dimensional joint density of $\vecXX{t}{(\mathcal{C})}$ \emph{and} end-point $\bm{y}^{(\mathcal{C})}$ conditional on $\vecX{s}{(\mathcal{C})}$, which we denote as $p_{1}$, then
\begin{equation*}
    -2 \log p_{1} = D_{1} + D_{2},
\end{equation*}
where $D_{1}$ is the log-density of $\bm{y}^{(\mathcal{C})}$ conditional on $\vecX{s}{(\mathcal{C})}$ and given by
\begin{equation*}
    D_{1} = \sum_{c\in\mathcal{C}} \frac{(\bm{y}^{(\mathcal{C})}-\bm{x}_{s}^{(c)})^{\intercal}\mathbf{\Lambda}_{c}^{-1}(\bm{y}^{(\mathcal{C})}-\bm{x}_{s}^{(c)})}{T-s} + k_{1}
\end{equation*}
where $k_{1}$ is a constant; $D_{2}$ is the log-density of $\vecXX{t}{(\mathcal{C})}$ conditional on $\vecX{s}{(\mathcal{C})}$ and $\bm{y}^{(\mathcal{C})}$ (which is simply the log-density of $|\mathcal{C}|$ Brownian bridges with respective covariance matrices $\mathbf{\Lambda}_{c}$ for $c\in\mathcal{C}$), given by
\begin{equation*}
    D_{2} = \sum_{c\in\mathcal{C}} \frac{T-s}{(t-s)(T-t)} \left[\bm{x}_{t}^{(c)} - \frac{t-s}{T-s}\bm{y}^{(\mathcal{C})} - \frac{T-t}{T-s}\bm{x}_{s}^{(c)}\right]^{\intercal}\mathbf{\Lambda}_{c}^{-1}\left[\bm{x}_{t}^{(c)} - \frac{t-s}{T-s}\bm{y}^{(\mathcal{C})} - \frac{T-t}{T-s}\bm{x}_{s}^{(c)}\right] + k_{2},
\end{equation*}
where $k_{2}$ is a constant. We therefore have
\begin{align*}
    -2 \log p_{1}
    & = \frac{(\bm{y}^{(\mathcal{C})}-\tilde{\bm{x}}_{s}^{(\mathcal{C})})^{\intercal}\mathbf{\Lambda}_{\mathcal{C}}^{-1}(\bm{y}^{(\mathcal{C})}-\tilde{\bm{x}}_{s}^{(\mathcal{C})})}{T-s} \nonumber \\
    & \qquad + \sum_{c\in\mathcal{C}} \left[ \frac{t-s}{(T-t)(T-s)} {\bm{y}^{(\mathcal{C})}}^{\intercal}\mathbf{\Lambda}_{c}^{-1}\bm{y}^{(\mathcal{C})} - \frac{2}{T-t}{\bm{y}^{(\mathcal{C})}}^{\intercal}\mathbf{\Lambda}_{c}^{-1}\bm{x}_{t}^{(c)} + \frac{2}{T-s}{\bm{y}^{(\mathcal{C})}}^{\intercal}\mathbf{\Lambda}_{c}^{-1}\bm{x}_{s}^{(c)}\right] \nonumber \\
    & \qquad + \sum_{c\in\mathcal{C}} \frac{T-s}{(t-s)(T-t)} \left[\bm{x}_{t}^{(c)} - \frac{T-t}{T-s}\bm{x}_{s}^{(c)}\right]^{\intercal}\mathbf{\Lambda}_{c}^{-1}\left[\bm{x}_{t}^{(c)} - \frac{T-t}{T-s}\bm{x}_{s}^{(c)}\right] + k_{3} \nonumber \\
    & = \frac{1}{T-s} \left[{\bm{y}^{(\mathcal{C})}}^{\intercal}\mathbf{\Lambda}_{\mathcal{C}}^{-1}\bm{y}^{(\mathcal{C})}-2{\bm{y}^{(\mathcal{C})}}^{\intercal}\mathbf{\Lambda}_{\mathcal{C}}^{-1}\tilde{\bm{x}}_{s}^{(\mathcal{C})}\right] \nonumber \\
    & \qquad + \left[ \frac{t-s}{(T-t)(T-s)} {\bm{y}^{(\mathcal{C})}}^{\intercal}\mathbf{\Lambda}_{\mathcal{C}}^{-1}\bm{y}^{(\mathcal{C})} - \frac{2}{T-t}{\bm{y}^{(\mathcal{C})}}^{\intercal}\mathbf{\Lambda}_{\mathcal{C}}^{-1}\tilde{\bm{x}}_{t}^{(\mathcal{C})} + \frac{2}{T-s}{\bm{y}^{(\mathcal{C})}}^{\intercal}\mathbf{\Lambda}_{\mathcal{C}}^{-1}\tilde{\bm{x}}_{s}^{(\mathcal{C})}\right] \nonumber \\
    & \qquad + \sum_{c\in\mathcal{C}} \frac{T-s}{(t-s)(T-t)} \left[\bm{x}_{t}^{(c)} - \frac{T-t}{T-s}\bm{x}_{s}^{(c)}\right]^{\intercal}\mathbf{\Lambda}_{c}^{-1}\left[\bm{x}_{t}^{(c)} - \frac{T-t}{T-s}\bm{x}_{s}^{(c)}\right] + k_{4} \nonumber \\
    & = \left[\frac{1}{T-s}+\frac{t-s}{(T-t)(T-s)}\right] {\bm{y}^{(\mathcal{C})}}^{\intercal}\mathbf{\Lambda}_{\mathcal{C}}^{-1}\bm{y}^{(\mathcal{C})} - \frac{2}{T-t}{\bm{y}^{(\mathcal{C})}}^{\intercal}\mathbf{\Lambda}_{\mathcal{C}}^{-1}\tilde{\bm{x}}_{t}^{(\mathcal{C})} \nonumber \\
    & \qquad + \sum_{c\in\mathcal{C}} \frac{T-s}{(t-s)(T-t)} \left[\bm{x}_{t}^{(c)} - \frac{T-t}{T-s}\bm{x}_{s}^{(c)}\right]^{\intercal}\mathbf{\Lambda}_{c}^{-1}\left[\bm{x}_{t}^{(c)} - \frac{T-t}{T-s}\bm{x}_{s}^{(c)}\right] + k_{4} \nonumber \\
    & = \frac{1}{T-t} \left[ {\bm{y}^{(\mathcal{C})}}^{\intercal}\mathbf{\Lambda}_{\mathcal{C}}^{-1}\bm{y}^{(\mathcal{C})} - 2{\bm{y}^{(\mathcal{C})}}^{\intercal}\mathbf{\Lambda}_{\mathcal{C}}^{-1}\tilde{\bm{x}}_{t}^{(\mathcal{C})} \right] \nonumber \\
    & \qquad + \sum_{c\in\mathcal{C}} \frac{T-s}{(t-s)(T-t)} \left[\bm{x}_{t}^{(c)} - \frac{T-t}{T-s}\bm{x}_{s}^{(c)}\right]^{\intercal}\mathbf{\Lambda}_{c}^{-1}\left[\bm{x}_{t}^{(c)} - \frac{T-t}{T-s}\bm{x}_{s}^{(c)}\right] +k_{4} \nonumber \\
    & = \frac{1}{T-t} \left[ (\bm{y}^{(\mathcal{C})}-\tilde{\bm{x}}_{t}^{(\mathcal{C})})^{\intercal}\mathbf{\Lambda}_{\mathcal{C}}^{-1}(\bm{y}^{(\mathcal{C})}-\tilde{\bm{x}}_{t}^{(\mathcal{C})}) \right] - \frac{1}{T-t}\tilde{\bm{x}}_{t}^{(\mathcal{C})^{\intercal}}\mathbf{\Lambda}_{\mathcal{C}}^{-1}\tilde{\bm{x}}_{t}^{(\mathcal{C})} \nonumber \\
    & \qquad + \sum_{c\in\mathcal{C}} \frac{T-s}{(t-s)(T-t)} \left[\bm{x}_{t}^{(c)} - \frac{T-t}{T-s}\bm{x}_{s}^{(c)}\right]^{\intercal}\mathbf{\Lambda}_{c}^{-1}\left[\bm{x}_{t}^{(c)} - \frac{T-t}{T-s}\bm{x}_{s}^{(c)}\right] + k_{4},
\end{align*}
where $k_{3}$ and $k_{4}$ are constants, and $\mathbf{\Lambda}_{\mathcal{C}}:=\left(\sum_{c\in\mathcal{C}} \mathbf{\Lambda}_{c}^{-1}\right)^{-1}$.

Next, we integrate out $\bm{y}^{(\mathcal{C})}$ to obtain the $d|\mathcal{C}|$-dimensional density of $\vecXX{t}{(\mathcal{C})}$ conditional on $\vecX{s}{(\mathcal{C})}$, which we denote $p_{2}$:
\begin{align*}
    - 2 \log p_{2}
    & = - \frac{1}{T-t}\tilde{\bm{x}}_{t}^{(\mathcal{C})^{\intercal}}\mathbf{\Lambda}_{\mathcal{C}}^{-1}\tilde{\bm{x}}_{t}^{(\mathcal{C})} + \sum_{c\in\mathcal{C}} \frac{T-s}{(t-s)(T-t)} \left[\bm{x}_{t}^{(c)} - \frac{T-t}{T-s}\bm{x}_{s}^{(c)}\right]^{\intercal}\mathbf{\Lambda}_{c}^{-1}\left[\bm{x}_{t}^{(c)} - \frac{T-t}{T-s}\bm{x}_{s}^{(c)}\right] + k_{5} \nonumber \\
    & = - \frac{1}{T-t}\tilde{\bm{x}}_{t}^{(\mathcal{C})^{\intercal}}\mathbf{\Lambda}_{\mathcal{C}}^{-1}\tilde{\bm{x}}_{t}^{(\mathcal{C})} + \sum_{c\in\mathcal{C}} \frac{T-s}{(t-s)(T-t)} \left[{\bm{x}_{t}^{(c)}}^{\intercal}\mathbf{\Lambda}_{c}^{-1}\bm{x}_{t}^{(c)} - 2\left(\frac{T-t}{T-s}\right){\bm{x}_{t}^{(c)}}^{\intercal}\mathbf{\Lambda}_{c}^{-1}\bm{x}_{s}^{(c)}\right] + k_{6},
\end{align*}
where $k_{5}$ and $k_{6}$ are constants. Noting that
\begin{align*}
    \tilde{\bm{x}}_{t}^{(\mathcal{C})^{\intercal}}\mathbf{\Lambda}_{\mathcal{C}}^{-1}\tilde{\bm{x}}_{t}^{(\mathcal{C})}
    & = \left[ \left(\sum_{c\in\mathcal{C}} \mathbf{\Lambda}_{c}^{-1}\right)^{-1}\left(\sum_{c\in\mathcal{C}} \mathbf{\Lambda}_{c}^{-1}\bm{x}_{t}^{(c)}\right) \right]^{\intercal} \left(\sum_{c\in\mathcal{C}} \mathbf{\Lambda}_{c}^{-1}\right) \left[ \left(\sum_{c\in\mathcal{C}} \mathbf{\Lambda}_{c}^{-1}\right)^{-1}\left(\sum_{c\in\mathcal{C}} \mathbf{\Lambda}_{c}^{-1}\bm{x}_{t}^{(c)}\right) \right] \nonumber \\
    & = \left(\sum_{c\in\mathcal{C}} \mathbf{\Lambda}_{c}^{-1}\bm{x}_{t}^{(c)}\right)^{\intercal} \mathbf{\Lambda}_{\mathcal{C}} \left(\sum_{c\in\mathcal{C}} \mathbf{\Lambda}_{c}^{-1}\bm{x}_{t}^{(c)}\right) \nonumber \\
    & = \sum_{i,j\in\mathcal{C}} {\bm{x}_{t}^{(i)}}^{\intercal} \left( \mathbf{\Lambda}_{i}^{-1} \mathbf{\Lambda}_{\mathcal{C}} \mathbf{\Lambda}_{j}^{-1} \right) \bm{x}_{t}^{(j)}.
\end{align*}
So we have,
\begin{align*}
    - 2 \log p_{2}
    & = - \frac{1}{T-t} \sum_{i,j\in\mathcal{C}} {\bm{x}_{t}^{(i)}}^{\intercal} \left( \mathbf{\Lambda}_{i}^{-1} \mathbf{\Lambda}_{\mathcal{C}} \mathbf{\Lambda}_{j}^{-1} \right) \bm{x}_{t}^{(j)} + \frac{T-s}{(t-s)(T-t)} \sum_{c\in\mathcal{C}} {\bm{x}_{t}^{(c)}}^{\intercal}\mathbf{\Lambda}_{c}^{-1}\bm{x}_{t}^{(c)} \nonumber \\
    & \qquad - \frac{2}{t-s}\sum_{c\in\mathcal{C}} {\bm{x}_{t}^{(c)}}^{\intercal}\mathbf{\Lambda}_{c}^{-1}\bm{x}_{s}^{(c)} + k_{6} \nonumber \\
    & = \frac{T-s}{(t-s)(T-t)} \sum_{c\in\mathcal{C}} {\bm{x}_{t}^{(c)}}^{\intercal}\mathbf{\Lambda}_{c}^{-1}\bm{x}_{t}^{(c)} - \frac{1}{T-t}\sum_{c\in\mathcal{C}} {\bm{x}_{t}^{(c)}}^{\intercal} \left( \mathbf{\Lambda}_{c}^{-1} \mathbf{\Lambda}_{\mathcal{C}} \mathbf{\Lambda}_{c}^{-1} \right) \bm{x}_{t}^{(c)} \nonumber \\
    & \qquad -\frac{1}{T-t} \sum_{\substack{i,j\in\mathcal{C} \\ i \neq j}} {\bm{x}_{t}^{(i)}}^{\intercal} \left( \mathbf{\Lambda}_{i}^{-1} \mathbf{\Lambda}_{\mathcal{C}} \mathbf{\Lambda}_{j}^{-1} \right) \bm{x}_{t}^{(j)} - \frac{2}{t-s}\sum_{c\in\mathcal{C}} {\bm{x}_{t}^{(c)}}^{\intercal}\mathbf{\Lambda}_{c}^{-1}\bm{x}_{s}^{(c)} + k_{6} \nonumber \\
    & = \vecX{t}{\intercal} \bm{V}_{s,t}^{-1} \vecX{t}{(\mathcal{C})} - \frac{2}{t-s} \vecX{t}{\intercal} \bm{L}^{-1} \vecX{s}{(\mathcal{C})} + k_{6}
\end{align*}
where
\begin{equation}
    \label{eq:V_inv}
    \bm{V}_{s,t}^{-1} =
    \begin{pmatrix}
    \bm{\Sigma}_{11}^{-1} & \bm{\Sigma}_{12}^{-1} & \dots & \bm{\Sigma}_{1|\mathcal{C}|}^{-1} \\
    \bm{\Sigma}_{21}^{-1} & \bm{\Sigma}_{22}^{-1} & \dots & \bm{\Sigma}_{2|\mathcal{C}|}^{-1} \\
    \vdots & \vdots & \ddots & \vdots \\
    \bm{\Sigma}_{|\mathcal{C}|1}^{-1} & \bm{\Sigma}_{|\mathcal{C}|2}^{-1} & \dots & \bm{\Sigma}_{|\mathcal{C}||\mathcal{C}|}^{-1}
    \end{pmatrix} \in \mathbb{R}^{|\mathcal{C}|d \times |\mathcal{C}|d},
\end{equation}
with
\begin{align*}
    \bm{\Sigma}_{ii}^{-1} & = \frac{T-s}{(t-s)(T-t)} \mathbf{\Lambda}_{c_{i}}^{-1} - \frac{1}{T-t} \left(\mathbf{\Lambda}_{c_{i}}^{-1}\mathbf{\Lambda}_{\mathcal{C}}\mathbf{\Lambda}_{c_{i}}^{-1}\right) \in \mathbb{R}^{d \times d}, \\
    \mathbf{\Sigma}_{ij}^{-1} & = - \frac{1}{T-t} \left( \mathbf{\Lambda}_{c_{i}}^{-1}\mathbf{\Lambda}_{\mathcal{C}}\mathbf{\Lambda}_{c_{j}}^{-1} \right) \in \mathbb{R}^{d \times d},
\end{align*}
for $i,j=1,\dots,|\mathcal{C}|$, and
\begin{equation*}
    \bm{L}^{-1} = 
    \begin{pmatrix}
    \mathbf{\Lambda}_{c_{1}}^{-1} & \bm{0}_{d \times d} & \dots & \bm{0}_{d \times d} \\
    \bm{0}_{d \times d} & \mathbf{\Lambda}_{c_{2}}^{-1} & \dots & \bm{0}_{d \times d} \\
    \vdots & \ddots & \vdots & \vdots \\
    \bm{0}_{d \times d} & \bm{0}_{d \times d} & \dots & \mathbf{\Lambda}_{c_{|\mathcal{C}|}}^{-1}
    \end{pmatrix} \in \mathbb{R}^{|\mathcal{C}|d \times |\mathcal{C}|d},
\end{equation*}
where $\bm{0}_{d \times d}$ is the $d \times d$ with all elements zero. We finally complete the square to get
\begin{equation*}
    - 2 \log p_{2} = \vecX{t}{(\mathcal{C})} \bm{V}_{s,t}^{-1} \vecX{t}{(\mathcal{C})} - 2 \vecX{t}{(\mathcal{C})} \bm{V}_{s,t}^{-1} \vecM{s,t}{(\mathcal{C})} + k_{6},
\end{equation*}
where 
\begin{equation*}
    \vecM{s,t}{(\mathcal{C})} = \frac{\bm{V}_{s,t}\bm{L}^{-1}\vecX{s}{(\mathcal{C})}}{t-s}.
\end{equation*}
Inverting $\bm{V}_{s,t}^{-1}$ in \eqref{eq:V_inv}, we obtain \eqref{eq:V_GBF} and subsequently we can get the expression for $\bm{M}_{s,t}^{(c)}$ in \eqref{eq:M_GBF} to prove the statement in part \ref{prop:proposal_simulation_GBF:a} of \thmref{theorem:fusion_measure_GBF}.

For part \ref{prop:proposal_simulation_GBF:b}, for $c\in\mathcal{C}$, the law of $\{\bm{X}_{t}^{(c)}, t\in(0,T)\}$ conditional on endpoints $\bm{x}_{0}^{(c)}$ and $\bm{y}^{(\mathcal{C})}$ is that of a Brownian bridge. This statement in the theorem holds from the standard properties of Brownian bridges (with covariance matrix $\mathbf{\Lambda}_{c}$). In particular, considering the distribution of $\bm{X}_{q}^{(c)}$ at an intermediate point $q\in(s,t)$ given the positions $\bm{X}_{s}^{(c)}=\bm{x}_{s}^{(c)}$ and $\bm{X}_{t}^{(c)}=\bm{x}_{t}^{(c)}$ at times $s$ and $t$ respectively, then we have
\begin{align*}
    & \mathbb{P}\left(\bm{X}_{q} = \bm{w} \middle| \bm{X}_{s}^{(c)} = \bm{x}_{s}^{(c)}, \bm{X}_{t}^{(c)} = \bm{x}_{t}^{(c)}\right) \\
    & \qquad \propto \mathbb{P}\left(\bm{X}_{t}^{(c)}=\bm{x}_{t}^{(c)}\middle|\bm{X}_{s}^{(c)}=\bm{x}_{s}^{(c)}, \bm{X}_{q}=\bm{w}\right) \cdot \mathbb{P}\left(\bm{X}_{q}=\bm{w}\middle|\bm{X}_{s}^{(c)}=\bm{x}_{s}^{(c)}\right) \\
    & \qquad \propto \mathbb{P}\left(\bm{X}_{t}^{(c)}=\bm{x}_{t}^{(c)}\middle|\bm{X}_{q}=\bm{w}\right) \cdot \mathbb{P}\left(\bm{X}_{q}=\bm{w}\middle|\bm{X}_{s}^{(c)}=\bm{x}_{s}^{(c)}\right) \\
    & \qquad \propto \exp\left(-\frac{(\bm{x}_{t}^{(c)}-\bm{w})^{\intercal}\mathbf{\Lambda}_{c}^{-1}(\bm{x}_{t}^{(c)}-\bm{w})}{2(t-q)}\right) \cdot \exp\left(-\frac{(\bm{w}-\bm{x}_{s}^{(c)})^{\intercal}\mathbf{\Lambda}_{c}^{-1}(\bm{w}-\bm{x}_{s}^{(c)})}{2(q-s)}\right), \nonumber
\end{align*}
and hence we arrive at the result in the statement. \hfill $\blacksquare$

\section{Proof of \texorpdfstring{\propositionref{prop:bounds}}{Proposition 3}}
\label{app:bounds}

\proof First note that we can rewrite \eqref{eq:phi} as follows,
\begin{equation}
    \label{eq:phi_expanded}
    \phi_{c} \left( \bm{x} \right) = \frac{1}{2} \left( \norm{\mathbf{\Lambda}_{c}^{\frac{1}{2}} \nabla \log f_{c} \left( \bm{x} \right)}^{2} + \Tr(\mathbf{\Lambda}_{c} \nabla^{2} \log f_{c} \left( \bm{x} \right)) \right).
\end{equation}
Let $R_{c} := R_{c} \big(\bm{X}_{[0,T]}^{(c)}\big)$ denote a compact subset of $\mathbb{R}^{d}$ for which $\bm{X}_{t}^{(c)}$ is constrained in time $[0,T]$ for $c\in\mathcal{C}$, then to bound the first term in \eqref{eq:phi_expanded}, we first use the triangle inequality by noting
\begin{align}
    \max_{\bm{x} \in R_{c}} \norm{\mathbf{\Lambda}_{c}^{\frac{1}{2}} \nabla \log f_{c}(\bm{x})}
    & = \max_{\bm{x} \in R_{c}} \norm{\mathbf{\Lambda}_{c}^{\frac{1}{2}} \nabla \log f_{c} (\hat{\bm{x}}^{(c)}) + \mathbf{\Lambda}_{c}^{\frac{1}{2}} \left( \nabla \log f_{c} (\bm{x}) - \nabla \log f_{c} (\hat{\bm{x}}^{(c)}) \right)} \nonumber \\
    & \leq \norm{\mathbf{\Lambda}_{c}^{\frac{1}{2}} \nabla \log f_{c} (\hat{\bm{x}}^{(c)})} + \max_{\bm{x} \in R_{c}} \norm{\mathbf{\Lambda}_{c}^{\frac{1}{2}} \left( \nabla \log f_{c} (\bm{x}) - \nabla \log f_{c} (\hat{\bm{x}}^{(c)}) \right)}, \label{eq:bounds:first_term}
\end{align}
where $\hat{\bm{x}}^{(c)}$ is a user-specified point in $\mathbb{R}^{d}$. Focusing now on bounding the second term of \eqref{eq:bounds:first_term}, then we express this as a line integral between $\bm{x}$ and $\hat{\bm{x}}^{(c)}$ so
\begin{equation*}
    \max_{\bm{x} \in R_{c}} \norm{\mathbf{\Lambda}_{c}^{\frac{1}{2}} \left( \nabla \log f_{c} (\bm{x}) - \nabla \log f_{c} (\hat{\bm{x}}^{(c)}) \right)} = \max_{\bm{x} \in R_{c}} \norm{\mathbf{\Lambda}_{c}^{-\frac{1}{2}} \int_{0}^{\norm{\bm{x}-\hat{\bm{x}}^{(c)}}} \mathbf{\Lambda}_{c} \nabla^{2} \log f(\bm{x}+u\bm{n}) \bm{n} \ddspace{u}},
\end{equation*}
where $\bm{u}=\bm{x}+u\bm{n}$, where $\bm{n}$ is a unit-vector. We have
\begin{align}
    \max_{\bm{x} \in R_{c}} \norm{\mathbf{\Lambda}_{c}^{-\frac{1}{2}} \int_{0}^{\norm{\bm{x}-\hat{\bm{x}}^{(c)}}} \mathbf{\Lambda}_{c} \nabla \log f(\bm{x}+u\bm{n}) \bm{n} \ddspace{u}}
    & \leq \max_{\bm{x} \in R_{c}} \norm{\mathbf{\Lambda}_{c}^{-\frac{1}{2}} \left(\bm{x}-\hat{\bm{x}}^{(c)}\right)} \nonumber \\
    & \qquad \cdot \sup_{\bm{n}; \bm{x} \in R_{c}} \norm{\mathbf{\Lambda}_{c} \nabla^{2} \log f(\bm{x}+u\bm{n})\bm{n}} \nonumber \\
    & \leq \max_{\bm{x} \in R_{c}} \norm{\mathbf{\Lambda}_{c}^{-\frac{1}{2}} \left(\bm{x}-\hat{\bm{x}}^{(c)}\right)} \cdot P^{\mathbf{\Lambda}_{c}}, \label{eq:bounds:first_term:second_term}
\end{align}
where $P^{\mathbf{\Lambda}_{c}}$ is defined in \eqref{eq:P}. Putting together \eqref{eq:bounds:first_term} and \eqref{eq:bounds:first_term:second_term}, we have
\begin{equation*}
    \max_{\bm{x} \in R_{c}} \norm{\mathbf{\Lambda}_{c}^{\frac{1}{2}} \nabla \log f_{c} (\bm{x})} \leq \norm{\mathbf{\Lambda}_{c}^{\frac{1}{2}} \nabla \log f_{c} \left(\hat{\bm{x}}^{(c)}\right)} + \max_{\bm{x} \in R_{c}} \norm{\mathbf{\Lambda}_{c}^{-\frac{1}{2}} \left( \bm{x} - \hat{\bm{x}}^{(c)} \right)} \cdot P^{\mathbf{\Lambda}_{c}}.
\end{equation*}
Since for a matrix $A \in \mathbb{R}^{d}$, $\Tr(A) \leq d \cdot \gamma(A)$, we can bound the second term in \eqref{eq:phi_expanded} as follows:
\begin{equation*}
    \max_{\bm{x} \in R_{c}} \abs{\Tr(\mathbf{\Lambda}_{c} \nabla^{2} \log f_{c} (\bm{x}))} \leq d \cdot P^{\mathbf{\Lambda}_{c}},
\end{equation*}
and hence we can bound $\phi_{c}$ as follows:
\begin{equation*}
     \max_{\bm{x} \in R_{c}} \abs{\phi_{c} \left( \bm{x} \right)} \leq \frac{1}{2} \left[ \left( \norm{\mathbf{\Lambda}_{c}^{\frac{1}{2}} \nabla \log f_{c} \left( \hat{\bm{x}}^{(c)} \right)} + \max_{\bm{x} \in R_{c}} \norm{\mathbf{\Lambda}_{c}^{-\frac{1}{2}}  \left( \bm{x} - \hat{\bm{x}}^{(c)} \right)} \cdot P^{\mathbf{\Lambda}_{c}} \right)^{2} + d \cdot P^{\mathbf{\Lambda}_{c}} \right].
\end{equation*}
Noting that in \eqref{eq:phi_expanded} that the first term is squared, then the lower and upper bounds of $\phi_{c}\left(\bm{x}\right)$ for $\bm{x} \in R_{c}$ are given by \eqref{eq:lower_bound} and \eqref{eq:upper_bound} respectively. \hfill $\blacksquare$

\section{Proof of \texorpdfstring{\thmref{theorem:unbiased_estimator_GBF}}{Theorem 4}} \label{app:unbiased_estimator_GBF_proof}

\proof Following in the style of \citet{Beskos_et_al_2006b, Beskos_et_al_2008, Fearnhead_2008} and \citet[Appendix B]{dai_et_al_2023}, for $j=1,\dots,n$, we have
\begin{align*}
    & \mathbb{E}_{\bar{\mathcal{R}}}\mathbb{E}_{\bar{\mathbb{W}}|\bar{\mathcal{R}}}\mathbb{E}_{\mathbb{\bar{\mathbb{K}}}}\mathbb{E}_{\bar{\mathbb{U}}}\left[a_{j}\tilde{\rho}_{j}\right] \\
    & \qquad = \mathbb{E}_{\bar{\mathcal{R}}}\mathbb{E}_{\bar{\mathbb{W}}|\bar{\mathcal{R}}}\mathbb{E}_{\mathbb{\bar{\mathbb{K}}}}\mathbb{E}_{\bar{\mathbb{U}}}\left[\prod_{c\in\mathcal{C}} \left( \frac{\Delta_{j}^{\kappa_{c}} \cdot e^{-(U_{j}^{(c)}-\mathbf{\Phi}_{c})\Delta_{j}}}{\kappa_{c}! \cdot p \left( \kappa_{c} | R_{c} \right)} \prod_{k_{c}=1}^{\kappa_{c}} \left(U_{j}^{(c)} - \phi_{c} \left( \bm{X}_{\xi_{c, k_{c}}}^{(c)} \right) \right) \right)\right] \\
    & \qquad = \mathbb{E}_{\bar{\mathcal{R}}}\mathbb{E}_{\bar{\mathbb{W}}|\bar{\mathcal{R}}}\mathbb{E}_{\mathbb{\bar{\mathbb{K}}}}\left[\prod_{c\in\mathcal{C}} \left( \frac{\Delta_{j}^{\kappa_{c}} \cdot e^{-(U_{j}^{(c)}-\mathbf{\Phi}_{c})\Delta_{j}}}{\kappa_{c}! \cdot p \left( \kappa_{c} | R_{c} \right)} \cdot \left[\int_{t_{j-1}}^{t_{j}} \frac{U_{j}^{(c)}-\phi_{c}\left(\bm{X}_{t}^{(c)}\right)}{\Delta_{j}}\ddspace{t}\right]^{\kappa_{c}} \right)\right] \\
    & \qquad = \mathbb{E}_{\bar{\mathcal{R}}}\mathbb{E}_{\bar{\mathbb{W}}|\bar{\mathcal{R}}}\left[\prod_{c\in\mathcal{C}} \left( \sum_{k_{c}=0}^{\infty} \frac{\Delta_{j}^{k_{c}} \cdot e^{-(U_{j}^{(c)}-\mathbf{\Phi}_{c})\Delta_{j}}}{k_{c}! \cdot p \left( k_{c} | R_{c} \right)} \cdot \left[\int_{t_{j-1}}^{t_{j}} \frac{U_{j}^{(c)}-\phi_{c}\left(\bm{X}_{t}^{(c)}\right)}{\Delta_{j}}\ddspace{t}\right]^{k_{c}} \right)\right] \\
    & \qquad = \mathbb{E}_{\bar{\mathcal{R}}}\mathbb{E}_{\bar{\mathbb{W}}|\bar{\mathcal{R}}}\left[\prod_{c\in\mathcal{C}} e^{-(U_{j}^{(c)}-\mathbf{\Phi}_{c})\Delta_{j}} \cdot \left( \sum_{k_{c}=0}^{\infty} \frac{\Delta_{j}^{k_{c}}}{k_{c}! \cdot p \left( k_{c} | R_{c} \right)} \cdot \left[\int_{t_{j-1}}^{t_{j}} \frac{U_{j}^{(c)}-\phi_{c}\left(\bm{X}_{t}^{(c)}\right)}{\Delta_{j}}\ddspace{t}\right]^{k_{c}} \right)\right] \\
    & \qquad = \mathbb{E}_{\bar{\mathcal{R}}}\mathbb{E}_{\bar{\mathbb{W}}|\bar{\mathcal{R}}}\left[\prod_{c\in\mathcal{C}} e^{-(U_{j}^{(c)}-\mathbf{\Phi}_{c})\Delta_{j}} \cdot \exp \left\{ \int_{t_{j-1}}^{t_{j}} \left(U_{j}^{(c)}-\phi_{c}\left(\bm{X}_{t}^{(c)}\right)\right) \ddspace{t} \right\} \right] \\
    & \qquad = \prod_{c\in\mathcal{C}} \mathbb{E}_{\mathbb{W}_{\mathbf{\Lambda}_{c},j}} \left[ \exp \left\{ -\int_{t_{j-1}}^{t_{j}} \left( \phi_{c} \left( \bm{X}_{t}^{(c)} \right) - \mathbf{\Phi}_{c} \right) \ddspace{t} \right\} \right] =: \rho_{j},
\end{align*}
and hence $a_{j}\tilde{\rho}_{j}$ is an unbiased estimator for $\rho_{j}$. \hfill $\blacksquare$

\section{Unbiased Estimation of \texorpdfstring{$\rho_{j}$}{rho j}}
\label{app:unbiased_estimators}

Computing $\tilde{\rho}_{1}^{(a)}$ and $\tilde{\rho}_{1}^{(b)}$ by means of \emph{layer information} in the case where $\mathbf{\Lambda}_{c} = \mathbb{I}_{d}$ is detailed explicitly in \citet[Algorithm 4, Appendix B]{dai_et_al_2023}. In the case where $\mathbf{\Lambda}_{c} \neq \mathbb{I}_{d}$, we simulate layers by appealing to a suitable transformation. In particular, we transform the start and end points of the Brownian bridge with transformation matrix $\mathbf{\Lambda}_{c}^{-\frac{1}{2}}$, letting $\bm{z}_{j-1}^{(c)} := \mathbf{\Lambda}_{c}^{-\frac{1}{2}} \bm{x}_{j-1}^{(c)}$ and $\bm{z}_{j}^{(c)} := \mathbf{\Lambda}_{c}^{-\frac{1}{2}} \bm{x}_{j}^{(c)}$. The resulting Brownian bridge sample path, $\bm{z}_{t}^{(c)} := \mathbf{\Lambda}_{c}^{-\frac{1}{2}} \bm{X}_{t}^{(c)}$, has identity covariance structure and thus we can use existing methods for simulating \emph{layered} Brownian bridge sample paths $\bm{z}_{t}^{(c)}$ with law $\mathbb{W}_{\mathbb{I}_{d}}$ from $\bm{z}_{j-1}^{(c)}$ to $\bm{z}_{j}^{(c)}$. By finding a bounding hyper cube for the reverse transformed bounds, we are able to find appropriate layer information for the case $\mathbf{\Lambda}_{c} \neq \mathbb{I}_{d}$. We are now able with minimal modification to apply the approach of \citet{dai_et_al_2023}, as given in Algorithm \ref{alg:unbiased_estimator_rho_j}.
\begin{algorithm}
    \caption{Simulating $\tilde{\rho}_{j}$.}
    \label{alg:unbiased_estimator_rho_j}
    \begin{enumerate}
        \item For $c\in\mathcal{C}$
        \begin{enumerate}
            \item $\bm{z}_{j-1}^{(c)}, \bm{z}_{j}^{(c)}$: Transform the path, setting $\bm{z}_{j-1}^{(c)} := \mathbf{\Lambda}_{c}^{-\frac{1}{2}} \bm{x}_{j-1}^{(c)}$,  and $\bm{z}_{j}^{(c)} := \mathbf{\Lambda}_{c}^{-\frac{1}{2}} \bm{x}_{j}^{(c)}$.
            \item $R_{c}$: Set $R_{c} := \mathbf{\Lambda}_{c}^{\frac{1}{2}} R_{c}^{(z)}$, where $R_{c}^{(z)} \sim \mathcal{R}_{c}^{(z)}$ as per \citet[Algorithm 14]{Pollock_et_al_2016}. \label{alg:unbiased_estimator_rho_j:R_c}
            \item $L_{X}^{(c)}, U_{X}^{(c)}$: Compute lower and upper bounds, $L_{X}^{(c)}$ and $U_{X}^{(c)}$, of $\phi_{c}(\bm{x})$ for $\bm{x} \in R_{c}$ (as per \eqref{eq:lower_bound} and \eqref{eq:upper_bound}, or otherwise). \label{alg:unbiased_estimator_rho_j:compute_phi_bounds}
            \item $p_{c}$: Choose $p(\cdot | R_{c})$ using either GPE-1 (\condref{cond:GPE1}) or GPE-2 (\condref{cond:GPE2}).
            \item $\kappa_{c}, \xi$: Simulate $\kappa_{c} \sim p(\cdot | R_{c})$, and simulate $\xi_{c,1}, \dots, \xi_{c,\kappa_{c}} \sim \mathcal{U}[t_{j-1},t_{j}]$.
            \item $\bm{z}^{(c)}$: Simulate $\bm{z}_{\xi_{c,1}}^{(c)}, \dots, \bm{z}_{\xi_{c,\kappa_{c}}}^{(c)} \sim \mathbb{W}_{\mathbb{I}_{d}} | R_{c}^{(z)}$ as per \citet[Algorithm 15]{Pollock_et_al_2016}.
            \item $\bm{X}^{(c)}$: Reverse transform the path, setting $\bm{X}_{\xi_{c,k_{c}}}^{(c)} = \mathbf{\Lambda}_{c}^{\frac{1}{2}} \bm{z}_{\xi_{c,k_{c}}}^{(c)}$ for $k_{c} \in \{1,\dots,\kappa_{c}\}$.
        \end{enumerate}
        \item Output:
        \begin{equation*}
            \tilde{\rho}_{j} = \prod_{c\in\mathcal{C}} \left[ \frac{\Delta_{j}^{\kappa_{c}} \cdot e^{-U_{j}^{(c)}\Delta_{j}}}{\kappa_{c}! \cdot p \left( \kappa_{c} | R_{c} \right)} \prod_{k_{c}=1}^{\kappa_{c}} \left(U_{j}^{(c)} - \phi_{c} \left( \bm{X}_{\xi_{c, k_{c}}}^{(c)} \right) \right) \right].
        \end{equation*}
    \end{enumerate}
\end{algorithm}

\section{Proof of \texorpdfstring{\thmref{theorem:T_guidance}}{Theorem 11}, \texorpdfstring{\cororef{corollary:T_guidance}}{Corollary 12}, \texorpdfstring{\thmref{theorem:mesh_guidance}}{Theorem 14}, \texorpdfstring{\propositionref{prop:k4_choice}}{Proposition 18}} \label{app:guidance_proofs}

\proof (\thmref{theorem:T_guidance}) Considering the initial conditional effective sample size, $\CESS{0}$, we have
\begin{align}
    N^{-1}\CESS{0}
    & := N^{-1} \left[ \frac{\left(\sum_{i=1}^{N} \rho_{0,i}\right)^{2}}{\sum_{i=1}^{N} \rho_{0,i}^{2}} \right]
    \rightarrow \frac{\left(\mathbb{E}[\rho_{0,i}]\right)^{2}}{\mathbb{E}[\rho_{0,i}^{2}]} \nonumber \\
    & = \frac{\mathbb{E}\left[\exp \left\{ - \sum_{c\in\mathcal{C}} \frac{(\tilde{\bm{x}}_{0}^{(\mathcal{C})}-\bm{x}_{0}^{(c)})^{\intercal} \mathbf{\Lambda}_{c}^{-1} (\tilde{\bm{x}}_{0}^{(\mathcal{C})}-\bm{x}_{0}^{(c)})}{2T} \right\}\right]^{2}}{\mathbb{E}\left[\exp \left\{ - \sum_{c\in\mathcal{C}} \frac{(\tilde{\bm{x}}_{0}^{(\mathcal{C})}-\bm{x}_{0}^{(c)})^{\intercal} \mathbf{\Lambda}_{c}^{-1} (\tilde{\bm{x}}_{0}^{(\mathcal{C})}-\bm{x}_{0}^{(c)})}{T} \right\}\right]} \nonumber \\
    & = \frac{\mathbb{E}\left[e^{-\frac{|\mathcal{C}|\sigma^{2}}{2T}}\right]^{2}}{\mathbb{E}\left[e^{-\frac{|\mathcal{C}|\sigma^{2}}{T}}\right]},
\end{align}
where $\sigma^{2}:=\frac{1}{|\mathcal{C}|}\sum_{c\in\mathcal{C}}(\tilde{\bm{x}}_{0}^{(\mathcal{C})}-\bm{x}_{0}^{(c)})^{\intercal} \mathbf{\Lambda}_{c}^{-1} (\tilde{\bm{x}}_{0}^{(\mathcal{C})}-\bm{x}_{0}^{(c)})$ where $\bm{x}_{0}^{(c)} \sim \mathcal{N}_{d}(\bm{a}_{c}, \frac{b|\mathcal{C}|}{m}\mathbf{\Lambda}_{c})$. To get an expression for $N^{-1}\CESS{0}$, we begin by obtaining the moment generating function (mgf) for $\sigma^{2}$. First note
\begin{align}
    \frac{1}{|\mathcal{C}|} \sum_{c\in\mathcal{C}} (\bm{x}_{0}^{(c)}-\tilde{\bm{a}})^{\intercal} \mathbf{\Lambda}_{c}^{-1} (\bm{x}_{0}^{(c)}-\tilde{\bm{a}})
    & = \sigma^{2} + \frac{1}{|\mathcal{C}|} \sum_{c\in\mathcal{C}} (\tilde{\bm{x}}_{0}^{(\mathcal{C})}-\tilde{\bm{a}})^{\intercal} \mathbf{\Lambda}_{c}^{-1} (\tilde{\bm{x}}_{0}^{(\mathcal{C})}-\tilde{\bm{a}}). \label{eq:mgf_sigma_2}
\end{align}
Considering the term $\frac{1}{|\mathcal{C}|} \sum_{c\in\mathcal{C}} (\bm{x}_{0}^{(c)}-\tilde{\bm{a}})^{\intercal} \mathbf{\Lambda}_{c}^{-1} (\bm{x}_{0}^{(c)}-\tilde{\bm{a}})$ and letting $\bm{Y}_{c}:=\mathbf{\Lambda}_{c}^{-\frac{1}{2}}(\bm{x}_{0}^{(c)}-\tilde{\bm{a}})$, then $\bm{Y}_{c}$ has mean $\mathbf{\Lambda}_{c}^{-\frac{1}{2}}(\bm{a}_{c}-\tilde{\bm{a}})$ and variance $\frac{b|\mathcal{C}|}{m}\mathbb{I}_{d}$.
Hence $\sqrt{\frac{m}{b|\mathcal{C}|}}\bm{Y}_{c}$ has mean $\sqrt{\frac{m}{b|\mathcal{C}|}}\mathbf{\Lambda}_{c}^{-\frac{1}{2}}(\bm{a}_{c}-\tilde{\bm{a}})$ and variance $\mathbb{I}_{d}$, and so let
\begin{align*}
    \lambda
    & = \sum_{c\in\mathcal{C}} \norm{\sqrt{\frac{m}{b|\mathcal{C}|}}\mathbf{\Lambda}_{c}^{-\frac{1}{2}}(\bm{a}_{c}-\tilde{\bm{a}})}^{2} = \frac{m}{b|\mathcal{C}|} \sum_{c\in\mathcal{C}} (\bm{a}_{c}-\tilde{\bm{a}})^{\intercal} \mathbf{\Lambda}_{c}^{-1} (\bm{a}_{c}-\tilde{\bm{a}}) = \frac{m}{b} \sigma_{\bm{a}}^{2},
\end{align*}
then $\frac{m}{b|\mathcal{C}|} \sum_{c\in\mathcal{C}} \norm{\bm{Y}_{c}}^{2} \sim \chi^{2}(|\mathcal{C}|d, \lambda)$ distribution (i.e.\ $\frac{m}{b|\mathcal{C}|} \sum_{c\in\mathcal{C}} (\bm{x}_{0}^{(c)}-\tilde{\bm{a}})^{\intercal} \mathbf{\Lambda}_{c}^{-1} (\bm{x}_{0}^{(c)}-\tilde{\bm{a}})$ has a non-central $\chi^{2}(|\mathcal{C}|d, \lambda)$ distribution) with mgf
\begin{equation}
    \label{eq:mgf_1}
    M_{1}(s) := \frac{\exp \left\{\frac{\lambda s}{1-2s}\right\}}{(1-2s)^{\frac{|\mathcal{C}|d}{2}}}.
\end{equation}
Secondly, consider $\frac{1}{|\mathcal{C}|} \sum_{c\in\mathcal{C}} (\tilde{\bm{x}}_{0}^{(\mathcal{C})}-\tilde{\bm{a}})^{\intercal} \mathbf{\Lambda}_{c}^{-1} (\tilde{\bm{x}}_{0}^{(\mathcal{C})}-\tilde{\bm{a}}) = \frac{1}{|\mathcal{C}|} (\tilde{\bm{x}}_{0}^{(\mathcal{C})}-\tilde{\bm{a}})^{\intercal}\mathbf{\Lambda}_{\mathcal{C}}^{-1}(\tilde{\bm{x}}_{0}^{(\mathcal{C})}-\tilde{\bm{a}})$, where $\mathbf{\Lambda}_{\mathcal{C}}^{-1}:=\sum_{c\in\mathcal{C}}\mathbf{\Lambda}_{c}^{-1}$. Then since $\tilde{\bm{x}}_{0}^{(\mathcal{C})} \sim \mathcal{N}_{d}(\tilde{\bm{a}}, \frac{b|\mathcal{C}|}{m} \mathbf{\Lambda}_{\mathcal{C}})$, then $\bm{Z}:=\sqrt{\frac{m}{b|\mathcal{C}|}}\mathbf{\Lambda}^{-\frac{1}{2}}(\tilde{\bm{x}}_{0}^{(\mathcal{C})}-\tilde{\bm{a}}) \sim \mathcal{N}_{d}(\bm{0},\mathbb{I}_{d})$ and so $\norm{\bm{Z}}^{2} \sim \chi^{2}_{d}$ (i.e.\ $\frac{m}{b|\mathcal{C}|} \sum_{c\in\mathcal{C}} (\tilde{\bm{x}}_{0}^{(\mathcal{C})}-\tilde{\bm{a}})^{\intercal} \mathbf{\Lambda}_{c}^{-1} (\tilde{\bm{x}}_{0}^{(\mathcal{C})}-\tilde{\bm{a}})$ has $\chi^{2}_{d}$ distribution) with mgf
\begin{equation}
    \label{eq:mgf_2}
    M_{2}(s):=(1-2s)^{-\frac{d}{2}}.
\end{equation}
From \eqref{eq:mgf_sigma_2}, we have
\begin{align*}
    \sigma^{2} 
    & = \frac{b}{m} \underbrace{\left[ \frac{m}{b|\mathcal{C}|} \sum_{c\in\mathcal{C}} (\bm{x}_{0}^{(c)}-\tilde{\bm{a}})^{\intercal} \mathbf{\Lambda}_{c}^{-1} (\bm{x}_{0}^{(c)}-\tilde{\bm{a}})\right]}_{\sim \chi^{2}(|\mathcal{C}|d,\lambda)} - \frac{b}{m} \underbrace{\left[\frac{m}{b|\mathcal{C}|} \sum_{c\in\mathcal{C}} (\tilde{\bm{x}}_{0}^{(\mathcal{C})}-\tilde{\bm{a}})^{\intercal} \mathbf{\Lambda}_{c}^{-1} (\tilde{\bm{x}}_{0}^{(\mathcal{C})}-\tilde{\bm{a}})\right]}_{\sim \chi^{2}_{d}},
\end{align*}
where $\lambda=\frac{m\sigma_{\bm{a}}^{2}}{b}$. Therefore, using \eqref{eq:mgf_1} and \eqref{eq:mgf_2}, the mgf for $\sigma^{2}$ is given by
\begin{align}
    M_{\sigma^{2}}(s)
    & = \frac{M_{1}(\frac{sb}{m})}{M_{2}(\frac{sb}{m})}
    = \exp\left\{\frac{m\sigma_{\bm{a}}^{2}s}{m-2sb} \right\} \cdot \left(1-2\frac{sb}{m}\right)^{-\frac{(|\mathcal{C}|-1)d}{2}}, \text{ where } \frac{sb}{m} < \frac{1}{2}.
\end{align}
Given the mgf of $\sigma^{2}$, then 
\begin{align*}
    N^{-1}\CESS{0}
    & \rightarrow \frac{\mathbb{E}\left[e^{-\frac{|\mathcal{C}|\sigma^{2}}{2T}}\right]^{2}}{\mathbb{E}\left[e^{-\frac{|\mathcal{C}|\sigma^{2}}{T}}\right]} = \frac{M_{\sigma^{2}}\left(-\frac{|\mathcal{C}|}{2T}\right)^{2}}{M_{\sigma^{2}}\left(-\frac{|\mathcal{C}|}{T}\right)} \\
    & = \frac{\left[\exp\left\{\frac{m\sigma_{\bm{a}}^{2}\left(-\frac{|\mathcal{C}|}{2T}\right)}{m-2\left(-\frac{|\mathcal{C}|b}{2T}\right)} \right\} \cdot \left(1-2\left(-\frac{|\mathcal{C}|}{2T}\right)\frac{b}{m}\right)^{-\frac{(|\mathcal{C}|-1)d}{2}}\right]^{2}}{\exp\left\{\frac{m\sigma_{\bm{a}}^{2}\left(-\frac{|\mathcal{C}|}{T}\right)}{m-2\left(-\frac{|\mathcal{C}|b}{T}\right)} \right\} \cdot \left(1-2\left(-\frac{|\mathcal{C}|}{T}\right)\frac{b}{m}\right)^{-\frac{(|\mathcal{C}|-1)d}{2}}} \\
    & = \frac{\exp\left\{-\frac{m\sigma_{\bm{a}}^{2}\left(\frac{|\mathcal{C}|}{T}\right)}{m+\frac{|\mathcal{C}|b}{T}} \right\} \cdot \left(1+\frac{|\mathcal{C}|b}{Tm}\right)^{-(|\mathcal{C}|-1)d}}{\exp\left\{-\frac{m\sigma_{\bm{a}}^{2}\left(\frac{|\mathcal{C}|}{T}\right)}{m+2\left(\frac{|\mathcal{C}|b}{T}\right)} \right\} \cdot \left(1+2\left(\frac{|\mathcal{C}|b}{Tm}\right)\right)^{-\frac{(|\mathcal{C}|-1)d}{2}}} \\
    & = \exp\left\{-\frac{\sigma_{\bm{a}}^{2}}{\frac{T}{|\mathcal{C}|}+\frac{b}{m}}\right\} \cdot \exp\left\{\frac{\sigma_{\bm{a}}^{2}}{\frac{T}{|\mathcal{C}|}+\frac{2b}{m}}\right\} \cdot \left[\frac{\left(1+\frac{|\mathcal{C}|b}{Tm}\right)^{2}}{1+2\left(\frac{|\mathcal{C}|b}{Tm}\right)}\right]^{-\frac{(|\mathcal{C}|-1)d}{2}} \\
    & = \exp\left\{-\frac{\sigma_{\bm{a}}^{2}\left(\frac{b}{m}\right)}{\left(\frac{T}{|\mathcal{C}|}+\frac{b}{m}\right)\left(\frac{T}{|\mathcal{C}|}+\frac{2b}{m}\right)}\right\} \cdot \left[ 1 + \frac{\left(\frac{|\mathcal{C}|b}{Tm}\right)^{2}}{1+\frac{2|\mathcal{C}|b}{Tm}} \right]^{-\frac{(|\mathcal{C}|-1)d}{2}},
\end{align*}
and so \thmref{theorem:T_guidance} immediately follows. \hfill $\blacksquare$

\proof (\cororef{corollary:T_guidance}) Under \condref{cond:SH},  $\sigma_{\bm{a}}^{2}=\frac{(|\mathcal{C}|-1)\lambda}{m}<\frac{|\mathcal{C}|\lambda}{m}$, so for the first term in \eqref{eq:CESS_0},
\begin{align}
    \exp\left\{-\frac{\sigma_{\bm{a}}^{2}\left(\frac{b}{m}\right)}{\left(\frac{T}{|\mathcal{C}|}+\frac{b}{m}\right)\left(\frac{T}{|\mathcal{C}|}+\frac{2b}{m}\right)}\right\}
    & \geq \exp \left\{ - \frac{\sigma_{\bm{a}}^{2}b|\mathcal{C}|^{2}}{T^{2}m} \right\} \nonumber \\
    & \geq \exp \left\{ - \frac{b^{2}|\mathcal{C}|^{3}\lambda}{T^{2}m^{2}} \right\} \nonumber \\
    & \geq \exp \left\{ - \frac{\lambda}{k_{1}^{2}} \right\}, \label{eq:T_guidance_first_term_bound_SH}
\end{align}
where $T \geq \frac{b|\mathcal{C}|^{3/2}k_{1}}{m}$ for some constant $k_{1}>0$, and for the second term in \eqref{eq:CESS_0}, then
\begin{align}
     \left[ 1 + \frac{\left(\frac{|\mathcal{C}|b}{Tm}\right)^{2}}{1+\frac{2|\mathcal{C}|b}{Tm}} \right]^{-\frac{(|\mathcal{C}|-1)d}{2}}
     & \geq \left[ \exp\left\{\frac{\left(\frac{|\mathcal{C}|b}{Tm}\right)^{2}}{1+\frac{2|\mathcal{C}|b}{Tm}}\right\} \right]^{-\frac{(|\mathcal{C}|-1)d}{2}} \nonumber \\
     & = \exp \left\{ - \frac{\left(\frac{|\mathcal{C}|b}{Tm}\right)^{2}(|\mathcal{C}|-1)d}{2(1+\frac{2C}{Tm})}\right\} \nonumber \\
     & \geq \exp \left\{ -\frac{\left(\frac{|\mathcal{C}|^{3}b^{2}}{T^{2}m^{2}}\right)d}{2} \right\} \nonumber \\
     & \geq \exp \left\{ - \frac{d}{2k_{1}^{2}} \right\}, \label{eq:T_guidance_second_term_bound}
\end{align}
with $T \geq \frac{b|\mathcal{C}|^{3/2}k_{1}}{m}$. Hence, under \condref{cond:SH} and choosing $T \geq \frac{b|\mathcal{C}|^{3/2}k_{1}}{m}$, combining the bounds from \eqref{eq:T_guidance_first_term_bound_SH} and \eqref{eq:T_guidance_second_term_bound} gives \eqref{eq:CESS_0_SH}. Under \condref{cond:SSH}, $\sigma_{\bm{a}}^{2}=b\gamma$, if we assume $T \geq \frac{b|\mathcal{C}|^{3/2}k_{1}}{m}$ for some constant $k_{1}>0$, and $T \geq |\mathcal{C}|^{\frac{1}{2}}k_{2}$ for some constant $k_{2}>0$, then
\begin{equation*}
    \left(\frac{T}{|\mathcal{C}|}+\frac{b}{m}\right)\left(\frac{T}{|\mathcal{C}|}+\frac{2b}{m}\right) \geq \frac{T^{2}}{|\mathcal{C}|^{2}} \geq \frac{bk_{1}k_{2}}{m},
\end{equation*}
and so we have
\begin{equation}
    \exp\left\{-\frac{\sigma_{\bm{a}}^{2}\left(\frac{b}{m}\right)}{\left(\frac{T}{|\mathcal{C}|}+\frac{b}{m}\right)\left(\frac{T}{|\mathcal{C}|}+\frac{2b}{m}\right)}\right\} \geq \exp \left\{ - \frac{\frac{b^{2}\gamma}{m}}{\frac{bk_{1}k_{2}}{m}} \right\} = \exp \left\{ - \frac{b\gamma}{k_{1}k_{2}} \right\}. \label{eq:T_guidance_first_term_bound_SSH}
\end{equation}
Hence, under \condref{cond:SSH} and choosing $T$ such that $T \geq \frac{b|\mathcal{C}|^{3/2}k_{1}}{m}$ and $T \geq |\mathcal{C}|^{\frac{1}{2}}k_{2}$, we can combine the bounds from \eqref{eq:T_guidance_first_term_bound_SSH} and \eqref{eq:T_guidance_second_term_bound} to obtain the bound in \eqref{eq:CESS_0_SSH}. \hfill $\blacksquare$

\proof (\thmref{theorem:mesh_guidance}) As $N\rightarrow\infty$, we have
\begin{align*}
    N^{-1}\CESS{j}
    & := N^{-1} \left[\frac{\left(\sum_{i=1}^{N}\tilde{\rho}_{j,i}\right)^{2}}{\sum_{i=1}^{N} \tilde{\rho}_{j,i}^{2}}\right]
    = \left[\frac{\left(N^{-1} \sum_{i=1}^{N}a_{j}\tilde{\rho}_{j,i}\right)^{2}}{N^{-1} \sum_{i=1}^{N} \left(a_{j}\tilde{\rho}_{j,i}\right)^{2}}\right] \rightarrow \frac{\mathbb{E}\left[a_{j}\tilde{\rho}_{j}\right]^{2}}{\mathbb{E}\left[\left(a_{j}\tilde{\rho_{j}}\right)^{2}\right]},
\end{align*}
where $a_{j}:=\exp\{\sum_{c\in\mathcal{C}}\mathbf{\Phi}_{c}\Delta_{j}\}$. Since $a_{j}\tilde{\rho}_{j}$ is an unbiased estimate of $\rho_{j}$ (see \thmref{theorem:unbiased_estimator_GBF}), then
\begin{align*}
    \mathbb{E}\left[a_{j}\tilde{\rho}_{j}\right]
    & = \prod_{c\in\mathcal{C}} \mathbb{E}_{\mathbb{W}_{\mathbf{\Lambda}_{c},j}} \left( \exp \left\{ -\int_{t_{j-1}}^{t_{j}} \left( \phi_{c} \left( \bm{X}_{t}^{(c)} \right) - \mathbf{\Phi}_{c} \right) \right\} \right) \\
    & = \mathbb{E}_{\bar{\mathbb{W}}_{\mathbf{\Lambda}}} \left( \exp \left\{ -\sum_{c\in\mathcal{C}} \int_{t_{j-1}}^{t_{j}} \phi_{c} \left( \bm{X}_{t}^{(c)} \right) \right\} \right) \cdot a_{j}
\end{align*}
where $\bar{\mathbb{W}}_{\mathbf{\Lambda}}$ denotes the law of the collection of Brownian bridges $\{\mathbb{W}_{\mathbf{\Lambda}_{c},j}: c\in\mathcal{C}\}$ for each $j$. Note that under the optimal distribution for $p(\kappa_{c}|R_{c})$ (a Poisson distribution with intensity given in \eqref{eq:optimal_lambda_c}), then $\mathbb{E}\left[\left(a_{j}\tilde{\rho}_{j}\right)^{2}\right] \leq 1$ \citep{Fearnhead_2008, dai_et_al_2023}, so
\begin{equation*}
    \lim_{N\rightarrow\infty} N^{-1}\CESS{j} \geq \mathbb{E}\left[a_{j}\tilde{\rho}_{j}\right]^{2} = \left[\mathbb{E}_{\bar{\mathbb{W}}_{\mathbf{\Lambda}}}\left(\exp\left\{-\sum_{c\in\mathcal{C}} \int_{t_{j-1}}^{t_{j}} \phi_{c} \left( \bm{X}_{t}^{(c)} \right)\right\}\right)\right]^{2} \cdot a_{j}^{2}.
\end{equation*}
If $f_{c} \sim \mathcal{N}_{d}(\bm{a}_{c}, \frac{b|\mathcal{C}|}{m}\mathbf{\Lambda}_{c})$, then $\phi_{c}(\bm{x})=\frac{1}{2} \left(\left(\frac{m}{b|\mathcal{C}|}\right)^{2}(\bm{x}-\bm{a}_{c})^{\intercal}\mathbf{\Lambda}_{c}^{-1}(\bm{x}-\bm{a}_{c}) - \frac{md}{b|\mathcal{C}|} \right)$ which has global lower bound $\mathbf{\Phi}_{c}=-\frac{1}{2}\left(\frac{md}{b|\mathcal{C}|}\right)$ (since the minimum of $\phi_{c}$ occurs at the mean, $\bm{a}_{c}$). Then by considering small intervals $(t_{j-1},t_{j})$ and taking the limit of $\Delta_{j}:=t_{j}-t_{j-1} \rightarrow 0$, then
\begin{align*}
    & \lim_{\Delta_{j}\rightarrow 0} \lim_{N\rightarrow\infty} N^{-1}\CESS{j} \nonumber \\
    & \geq \lim_{\Delta_{j}\rightarrow 0} \left[\mathbb{E}\left(\mathbb{E}\left\{\mathbb{E}\left(\exp\left\{-\sum_{c\in\mathcal{C}}\int_{t_{j-1}}^{t_{j}} \phi_{c}\left(\bm{X}_{t}^{(c)}\right)\dd t\right\} \middle| \bm{\xi}_{j}, \vecX{j-1}{(\mathcal{C})} \right) \middle| \vecX{j-1}{(\mathcal{C})}\right\}\right)\right]^{2} \cdot a_{j}^{2} \nonumber \\
    & \geq \lim_{\Delta_{j}\rightarrow 0} \left[\mathbb{E}\left(\mathbb{E}\left\{\mathbb{E}\left(\exp\left\{-\frac{\Delta_{j}}{2} \sum_{c\in\mathcal{C}} \left(\frac{m}{b|\mathcal{C}|}\right)^{2}(\bm{x}_{j}^{(c)}-\bm{a}_{c})^{\intercal}\mathbf{\Lambda}_{c}^{-1}(\bm{x}_{j}^{(c)}-\bm{a}_{c})\right\} \middle| \bm{\xi}_{j}, \vecX{j-1}{(\mathcal{C})} \right) \middle| \vecX{j-1}{(\mathcal{C})}\right\}\right)\right]^{2} \nonumber \\
    & \geq \left[\mathbb{E}\left(\mathbb{E}\left\{\lim_{\Delta_{j}\rightarrow 0}\mathbb{E}\left(\exp\left\{-\frac{\Delta_{j}}{2} \sum_{c\in\mathcal{C}} \left(\frac{m}{b|\mathcal{C}|}\right)^{2}(\bm{x}_{j}^{(c)}-\bm{a}_{c})^{\intercal}\mathbf{\Lambda}_{c}^{-1}(\bm{x}_{j}^{(c)}-\bm{a}_{c})\right\} \middle| \bm{\xi}_{j}, \vecX{j-1}{(\mathcal{C})} \right) \middle| \vecX{j-1}{(\mathcal{C})}\right\}\right)\right]^{2},
\end{align*}
(by using a trapezoidal rule approximation of the integral and exploiting the use of small intervals) where $\lim_{\Delta_{j}\rightarrow 0}$ and expectations are exchanged using the dominated convergence theorem (as the exponential term is bounded above by $1$ and its expectation exists \citep[Appendix C]{dai_et_al_2023}).

From \eqref{eq:GBF_propagate_mod} in \cororef{corollary:GBF_propagate_mod}, we note that $\bm{x}_{j}^{(c)}$ only depends $\bm{x}_{j-1}^{(c)}$ through $\bm{\xi}_{j}$ and $\bm{\zeta}_{j}^{(c)}$ for all $c\in\mathcal{C}$, and we have
\begin{equation*}
    \left. \bm{x}_{j}^{(c)} \middle| \bm{\xi}_{j}, \vecX{j-1}{(\mathcal{C})} \right. \sim \mathcal{N}_{d}\left(\mathbb{E}\left[\bm{x}_{j}^{(c)} \middle| \bm{\xi}_{j},  \vecX{j-1}{(\mathcal{C})}\right], \frac{T-t_{j}}{T-t_{j-1}} \Delta_{j} \mathbf{\Lambda}_{c}\right),
\end{equation*}
and consequently, 
\begin{equation*}
    \left( \frac{T-t_{j}}{T-t_{j-1}}\Delta_{j} \right)^{-1} \sum_{c\in\mathcal{C}} (\bm{x}_{j}^{(c)}-\bm{a}_{c})^{\intercal}\mathbf{\Lambda}_{c}^{-1}(\bm{x}_{j}^{(c)}-\bm{a}_{c}) \sim \chi^{2}(|\mathcal{C}|d, \lambda_{j}^{\prime}),
\end{equation*}
with moment generating function $M_{j}(s):=\exp\left\{\frac{\lambda_{j}^{\prime}s}{1-2s}\right\}\cdot(1-2s)^{-\frac{|\mathcal{C}|d}{2}}$, where
\begin{align*}
    \lambda_{j}^{\prime}
    & = \left(\frac{T-t_{j}}{T-t_{j-1}}\Delta_{j}\right)^{-1} \sum_{c\in\mathcal{C}} \left(\mathbb{E}\left[\bm{x}_{j}^{(c)}\middle|\bm{\xi}_{j},\vecX{j-1}{(\mathcal{C})}\right]-\bm{a}_{c}\right)^{\intercal}\mathbf{\Lambda}_{c}^{-1}\left(\mathbb{E}\left[\bm{x}_{j}^{(c)}\middle|\bm{\xi}_{j},\vecX{j-1}{(\mathcal{C})}\right]-\bm{a}_{c}\right) \\
    & = \left(\frac{T-t_{j}}{T-t_{j-1}}\Delta_{j}\right)^{-1} |\mathcal{C}| \sigma_{t_{j}}^{2},
\end{align*}
with
\begin{equation*}
    \sigma_{t_{j}}^{2} := \frac{1}{|\mathcal{C}|} \sum_{c\in\mathcal{C}} \left(\mathbb{E}\left[\bm{x}_{j}^{(c)}\middle|\bm{\xi}_{j},\vecX{j-1}{(\mathcal{C})}\right]-\bm{a}_{c}\right)^{\intercal}\mathbf{\Lambda}_{c}^{-1}\left(\mathbb{E}\left[\bm{x}_{j}^{(c)}\middle|\bm{\xi}_{j},\vecX{j-1}{(\mathcal{C})}\right]-\bm{a}_{c}\right).
\end{equation*}
Letting $s=-\frac{1}{2}\left(\frac{m}{b|\mathcal{C}|}\right)^{2}\left(\frac{T-t_{j}}{T-t_{j-1}}\right)\Delta_{j}^{2}$, then
\begin{align*}
    \lim_{\Delta_{j}\rightarrow 0} \lim_{N\rightarrow\infty} N^{-1}\CESS{j}
    & \geq \left[\mathbb{E}\left(\mathbb{E}\left\{\lim_{\Delta_{j}\rightarrow 0} \exp\left\{\frac{\lambda_{j}^{\prime}s}{1-2s}\right\} \middle| \vecX{j-1}{(\mathcal{C})} \right\}\right)\right]^{2} \cdot (1-2s)^{-|\mathcal{C}|d} \\
    & \geq \left[\mathbb{E}\left(\mathbb{E}\left\{\lim_{\Delta_{j}\rightarrow 0} \exp\left\{\frac{-\frac{1}{2}\left(\frac{m^{2}}{b^{2}C}\right)\sigma_{t_{j}}^{2}\Delta_{j}}{1-2s}\right\} \middle| \vecX{j-1}{(\mathcal{C})} \right\}\right)\right]^{2} \cdot (1-2s)^{-|\mathcal{C}|d}.
\end{align*}
From \eqref{eq:GBF_propagate_mod}, we have
\begin{equation*}
    \mathbb{E}\left[\bm{x}_{j}^{(c)}\middle|\bm{\xi}_{j},\vecX{j-1}{(\mathcal{C})}\right] = \left[\frac{\Delta_{j}^{2}}{T-t_{j-1}}\right]^{\frac{1}{2}} \bm{\xi}_{j} + \frac{T-t_{j}}{T-t_{j-1}} \bm{x}_{j-1}^{(c)} + \frac{t_{j}-t_{j-1}}{T-t_{j-1}} \tilde{\bm{x}}_{j-1},
\end{equation*}
and so we have $\lim_{\Delta_{j}\rightarrow 0} \sigma_{t_{j}}^{2}=:\nu_{j}$ where $\nu_{j}$ is given in \eqref{eq:nu_j}.
Using Jensen's inequality, we can get
\begin{align}
    \lim_{\Delta_{j}\rightarrow 0} \lim_{N\rightarrow\infty} N^{-1}\CESS{j} 
    & \geq \lim_{\Delta_{j}\rightarrow 0} \left[ \exp\left\{\frac{-\frac{1}{2}\mathbb{E}\left[\nu_{j}\right]\left(\frac{m^{2}}{b^{2}|\mathcal{C}|}\right)\Delta_{j}}{1-2s}\right\} \right]^{2} \cdot (1-2s)^{-|\mathcal{C}|d} \nonumber \\
    & \geq \lim_{\Delta_{j}\rightarrow 0} \exp\left\{\frac{-\mathbb{E}\left[\nu_{j}\right]\left(\frac{m^{2}}{b^{2}|\mathcal{C}|}\right)\Delta_{j}}{1-2s}\right\} \cdot (1-2s)^{-|\mathcal{C}|d}. \label{eq:CESS_j}
\end{align}
Consider the first term in \eqref{eq:CESS_j}, then taking the limit $\Delta_{j} \rightarrow 0$ implies that $s \rightarrow 0$, and if $\Delta_{j} \leq \frac{b^{2}|\mathcal{C}|k_{3}}{\mathbb{E}\left[\nu_{j}\right]m^{2}}$ for some $k_{3}>0$, then
\begin{equation}
    \exp\left\{\frac{-\mathbb{E}\left[\nu_{j}\right]\left(\frac{m^{2}}{b^{2}|\mathcal{C}|}\right)\Delta_{j}}{1-2s}\right\} \geq \exp\left\{-k_{3}\right\}. \label{eq:mesh_guidance_first_term_bound}
\end{equation}
Similarly for the second term in \eqref{eq:CESS_j}, if $\Delta_{j} \leq \left(\frac{b^{2}|\mathcal{C}|k_{4}}{2m^{2}d}\right)^{\frac{1}{2}}$, we have
\begin{align}
    (1-2s)^{-|\mathcal{C}|d}
    & \geq \exp\left\{4s|\mathcal{C}|d\right\} \nonumber \\
    & = \exp\left\{4|\mathcal{C}|d\left(-\frac{1}{2}\left(\frac{m}{b|\mathcal{C}|}\right)^{2}\left(\frac{T-t_{j}}{T-t_{j-1}}\right)\Delta_{j}^{2}\right)\right\} \nonumber \\
    & = \exp\left\{-2\left(\frac{m^{2}}{b^{2}|\mathcal{C}|}\right)d\Delta_{j}^{2}\right\} \geq \exp\left\{-k_{4}\right\}. \label{eq:mesh_guidance_second_term_bound}
\end{align}

Combining the bounds in \eqref{eq:mesh_guidance_first_term_bound} and \eqref{eq:mesh_guidance_second_term_bound}, and taking the limit $\Delta_{j}\rightarrow 0$ over sequences of $t_{j}-t_{j-1}\rightarrow 0$, with \eqref{eq:mesh_guidance}, we arrive at the result given in the theorem. \hfill $\blacksquare$

\proof (\propositionref{prop:k4_choice} Using \thmref{theorem:mesh_guidance}, then for iteration $j$, we want to choose $\exp\{-k_{3,j}-k_{4,j}\}=\zeta^{\prime}\in(0,1)$, and so $k_{3,j}=-\log(\zeta^{\prime})-k_{4,j}$. By substituting this into \eqref{eq:mesh_guidance}, we can choose the mesh size as
\begin{equation}
    \label{eq:mesh_guidance_one_parameter}
    \tilde{\Delta}_{j} = \min\left\{ \frac{b^{2}|\mathcal{C}|[-\log(\zeta^{\prime})-k_{4,j}]}{\mathbb{E}[\nu_{j}]m^{2}}, \left(\frac{b^{2}|\mathcal{C}|k_{4,j}}{2m^{2}d}\right)^{\frac{1}{2}}\right\},
\end{equation}
where $k_{4,j}<-\log(\zeta^{\prime})$ (in order to ensure that $k_{3,j}>0$). Here, we want the largest interval which satisfies $N^{-1}\CESS{j}\geq\zeta^{\prime}$. This corresponds to choosing $k_{4,j}$ with
\begin{align}
    \frac{b^{2}|\mathcal{C}|[-\log(\zeta^{\prime})-k_{4,j}]}{\mathbb{E}[\nu_{j}]m^{2}} & = \left(\frac{b^{2}|\mathcal{C}|k_{4,j}}{2m^{2}d}\right)^{\frac{1}{2}} \nonumber \\
    \implies \frac{b^{4}|\mathcal{C}|^{2}[-\log(\zeta^{\prime})-k_{4,j}]^{2}}{\mathbb{E}[\nu_{j}]^{2}m^{4}} & = \frac{b^{2}|\mathcal{C}|k_{4,j}}{2m^{2}d} \nonumber \\
    \implies [-\log(\zeta^{\prime})-k_{4,j}]^{2} & = \frac{\mathbb{E}[\nu_{j}]^{2}m^{2}}{2b^{2}|\mathcal{C}|d}k_{4,j} \nonumber \\
    \implies \log(\zeta^{\prime})^{2}+2k_{4,j}\log(\zeta^{\prime})+k_{4,j}^{2} & = \frac{\mathbb{E}[\nu_{j}]^{2}m^{2}}{2b^{2}|\mathcal{C}|d}k_{4,j} \nonumber \\
    \implies k_{4,j}^{2} + \left(2\log(\zeta^{\prime})-\frac{\mathbb{E}[\nu_{j}]^{2}m^{2}}{2b^{2}|\mathcal{C}|d}\right)k_{4,j}+\log(\zeta^{\prime})^{2} & = 0. \label{eq:k4_quadratic}
\end{align}
Applying the quadratic formula to solve \eqref{eq:k4_quadratic} gives
\begin{equation*}
    k_{4,j} = \frac{\left(\frac{\mathbb{E}[\nu_{j}]^{2}m^{2}}{2b^{2}|\mathcal{C}|d}-2\log(\zeta^{\prime})\right) \pm \sqrt{\left(2\log(\zeta^{\prime})-\frac{\mathbb{E}[\nu_{j}]^{2}m^{2}}{2b^{2}|\mathcal{C}|d}\right)^{2}-4\log(\zeta^{\prime})^{2}}}{2}.
\end{equation*}
Note that we have the constraints that $0<k_{4,j}<-\log(\zeta^{\prime})$, and since from \eqref{eq:k4_quadratic}, we have
\begin{equation*}
    k_{4,j}^{2}+\left(2\log(\zeta^{\prime})-\frac{\mathbb{E}[\nu_{j}]^{2}m^{2}}{2b^{2}|\mathcal{C}|d}\right)k_{4,j}=-\log(\zeta^{\prime})^{2},
\end{equation*}
then we will always choose the smaller root and arrive at the statement of the theorem. \hfill $\blacksquare$

\section{Practical implementation considerations} \label{app:practical_considerations}

In many practical settings there will be additional constraints which require us to modify \algoref{alg:GBF} appropriately. Examples include settings where latency between cores is problematic, or in scenarios where functional evaluations of the sub-posterior densities $f_{c}$ are not available. In this section, consider several modifications to \algoref{alg:GBF} to make it more amenable to certain application areas. To clarify, the implementation of our methodology in examples presented in \secref{sec:dc_gbf_examples} do not exploit these modifications that we present below.

\subsection{Reducing communication between the cores} \label{sec:GBF_guidance:communication}

For our \gbfa approach, we highlight two steps where communication between cores could be reduced. In particular, it is possible to limit the amount of communication necessary when initialising the particle set, and also when we propagate the particles in the iterative steps of the algorithm. In a distributed/parallel setting, it is desirable to reduce the number of communication between cores since there is a latency penalty for each communication leading to a more computationally expensive algorithm.

In \algstepref{alg:GBF}{alg:GBF:initial_weights}, the particles are composed by pairing the sub-posterior draws index-wise to obtain $\{\vecX{0,i}{(\mathcal{C})}\}_{i=1}^{M}$ which requires a communication between the cores. To fully initialise the algorithm, we must assign importance weights to the particles which requires an additional two communications between the cores; namely a communication back to the individual cores to provide the weighted mean of the particles $\tilde{\bm{x}}_{0,i}$, and a communication between the cores to compute $\rho_{0,i}(\vecX{0,\cdot}{(\mathcal{C})})$ (since \eqref{eq:rho_0_gbf} can be decomposed into a product of $|\mathcal{C}|$ terms corresponding to the individual contributions from each sub-posterior. Following the approach of \citet[Section 3.7.1]{dai_et_al_2023}, let $\tilde{\theta}\in\mathbb{R}^{d}$ be a weighted average of approximate modes (or means) of each sub-posterior. Noting that this can be computed in a single pre-processing step prior to initialisation, then we can modify the proposal mechanism for the initial draw to be from the density
\begin{equation}
    \label{eq:f_c_tilde_GBF}
    \tilde{f}_{c}\left(\bm{x}_{0}^{(c)}\right) \propto \exp\left\{-\frac{(\bm{x}_{0}^{(c)}-\tilde{\bm{\theta}})^{\intercal}\mathbf{\Lambda}_{c}^{-1}(\bm{x}_{0}^{(c)}-\tilde{\bm{\theta}})}{2T}\right\} \cdot f_{c}\left(\bm{x}_{0}^{(c)}\right),
\end{equation}
then by modifying the algorithm by replacing $\rho_{0}$ with
\begin{equation}
    \label{eq:varrho_0_GBF}
    \tilde{\varrho}_{0}:=\exp\left\{\frac{(\tilde{\bm{x}}_{0}^{(\mathcal{C})}-\tilde{\bm{\theta}})^{\intercal}\mathbf{\Lambda}_{\mathcal{C}}^{-1}(\tilde{\bm{x}}_{0}^{(\mathcal{C})}-\tilde{\bm{\theta}})}{2T}\right\},
\end{equation}
where $\mathbf{\Lambda}_{\mathcal{C}}^{-1}:=(\sum_{c\in\mathcal{C}}\mathbf{\Lambda}_{c}^{-1})$, we can see that
\begin{equation*}
    \tilde{\varrho}_{0}\left(\vecX{0}{(\mathcal{C})}\right) \cdot \prod_{c\in\mathcal{C}} \tilde{f}_{c}\left(\bm{x}_{0}^{(c)}\right) \propto \rho_{0}\left(\vecX{0}{(\mathcal{C})}\right) \cdot \prod_{c\in\mathcal{C}} f_{c}\left(\bm{x}_{0}^{(c)}\right).
\end{equation*}

Since we subsequently re-normalise the importance weights, we do not need to compute any constant of proportionality for $\tilde{\varrho}_{0}$. Adopting this approach means that we can sample from $\tilde{f}_{c}$ on each core independently and evaluate the modified importance weight without any further communication between the cores. This therefore reduces the number of communications required to initialise the particle set from three (in the original formulation) to two (since this approach does require one communication in order to compute $\tilde{\theta}$). The modified initialisation is summarised in \algoref{alg:GBF_initialisation_mod}.

\begin{algorithm}
    \caption{Particle set initialisation modification (to replace \algstepref{alg:GBF}{alg:GBF:initial_weights}).}
    \label{alg:GBF_initialisation_mod}
    1(b) For $k$ in $1$ to $M$,
    \begin{enumerate}[(i)]
        \item \textbf{$\vecX{0,k}{(\mathcal{C})}$}: For $c\in\mathcal{C}$, simulate $\bm{x}_{0,k}^{(c)} \sim \tilde{f}_{c}$ \eqref{eq:f_c_tilde_GBF}. Set $\vecX{0,k}{(\mathcal{C})}:=(\bm{x}_{0,k}^{(c_{1})},\dots,\bm{x}_{0,k}^{(c_{|\mathcal{C}|})})$.
        \item Compute un-normalised weight
        $w^{(\mathcal{C})\prime}_{0,k} := \big(\prod_{c\in\mathcal{C}} w_{k}^{(c)}\big) \cdot \tilde{\varrho}_{0}(\vecX{0,k}{(\mathcal{C})})$ as per \eqref{eq:varrho_0_GBF}.
    \end{enumerate}
\end{algorithm}

There is also scope to reduce the number of communications required to propagate the particle set in \algstepref{alg:GBF}{alg:GBF:propagate}. To propagate the particles, there is a communication between the cores in order to compute $\vecM{j}{(\mathcal{C})}:=\vecM{t_{j-1},t_{j}}{(\mathcal{C})}$ as per \eqref{eq:M_GBF} since this requires the current position of each of the $|\mathcal{C}|$ trajectories. Once we have computed this and propagated the samples, a further communication back to the cores would be necessary so that each core can compute their contribution to the $\tilde{\rho}_{j}$ importance weight. Alternatively, we can utilise \cororef{corollary:GBF_propagate_mod} so that each of the $|\mathcal{C}|$ processes can propagate their own individual particles to compose $\vecX{j}{(\mathcal{C})}$.

\begin{corollary}
    \label{corollary:GBF_propagate_mod}
    Simulating $\vecX{j}{(\mathcal{C})}\sim\mathcal{N}_{d}\left(\vecM{j}{(\mathcal{C})}, \bm{V}_{j}\right)$, the required transition from $\vecX{j-1}{(\mathcal{C})}$ to $\vecX{j}{(\mathcal{C})}$ in \algstepref{alg:GBF}{alg:GBF:propagate}, can be expressed as
    \begin{equation}
        \label{eq:GBF_propagate_mod}
        \bm{x}_{j}^{(c)} = \left[\frac{\Delta_{j}^{2}}{T-t_{j-1}}\right]^{\frac{1}{2}} \bm{\xi}_{j} + \left[\frac{T-t_{j}}{T-t_{j-1}}\Delta_{j}\right]^{\frac{1}{2}} \bm{\eta}_{j}^{(c)} + \bm{M}_{j}^{(c)},
    \end{equation}
    where $\bm{\xi}_{j} \sim \mathcal{N}_{d}(\bm{0}, \mathbf{\Lambda}_{\mathcal{C}})$, $\bm{\eta}_{j}^{(c)} \sim \mathcal{N}_{d}(\bm{0}, \mathbf{\Lambda}_{c})$ and $\bm{M}_{j}^{(c)}$ is the sub-vector of $\vecM{j}{(\mathcal{C})}$ corresponding to the $c$th component given by \eqref{eq:M_GBF}.
\end{corollary}

\proof From \propositionref{prop:proposal_simulation_GBF:a}, we have $\vecX{j}{(\mathcal{C})}\sim\mathcal{N}_{d}\left(\vecM{j}{(\mathcal{C})}, \bm{V}_{j}\right)$ where $\vecM{j}{(\mathcal{C})}:=\vecM{t_{j-1},t_{j}}{(\mathcal{C})}$ is given by \eqref{eq:M_GBF} and $\bm{V}_{j}:=\bm{V}_{t_{j-1},t_{j}}$ is given by \eqref{eq:V_GBF}.
From \eqref{eq:GBF_propagate_mod}, the mean and covariance matrix of $\vecX{j}{(\mathcal{C})}$ given $\vecX{j}{(\mathcal{C})}$ are also given by $\vecM{j}{(\mathcal{C})}$ and $\bm{V}_{j}$ as required. \hfill $\blacksquare$

By using \cororef{corollary:GBF_propagate_mod}, we can see that the interaction between the $|\mathcal{C}|$ trajectories occurs through their weighted mean $\tilde{\bm{x}}_{j-1}$ at the previous iteration. This can be computed at the previous iteration, and we can communicate this along with the common Gaussian vector $\bm{\xi}_{j}$ at the same time. This therefore removes an unnecessary additional communication between the cores at every iteration, resulting in a much more efficient algorithm if latency is a concern. This approach is presented in \algoref{alg:GBF_propagate_mod}.
\begin{algorithm}
    \caption{Particle set propagation modification (to replace \algstepref{alg:GBF}{alg:GBF:propagate}).}
    \label{alg:GBF_propagate_mod}
    2(b)i.
    \begin{enumerate}[(A)]
        \item For $c\in\mathcal{C}$, simulate $\bm{x}_{j,i}^{(c)}|(\tilde{\bm{x}}_{j-1,i}, \bm{x}_{j-1,i}^{(c)})$ in \eqref{eq:GBF_propagate_mod}.
        \item  Set $\vecX{j,i}{(\mathcal{C})}:=(\bm{x}_{j,i}^{(c_{1})},\dots,\bm{x}_{j,i}^{(c_{|\mathcal{C}|})})$ and compute $\tilde{\bm{x}}_{j,i}:=(\sum_{c\in\mathcal{C}}\mathbf{\Lambda}_{c}^{-1})^{-1}(\sum_{c\in\mathcal{C}}\mathbf{\Lambda}_{c}^{-1}\bm{x}_{j,i}^{(c)})$.
    \end{enumerate}
\end{algorithm}

\subsection{Alternative methods for updating the particle set weights} \label{sec:GBF_guidance_unbiased_estimators}

In this paper, we have assumed that we have been able to compute functionals of each sub-posterior $f_{c}$ for $c\in\mathcal{C}$, however there are many settings where it may be impractical or infeasible to do so. This may be case if there is some form of intractability of the sub-posteriors (see for instance \citet{Andrieu_Roberts_2009}), or maybe the evaluation of such quantities may be simply too computationally expensive (for instance in large data settings \citep{Pollock_et_al_2020, Bouchard_et_al_2018, Bierkens_et_al_2019, dai_et_al_2023}). In these settings, we no longer are able to evaluate $\phi_{c}$ in \eqref{eq:phi} which is necessary to update the particle weights in the iterative steps of the BF algorithm. However, it is possible to consider alternative unbiased estimators for $\tilde{\rho}_{j}$ in \stepref{alg:GBF:update_weight}.
\begin{corollary}
    \label{corollary:GBF_alt_unbiased_estimator}
    \citep[Corollary 3]{dai_et_al_2023}
    The estimator
    \begin{equation}
        \tilde{\varrho}_{j} \left(\vecX{j-1}{(\mathcal{C})}, \vecX{j}{(\mathcal{C})}\right) := \prod_{c\in\mathcal{C}} \left( \frac{\Delta_{j}^{\kappa_{c}} \cdot e^{-\bar{U}_{X}^{(c)}\Delta_{j}}}{\kappa_{c}! \cdot p \left( \kappa_{c} | R_{c} \right)} \prod_{k_{c}=1}^{\kappa_{c}} \left(\bar{U}_{X}^{(c)} - \tilde{\phi}_{c} \left( \bm{X}_{\xi_{c, k_{c}}}^{(c)} \right) \right) \right),
    \end{equation}
    where $\tilde{\phi}_{c}$ is an unbiased estimator of $\phi_{c}$ and $\bar{U}_{j}^{(c)}$ is a constant such that $\tilde{\phi}_{c}(\bm{x}) \leq \bar{U}_{j}^{(c)}$ for $\bm{x} \in R_{c}$.
\end{corollary}

\proof This follows directly from \thmref{theorem:unbiased_estimator_GBF}. \hfill $\blacksquare$

The estimator $\tilde{\varrho}_{j}$ in \cororef{corollary:GBF_alt_unbiased_estimator} can therefore be used as a substitute for $\tilde{\rho}_{j}$ in \algstepref{alg:GBF}{alg:GBF:update_weight}. However, we must be careful in constructing $\tilde{\varrho}_{j}$ since its introduction typically increases the variance of the estimator which ultimately causes higher variance in the particle set weights in the BF algorithm. In particular, by using \cororef{corollary:GBF_alt_unbiased_estimator}, the number of expected functional evaluations will change from $K$ to $K^{\prime}$ and so we must consider the growth in the ratio $K^{\prime}/K$ as $m_{c}\rightarrow\infty$ \citep{Pollock_et_al_2020, dai_et_al_2023}. However, as noted, introducing an alternative unbiased estimator may be necessary to apply the BF approach to some settings.

For instance, consider the example setting provided in \citet[Appendix E]{dai_et_al_2023}, where we have a large number of data points associated to each sub-posterior (i.e we have $m_{c}\gg1$ data points for core $c\in\mathcal{C}$) then computing $\phi_{c}$ in \eqref{eq:phi} is an expensive $\mathcal{O}(m_{c})$ operation. However, since $\phi_{c}$ is linear in terms terms of $\nabla \log f_{c}(\bm{x})$ and $\nabla^{2} \log f_{c}(\bm{x})$, it is simple to construct an unbiased estimator $\tilde{\phi}_{c}$ for $\phi_{c}$. In the setting, we also assume the sub-posteriors admit a structure with conditional independence and can be factorised as follows,
\begin{equation}
    f_{c}(\bm{x}) \propto \prod_{i=1}^{m_{c}} l_{i,c}(\bm{x}).
\end{equation}

Then since $\phi_{c}$ is linear in terms of $\nabla \log l_{i,c}(\bm{x})$ and $\nabla \log l_{i,c}(\bm{x})$, then we could use the following naive unbiased estimator for $\phi_{c}^{dl}$:
\begin{equation}
    \label{eq:tilde_phi_c_naive_GBF}
    \tilde{\phi}_{c}(\bm{x}) = \frac{m_{c}}{2} \left( \nabla \log l_{I,c}(\bm{x}^{*})^{\intercal} \mathbf{\Lambda}_{c} \nabla \log l_{J,c}(\bm{x}^{*}) + \Tr(\mathbf{\Lambda}_{c} \nabla^{2} \log l_{I,c}(\bm{x}^{*})) \right),
\end{equation}
where $I,J\overset{\iid}{\sim}\mathcal{U}\{1,\dots,m_{c}\}$. Although using such an estimator has the advantage of having $\mathcal{O}(1)$ cost when evaluating, this comes at the cost of an $\mathcal{O}(m_{c})$ inflation in the expected number of evaluations when evaluating $\tilde{\varrho}_{j}$ over $\tilde{\rho}_{j}$. However, following the approach of \citet[Section 4]{Pollock_et_al_2020} and \citet[Appendix E]{dai_et_al_2023}, we first want to suitable choose some \emph{control variates} to construct our estimator, and compute $\nabla \log f_{c}$ and $\nabla^{2} \log f_{c}$ at points \emph{close} to either the mode of the sub-posterior, $\hat{\bm{x}}_{c}$, or the mode of the target posterior $\hat{\bm{x}}$ (where close means within $\mathcal{O}(m_{c}^{-\frac{1}{2}})$ of the true respective modes). Computing these control variates will typically be one-time $\mathcal{O}(m_{c})$ computations.

Let 
\begin{align}
    \tilde{\alpha}_{I,c}(\bm{x}) & := n\cdot[\nabla \log l_{I,c}(\bm{x}) - \nabla \log l_{I,c}(\bm{x}^{*})], \\
    \tilde{H}_{I,c}(\bm{x}) & := n\cdot[\nabla^{2} \log l_{I,c}(\bm{x}) - \nabla^{2} \log l_{I,c}(\bm{x}^{*})],
\end{align}
then since $\log f_{c}(\bm{x}) = \sum_{i=1}^{m_{c}} \log l_{i,c}(\bm{x})$, we have
\begin{equation}
    \mathbb{E}_{\mathcal{A}}\left[ \tilde{\alpha}_{I,c}(\bm{x}) \right] = \alpha_{c}(\bm{x}), \quad \mathbb{E}_{\mathcal{A}}\left[ \tilde{H}_{I,c}(\bm{x}) \right] = H_{c}(\bm{x}).
\end{equation}
where $\alpha_{c}(\bm{x}):=\nabla \log f_{c}(\bm{x}) - \nabla \log f_{c}(\bm{x}^{*})$ and $H_{c}(\bm{x}):=\nabla^{2} \log f_{c}(\bm{x}) - \nabla^{2} \log f_{c}(\bm{x}^{*})$ and $\mathcal{A}$ is the law of $I\sim\mathcal{U}\{1,\dots,n\}$.

Noting that $\phi_{c}(\bm{x})$ in \eqref{eq:phi} can be re-expressed as
\begin{equation}
    \phi_{c}(\bm{x}) = \frac{1}{2} \left[ \alpha_{c}(\bm{x})^{\intercal}\mathbf{\Lambda}_{c}(2\nabla\log f_{c}(\bm{x}^{*}) + \alpha_{c}(\bm{x})) + \Tr(\mathbf{\Lambda}_{c}H_{c}(\bm{x})) \right] + C^{*},
\end{equation}
where $C^{*}:=\frac{1}{2} \left( \nabla \log f_{c}(\bm{x}^{*})^{\intercal} \mathbf{\Lambda}_{c} \nabla \log f_{c}(\bm{x}^{*}) + \Tr(\mathbf{\Lambda}_{c} \nabla^{2} \log f_{c}(\bm{x}^{*})) \right)$, then this leads to the following unbiased estimator for $\phi_{c}$:
\begin{equation}
    \tilde{\phi}_{c}(\bm{x}) := \frac{1}{2} \left[ \alpha_{I,c}(\bm{x})^{\intercal}(2\nabla\log f_{c}(\bm{x}^{*}) + \alpha_{J,c}(\bm{x})) + \Tr(\mathbf{\Lambda}_{c}\tilde{H}_{I,c}(\bm{x})) \right] + C^{*},
\end{equation}
where $I,J\overset{\iid}{\sim}\mathcal{U}\{1,\dots,m_{c}\}$, i.e.\ if now we let $\mathcal{A}$ be the law of $I,J\overset{\iid}{\sim}\mathcal{U}\{1,\dots,m_{c}\}$, we have $\mathbb{E}_{\mathcal{A}}\left[\tilde{\phi}_{c}(\bm{x})\right] = \phi_{c}(\bm{x})$.

Here the evaluations of the constants $\norm{\nabla \log f_{c}(\bm{x}^{*})}^{2}$, $\Tr(\nabla^{2} \log f_{c}(\bm{x}^{*}))$ are of $\mathcal{O}(m_{c})$ cost, but they only need to be computed once prior to calling \algoref{alg:GBF}. The unbiased estimator $\tilde{\phi}_{c}(\bm{x})$ uses only double draws from $\{1,\dots,m_{c}\}$, although \citet{Pollock_et_al_2020} notes that it would be possible to replace this by averaging over multiple draws (sampling from $\{1,\dots,m_{c}\}$ with replacement) which could have advantages of reducing the variance of the estimator at the cost of increasing the number of data points to evaluate at.


\section{Simulation studies} \label{app:dc_gbf_simulation_studies}

In this section we study empirically the performance of our Fusion algorithms (Sections \ref{sec:GBF} and \ref{sec:dc_gbf}), and selection of tuning parameters ($T$, $n$ and $\mathcal{P}$ as discussed in \secref{sec:GBF_guidance}) in our two idealised key settings---the $\SH{\lambda}$ setting (\condref{cond:SH}) and $\SSH{\gamma}$ setting (\condref{cond:SSH}) described in \secref{sec:GBF_guidance}. We do this in Sections \ref{subsec:GBF_similar_means} and \ref{subsec:GBF_dissimilar_means} respectively. For simplicity, here we focus on BF and \gbfa, noting that \gbfa is simply \hgbfa with a \emph{fork-and-join tree} hierarchy (as in \figref{fig:fork_and_join}). Finally, in \secref{subsec:GBF_dimension_study} we compare the performance of Fusion methodologies (including \hgbfa with a \emph{balanced-binary} tree hierarchy) with increasing dimensionality. In \secref{sec:dc_gbf_examples}, we consider more substantive examples using real data. Note that the earlier Bayesian Fusion approach is simply a special case of our \gbfa approach with $\mathbf{\Lambda}_{c}=\mathbb{I}_{d}$ for $c\in\mathcal{C}$, and so comparison with this work is straight-forward. 

To compare the performance of different approaches we consider their computational cost (both the total run time, and $n$ which represents the number of iterations of \algstepref{alg:GBF}{alg:GBF:iterations} and so is a proxy for the amount of communication between cores), and \emph{Integrated Absolute Distance (IAD)} defined in \eqref{eq:IAD}. 

Throughout this section we use the GPE-2 estimator of $\rho_{j}$ as given in \defnref{cond:GPE2}, and use the Trapezoidal rule to estimate the mean $\gamma_{c}$ in \eqref{eq:NB_mean} and set $\beta_{c}=10$ for $c\in\mathcal{C}$. Code to run these simulation studies can be found at \href{https://github.com/rchan26/DCFusion}{\texttt{https://github.com/rchan26/DCFusion}}.

\subsection{Sub-posterior Homogeneity} \label{subsec:GBF_similar_means}

We first study the guidance developed for $T$ and $\mathcal{P}$ in \secref{sec:GBF_guidance} for \gbfa (\algoref{alg:GBF}) in the $\SH{\lambda}$ setting of \condref{cond:SH}. Recall, this is the setting in which we are combining homogeneous sub-posteriors, and would naturally arise if a dataset was split randomly across several cores. To study this setting, we consider the idealised scenario of combining $C=10$ bi-variate Gaussian sub-posteriors, with a range of data sizes from $m=1000$ to $m=40000$, which have been randomly split across the $C=10$ cores. In particular, each sub-posterior has mean $\bm{0}=(0,0)$ and variance $\frac{C}{m}\mathbf{\Sigma}$, where $\mathbf{\Sigma}=\begin{pmatrix} 1 & \rho \\ \rho & 1 \end{pmatrix}$ with $\rho=0.9$. For this example, we apply both BF and \gbfa with a fixed particle set size of $N=10000$. 

To verify the guidance for $T$ and $\mathcal{P}$, we consider varying $T$ and $\mathcal{P}$ with increasing data size $m$, and the impact this has on $\CESS{0}$ and $\CESS{j}$ (for $j\in\{1,\dots,n\}$). We consider the four following choices of $T$ and $\mathcal{P}$:
\begin{enumerate}
    \item a fixed choice of $T$ and $n$ to obtain $\mathcal{P}$ (for \gbfa, $T=1$ and $n=5$, and for BF, $T=0.005$ and $n=5$),
    \item using the recommended $T$ from \secref{subsec:GBF_T_guidance} and fixed $n=5$ to obtain $\mathcal{P}$,
    \item using the recommended $T$ and $\mathcal{P}$ using a regular mesh (as outlined in \algoref{alg:regular_mesh} and \secref{subsubsec:regular_mesh}),
    \item using the recommended $T$ and $\mathcal{P}$ using an adaptive mesh (as outlined in \algoref{alg:adaptive_mesh} and \secref{subsubsec:adaptive_mesh}).
\end{enumerate}

In implementing the BF and \gbfa, we set our lower tolerable bounds for the initial ($\CESS{0}$) and the iterative ($\CESS{j}$) conditional effective sample sizes to be $0.5N$ (i.e.\ we set $\zeta=\zeta^{\prime}=0.5$), resampling if $\ESS$ falls below $0.5N$. To summarise how one might practically use our guidance to choose $T$ and $\mathcal{P}$, we present our approach in \remref{remark:GBF_guidance_similar_means}.

\begin{remark}
    \label{remark:GBF_guidance_similar_means}
    We set the tuning parameters for BF and \gbfa (for the $\SH{\lambda}$ setting of \secref{subsec:GBF_similar_means}) as follows:
    \begin{enumerate}
        \item Following the guidance outlined in \remref{remark:T_guidance_k1_k2}, and with $\zeta=0.5, \lambda = 1$ and $d=2$, we have $k_{1}=\sqrt{-\frac{(\lambda+\frac{d}{2})}{\log(\zeta)}} \approx 1.7$. For \gbfa, $\mathbf{\Lambda}_{c}$ is the estimated covariance matrices for sub-posterior $c\in\{1,\dots,C\}$, so $b=\frac{m}{C}$ (see \remref{remark:b_choice_GBF}), and we choose $T=C^{\frac{1}{2}}k_{1}$. For BF, $\mathbf{\Lambda}_{c}=\mathbb{I}_{d}$ for $c\in\{1,\dots,C\}$, so we have $b=1$ and so we choose $T=C^{3/2}k_{1}/m$.
        \item When using the regular mesh, we use \algoref{alg:regular_mesh} to obtain $\mathcal{P}$. First let $\zeta^{\prime}=0.5$ then for \gbfa we have $b=\frac{m}{C}$, and so $\Delta_{j}=\Delta=\sqrt{\frac{k_{4}}{2Cd}}$ for each $j$, where $k_{4}$ is computed as per \eqref{eq:k4_choice} and computing an estimate of the supremum of $\widehat{\mathbb{E}[\nu_{j}]}$ as per \eqref{eq:E_nu_j_max}. For BF, $b=1$, so $\Delta_{j}=\Delta=\sqrt{\frac{Ck_{4}}{2m^{2}d}}$ for each $j$.
        \item When using the adaptive mesh, we use \algoref{alg:adaptive_mesh} to obtain $\Delta_{j}$ recursively at each iteration to construct $\mathcal{P}$. We let $\zeta^{\prime}=0.5$ and for \gbfa (where $b=\frac{m}{C}$) we compute $t_{j}=\min\{T,t_{j-1}+\Delta_{j}\}$ where $\Delta_{j}=\sqrt{\frac{k_{4}}{2Cd}}$ at each iteration of \algoref{alg:GBF}, until we have $t_{j}=T$. For the standard BF approach, note that $b=1$ so we must compute $\Delta_{j}=\sqrt{\frac{Ck_{4}}{2m^{2}d}}$ instead at each iteration. 
    \end{enumerate}
\end{remark}

The conditional effective sample size of the \gbfa and (standard) BF approaches with increasing data size in this $\SH{\lambda}$ setting are shown in \figref{fig:bivG_similar_means}. 

First considering the results from fixing $T$ and $n$ in \figref{fig:bivG_similar_means_a}, we can see that BF lacks robustness with increasing data size. Here $\CESS{0}$ improves with increasing data size ($m$), which is due to the sub-posteriors becoming increasingly similar with $m$ in this idealised scenario. However, as we increase $m$ the fixed choice for $T$ (and hence the size of the intervals) becomes increasingly inappropriate for the sub-posteriors, which leads to a degradation in average $\CESS{j}$. In contrast, \gbfa incorporates global information about the sub-posteriors (i.e.\ the variance of the sub-posteriors), so there is no change in performance with $m$. Here there is a trade-off with the choice of $T$: a small $T$ leads to poor behaviour on initialisation (i.e.\ low $\CESS{0}$), but good behaviour at each iteration (i.e.\ high average $\CESS{j}$). 

Considering \figref{fig:bivG_similar_means_b}, we see that scaling $T$ following the guidance developed in \secref{subsec:GBF_T_guidance} immediately stabilises $\CESS{0}$, although $\CESS{j}$ performance is still poor ($n$ is too small). In \figref{fig:bivG_similar_means_c} and \figref{fig:bivG_similar_means_d}, we see that utilising both the guidance for $T$ and the mesh $\mathcal{P}$ drastically improves the performance of both BF and \gbfa. In both cases \gbfa outperforms BF: it achieves higher average $\CESS{j}$, and the variance of $\CESS{j}$ is lower. Given BF is a special case of \gbfa, this improvement can be ascribed to the use of estimated covariance matrices for $\mathbf{\Lambda}_{c}$. In particular, this choice leads to a lower variance unbiased estimator for $\rho_{j}$, and an improved proposal $h^{bf}$ \eqref{eq:h_GBF} for $g^{bf}$ \eqref{eq:g_GBF}.

From \figref{fig:bivG_similar_means_IAD} we see that with BF that without our guidance on $T$ and $\mathcal{P}$, average IAD is poor, and the variance of the IAD is very large. In contrast, \gbfa with our guidance is robust across the different scenarios. Comparing the regular and adaptive meshes simply using $\CESS{0}$ and $\CESS{j}$ would imply that the regular mesh is performing better (since it has slightly better $\CESS{j}$), however the adaptive mesh is slightly more computationally efficient as shown by having a smaller mesh size, $n$, (illustrated in \figref{fig:bivG_similar_means_mesh_size}) and having a faster algorithm run-time (illustrated in \figref{fig:bivG_similar_means_time}). By looking at the IAD obtained for these approaches, we can see that we are able to obtain similar performance at a lower cost with the adaptive mesh construction.

\begin{figure}[htp]
    \centering
    \begin{subfigure}[b]{\textwidth}
        \setcounter{subfigure}{0}
        \renewcommand\thesubfigure{\roman{subfigure}}
        \begin{subfigure}[b]{0.45\textwidth}
            \centering
            \includegraphics[width=\textwidth]{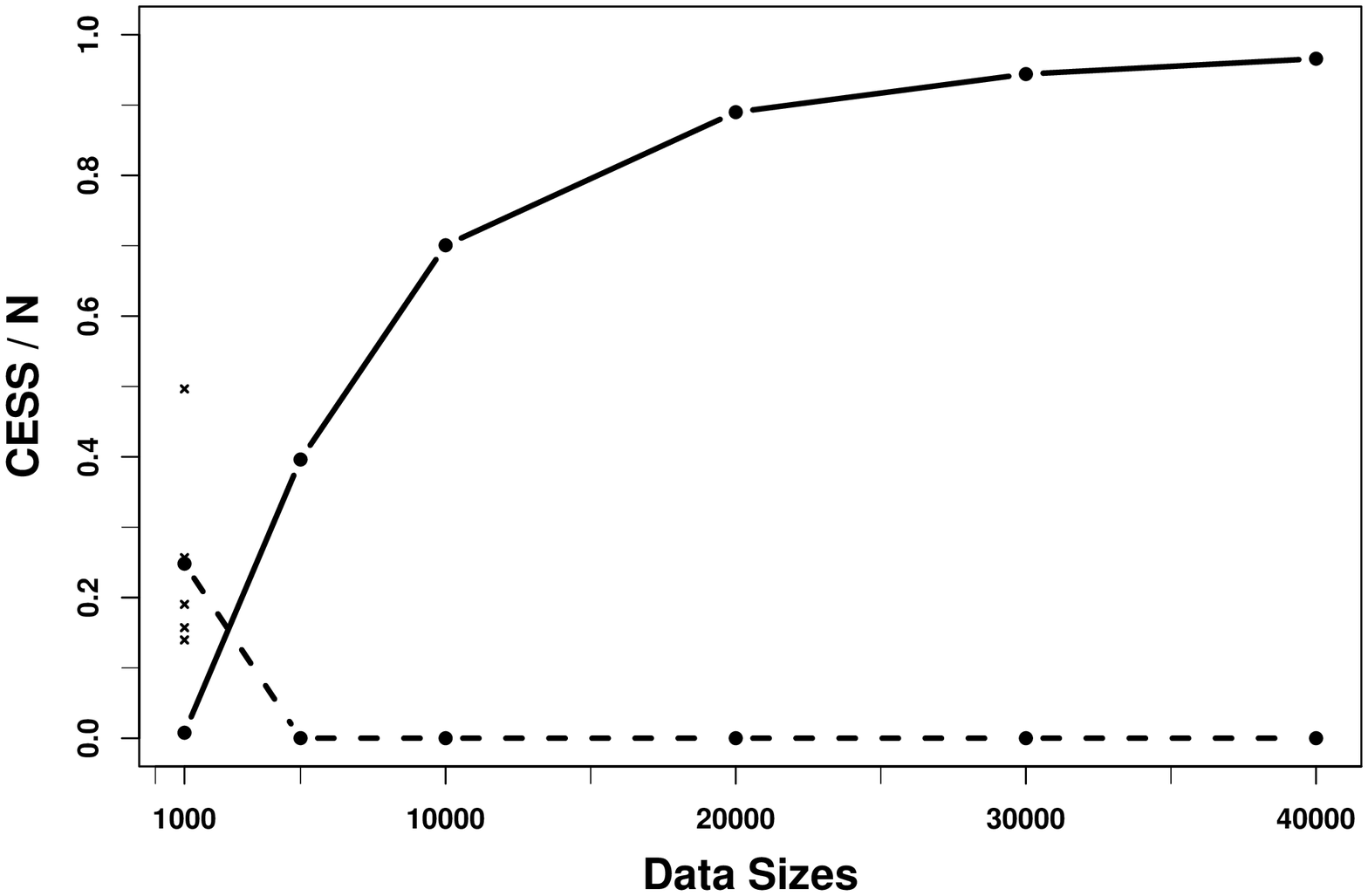}
            \caption{BF}
            \label{fig:bivG_similar_means_vanilla_a}
        \end{subfigure}%
        \hfill
        \begin{subfigure}[b]{0.45\textwidth}  
            \centering 
            \includegraphics[width=\textwidth]{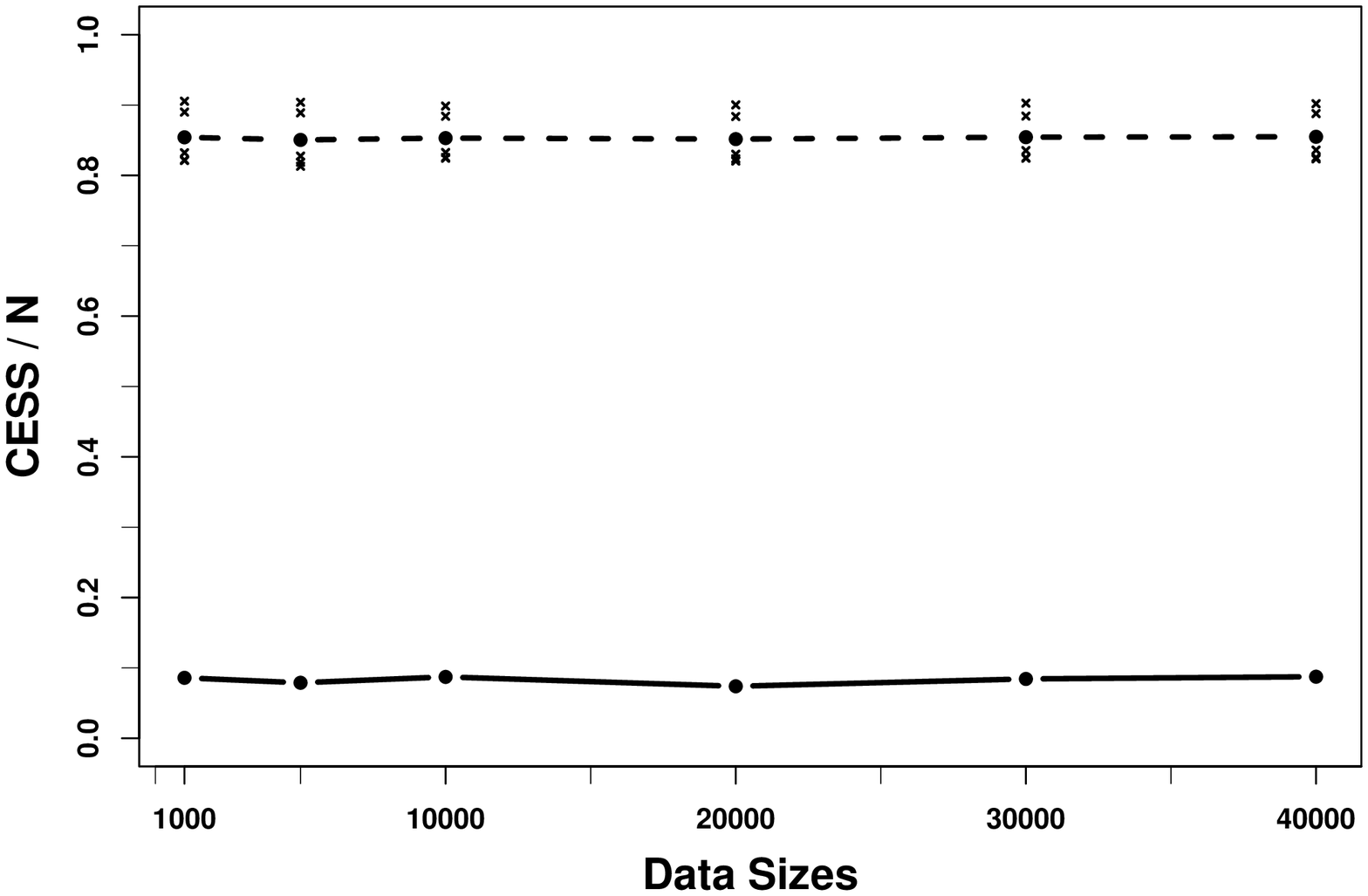}
            \caption{\gbfa}
            \label{fig:bivG_similar_means_generalised_a}
        \end{subfigure}%
        \setcounter{subfigure}{0}
        \renewcommand\thesubfigure{\alph{subfigure}}
        \caption{Fixed user-specified $T$ and $n$.}
        \label{fig:bivG_similar_means_a}
    \end{subfigure}%
    \vskip\baselineskip
    \begin{subfigure}[b]{\textwidth}
        \setcounter{subfigure}{0}
        \renewcommand\thesubfigure{\roman{subfigure}}
        \begin{subfigure}[b]{0.45\textwidth}
            \centering
            \includegraphics[width=\textwidth]{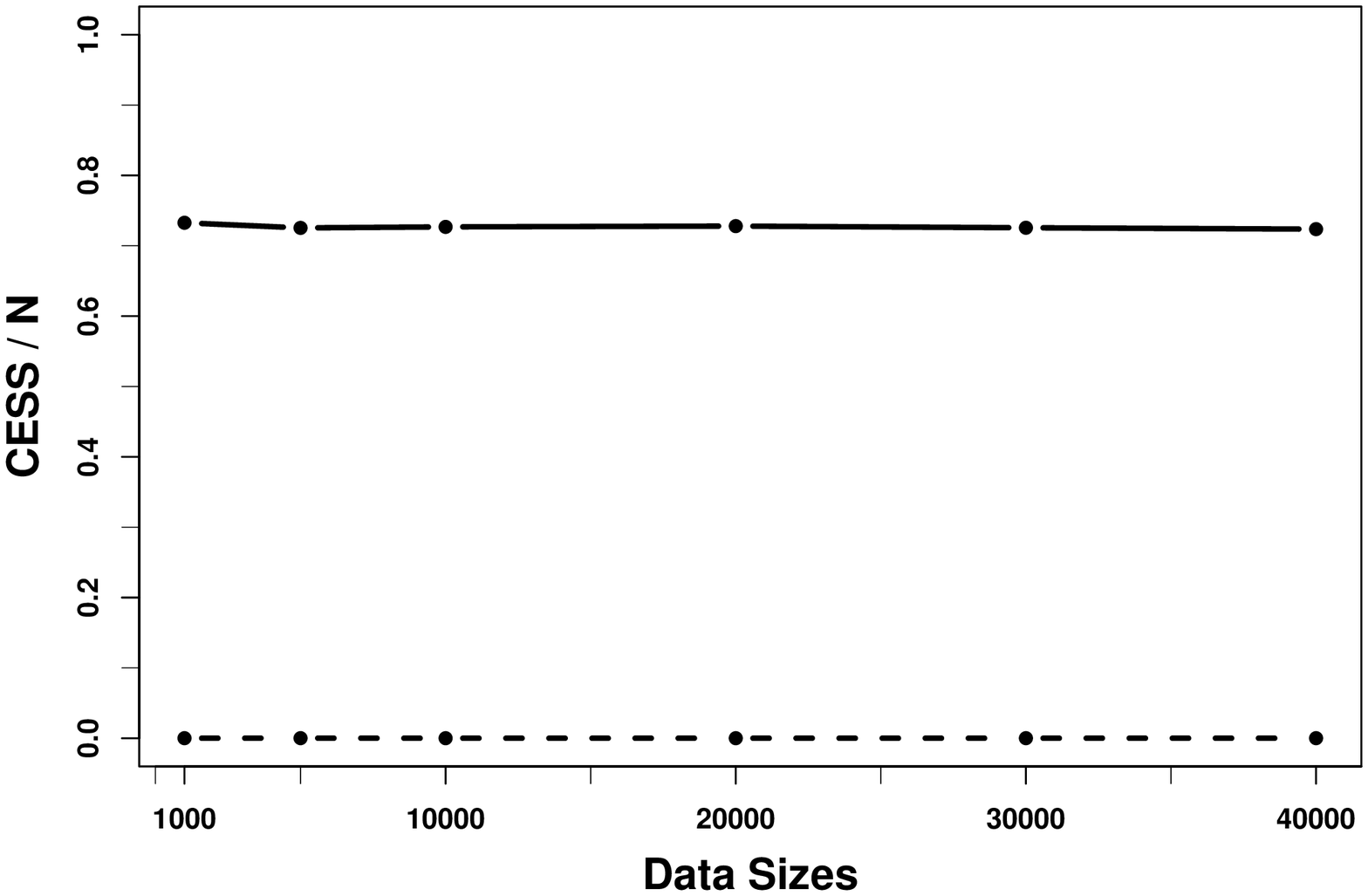}
            \caption{BF}
            \label{fig:bivG_similar_means_vanilla_b}
        \end{subfigure}%
        \hfill
        \begin{subfigure}[b]{0.45\textwidth}  
            \centering 
            \includegraphics[width=\textwidth]{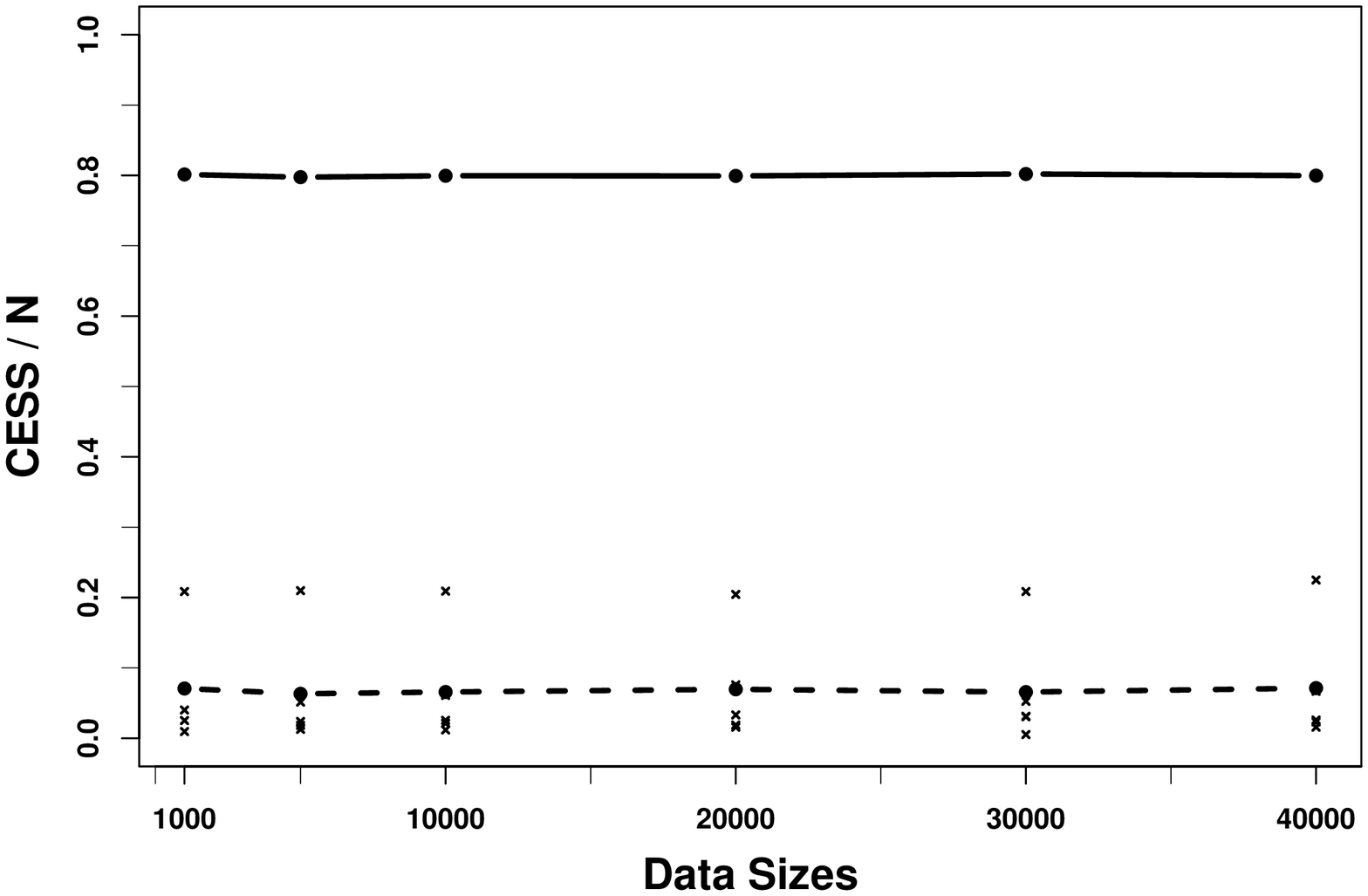}
            \caption{\gbfa}
            \label{fig:bivG_similar_means_generalised_b}
        \end{subfigure}%
        \setcounter{subfigure}{1}
        \renewcommand\thesubfigure{\alph{subfigure}}
        \caption{Recommended $T$ and fixed $n$.}
        \label{fig:bivG_similar_means_b}
    \end{subfigure}%

    \caption{Bivariate Gaussian example in $\SH{\lambda}$ setting with increasing data size. In Figures \ref{fig:bivG_similar_means_a}, \ref{fig:bivG_similar_means_b}, \ref{fig:bivG_similar_means_c}, \ref{fig:bivG_similar_means_d} solid lines denote initial CESS ($\text{CESS}_{0}$), and dotted lines denote averaged CESS in subsequent iterations $(\frac{1}{n}\sum_{j=1}^{n}\text{CESS}_{j})$, and crosses denote $\text{CESS}_{j}$ for each $j\in\{1,\dots,n\}$.}
    \label{fig:bivG_similar_means}    
\end{figure}

\begin{figure}[p]
    \centering
    \addtocounter{figure}{-1}
    \begin{subfigure}[b]{\textwidth}
        \setcounter{subfigure}{0}
        \renewcommand\thesubfigure{\roman{subfigure}}
        \begin{subfigure}[b]{0.45\textwidth}
            \centering
            \includegraphics[width=\textwidth]{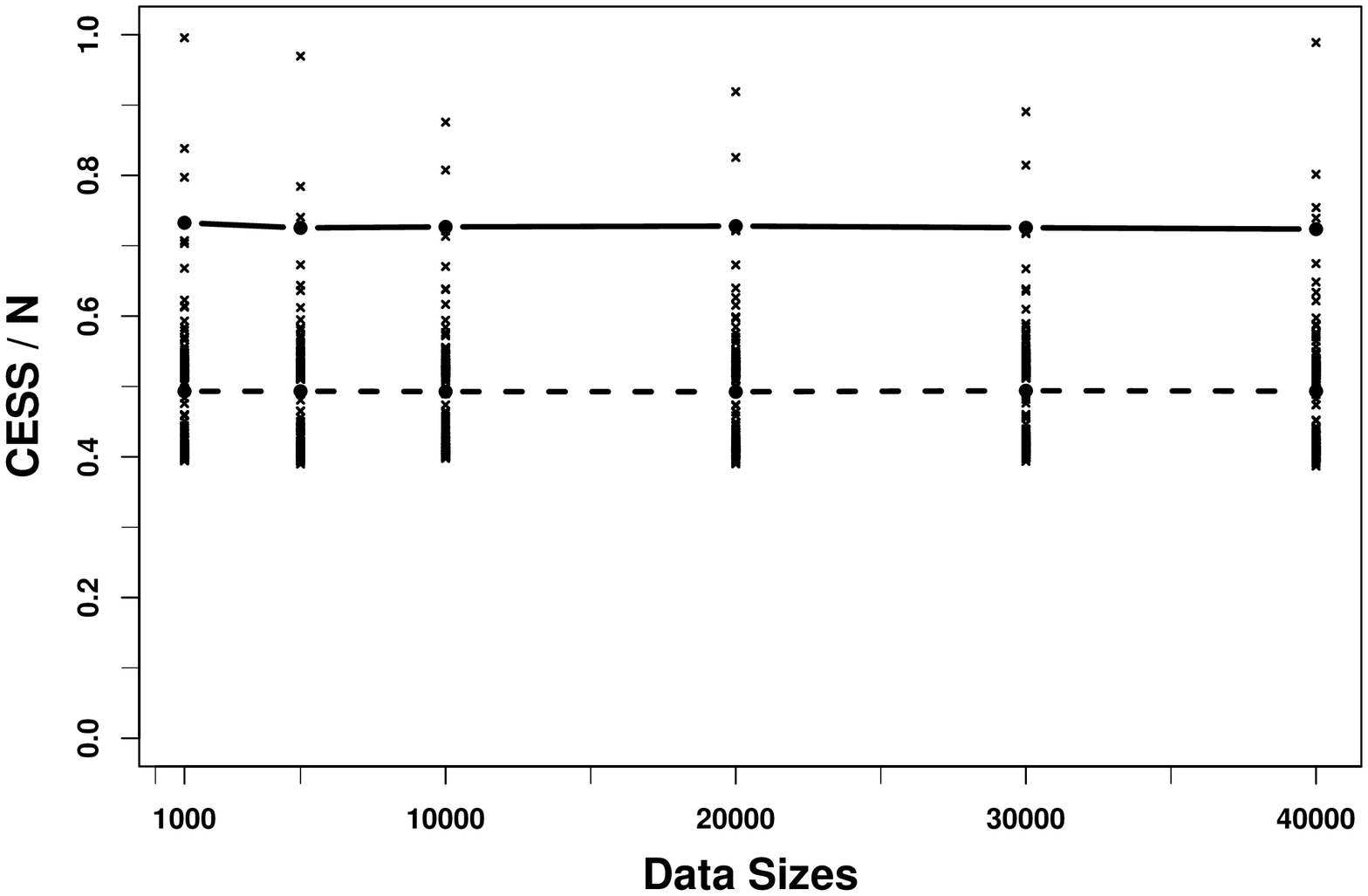}
            \caption{BF}
            \label{fig:bivG_similar_means_vanilla_c}
        \end{subfigure}%
        \hfill
        \begin{subfigure}[b]{0.45\textwidth}  
            \centering 
            \includegraphics[width=\textwidth]{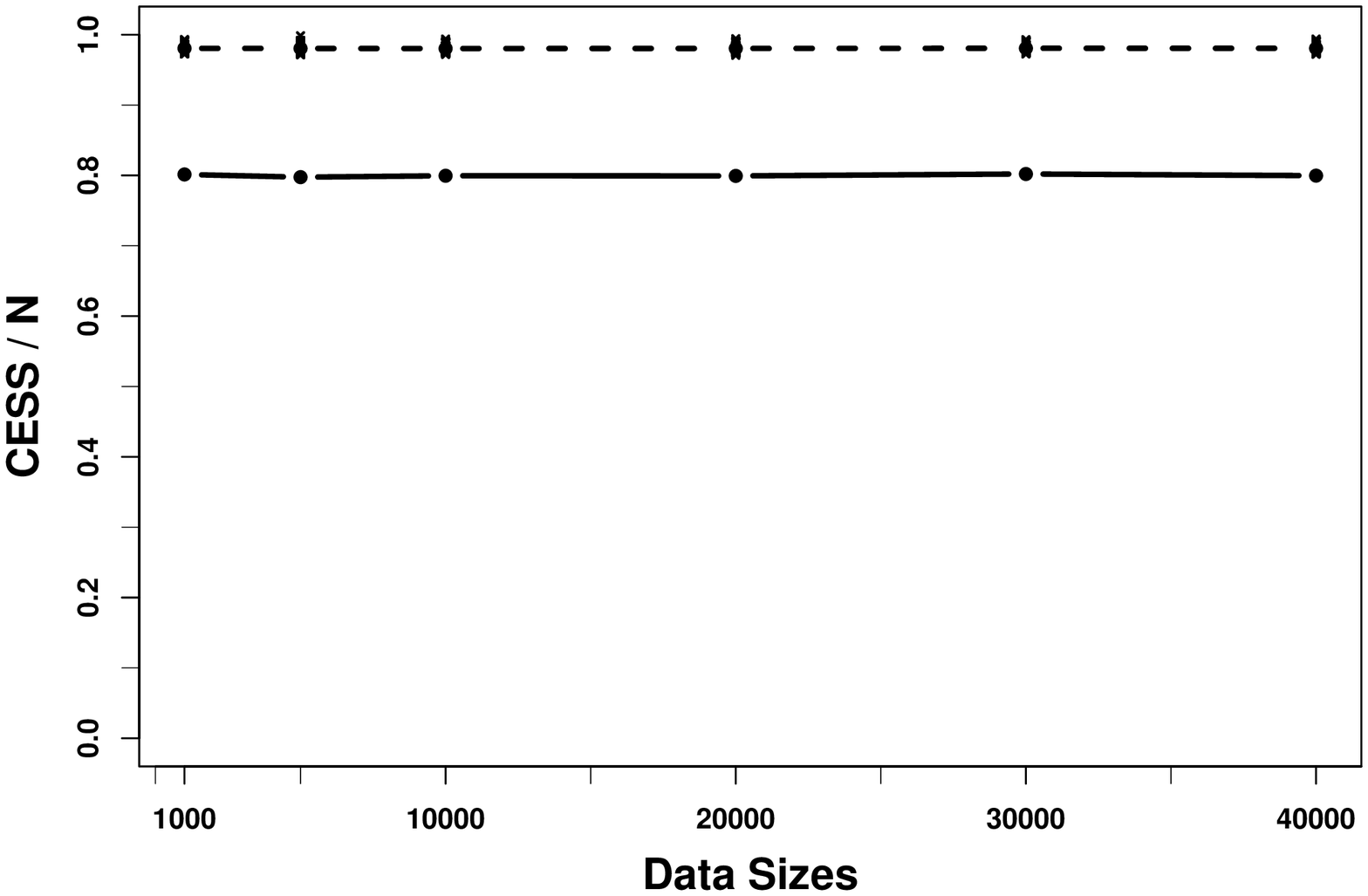}
            \caption{\gbfa}
            \label{fig:bivG_similar_means_generalised_c}
        \end{subfigure}%
        \setcounter{subfigure}{2}
        \renewcommand\thesubfigure{\alph{subfigure}}
        \caption{Recommended $T$ and recommended regular mesh $\mathcal{P}$.}
        \label{fig:bivG_similar_means_c}
    \end{subfigure}%
    \vskip\baselineskip
    \begin{subfigure}[b]{\textwidth}
        \setcounter{subfigure}{0}
        \renewcommand\thesubfigure{\roman{subfigure}}
        \begin{subfigure}[b]{0.45\textwidth}
            \centering
            \includegraphics[width=\textwidth]{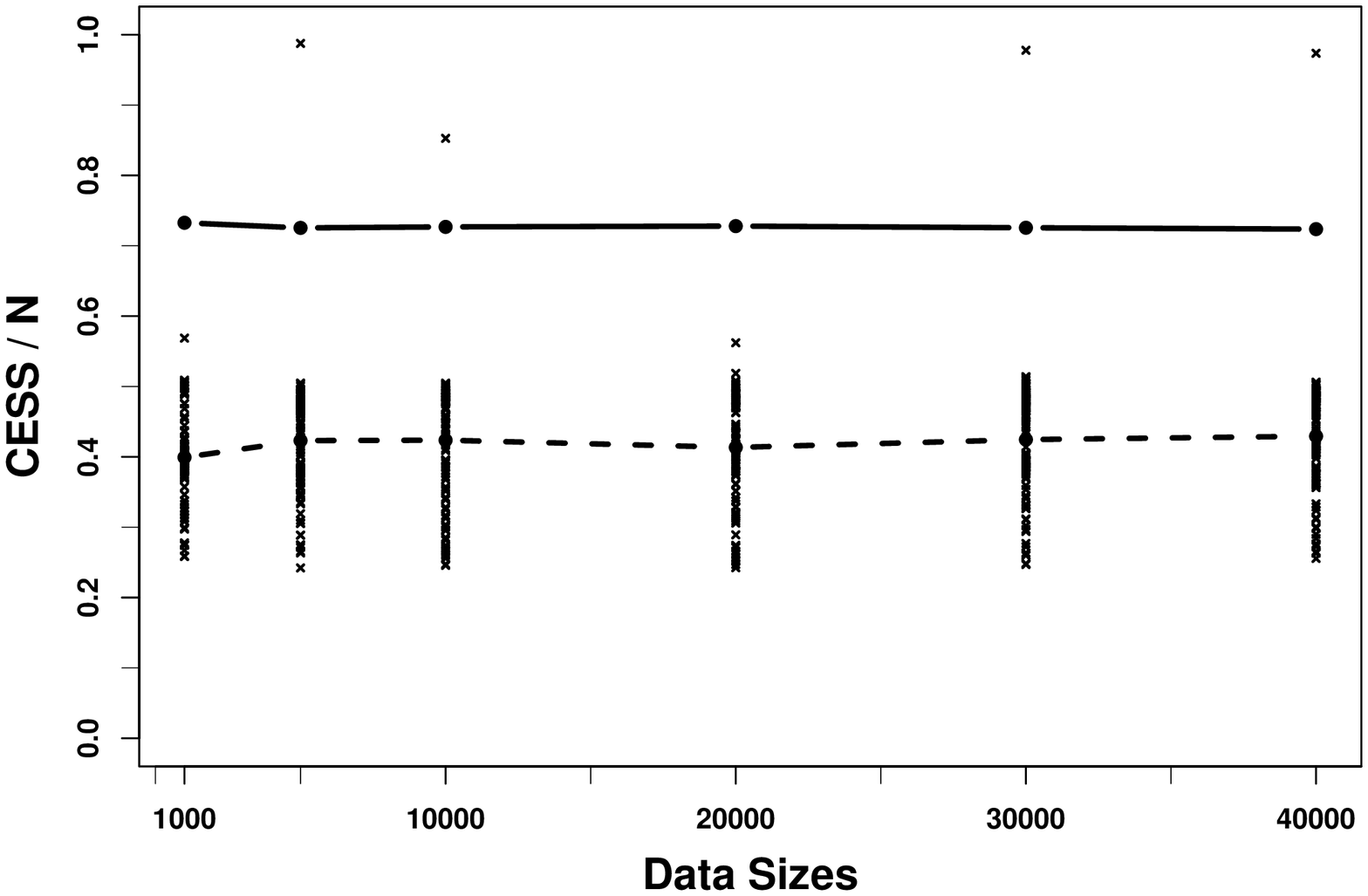}
            \caption{BF}
            \label{fig:bivG_similar_means_vanilla_d}
        \end{subfigure}%
        \hfill
        \begin{subfigure}[b]{0.45\textwidth}  
            \centering 
            \includegraphics[width=\textwidth]{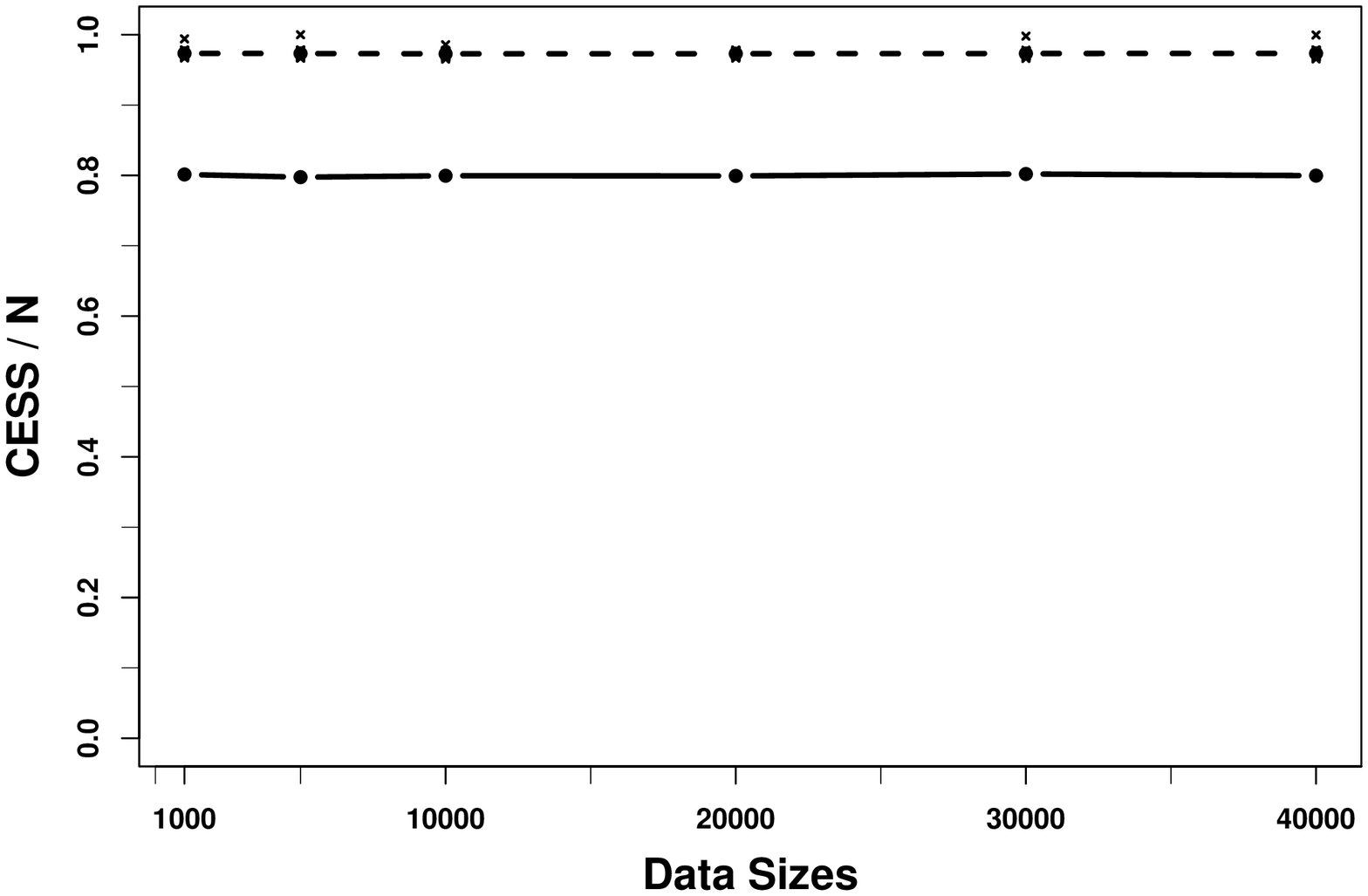}
            \caption{\gbfa}
            \label{fig:bivG_similar_means_generalised_d}
        \end{subfigure}%
        \setcounter{subfigure}{3}
        \renewcommand\thesubfigure{\alph{subfigure}}
        \caption{Recommended $T$ and recommended adaptive mesh $\mathcal{P}$.}
        \label{fig:bivG_similar_means_d}
    \end{subfigure}%
    \vskip\baselineskip
    \begin{subfigure}[b]{\textwidth}
        \setcounter{subfigure}{0}
        \renewcommand\thesubfigure{\roman{subfigure}}
        \begin{subfigure}[b]{0.45\textwidth}
            \centering
            \includegraphics[width=\textwidth]{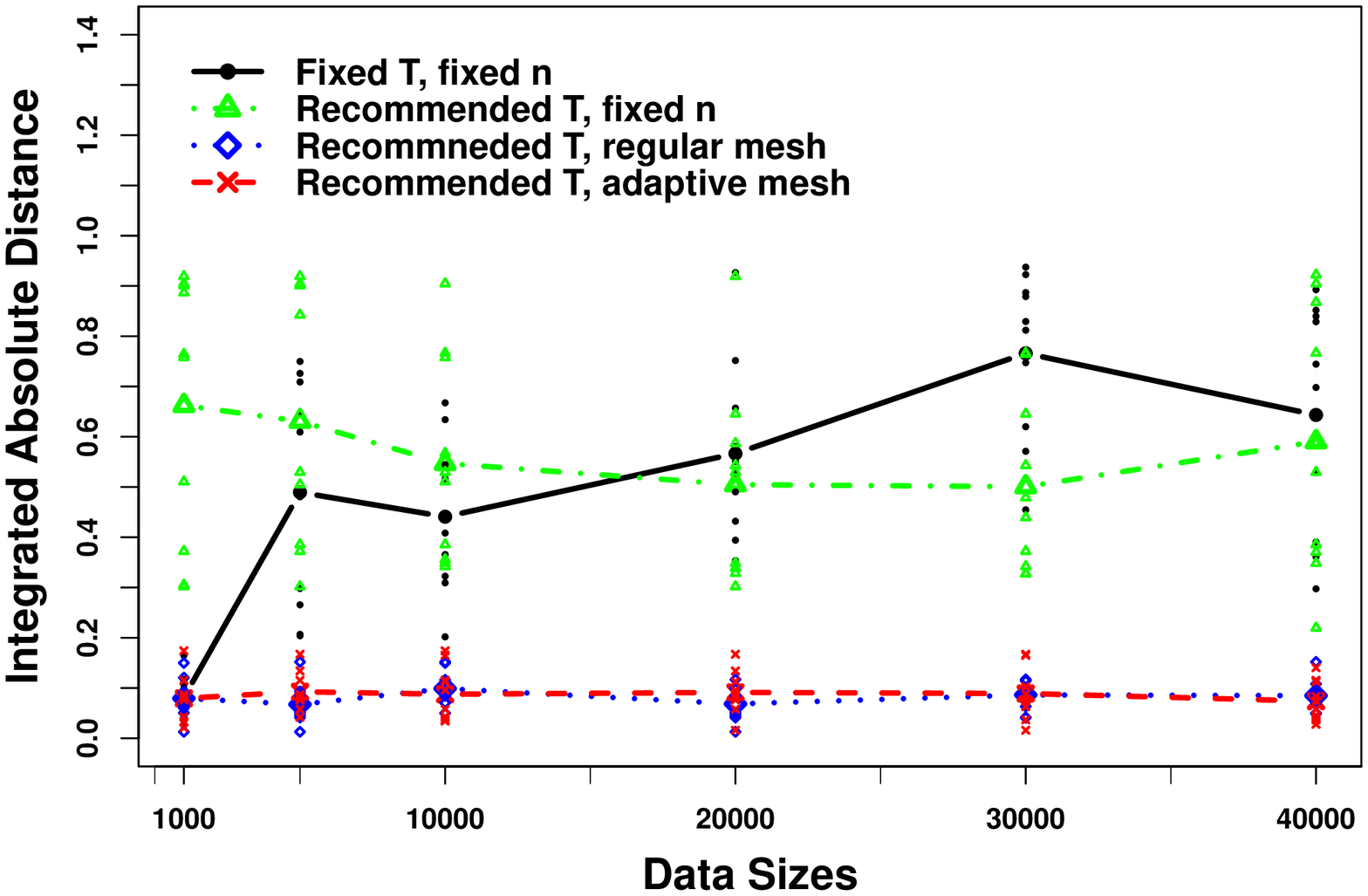}
            \caption{BF}
            \label{fig:bivG_similar_means_vanilla_IAD_average}
        \end{subfigure}%
        \hfill
        \begin{subfigure}[b]{0.45\textwidth}  
            \centering 
            \includegraphics[width=\textwidth]{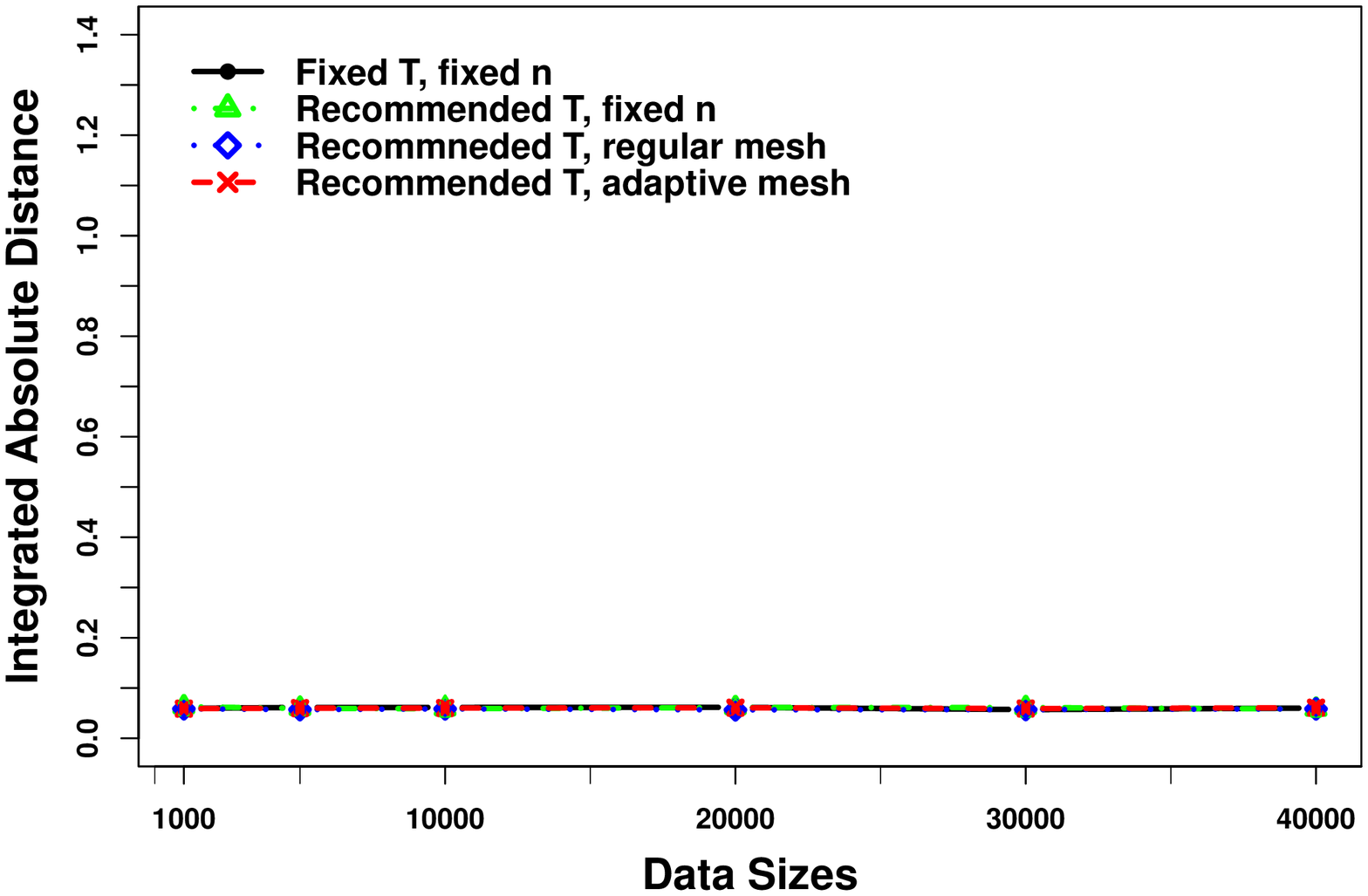}
            \caption{\gbfa}
            \label{fig:bivG_similar_means_generalised_IAD_average}
        \end{subfigure}%
        \setcounter{subfigure}{4}
        \renewcommand\thesubfigure{\alph{subfigure}}
        \caption{Integrated absolute distance: lines connect the mean IAD (averaged over ten runs) while the points denote the individual IAD achieved on each run.}
        \label{fig:bivG_similar_means_IAD}
    \end{subfigure}
        \caption{Bivariate Gaussian example in $\SH{\lambda}$ setting (continued).}
\end{figure}

\begin{figure}[p]
    \centering
    \addtocounter{figure}{-1}
    \begin{subfigure}[b]{\textwidth}
        \setcounter{subfigure}{0}
        \renewcommand\thesubfigure{\roman{subfigure}}
        \begin{subfigure}[b]{0.45\textwidth}
            \centering
            \includegraphics[width=\textwidth]{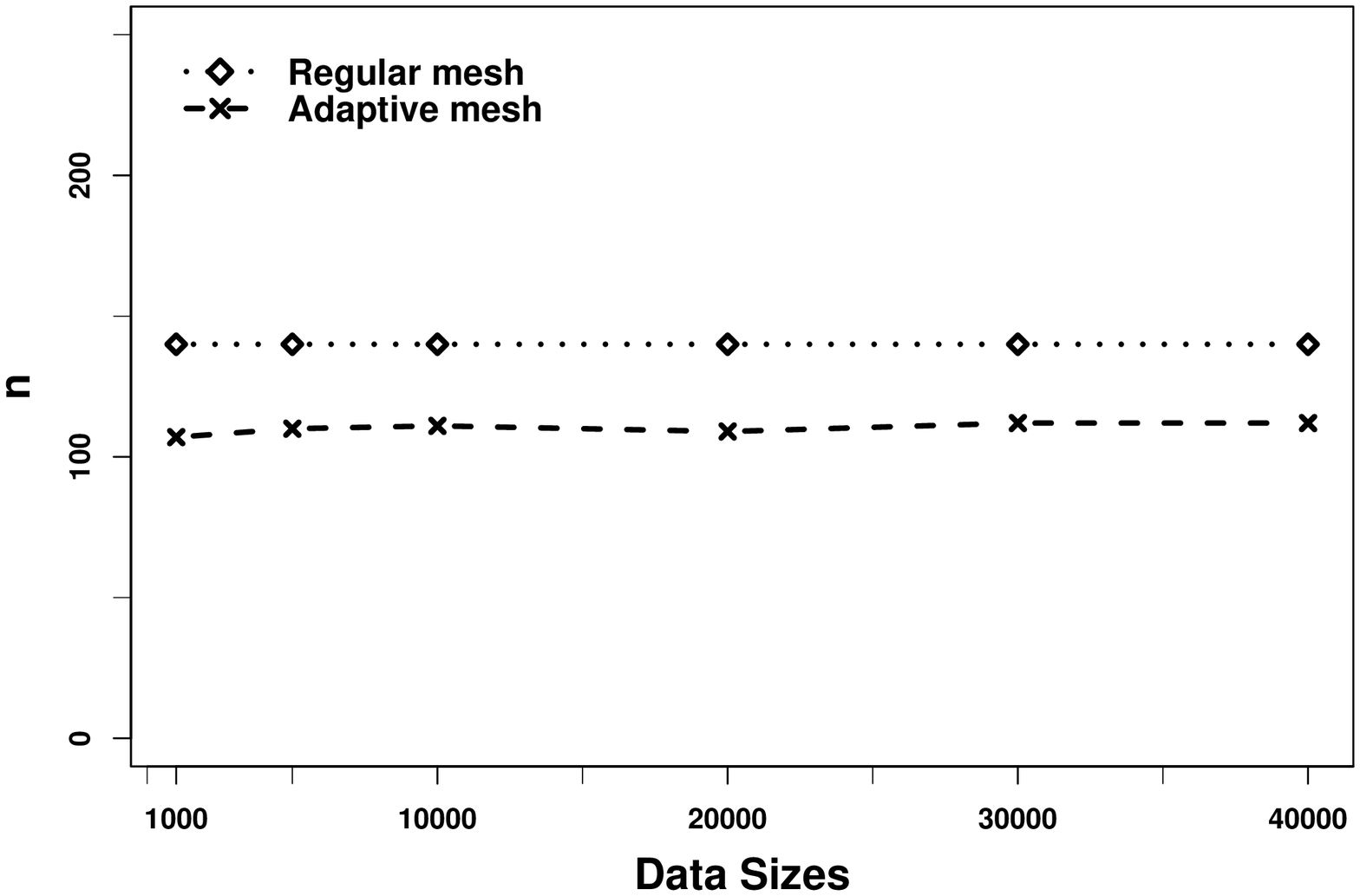}
            \caption{BF}
            \label{fig:bivG_similar_means_vanilla_mesh_size}
        \end{subfigure}%
        \hfill
        \begin{subfigure}[b]{0.45\textwidth}  
            \centering 
            \includegraphics[width=\textwidth]{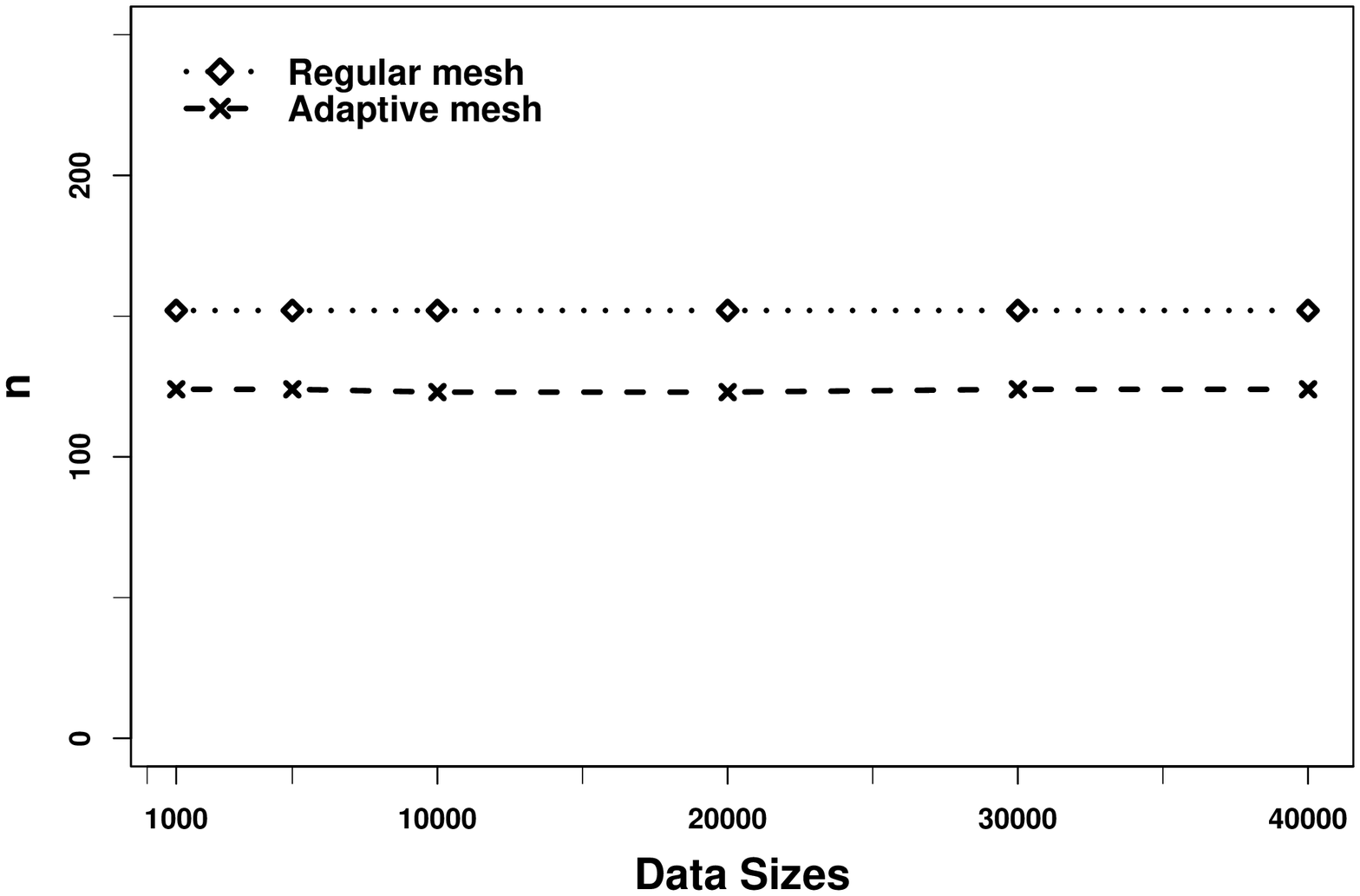}
            \caption{\gbfa}
            \label{fig:bivG_similar_means_generalised_mesh_size}
        \end{subfigure}%
        \setcounter{subfigure}{5}
        \renewcommand\thesubfigure{\alph{subfigure}}
        \caption{Comparison of mesh sizes between regular and adaptive schemes.}
        \label{fig:bivG_similar_means_mesh_size}
    \end{subfigure}%
    \vskip\baselineskip
    \begin{subfigure}[b]{\textwidth}
        \setcounter{subfigure}{0}
        \renewcommand\thesubfigure{\roman{subfigure}}
        \begin{subfigure}[b]{0.45\textwidth}
            \centering
            \includegraphics[width=\textwidth]{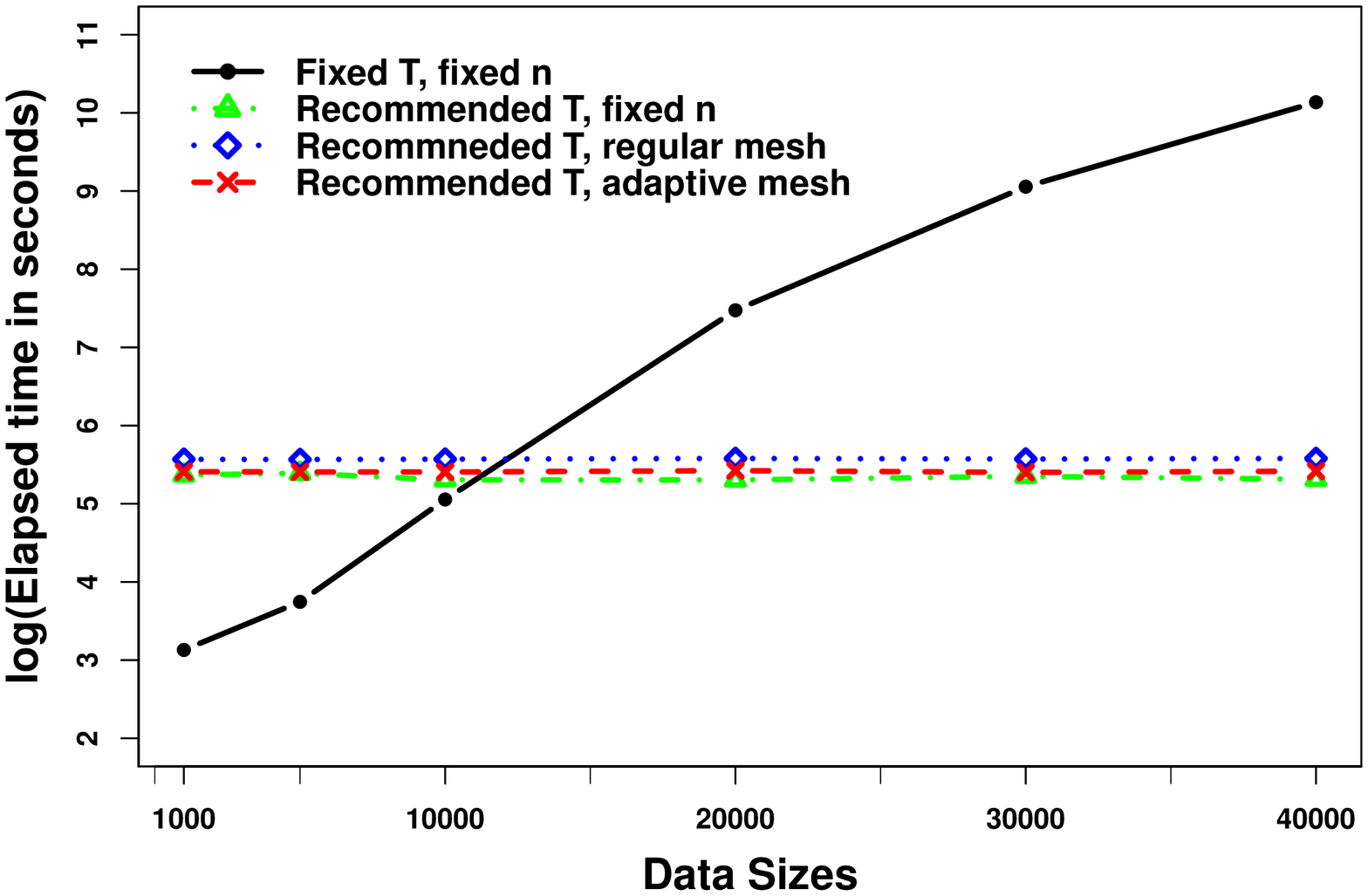}
            \caption{BF}
            \label{fig:bivG_similar_means_vanilla_time_average}
        \end{subfigure}%
        \hfill
        \begin{subfigure}[b]{0.45\textwidth}  
            \centering 
            \includegraphics[width=\textwidth]{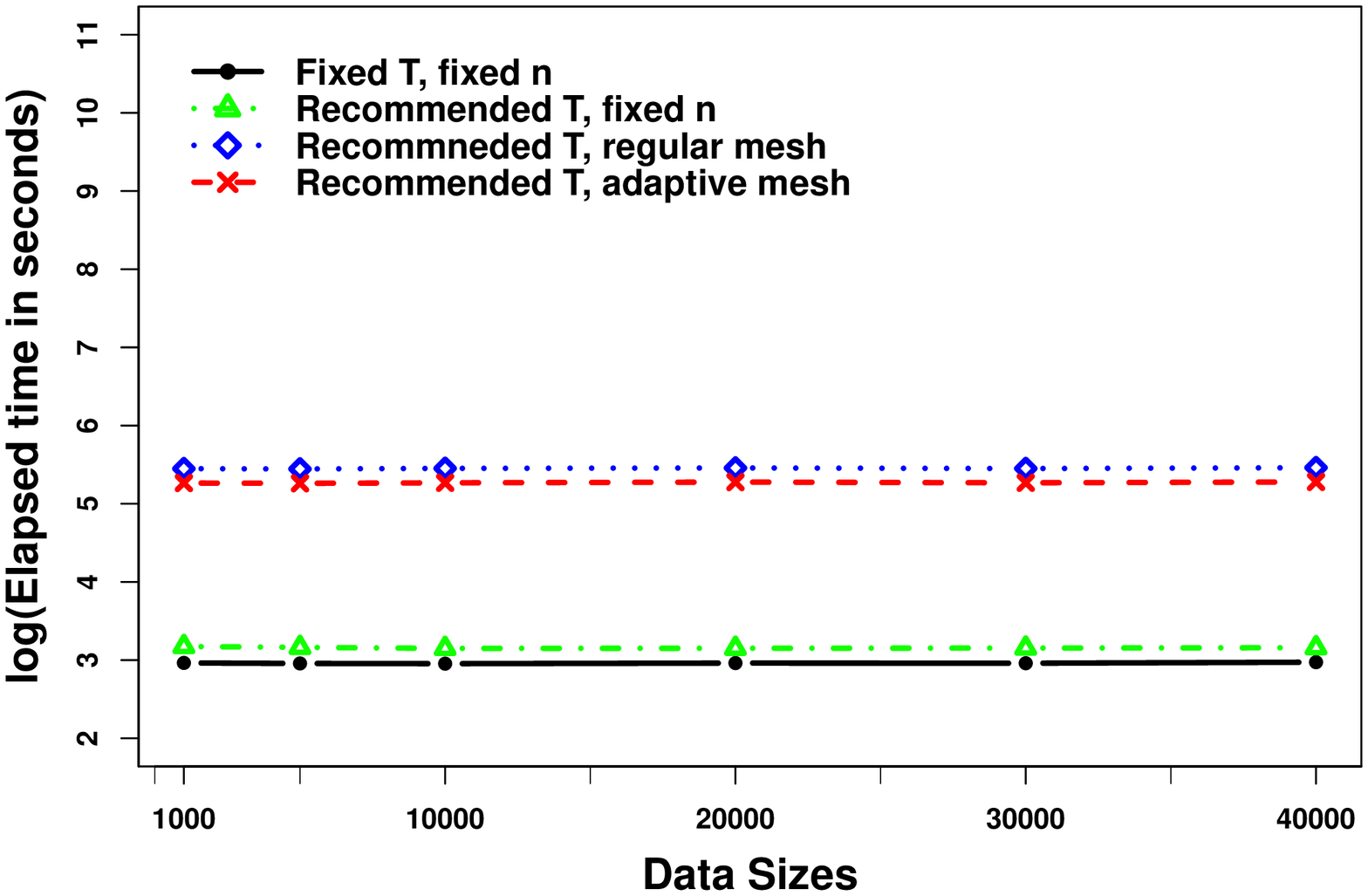}
            \caption{\gbfa}
            \label{fig:bivG_similar_means_generalised_time_average}
        \end{subfigure}%
        \setcounter{subfigure}{6}
        \renewcommand\thesubfigure{\alph{subfigure}}
        \caption{Mean average computational run-times (based on ten runs).}
        \label{fig:bivG_similar_means_time}
    \end{subfigure}%
        \caption{Bivariate Gaussian example in $\SH{\lambda}$ setting (continued).}
\end{figure}

\subsection{Sub-posterior Heterogeneity} \label{subsec:GBF_dissimilar_means}

Now we study the guidance for $T$ and $\mathcal{P}$ for \gbfa (\algoref{alg:GBF}) developed in \secref{sec:GBF_guidance} in the $\SSH{\gamma}$ setting of \condref{cond:SSH}. This represents the setting where sub-posterior heterogeneity does not decay with data size. Here, we consider the scenario of combining $C=2$ bi-variate Gaussian sub-posteriors, $f_{c} \sim \mathcal{N} \left(\bm{\mu}_{c}, \frac{2}{m}\mathbf{\Sigma}\right)$, where $\bm{\mu}_{1}=-(0.25, 0.25)$ and $\bm{\mu}_{2}=(0.25, 0.25)$ and $\mathbf{\Sigma}=\begin{pmatrix} 1 & \rho \\ \rho & 1 \end{pmatrix}$, with $\rho=0.9$. We again consider a range of data sizes, which ranges from $m=250$ to $m=2500$ and are randomly split between $C=2$ cores. We apply BF and \gbfa with a fixed particle set size of $N=10000$.

In this setting as $m$ increases the sub-posterior heterogeneity increases, which is a consequence of the sub-posteriors having diminishing overlapping support. In BF (where $\mathbf{\Lambda}_{1}=\mathbf{\Lambda}_{2}=\mathbb{I}_{d}$) this heterogeneity is not captured, and $\sigma_{\bm{a}}^{2}=0.125$ irrespective of $m$. By contrast, the generalised approach is able to capture the heterogeneity with $m$ with simply the inclusion of the estimated covariance matrices $\{\mathbf{\Lambda}_{c}\}_{c=1,2}$.

As with the previous example in \secref{subsec:GBF_similar_means}, we will investigate the effect of varying $T$ and $\mathcal{P}$ with $m$, and its impact upon $\CESS{0}$ and $\CESS{j}$. We consider the following choices for $T$ and $\mathcal{P}$:
\begin{enumerate}
    \item a fixed choice of $T$ and $n$ to obtain $\mathcal{P}$ (for \gbfa, $T=2$ and $n=5$, and for BF, $T=0.01$ and $n=5$),
    \item using the recommended $T$ from \secref{subsec:GBF_T_guidance} and fixed $n=5$ to obtain $\mathcal{P}$,
    \item using the recommended $T$ and $\mathcal{P}$ using a regular mesh (as outlined in \algoref{alg:regular_mesh} and \secref{subsubsec:regular_mesh}),
    \item using the recommended $T$ and $\mathcal{P}$ using a, adaptive mesh (as outlined in \algoref{alg:adaptive_mesh} and \secref{subsubsec:adaptive_mesh}).
\end{enumerate}

When applying the guidance, we set the lower tolerable bounds on the initial ($\CESS{0}$) and iterative ($\CESS{j}$) conditional effective sample sizes to be $0.5N$ (i.e.\ we set $\zeta=\zeta^{\prime}=0.5$), and re-sample if $\ESS$ drops below $0.5N$. Again, for helping with the practical interpretation of our extensive guidance for selecting $T$ and $\mathcal{P}$, we summarise our approach in \remref{remark:GBF_guidance_dissimilar_means}:
\begin{remark}
    \label{remark:GBF_guidance_dissimilar_means}
    We set the tuning parameters for BF and \gbfa (for the $\SSH{\gamma}$ setting of \secref{subsec:GBF_dissimilar_means}) as follows:
    \begin{enumerate}
        \item We follow the guidance outlined in \remref{remark:T_guidance_k1_k2}, noting that $\zeta=0.5$ and $d=2$. For \gbfa, $\mathbf{\Lambda}_{c=1,2}$ are the estimated covariance matrices for each of the sub-posteriors, so $b=\frac{m}{C}$ (see \remref{remark:b_choice_GBF}), and $\gamma=m\sigma_{\bm{a}}^{2}/C$ (where $\sigma_{\bm{a}}^{2}$ is estimated from the sub-posterior samples). Consequently, we can compute $k_{1}=k_{2}=\sqrt{-\frac{\left(\frac{\gamma m}{C}+\frac{d}{2}\right)}{\log(\zeta)}}$, and choose $T=C^{\frac{1}{2}}k_{1}$. For BF, $\mathbf{\Lambda}_{c=1,2}=\mathbb{I}_{d}$, so $b=1$ and $\gamma=\sigma_{\bm{a}}^{2}$, and so we can compute $k_{1}=\sqrt{-\frac{\left(\frac{\gamma m}{C}+\frac{d}{2}\right)}{\log(\zeta)}}$ and $k_{2}=\frac{Ck_{1}}{m}$, and choose $T=\frac{C^{3/2}k_{1}}{m}=C^{\frac{1}{2}}k_{2}$.
        \item When using the regular mesh, we use \algoref{alg:regular_mesh} to obtain $\mathcal{P}$. As $\zeta^{\prime}=0.5$, we have for \gbfa $b=\frac{m}{C}$, and so $\Delta_{j}=\Delta=\sqrt{\frac{k_{4}}{2Cd}}$ for each $j$ where $k_{4}$ is computed as per \eqref{eq:k4_choice} (with $\sup_j\widehat{\mathbb{E}[\nu_{j}]}$ computed as per \eqref{eq:E_nu_j_max}). For BF we have $b=1$, so $\Delta_{j}=\Delta=\sqrt{\frac{Ck_{4}}{2m^{2}d}}$ for each $j$.
        \item When using the adaptive mesh, we use \algoref{alg:adaptive_mesh} to obtain $\Delta_{j}$ recursively at each iteration to construct $\mathcal{P}$. With $\zeta^{\prime}=0.5$ for the \gbfa (where $b=\frac{m}{C}$) we compute $t_{j}=\min\{T,t_{j-1}+\Delta_{j}\}$ where $\Delta_{j}=\sqrt{\frac{k_{4}}{2Cd}}$ at each iteration of \algoref{alg:GBF} until we have $t_{j}=T$. For BF with $b=1$ we have instead $\Delta_{j}=\sqrt{\frac{Ck_{4}}{2m^{2}d}}$ at each iteration.
    \end{enumerate}
\end{remark}

CESS for BF and \gbfa with increasing $m$ in this $\SSH{\gamma}$ setting are shown in \figref{fig:bivG_dissimilar_means}. We can immediately see that the $\SSH{\gamma}$ setting is much more challenging than the idealised $\SH{\lambda}$ setting of \secref{subsec:GBF_similar_means} and \figref{fig:bivG_similar_means}, which is unsurprising as in this case the sub-posteriors are becoming increasingly mismatched as data size increases. 

In \figref{fig:bivG_dissimilar_means_a}, we see that fixing $T$ and $n$ is not ideal for either BF or \gbfa. As shown in \figref{fig:bivG_dissimilar_means_b}, there is a positive effect for both BF and \gbfa in using our recommended scaling of $T$ in the quality of the initialisation. In \figref{fig:bivG_dissimilar_means_c} and \figref{fig:bivG_dissimilar_means_d}, where both the guidance for $T$ and $\mathcal{P}$ are implemented, we see a substantial improvement in the performance of both approaches with respect to CESS, with again our new \gbfa approach outperforming BF.

In \figref{fig:bivG_dissimilar_means_c} we see that the use of a regular mesh in choosing $\mathcal{P}$, following our guidance, provides robust $\CESS{j}$ with low variance. Indeed, it appears to outperform the adaptive mesh approach for $\mathcal{P}$ (see \figref{fig:bivG_dissimilar_means_d}). However, as discussed in \secref{subsubsec:adaptive_mesh}, the regular mesh is overly conservative, and when we factor in the reduced number of iterations required in the adaptive case (\figref{fig:bivG_dissimilar_means_mesh_size}), along with the overall reduction in computational cost (\figref{fig:bivG_dissimilar_means_time}) for comparable IAD (\figref{fig:bivG_dissimilar_means_IAD}), we see that the use of an adaptive mesh is preferable. 
\begin{figure}[htp]
    \centering
    \begin{subfigure}[b]{\textwidth}
        \setcounter{subfigure}{0}
        \renewcommand\thesubfigure{\roman{subfigure}}
        \begin{subfigure}[b]{0.45\textwidth}
            \centering
            \includegraphics[width=\textwidth]{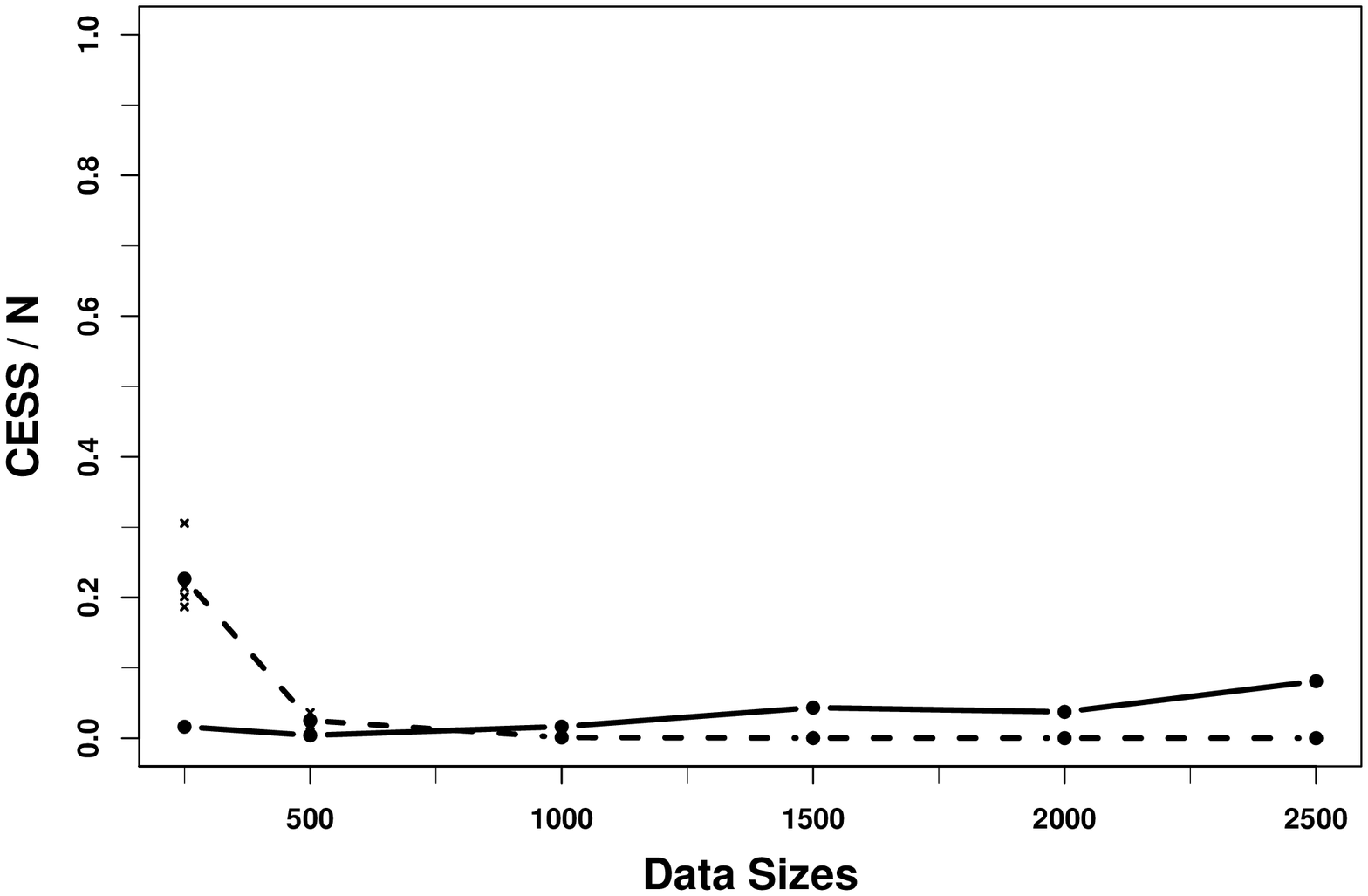}
            \caption{BF}
            \label{fig:bivG_dissimilar_means_vanilla_a}
        \end{subfigure}%
        \hfill
        \begin{subfigure}[b]{0.45\textwidth}  
            \centering 
            \includegraphics[width=\textwidth]{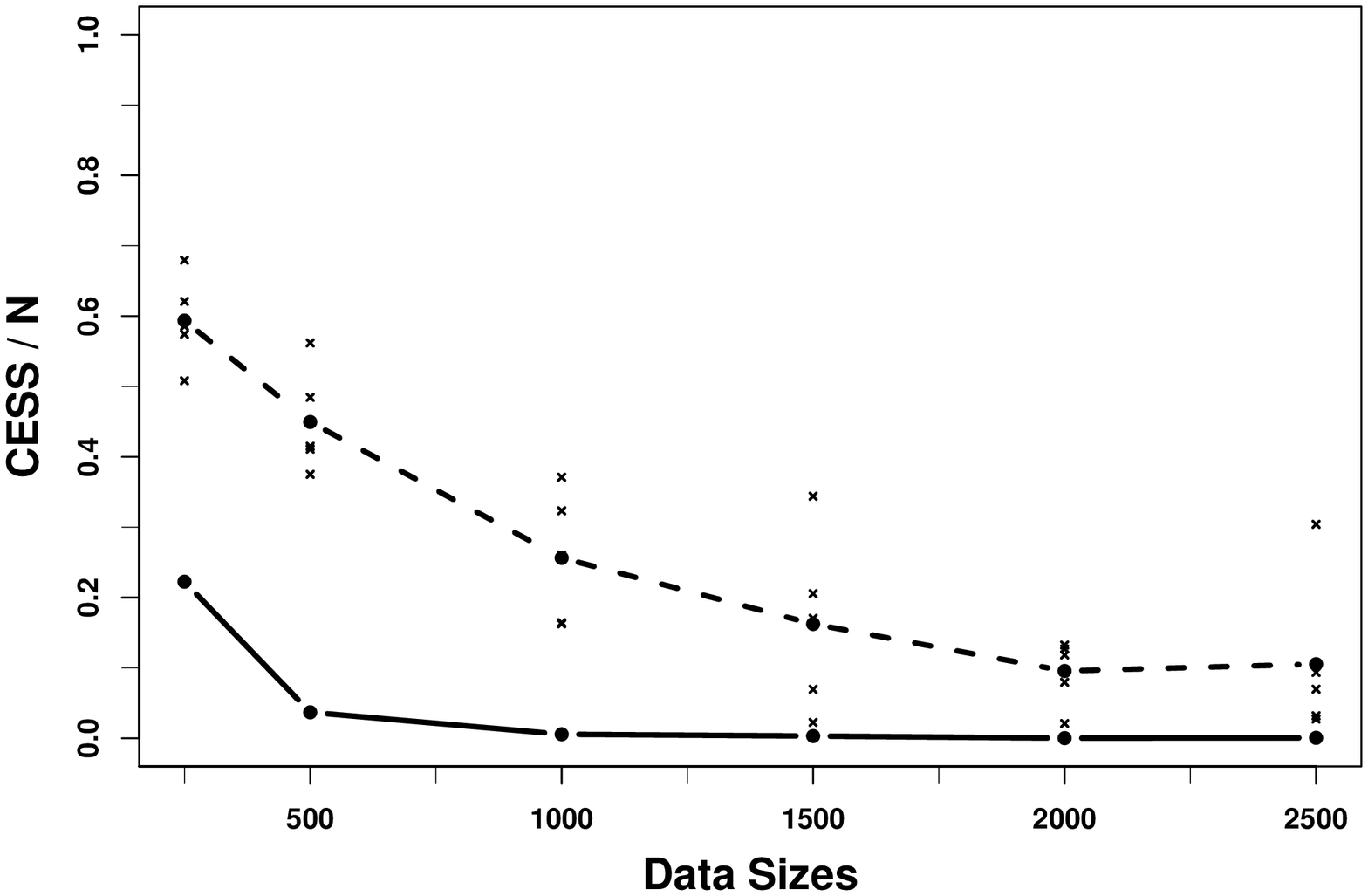}
            \caption{\gbfa}
            \label{fig:bivG_dissimilar_means_generalised_a}
        \end{subfigure}%
        \setcounter{subfigure}{0}
        \renewcommand\thesubfigure{\alph{subfigure}}
        \caption{Fixed user-specified $T$ and $n$.}
        \label{fig:bivG_dissimilar_means_a}
    \end{subfigure}%
    \vskip\baselineskip
    \begin{subfigure}[b]{\textwidth}
        \setcounter{subfigure}{0}
        \renewcommand\thesubfigure{\roman{subfigure}}
        \begin{subfigure}[b]{0.45\textwidth}
            \centering
            \includegraphics[width=\textwidth]{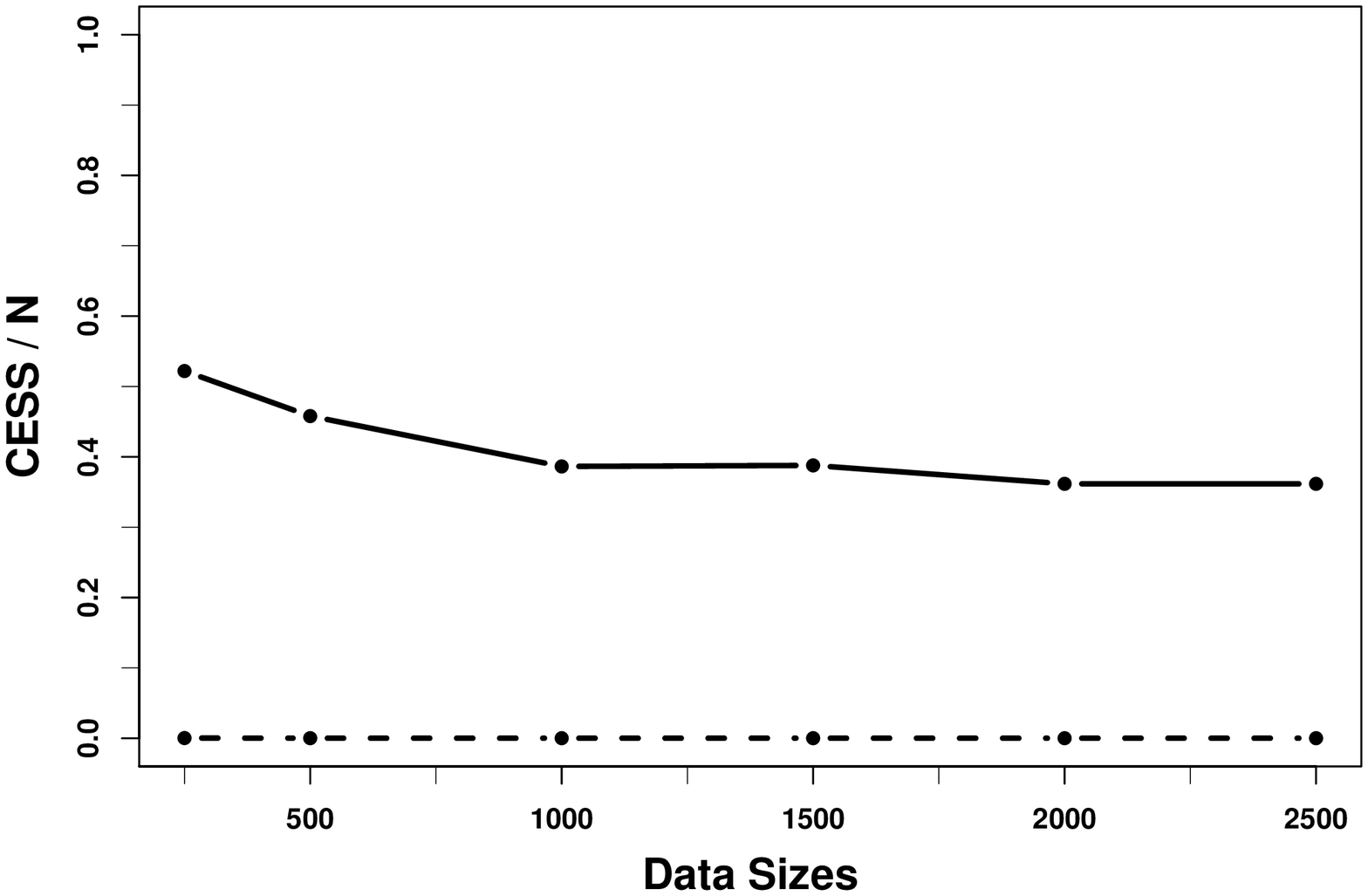}
            \caption{BF}
            \label{fig:bivG_dissimilar_means_vanilla_b}
        \end{subfigure}%
        \hfill
        \begin{subfigure}[b]{0.45\textwidth}  
            \centering 
            \includegraphics[width=\textwidth]{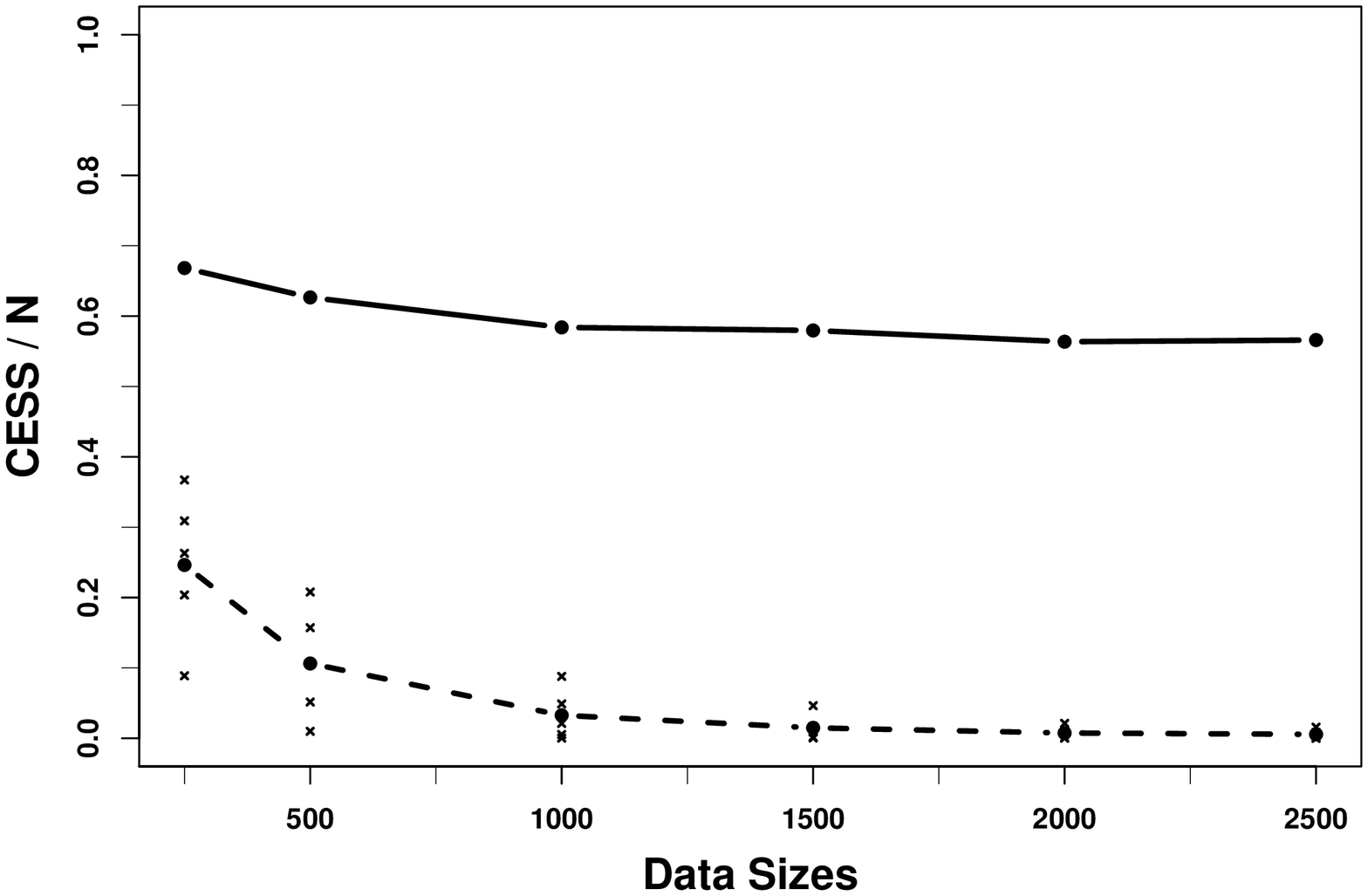}
            \caption{\gbfa}
            \label{fig:bivG_dissimilar_means_generalised_b}
        \end{subfigure}%
        \setcounter{subfigure}{1}
        \renewcommand\thesubfigure{\alph{subfigure}}
        \caption{Recommended $T$ and fixed $n$.}
        \label{fig:bivG_dissimilar_means_b}
    \end{subfigure}%
        \caption{Bivariate Gaussian example in $\SSH{\gamma}$ setting with increasing data size. In Figures \ref{fig:bivG_dissimilar_means_a}, \ref{fig:bivG_dissimilar_means_b}, \ref{fig:bivG_dissimilar_means_c}, \ref{fig:bivG_dissimilar_means_d} solid lines denote initial CESS ($\text{CESS}_{0}$), and dotted lines denote averaged CESS in subsequent iterations $(\frac{1}{n}\sum_{j=1}^{n}\text{CESS}_{j})$, and crosses denote $\text{CESS}_{j}$ for each $j\in\{1,\dots,n\}$.}
    \label{fig:bivG_dissimilar_means}
\end{figure}

\begin{figure}[p]
    \centering
    \addtocounter{figure}{-1}
    \begin{subfigure}[b]{\textwidth}
        \setcounter{subfigure}{0}
        \renewcommand\thesubfigure{\roman{subfigure}}
        \begin{subfigure}[b]{0.45\textwidth}
            \centering
            \includegraphics[width=\textwidth]{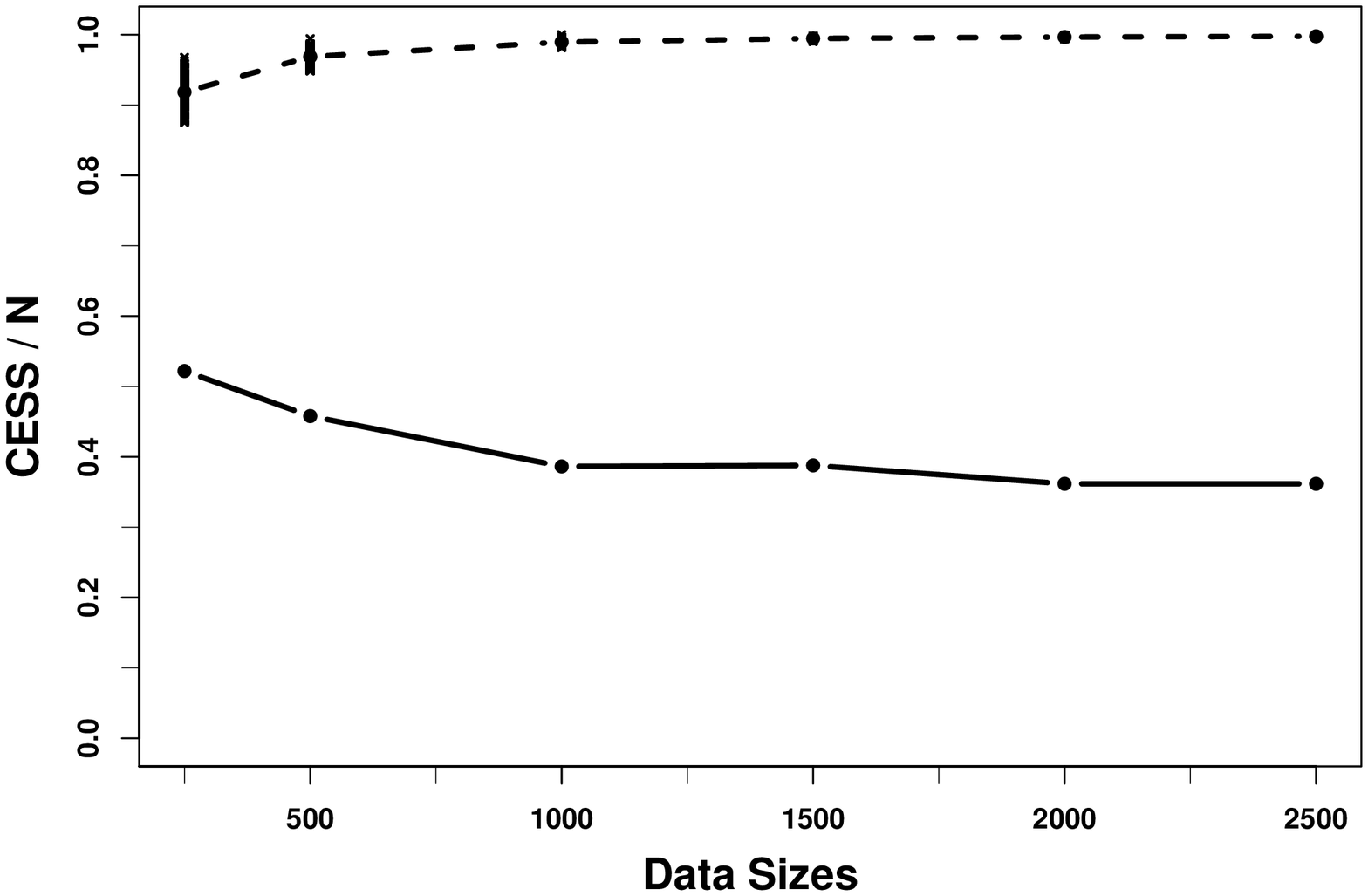}
            \caption{BF}
            \label{fig:bivG_dissimilar_means_vanilla_c}
        \end{subfigure}%
        \hfill
        \begin{subfigure}[b]{0.45\textwidth}  
            \centering 
            \includegraphics[width=\textwidth]{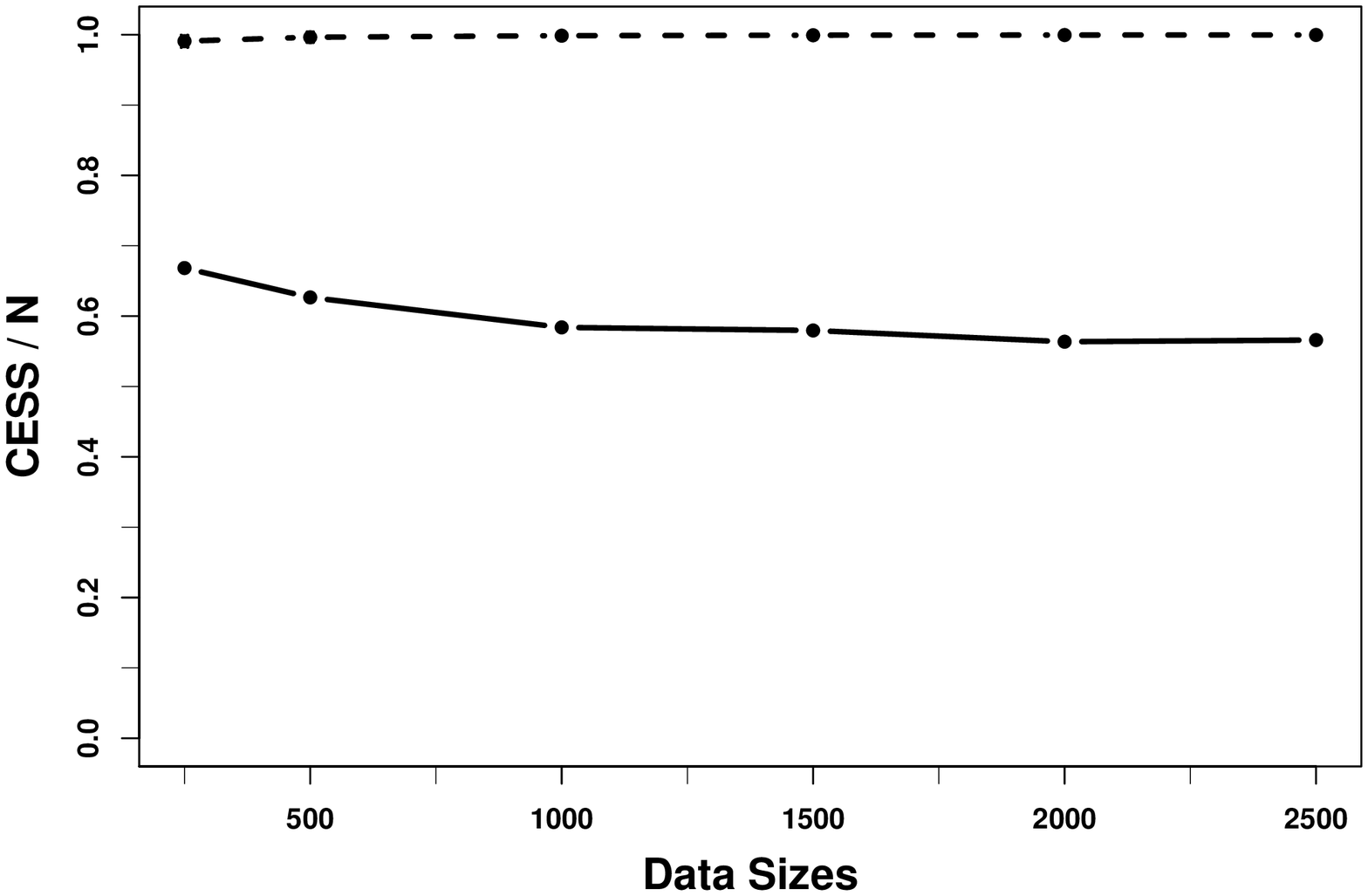}
            \caption{\gbfa}
            \label{fig:bivG_dissimilar_means_generalised_c}
        \end{subfigure}%
        \setcounter{subfigure}{2}
        \renewcommand\thesubfigure{\alph{subfigure}}
        \caption{Recommended $T$ and recommended regular mesh $\mathcal{P}$.}
        \label{fig:bivG_dissimilar_means_c}
    \end{subfigure}%
    \vskip\baselineskip
    \begin{subfigure}[b]{\textwidth}
        \setcounter{subfigure}{0}
        \renewcommand\thesubfigure{\roman{subfigure}}
        \begin{subfigure}[b]{0.45\textwidth}
            \centering
            \includegraphics[width=\textwidth]{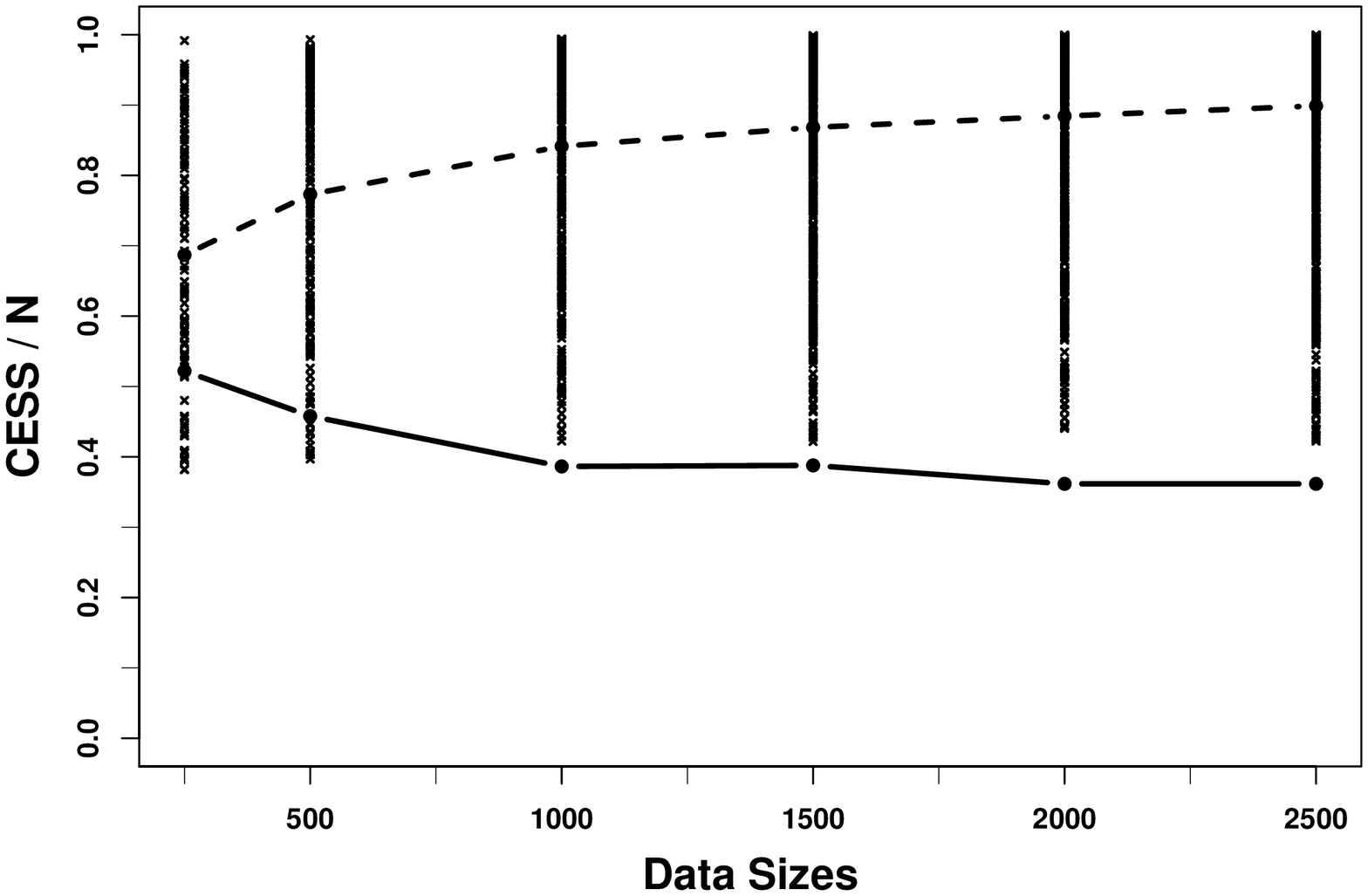}
            \caption{BF}
            \label{fig:bivG_dissimilar_means_vanilla_d}
        \end{subfigure}%
        \hfill
        \begin{subfigure}[b]{0.45\textwidth}  
            \centering 
            \includegraphics[width=\textwidth]{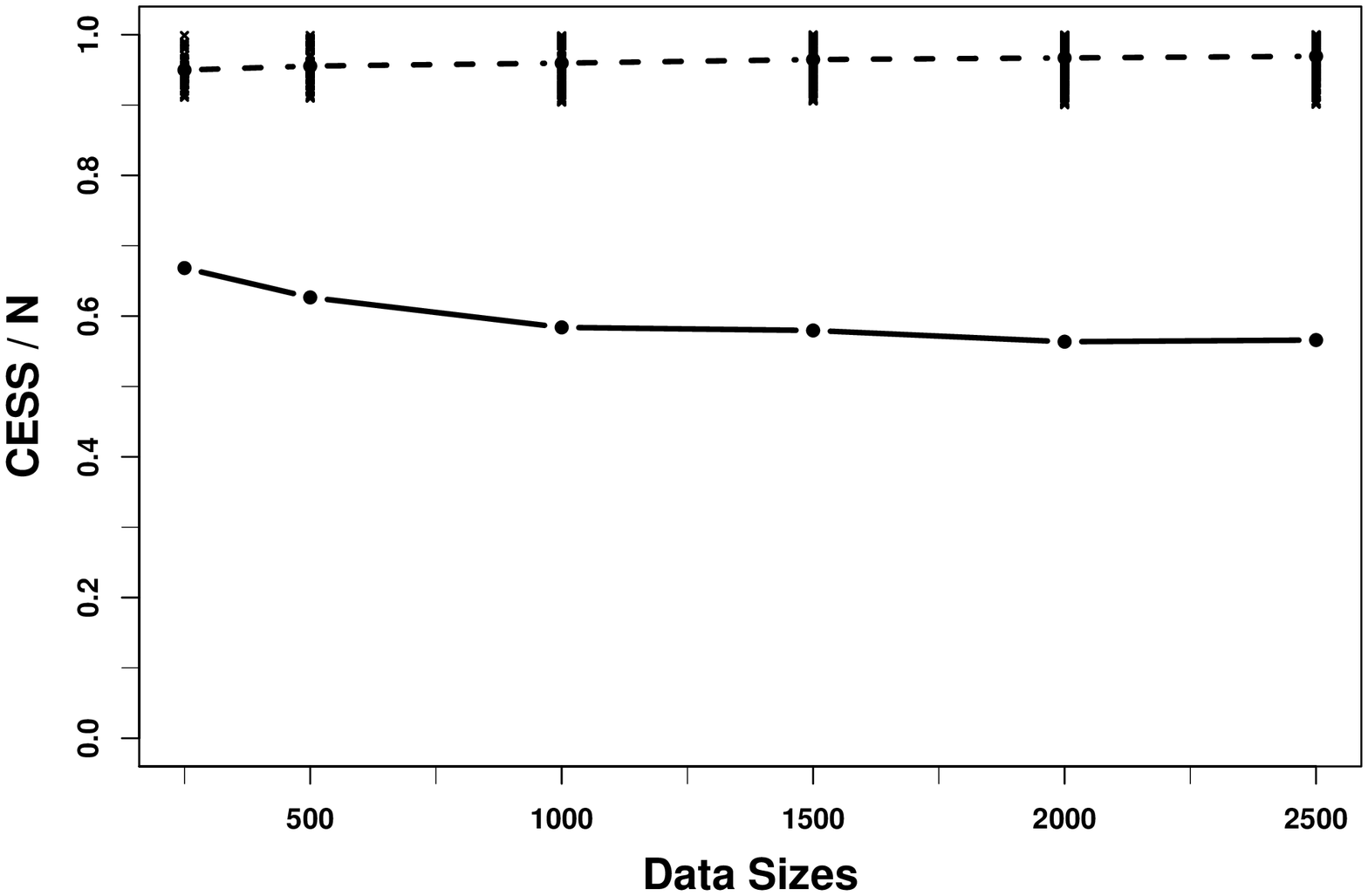}
            \caption{\gbfa}
            \label{fig:bivG_dissimilar_means_generalised_d}
        \end{subfigure}%
        \setcounter{subfigure}{3}
        \renewcommand\thesubfigure{\alph{subfigure}}
        \caption{Recommended $T$ and recommended adaptive mesh $\mathcal{P}$.}
        \label{fig:bivG_dissimilar_means_d}
    \end{subfigure}%
    \vskip\baselineskip
    \begin{subfigure}[b]{\textwidth}
        \setcounter{subfigure}{0}
        \renewcommand\thesubfigure{\roman{subfigure}}
        \begin{subfigure}[b]{0.45\textwidth}
            \centering
            \includegraphics[width=\textwidth]{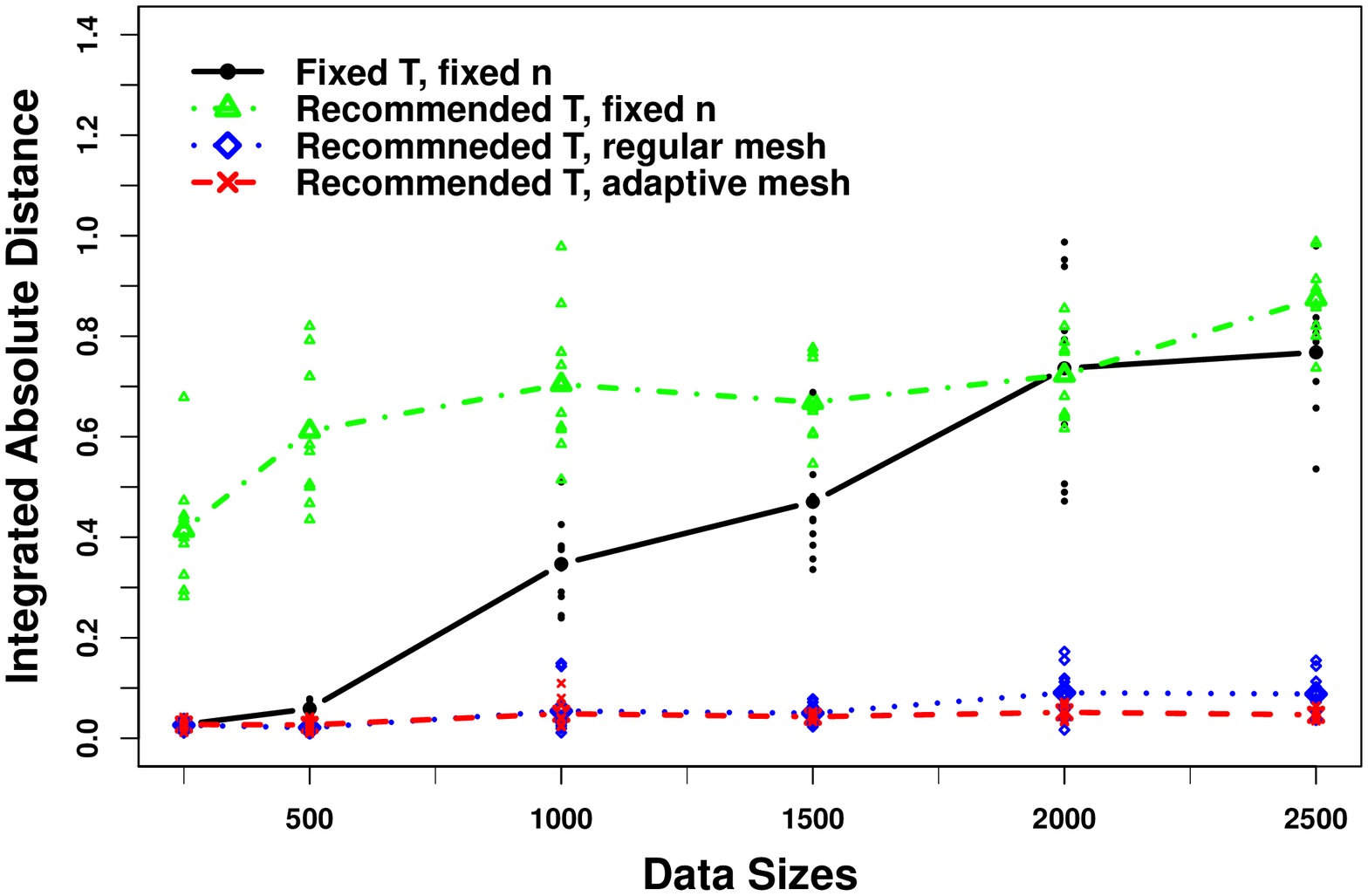}
            \caption{BF}
            \label{fig:bivG_dissimilar_means_vanilla_IAD_average}
        \end{subfigure}%
        \hfill
        \begin{subfigure}[b]{0.45\textwidth}  
            \centering 
            \includegraphics[width=\textwidth]{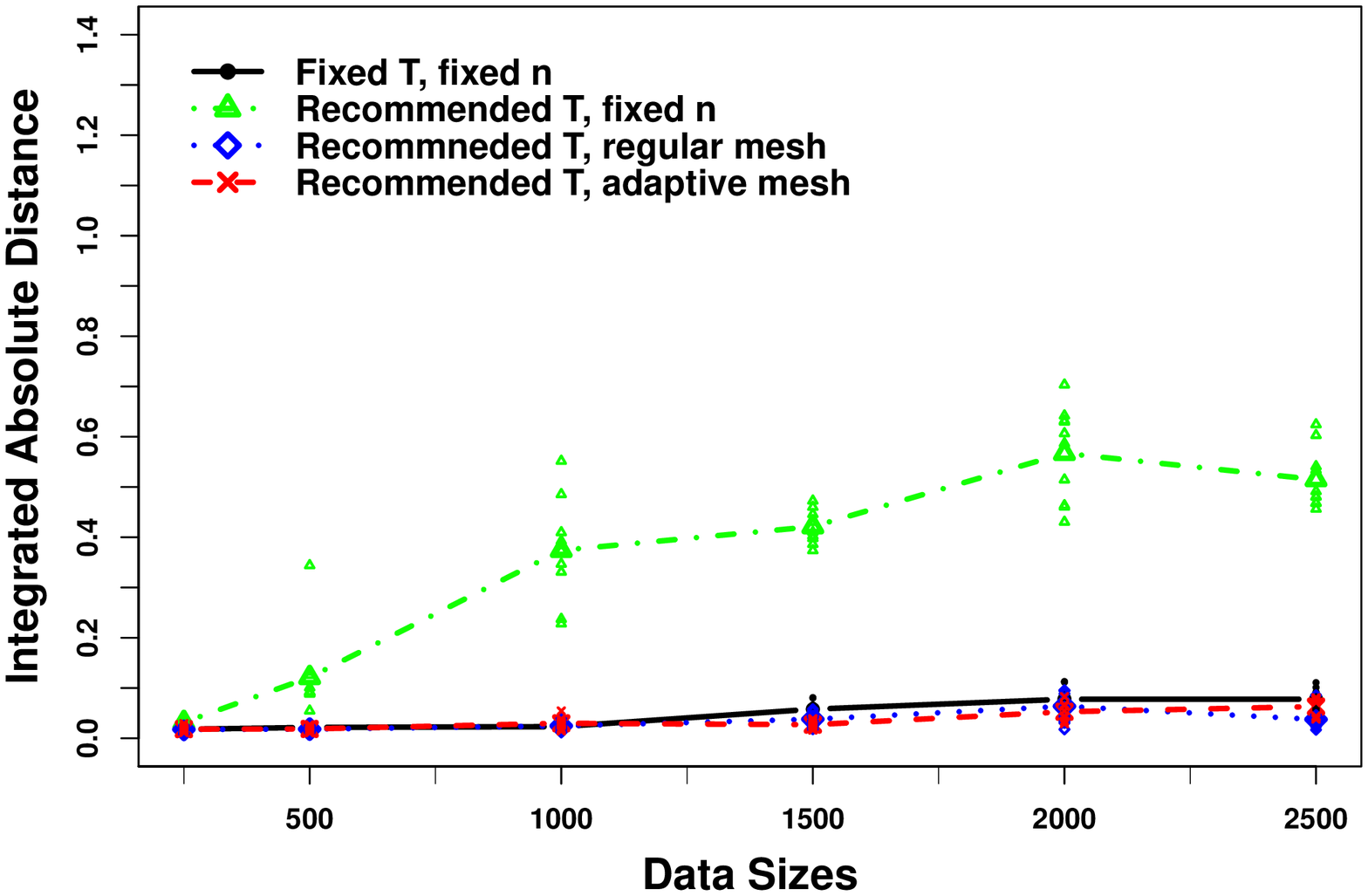}
            \caption{\gbfa}
            \label{fig:bivG_dissimilar_means_generalised_IAD_average}
        \end{subfigure}%
        \setcounter{subfigure}{4}
        \renewcommand\thesubfigure{\alph{subfigure}}
        \caption{Integrated absolute distance: lines connect the mean IAD (averaged over ten runs) while the points denote the individual IAD achieved on each run.}
        \label{fig:bivG_dissimilar_means_IAD}
    \end{subfigure}%
    \caption{Bivariate Gaussian Example in $\SSH{\gamma}$ setting (continued).}
\end{figure}

\begin{figure}[p]
    \centering
    \addtocounter{figure}{-1}
    \begin{subfigure}[b]{\textwidth}
        \setcounter{subfigure}{0}
        \renewcommand\thesubfigure{\roman{subfigure}}
        \begin{subfigure}[b]{0.45\textwidth}
            \centering
            \includegraphics[width=\textwidth]{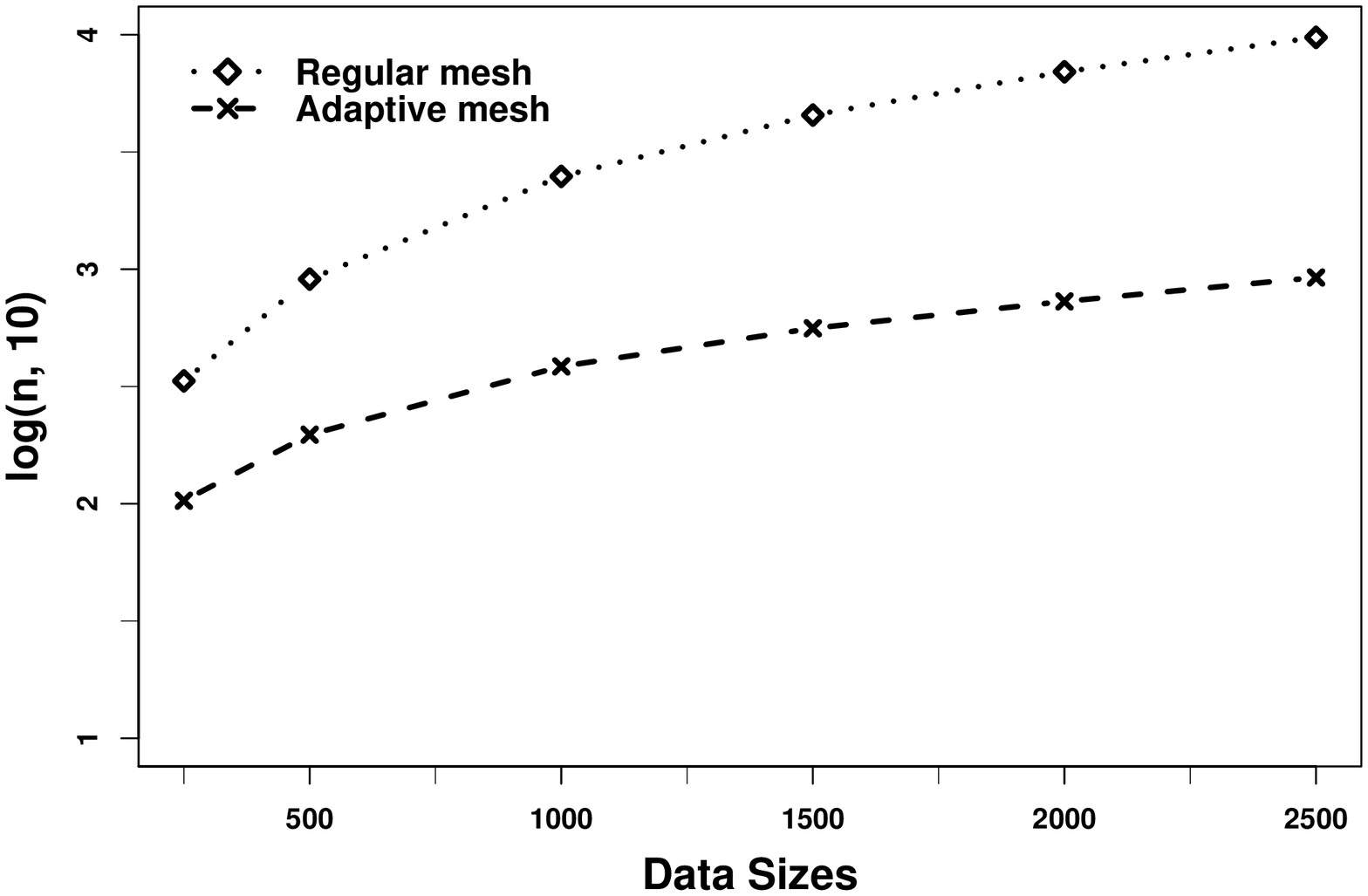}
            \caption{BF}
            \label{fig:bivG_dissimilar_means_vanilla_mesh_size}
        \end{subfigure}%
        \hfill
        \begin{subfigure}[b]{0.45\textwidth}  
            \centering 
            \includegraphics[width=\textwidth]{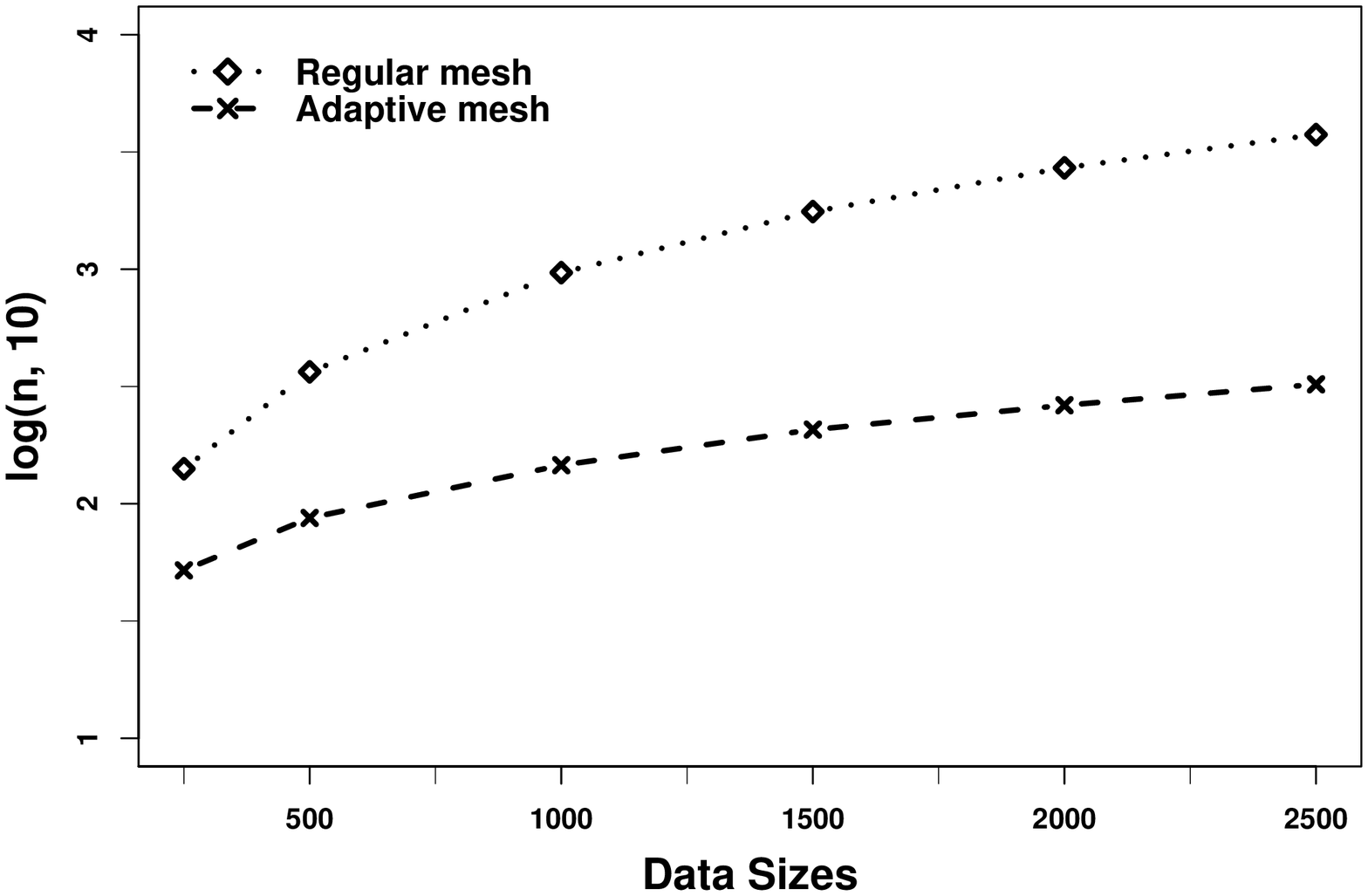}
            \caption{\gbfa}
            \label{fig:bivG_dissimilar_means_generalised_mesh_size}
        \end{subfigure}%
        \setcounter{subfigure}{5}
        \renewcommand\thesubfigure{\alph{subfigure}}
        \caption{Comparison of mesh sizes between regular and adaptive schemes.}
        \label{fig:bivG_dissimilar_means_mesh_size}
    \end{subfigure}%
    \vskip\baselineskip
    \begin{subfigure}[b]{\textwidth}
        \setcounter{subfigure}{0}
        \renewcommand\thesubfigure{\roman{subfigure}}
        \begin{subfigure}[b]{0.45\textwidth}
            \centering
            \includegraphics[width=\textwidth]{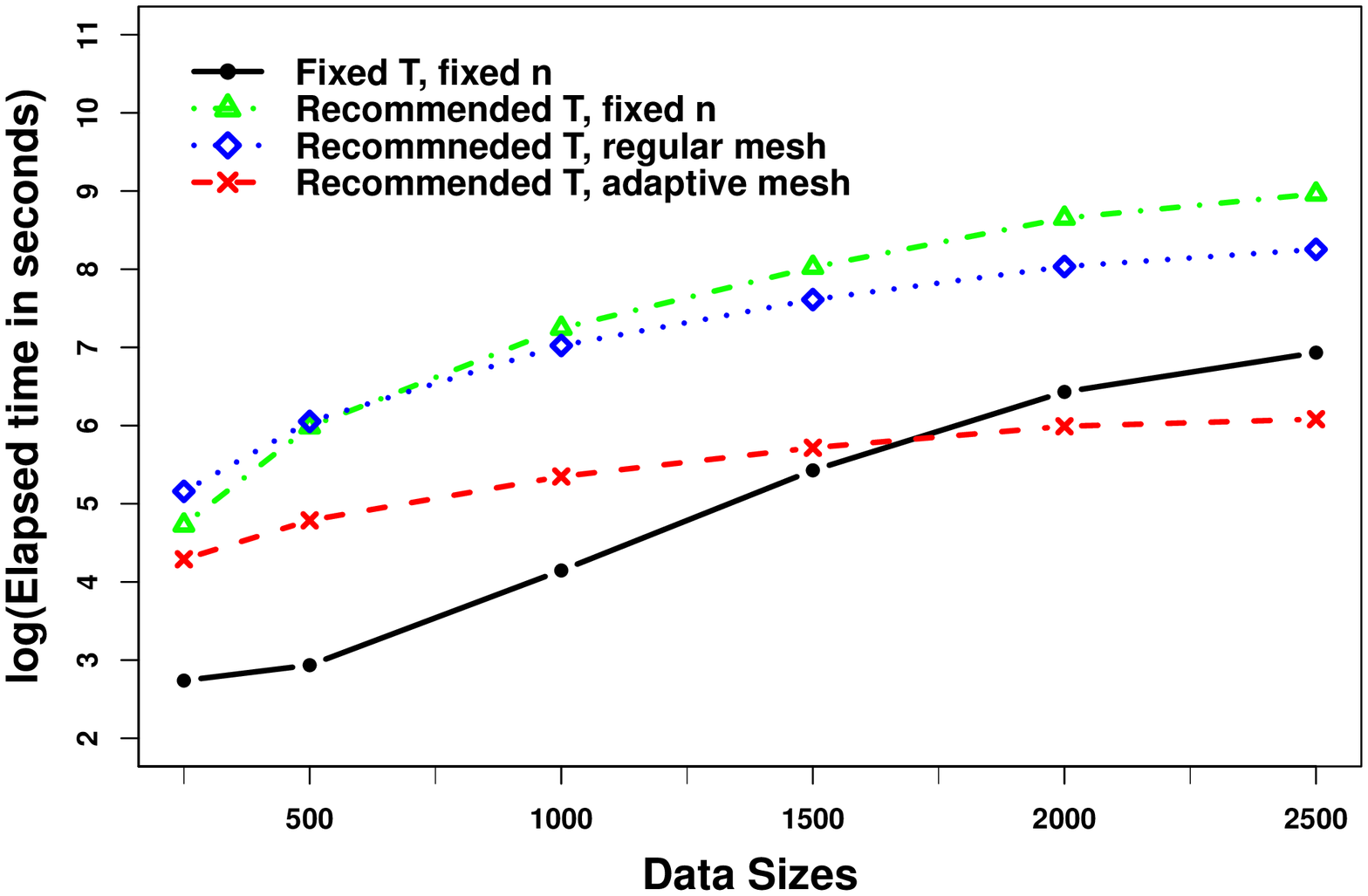}
            \caption{BF}
            \label{fig:bivG_dissimilar_means_vanilla_time_average}
        \end{subfigure}%
        \hfill
        \begin{subfigure}[b]{0.45\textwidth}  
            \centering 
            \includegraphics[width=\textwidth]{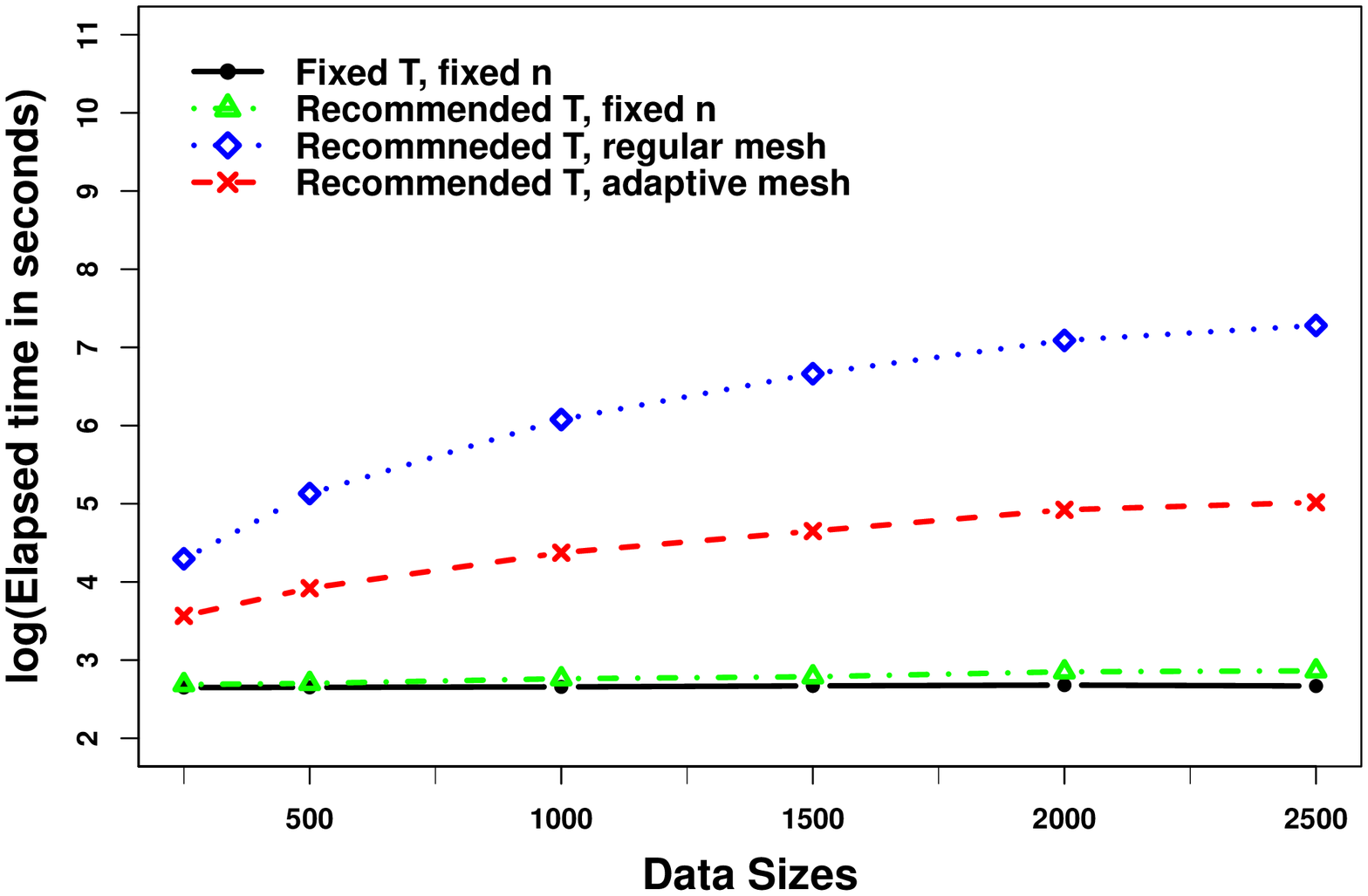}
            \caption{\gbfa}
            \label{fig:bivG_dissimilar_means_generalised_time_average}
        \end{subfigure}%
        \setcounter{subfigure}{6}
        \renewcommand\thesubfigure{\alph{subfigure}}
        \caption{Mean average computational run-times (based on ten runs).}
        \label{fig:bivG_dissimilar_means_time}
    \end{subfigure}%
    \caption{Bivariate Gaussian Example in $\SSH{\gamma}$ setting (continued).}    
\end{figure}

\clearpage
\subsection{Dimension Scaling} \label{subsec:GBF_dimension_study}

\begin{figure}[ht]
    \centering
    \begin{subfigure}[b]{0.49\textwidth}
        \centering
        \includegraphics[width=\textwidth]{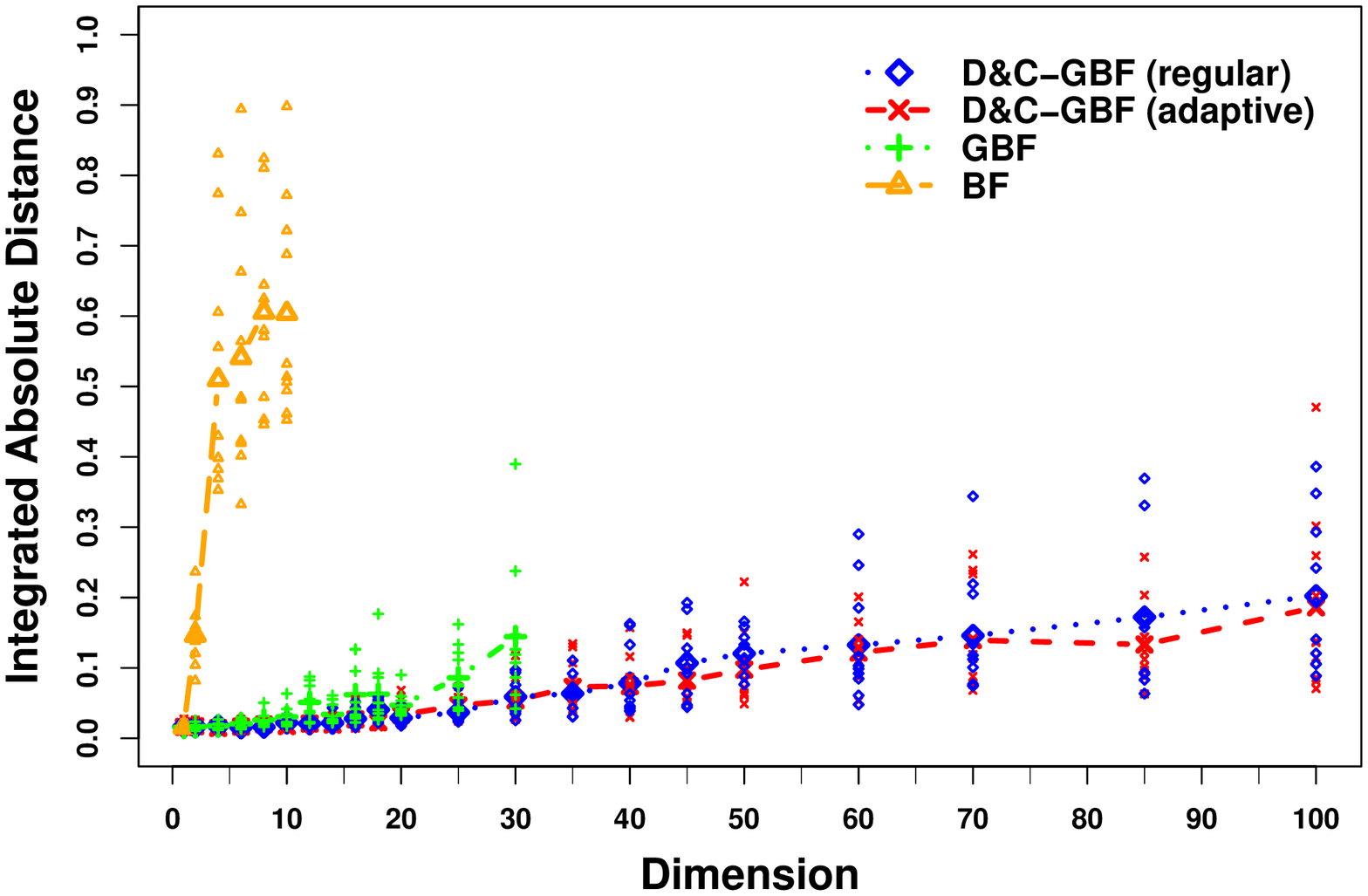}
        \caption{Integrated Absolute Distance.}
    \label{fig:dimension_study_IAD}
    \end{subfigure}
    \hfill
    \begin{subfigure}[b]{0.49\textwidth}
        \centering
        \includegraphics[width=\textwidth]{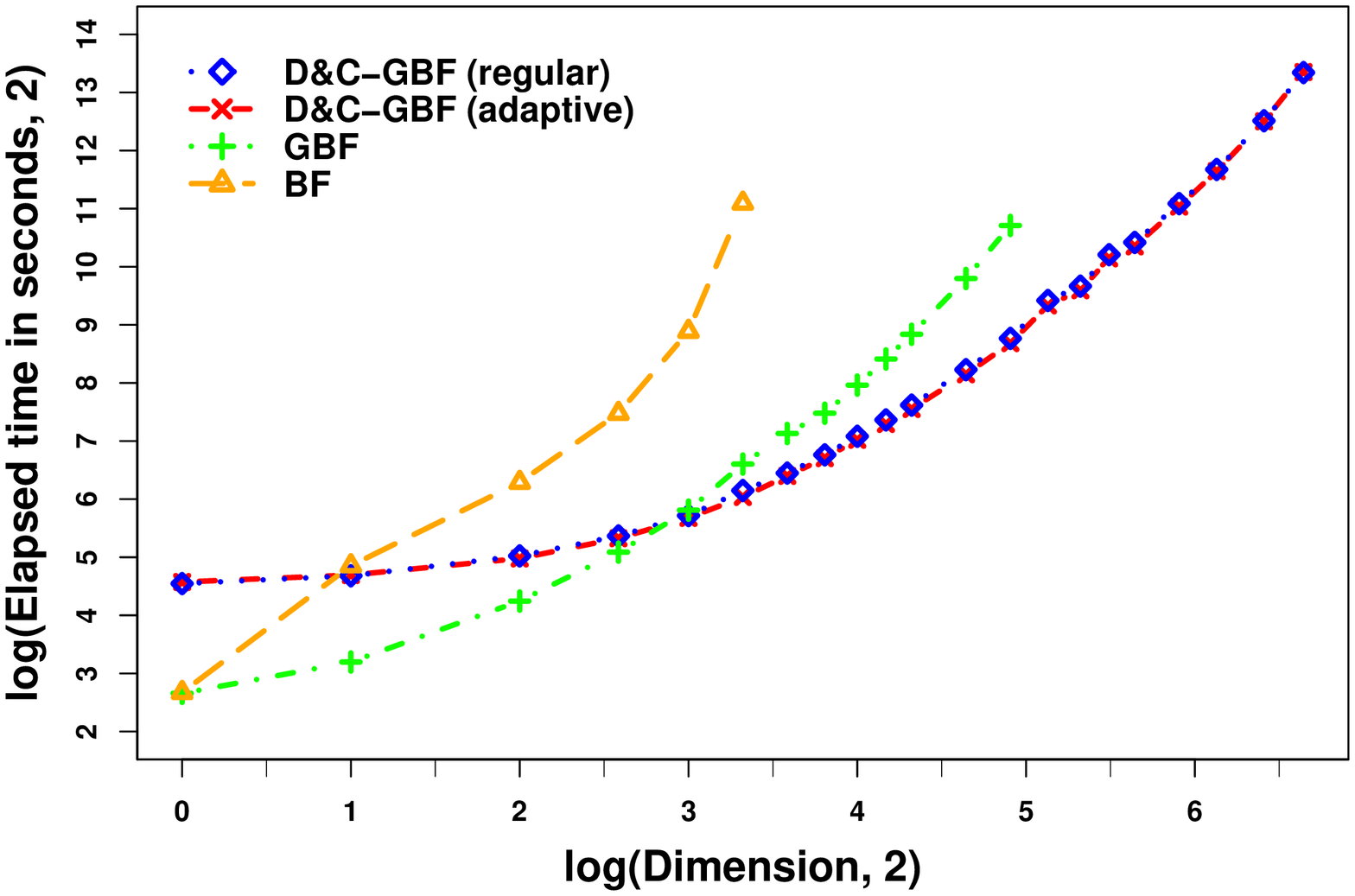}
        \caption{Computational cost.}
        \label{fig:dimension_study_time}
    \end{subfigure}
    \caption{Comparison of Fusion methodologies with increasing dimensionality (in the setting of \secref{subsec:GBF_dimension_study}). In \figref{fig:dimension_study_IAD}, lines connect the mean IAD (averaged over ten runs) while the points denote the individual IAD achieved on each run.}
    \label{fig:dimension_study}
\end{figure}

In this section we empirically study the performance of Fusion approaches (BF, \gbfa and \hgbfa) with increasing dimensionality. To do so we consider a $d$-dimensional multivariate Gaussian $f \propto \prod_{c=1}^{C} f_{c}$, where we let $C=8$ and $f_{c} \sim \mathcal{N}_{d}(\bm{0}, C\mathbf{\Sigma})$, and where
\begin{align*}
    \Sigma_{ii} = 1, & \qquad \text{for all } i \in\{1,\dots,d\}, \\
    \Sigma_{ij} = 0.9, & \qquad \text{for all } i \neq j, \, (i,j) \in \{1,\dots,d\},
\end{align*}
and simply vary $d$ (in steps from $d=1$ to $d=100$). For BF and \gbfa we use an adaptive mesh for $\mathcal{P}$, and for \hgbfa we consider both a regular and adaptive mesh for $\mathcal{P}$ with a balanced-binary tree hierarchy. In all cases we use the guidance developed in \secref{sec:GBF_guidance}. As we are in the $\SH{\lambda}$ setting (the true sub-posterior means are the same), we set $\lambda=1$. The lower bounds of the tolerable initial and iterative CESS are set to $0.05N$ (i.e.\ $\zeta=\zeta^{\prime}=0.05$) and we resample if the $\ESS$ drops below $0.5N$, where here we have $N=10000$. The results are presented in \figref{fig:dimension_study}.

As shown in \figref{fig:dimension_study_IAD}, the performance of all Fusion methods degrades with increasing dimensionality: both in terms of the average IAD and also the variance. As our target exhibits high correlation between components, BF struggles here even in low dimensions, whereas the \gbfa and \hgbfa approaches we have developed in this paper offer much better scaling with dimension. \hgbfa comfortably outperforms existing Fusion approaches for even moderate dimensionality in terms of IAD and computational cost. 

\section{Calculations for examples} \label{app:phi_calulations}

In this section, we provide the calculations necessary to implement the Fusion algorithms discussed in this paper. In particular, to implement \gbf (\secref{sec:GBF}) and \hgbf (\secref{sec:dc_gbf}), we must be able to compute $\phi_{c}$ given in \eqref{eq:phi}. This requires the computation of the first and second order derivatives of the log sub-posterior densities.

Furthermore, it is necessary to compute bounds of $\phi_{c}$. As noted in \secref{subsec:GBF_radonniko}, if it is not possible to (or simply difficult to) find tight bounds for $\phi_{c}$, we can use the general bounds given in \propositionref{prop:bounds}. To use these general bounds, we must find a upper bound on the matrix norm of $\mathbf{\Lambda}_{c}\nabla^{2}\log f_{c}(\bm{x})$ for $\bm{x} \in R_{c}$ (i.e.\ find $P^{\mathbf{\Lambda}_{c}}$ given in \eqref{eq:P}), which can be done by computing the matrix norm of the matrix which bounds the matrix $\mathbf{\Lambda}_{c}\nabla^{2}\log f_{c}(\bm{x})$ element-wise.

We note that in some cases, it may be easier to find the bound on the matrix norm of the Hessian of the \emph{transformed} sub-posterior, $f_{c}^{(z)}(\bm{z})$ where $\bm{z}:=\mathbf{\Lambda}_{c}^{-\frac{1}{2}}\bm{x}$. In particular, rather than finding a bound in \eqref{eq:P}, we can focus on finding the bound
\begin{equation}
    \label{eq:P_transformed}
    P^{\mathbf{\Lambda}_{c}} \geq \max_{\bm{z} \in R_{c}^{(z)}} \gamma\left(\nabla^{2} \log f_{c}^{(z)}(\bm{z})\right),
\end{equation}
which is equivalent to finding the bound in \eqref{eq:P}.

In \apxref{app:unbiased_estimators}, we detailed how we can simulate $\tilde{\rho}_{j}$. In particular, in \algoref{alg:unbiased_estimator_rho_j}, we perform a transformation on the space and in \stepref{alg:unbiased_estimator_rho_j:R_c}, we compute the layer information $R_{c}^{(z)}$ and so we can directly use this to find local element-wise bounds $\nabla^{2} \log f_{c}^{(z)}(\bm{z})$ for $\bm{z} \in R_{c}^{(z)}$. Therefore, to find $P^{\mathbf{\Lambda}_{c}}$, we just need to find bounds on the second order derivatives of the log-sub-posterior in the transformed space $\bm{z}:=\mathbf{\Lambda}_{c}^{-\frac{1}{2}}\bm{x}$ so that we can compute the matrix norm of the matrix which bounds $\nabla^{2} \log f_{c}^{(z)}(\bm{z})$ element-wise.

\subsection{Logistic Regression} \label{app:BLR_phi_calculations}

In \secref{subsec:log_reg}, we considered applying our Fusion methodologies to a logistic regression example with Gaussian prior distributions for the parameters. In particular, our sub-posterior densities were given by the posterior for Bayesian logistic regression with $\mathcal{N}_{d}(\mu_{j}, C\sigma_{\beta_{j}}^{2})$ prior for $\beta_{j}$ for $j=0,\dots,p$ is given by
\begin{equation}
    f_{c}(\bm{\beta}) := \pi(\bm{\beta} | \bm{y}) = \left[ \prod_{i=1}^{n} \frac{e^{X_{i}\bm{\beta} \cdot y_{i}}}{1+e^{X_{i}\bm{\beta}}} \right] \cdot \left[ \prod_{j=0}^{p} \frac{1}{\sqrt{2\pi C\sigma_{\beta_{j}}^{2}}} \exp \left( - \frac{(\beta_{j}-\mu_{j})^{2}}{2C\sigma_{\beta_{j}}^{2}} \right) \right]
\end{equation}
where $X\in\mathbb{R}^{n \times (p+1)}$ is the design matrix  so $X_{i} \bm{\beta} = \beta_{0} + \beta_{1} X_{i1} + \cdots \beta_{p} X_{ip}$. The log-posterior is given by
\begin{equation}
    \log f_{c}(\bm{\beta}) = \sum_{i=1}^{n} \left[ X_{i} \bm{\beta} \cdot y_{i} - \log(1+e^{\bm{\beta} X_{i}}) \right] - \sum_{j=0}^{p} \frac{(\beta_{j}-\mu_{j})^{2}}{2C\sigma_{\beta_{j}}^{2}} + \text{constant}.
\end{equation}
The first derivative of the log-posterior with respect to $\beta_{k}$ for $k=0,\dots,p$, is given by
\begin{align}
    \frac{\partial \log f_{c}(\bm{\beta})}{\partial \beta_{k}} 
	    & = \sum_{i=1}^{n} \left[ X_{ik} \cdot y_{i} - \frac{X_{ik} e^{X_{i} \bm{\beta}}}{1+e^{X_{i}\bm{\beta}}} \right] - \frac{(\beta_{k} - \mu_{k})}{C\sigma_{\beta_{k}}^{2}} \nonumber \\
	    & = \sum_{i=1}^{n} \left[ X_{ik} \cdot \left( y_{i} - \frac{1}{1+e^{-X_{i}\bm{\beta}}} \right) \right] - \frac{(\beta_{k} - \mu_{k})}{C\sigma_{\beta_{k}}^{2}}
\end{align}
and the second order derivatives of the log-posterior are given by
\begin{align}
    \frac{\partial^{2} \log f_{c}(\bm{\beta})}{\partial \beta_{k}^{2}} & = - \sum_{i=1}^{n} \frac{X_{ik}^{2} e^{X_{i}\bm{\beta}}}{(1+e^{X_{i}\bm{\beta}})^{2}} - \frac{1}{C\sigma_{\beta_{k}}^{2}}, \\
    \frac{\partial^{2} \log f_{c}(\bm{\beta})}{\partial \beta_{k} \partial \beta_{l}} & = - \sum_{i=1}^{n} \frac{X_{ik} X_{il} e^{X_{i} \bm{\beta}}}{(1+e^{X_{i} \bm{\beta}})^{2}} \text{ for } k \neq l,
\end{align}
for $k,l=0,\dots,p$. We can use these directly to compute $\phi_{c}$ given in \eqref{eq:phi}.

To compute the bounds of $\phi_{c}$, we can utilise the bounds provided in \propositionref{prop:bounds} (or in \eqref{eq:lower_bound} and \eqref{eq:upper_bound}). To do so, we must be able to compute an upper bound of the matrix norm $\mathbf{\Lambda}_{c}\nabla^{2}\log f_{c}(\bm{x})$ for $\bm{x}\in R_{c}$ where $R_{c}$ denotes the simulated layer information, i.e.\ to compute \eqref{eq:P}. While this can be done by computing the matrix norm of the matrix which bounds the matrix $\mathbf{\Lambda}_{c}\nabla^{2}\log f_{c}(\bm{x})$ element-wise, we noted above that it is typically easier to find bounds on the matrix norm of $\nabla^{2} \log f_{c}^{(z)}(\bm{z})$ where $\bm{z}:=\mathbf{\Lambda}_{c}^{-\frac{1}{2}}\bm{\beta}$, and instead we can focus on finding a bound in the transformed space, i.e.\ compute \eqref{eq:P_transformed}.

In this logistic regression setting, let $\bm{z} = \mathbf{\Lambda}_{c}^{-\frac{1}{2}} \bm{\beta}$ then the transformed posterior density is given by
\begin{equation}
    f_{c}^{(z)}(\bm{z}) = \pi(\bm{\beta}|X,\bm{y}) |J|,
\end{equation}
where $J = \mathbf{\Lambda}_{c}^{-\frac{1}{2}}$ is the Jacobian matrix with elements $J_{ij} = \frac{\partial z_{i}}{\partial \beta_{j}} = \mathbf{\Lambda}_{c,ij}^{-\frac{1}{2}}$. We have
\begin{equation}
    \log f_{c}^{(z)}(\bm{z}) = \log \pi(\bm{\beta}|X,\bm{y}) + \log |J|.
\end{equation}

Since $\bm{\beta} = \mathbf{\Lambda}_{c}^{\frac{1}{2}} \bm{z}$, we have
\begin{align}
    f_{c}^{(z)}(\bm{z}) 
    & := \pi(\bm{\beta}|X,\bm{y}) \cdot |\mathbf{\Lambda}_{c}^{-\frac{1}{2}}| \nonumber \\
    & = \left[ \prod_{i=1}^{n} \frac{e^{X_{i} \bm{\beta} \cdot y_{i}}}{1+e^{X_{i} \bm{\beta}}} \right] \cdot \left[ \prod_{j=0}^{p} \frac{1}{\sqrt{2 \pi C \sigma_{\beta_{j}}^{2}}} \exp \left( - \frac{(\beta_{j}-\mu_{j})^{2}}{2C\sigma_{\beta_{j}}^{2}} \right) \right] \cdot |\mathbf{\Lambda}_{c}^{-\frac{1}{2}}| \nonumber \\
    & = \left[ \prod_{i=1}^{n} \frac{e^{X_{i} (\mathbf{\Lambda}_{c}^{\frac{1}{2}} \bm{z}) \cdot y_{i}}}{1+e^{X_{i} (\mathbf{\Lambda}_{c}^{\frac{1}{2}} \bm{z})}} \right] \cdot \left[ \prod_{j=0}^{p} \frac{1}{\sqrt{2 \pi C \sigma_{\beta_{j}}^{2}}} \exp \left( - \frac{\left((\mathbf{\Lambda}_{c}^{\frac{1}{2}}\bm{z})_{j}-\mu_{j}\right)^{2}}{2C\sigma_{\beta_{j}}^{2}} \right) \right] \cdot |\mathbf{\Lambda}_{c}^{-\frac{1}{2}}|,
\end{align}
so
\begin{equation}
    \log f_{c}^{(z)}(\bm{z}) = \sum_{i=1}^{n} \left[ X_{i} (\mathbf{\Lambda}_{c}^{\frac{1}{2}} \bm{z}) \cdot y_{i} - \log(1+e^{X_{i} (\mathbf{\Lambda}_{c}^{\frac{1}{2}} \bm{z})}) \right] - \sum_{j=0}^{p} \frac{\left((\mathbf{\Lambda}_{c}^{\frac{1}{2}}\bm{z})_{j}-\mu_{j}\right)^{2}}{2C\sigma_{\beta_{j}}^{2}} + \text{constant}.
\end{equation}

We first note that since $\bm{\beta} = \mathbf{\Lambda}_{c}^{\frac{1}{2}} \bm{z}$, then $\beta_{i} = (\mathbf{\Lambda}_{c}^{\frac{1}{2}} \bm{z})_{i} = \sum_{k} \Lambda_{ik}^{\frac{1}{2}} z_{k}$. So we have
\begin{align}
    \frac{\partial (X_{i}\mathbf{\Lambda}_{c}^{\frac{1}{2}})\bm{z}}{\partial z_{k}}
    & = \frac{\partial}{\partial z_{k}} \sum_{j} X_{ij} \beta_{j} \nonumber \\
    & = \frac{\partial}{\partial z_{k}} \sum_{j} X_{ij} \left( \sum_{k} \Lambda_{jk}^{\frac{1}{2}} z_{k} \right) \nonumber \\
    & = \sum_{j} X_{ij} \Lambda_{jk}^{\frac{1}{2}} \nonumber \\
    & = (X\mathbf{\Lambda}_{c}^{\frac{1}{2}})_{ik} \label{eq:transformed_derivatives_1}
\end{align}
and also we have
\begin{align}
    \frac{\partial (\mathbf{\Lambda}_{c}^{\frac{1}{2}} \bm{z})_{i}}{\partial z_{k}} 
    & = \frac{\partial}{\partial z_{k}} \sum_{j} \Lambda_{ij}^{\frac{1}{2}} z_{j} = \Lambda_{ik}^{\frac{1}{2}} \label{eq:transformed_derivatives_2}
\end{align}

Using \eqref{eq:transformed_derivatives_1} and \eqref{eq:transformed_derivatives_2}, then the first derivative of the log transformed posterior with respect to $\beta_{k}$ for $k=0,\dots,p$, is given by
\begin{align}
    \frac{\partial \log f_{c}^{(z)}(\bm{z})}{\partial z_{k}} 
    & = \sum_{i=1}^{n} \left[ (X\mathbf{\Lambda}_{c}^{\frac{1}{2}})_{ik} \cdot \left( y_{i} - \frac{1}{1+e^{-(X_{i}\mathbf{\Lambda}_{c}^{\frac{1}{2}})\bm{z}}} \right) \right] - \sum_{j=0}^{p} \frac{\Lambda_{jk}^{\frac{1}{2}}\left((\mathbf{\Lambda}_{c}^{\frac{1}{2}}\bm{z})_{j}-\mu_{j}\right)}{C\sigma_{\beta_{j}}^{2}}. 
\end{align}

Then the second order derivatives are given by
\begin{align}
    \frac{\partial^{2} \log f_{c}^{(z)}(\bm{z})}{\partial z_{k} \partial z_{l}} 
    & = - \sum_{i=1}^{n} \frac{(X\mathbf{\Lambda}_{c}^{\frac{1}{2}})_{ik} (X\mathbf{\Lambda}_{c}^{\frac{1}{2}})_{il} e^{(X_{i}\mathbf{\Lambda}_{c}^{\frac{1}{2}})\bm{z}}}{\left(1+e^{(X_{i} \mathbf{\Lambda}_{c}^{\frac{1}{2}})\bm{z}}\right)^{2}} - \sum_{j=0}^{p} \frac{\Lambda_{jk}^{\frac{1}{2}} \Lambda_{jl}^{\frac{1}{2}}}{C \sigma_{\beta_{j}}^{2}},
\end{align}
for $k,l=0,\dots,p$.

To find bounds for $\phi_{c}$, we must now try to find bounds on the second derivatives given above and compute the matrix norm of the matrix made up of these bounds (which ultimately bounds $\nabla^{2} \log f_{c}^{(z)}(\bm{z})$ element-wise). For this example, we can find global and lower bounds of the second derivatives. Note however, we typically will expect better performance with the local bounds on $P^{\mathbf{\Lambda}_{c}}$ \eqref{eq:P_transformed} (as this will typically lead to the expected number of points we need to evaluate while performing Poisson thinning, $\kappa_{c}$, to be lower) despite these bounds being slightly more expensive to compute in practice.

\subsubsection{Global bounds of \texorpdfstring{$P^{\mathbf{\Lambda}_{c}}$}{the matrix norm}}

We first note that $\frac{e^{x}}{(1+e^{x})^{2}} \leq \frac{1}{4}$ for all $x$ (and this maximum occurs at $x=0$). We can utilise this to obtain a global bound:
\begin{equation}
    \sup\left[\left| \frac{\partial^{2} \log f_{c}^{(z)}(\bm{z})}{\partial z_{k} \partial z_{l}} \right| \right] = \sum_{i=1}^{n} \frac{|X\mathbf{\Lambda}_{c}^{\frac{1}{2}}|_{ik}\cdot|X\mathbf{\Lambda}_{c}^{\frac{1}{2}}|_{il}}{4} + \sum_{j=0}^{p} \frac{\Lambda_{jk}^{\frac{1}{2}} \Lambda_{jl}^{\frac{1}{2}}}{C \sigma_{\beta_{j}}^{2}}.
\end{equation}

\subsubsection{Local bounds of \texorpdfstring{$P^{\mathbf{\Lambda}_{c}}$}{the matrix norm}} \label{sec:log_reg_phi_bounds}

Local bounds can be obtained if we can find local bounds for 
\begin{equation}
    G_{1}(\bm{z}) := \frac{e^{(X_{i}\mathbf{\Lambda}_{c}^{\frac{1}{2}})\bm{z}}}{\left(1+e^{(X_{i} \mathbf{\Lambda}_{c}^{\frac{1}{2}})\bm{z}}\right)^{2}},
\end{equation}
for $i=1,\dots,n$. In that case, we have
\begin{equation}
    \sup_{\bm{z} \in R^{(z)}}\left[\left| \frac{\partial^{2} \log f_{c}^{(z)}(\bm{z})}{\partial z_{k} \partial z_{l}} \right| \right] = \sum_{i=1}^{n} \left[|X\mathbf{\Lambda}_{c}^{\frac{1}{2}}|_{ik}\cdot|X\mathbf{\Lambda}_{c}^{\frac{1}{2}}|_{il}\cdot\max_{\bm{z} \in R^{(z)}}\left\{G_{1}(\bm{z})\right\}\right] + \sum_{j=0}^{p} \frac{\Lambda_{jk}^{\frac{1}{2}} \Lambda_{jl}^{\frac{1}{2}}}{C \sigma_{\beta_{j}}^{2}}.
\end{equation}

To compute $\max_{\bm{z} \in R^{(z)}}\left\{G_{1}(\bm{z})\right\}$, see \secref{sec:NB_phi_local_bounds} and \algoref{alg:G_max} and set $r=1$.

\subsection{Robust Regression} \label{app:BRR_phi_calculations}

In \secref{subsec:robust_reg}, we considered a robust regression example (using a student-$t$ distribution) with Gaussian prior distributions for the parameters. In particular, our sub-posterior densities were given by the posterior for Bayesian robust regression with $\mathcal{N}_{d}(\mu_{j}, C\sigma_{\beta_{j}}^{2})$ prior for $\beta_{j}$ for $j=0,\dots,p$ is given by
\begin{align}
    f_{c}(\bm{\beta}) = \pi(\bm{\beta}|X,\bm{y}) 
    & := \left[ \prod_{i=1}^{n} \frac{\Gamma(\frac{\nu+1}{2})}{\Gamma(\frac{\nu}{2}) \sqrt{\pi\nu} \sigma} \left( 1 + \frac{1}{\nu} \left( \frac{y_{i}-X_{i}\bm{\beta}}{\sigma} \right)^{2} \right)^{-\left(\frac{\nu+1}{2}\right)} \right] \nonumber \\
    & \qquad \cdot \left[ \prod_{j=0}^{p} \frac{1}{\sqrt{2\pi C\sigma_{\beta_{j}}^{2}}} \exp \left( - \frac{(\beta_{j}-\mu_{j})^{2}}{2C\sigma_{\beta_{j}}^{2}} \right) \right].
\end{align}

The log-posterior is given by
\begin{equation}
    \log f_{c}(\bm{\beta})= -\left(\frac{\nu+1}{2}\right) \sum_{i=1}^{n} \log \left( 1+\frac{1}{\nu\sigma^{2}} \left(y_{i}-X_{i}\bm{\beta}\right)^{2} \right) - \sum_{j=0}^{p} \frac{(\beta_{j}-\mu_{j})^{2}}{2C\sigma_{\beta_{j}}^{2}} + \text{constant}.
\end{equation}

The first derivative of the log-posterior with respect to $\beta_{k}$ for $k=0,\dots,p$ is given by
\begin{align}
    \frac{\partial \log \pi(\bm{\beta}|X,\bm{y})}{\partial \beta_{k}}
    & = - \left(\frac{\nu+1}{2}\right) \sum_{i=1}^{n} \frac{-\frac{2X_{ik}}{\nu\sigma^{2}} (y_{i}-X_{i}\bm{\beta})}{1+\frac{1}{\nu\sigma^{2}}(y_{i}-X_{i}\bm{\beta})^{2}} - \frac{(\beta_{k}-\mu_{k})}{C\sigma_{\beta_{k}}^{2}} \nonumber \\
    & = (\nu+1) \sum_{i=1}^{n} \frac{X_{ik}(y_{i}-X_{i}\bm{\beta})}{\nu\sigma^{2}+(y_{i}-X_{i}\bm{\beta})^{2}} - \frac{(\beta_{k}-\mu_{k})}{C\sigma_{\beta_{k}}^{2}},
\end{align}
and the second order derivatives of the log-posterior are given by
\begin{align}
    \frac{\partial^{2} \log \pi(\bm{\beta}|X,\bm{y})}{\partial \beta_{k}^{2}} & = (\nu+1) \sum_{i=1}^{n} \frac{X_{ik}^{2} \left( (y_{i}-X_{i}\bm{\beta})^{2}-\nu\sigma^{2} \right)}{\left(\nu\sigma^{2}+(y_{i}-X_{i}\bm{\beta})^{2}\right)^{2}} - \frac{1}{C\sigma_{\beta_{k}}^{2}}, \\
    \frac{\partial^{2} \log \pi(\bm{\beta}|X,\bm{y})}{\partial \beta_{k} \partial \beta_{l}} & = (\nu+1) \sum_{i=1}^{n} \frac{X_{ik}X_{il} \left( (y_{i}-X_{i}\bm{\beta})^{2}-\nu\sigma^{2} \right)}{\left(\nu\sigma^{2}+(y_{i}-X_{i}\bm{\beta})^{2}\right)^{2}} \text{ for } k \neq l,
\end{align}
for $k,l=0,\dots,p$. We can use these derivatives directly to compute $\phi_{c}$ given in \eqref{eq:phi}.

Following in the same approach as \secref{sec:log_reg_phi_bounds}, we can compute the bounds of $\phi_{c}$ (in \eqref{eq:phi}) by utilising the bounds provided in \eqref{eq:lower_bound} and \eqref{eq:upper_bound}. As noted in \secref{sec:log_reg_phi_bounds}, we must be able to find an upper bound on the matrix norm of $\nabla^{2} \log f_{c}^{(z)}(\bm{z})$ where $\bm{z}:=\mathbf{\Lambda}_{c}^{-\frac{1}{2}}\bm{\beta}$, i.e.\ compute \eqref{eq:P_transformed}. To do so, we can compute the matrix norm of the matrix which bounds $\nabla^{2} \log f_{c}^{(z)}(\bm{z})$ element-wise. Now, let $\bm{z} = \mathbf{\Lambda}_{c}^{-\frac{1}{2}} \bm{\beta}$ then $f_{c}^{(z)}(\bm{z}) = \pi(\bm{\beta}|X,\bm{y}) |J|$, where $J = \mathbf{\Lambda}_{c}^{-\frac{1}{2}}$ is the Jacobian matrix, so we have
\begin{align}
    f_{c}^{(z)}(\bm{z}) 
    & := \pi(\bm{\beta}|X,\bm{y}) \cdot |\mathbf{\Lambda}_{c}^{-\frac{1}{2}}| \nonumber \\
    & = \left[ \prod_{i=1}^{n} \frac{\Gamma(\frac{\nu+1}{2})}{\Gamma(\frac{\nu}{2}) \sqrt{\pi\nu} \sigma} \left( 1 + \frac{1}{\nu} \left( \frac{y_{i}-X_{i}(\mathbf{\Lambda}_{c}^{\frac{1}{2}} \bm{z})}{\sigma} \right)^{2} \right)^{-\left(\frac{\nu+1}{2}\right)} \right] \nonumber \\
    & \qquad \cdot \left[ \prod_{j=0}^{p} \frac{1}{\sqrt{2\pi C\sigma_{\beta_{j}}^{2}}} \exp \left( - \frac{\left((\mathbf{\Lambda}_{c}^{\frac{1}{2}}\bm{z})_{j}-\mu_{j}\right)^{2}}{2C\sigma_{\beta_{j}}^{2}} \right) \right] \cdot |\mathbf{\Lambda}_{c}^{-\frac{1}{2}}|.
\end{align}
so
\begin{align}
    \log f_{c}^{(z)}(\bm{z}) 
    & = -\left(\frac{\nu+1}{2}\right) \sum_{i=1}^{n} \log \left( 1+\frac{1}{\nu\sigma^{2}} \left(y_{i}-X_{i}(\mathbf{\Lambda}_{c}^{\frac{1}{2}} \bm{z})\right)^{2} \right) \nonumber \\
    & \qquad - \sum_{j=0}^{p} \frac{\left((\mathbf{\Lambda}_{c}^{\frac{1}{2}}\bm{z})_{j}-\mu_{j}\right)^{2}}{2C\sigma_{\beta_{j}}^{2}} + \text{constant}.
\end{align}

Recall from \eqref{eq:transformed_derivatives_1} and \eqref{eq:transformed_derivatives_2}, then The first derivative of the log transformed posterior with respect to $\beta_{k}$ for $k=0,\dots,p$ is given by
\begin{align}
    \frac{\partial \log f_{c}^{(z)}(\bm{z})}{\partial z_{k}} 
    & = \left(\frac{\nu+1}{2}\right) \sum_{i=1}^{n} \frac{-\frac{2(X\mathbf{\Lambda}_{c}^{\frac{1}{2}})_{ik}}{\nu\sigma^{2}}\left(y_{i}-(X_{i}\mathbf{\Lambda}_{c}^{\frac{1}{2}}) \bm{z}\right)}{1+\frac{1}{\nu\sigma^{2}}\left(y_{i}-(X_{i}\mathbf{\Lambda}_{c}^{\frac{1}{2}})\bm{z}\right)^{2}} - \sum_{j=0}^{p} \frac{\Lambda_{jk}^{\frac{1}{2}}\left((\mathbf{\Lambda}_{c}^{\frac{1}{2}}\bm{z})_{j}-\mu_{j}\right)}{C\sigma_{\beta_{j}}^{2}} \nonumber \\
    & = (\nu+1) \sum_{i=1}^{n} \frac{(X\mathbf{\Lambda}_{c}^{\frac{1}{2}})_{ik}\left(y_{i}-(X_{i}\mathbf{\Lambda}_{c}^{\frac{1}{2}}) \bm{z}\right)}{\nu\sigma^{2}+\left(y_{i}-(X_{i}\mathbf{\Lambda}_{c}^{\frac{1}{2}}) \bm{z}\right)^{2}} - \sum_{j=0}^{p} \frac{\Lambda_{jk}^{\frac{1}{2}}\left((\mathbf{\Lambda}_{c}^{\frac{1}{2}}\bm{z})_{j}-\mu_{j}\right)}{C\sigma_{\beta_{j}}^{2}} 
\end{align}

Then the second order derivatives are given by
\begin{align}
    \frac{\partial^{2} \log f_{c}^{(z)}(\bm{z})}{\partial z_{k} \partial z_{l}} 
    & = (\nu+1) \sum_{i=1}^{n} \frac{(X\mathbf{\Lambda}_{c}^{\frac{1}{2}})_{ik}(X\mathbf{\Lambda}_{c}^{\frac{1}{2}})_{il} \left( \left(y_{i}-(X_{i}\mathbf{\Lambda}_{c}^{\frac{1}{2}})\bm{z}\right)^{2}-\nu\sigma^{2} \right)}{\left(\left(y_{i}-(X_{i}\mathbf{\Lambda}_{c}^{\frac{1}{2}})\bm{z}\right)^{2}+\nu\sigma^{2}\right)^{2}} - \sum_{j=0}^{p} \frac{\Lambda_{jk}^{\frac{1}{2}} \Lambda_{jl}^{\frac{1}{2}}}{C \sigma_{\beta_{j}}^{2}} \label{eq:RR_transformed_second_order_derivatives_2}
\end{align}
for $k,l=0,\dots,p$.

\subsubsection{Global bounds of \texorpdfstring{$P^{\mathbf{\Lambda}_{c}}$}{the matrix norm}}

To compute $P^{\mathbf{\Lambda}_{c}}$ for this example, first note that we can write 
\begin{equation}
    \frac{\partial^{2} \log f_{c}^{(z)}(\bm{z})}{\partial z_{k} \partial z_{l}} = (\nu+1) \sum_{i=1}^{n} (X\mathbf{\Lambda}_{c}^{\frac{1}{2}})_{ik}(X\mathbf{\Lambda}_{c}^{\frac{1}{2}})_{il} \left[ \frac{1}{E_{i}+b}-\frac{2b}{(E_{i}+b)^{2}} \right] - \sum_{j=0}^{p} \frac{\Lambda_{jk}^{\frac{1}{2}} \Lambda_{jl}^{\frac{1}{2}}}{C \sigma_{\beta_{j}}^{2}},
\end{equation}
for $k,l=0,\dots,p$, where $b=\nu\sigma^{2}$ and $E_{i}=\left(y_{i}-(X_{i}\mathbf{\Lambda}_{c}^{\frac{1}{2}})\bm{z}\right)^{2}$. Now let
\begin{equation*}
    K(E_{i})=\frac{1}{E_{i}+b}-\frac{2b}{(E_{i}+b)^{2}},
\end{equation*}
then the derivative is given by
\begin{equation*}
    K^{\prime}(E_{i}) = -\frac{1}{(E_{i}+b)^{2}}+\frac{4b}{(E_{i}+b)^{3}}.
\end{equation*}
Setting $K^{\prime}(E_{i})=0$ gives $E_{i}=3b$, and we have $K(E_{i}=3b)=\frac{1}{8b}$. So the supremum of the second derivative is given by
\begin{equation}
    \sup\left[\left| \frac{\partial^{2} \log f_{c}^{(z)}(\bm{z})}{\partial z_{k} \partial z_{l}} \right| \right] = \frac{(\nu+1)}{8\nu\sigma^{2}} \sum_{i=1}^{n} |X\mathbf{\Lambda}_{c}^{\frac{1}{2}}|_{ik}\cdot|X\mathbf{\Lambda}_{c}^{\frac{1}{2}}|_{il} - \sum_{j=0}^{p} \frac{\Lambda_{jk}^{\frac{1}{2}} \Lambda_{jl}^{\frac{1}{2}}}{C \sigma_{\beta_{j}}^{2}}.
\end{equation}

We can therefore use this to compute $P^{\mathbf{\Lambda}_{c}}$ to compute bounds for $\phi_{c}$ as per \eqref{eq:lower_bound} and \eqref{eq:upper_bound}.

\subsection{Negative Binomial Regression} \label{app:BNBR_phi_calculations}

In \secref{subsec:NB_reg}, we considered a negative Binomial regression example with Gaussian prior distributions for the parameters. In particular, our sub-posterior densities were given by the posterior density with $\mathcal{N}_{d}(\mu_{j},C\sigma_{\beta_{j}}^{2})$ priors for $\beta_{j}$ for $j=0,\dots,p$, is given by
\begin{align}
    f_{c}(\bm{\beta}) 
    & := \pi(\bm{\beta}|X,\bm{y}) \nonumber \\
    & = \left[ \prod_{i=1}^{n} \frac{\Gamma(y_{i}+r)}{y_{i}!\Gamma(r)} \left(\frac{\mu_{i}}{\mu_{i}+r}\right)^{y_{i}} \left(\frac{r}{\mu_{i}+r}\right)^{r} \right] \cdot \left[ \prod_{j=0}^{p} \frac{1}{\sqrt{2\pi C\sigma_{\beta_{j}}^{2}}} \exp \left( - \frac{(\beta_{j}-\mu_{j})^{2}}{2C\sigma_{\beta_{j}}^{2}} \right) \right] \nonumber \\
    & = \left[ \prod_{i=1}^{n} \frac{\Gamma(y_{i}+r)}{y_{i}!\Gamma(r)} \frac{\exp(X_{i}\bm{\beta} \cdot y_{i}) \cdot r^{r}}{(\exp(X_{i}\bm{\beta})+r)^{y_{i}+r}} \right] \cdot \left[ \prod_{j=0}^{p} \frac{1}{\sqrt{2\pi C\sigma_{\beta_{j}}^{2}}} \exp \left( - \frac{(\beta_{j}-\mu_{j})^{2}}{2C\sigma_{\beta_{j}}^{2}} \right) \right]
\end{align}

The log-posterior is given by
\begin{equation}
    \log f_{c}(\bm{\beta}) = \sum_{i=1}^{n} \left[ X_{i}\bm{\beta} \cdot y_{i} - (y_{i}+r)\log\left(\exp(X_{i}\bm{\beta})+r\right) \right] - \sum_{j=0}^{p} \frac{(\beta_{j}-\mu_{j})^{2}}{2C\sigma_{\beta_{j}}} + \text{constant}.
\end{equation}

The first order derivative of the log-posterior with respect to $\beta_{k}$ for $k=0,\dots,p$, is given by
\begin{align}
    \frac{\partial\log f_{c}(\bm{\beta})}{\partial \beta_{k}}
    & = \sum_{i=1}^{n} \left[ X_{ik}y_{i} - \frac{(y_{i}+r)X_{ik}\exp(X_{i}\bm{\beta})}{\exp(X_{i}\bm{\beta})+r} \right] - \frac{(\beta_{k}-\mu_{k})}{C\sigma_{\beta_{k}}^{2}}, \nonumber \\
    & = \sum_{i=1}^{n} \left[ X_{ik} \cdot \left( y_{i} - \frac{(y_{i}+r)\exp(X_{i}\bm{\beta})}{\exp(X_{i}\bm{\beta})+r} \right) \right] - \frac{(\beta_{k}-\mu_{k})}{C\sigma_{\beta_{k}}^{2}},
\end{align}
and the second order derivatives of the log-posterior are given by
\begin{align}
    \frac{\partial^{2}\log f_{c}(\bm{\beta})}{\partial\beta_{k}^{2}} & = - \sum_{i=1}^{n} \frac{(y_{i}+r) r X_{ik}^{2}\exp(X_{i}\bm{\beta})}{\left(\exp(X_{i}\bm{\beta})+r\right)^{2}} - \frac{1}{C\sigma_{\beta_{k}}^{2}}, \\
    \frac{\partial^{2}\log f_{c}(\bm{\beta})}{\partial\beta_{k}\partial\beta_{l}} & = - \sum_{i=1}^{n} \frac{(y_{i}+r) r X_{ik}X_{il}\exp(X_{i}\bm{\beta})}{\left(\exp(X_{i}\bm{\beta})+r\right)^{2}} \quad \text{ for } k \neq l,
\end{align}
for $k,l=0,\dots,p$. We can use these directly to compute $\phi_{c}$ given in \eqref{eq:phi}.

Following in the same approach as \secref{sec:log_reg_phi_bounds}, we can compute the bounds of $\phi_{c}$ (in \eqref{eq:phi}) by utilising the bounds provided in \eqref{eq:lower_bound} and \eqref{eq:upper_bound}. As noted in \secref{sec:log_reg_phi_bounds}, we must compute \eqref{eq:P_transformed}. To do so, we can compute the matrix norm of the matrix which bounds $\nabla^{2} \log f_{c}^{(z)}(\bm{z})$ element-wise. We have
\begin{align}
    f_{c}^{(z)}(\bm{z}) 
    & := \pi(\bm{\beta}|X,\bm{y}) \cdot |\mathbf{\Lambda}_{c}^{-\frac{1}{2}}| \nonumber \\
    & = \left[ \prod_{i=1}^{n} \frac{\Gamma(y_{i}+r)}{y_{i}!\Gamma(r)} \frac{\exp(X_{i}(\mathbf{\Lambda}_{c}^{\frac{1}{2}}\bm{z}) \cdot y_{i}) \cdot r^{r}}{\left(\exp(X_{i}(\mathbf{\Lambda}_{c}^{\frac{1}{2}}\bm{z}))+r\right)^{y_{i}+r}} \right] \nonumber \\
    & \qquad \cdot \left[ \prod_{j=0}^{p} \frac{1}{\sqrt{2\pi C\sigma_{\beta_{j}}^{2}}} \exp \left( - \frac{\left((\mathbf{\Lambda}_{c}^{\frac{1}{2}}\bm{z})_{j}-\mu_{j}\right)^{2}}{2C\sigma_{\beta_{j}}^{2}} \right) \right] \cdot |\mathbf{\Lambda}_{c}^{-\frac{1}{2}}|,
\end{align}
and
\begin{align}
    \log f_{c}^{(z)}(\bm{z}) 
    & = \sum_{i=1}^{n} \left[ (X_{i}\mathbf{\Lambda}_{c}^{\frac{1}{2}})\bm{z} \cdot y_{i} - (y_{i}+r)\log(\exp((X_{i}\mathbf{\Lambda}_{c}^{\frac{1}{2}})\bm{z})+r) \right] \nonumber \\
    & \qquad - \sum_{j=0}^{p} \frac{\left((\mathbf{\Lambda}_{c}^{\frac{1}{2}}\bm{z})_{j}-\mu_{j}\right)^{2}}{2C\sigma_{\beta_{j}}} + \text{constant}.
\end{align}
    
Recall from \eqref{eq:transformed_derivatives_1} and \eqref{eq:transformed_derivatives_2} that we have $\frac{\partial (X_{i}\mathbf{\Lambda}_{c}^{\frac{1}{2}})\bm{z}}{\partial z_{k}}=(X\mathbf{\Lambda}_{c}^{\frac{1}{2}})_{ik}$ and $\frac{\partial (\mathbf{\Lambda}_{c}^{\frac{1}{2}} \bm{z})_{i}}{\partial z_{k}}=\Lambda_{ik}^{\frac{1}{2}}$. Then first derivative of the log transformed posterior with respect to $\beta_{k}$ is given by
\begin{align}
    \frac{\partial \log f_{c}^{(z)}(\bm{z})}{\partial z_{k}}
    & = \sum_{i=1}^{n} \left[ (X\mathbf{\Lambda}_{c}^{\frac{1}{2}})_{ik} \cdot \left( y_{i} - \frac{(y_{i}+r)\exp((X_{i}\mathbf{\Lambda}_{c}^{\frac{1}{2}})\bm{z})}{\exp((X_{i}\mathbf{\Lambda}_{c}^{\frac{1}{2}})\bm{z})+r} \right) \right] \nonumber \\
    & \qquad - \sum_{j=0}^{p} \frac{\Lambda_{jk}^{\frac{1}{2}}\left((\mathbf{\Lambda}_{c}^{\frac{1}{2}}\bm{z})_{j}-\mu_{j}\right)}{C\sigma_{\beta_{j}}^{2}} 
\end{align}
and the second order derivatives are given by
\begin{align}
    \frac{\partial^{2} \log f_{c}^{(z)}(\bm{z})}{\partial z_{k} \partial z_{l}} & = - \sum_{i=1}^{n} \frac{(y_{i}+r)r(X\mathbf{\Lambda}_{c}^{\frac{1}{2}})_{ik}(X\mathbf{\Lambda}_{c}^{\frac{1}{2}})_{il}\exp((X_{i}\mathbf{\Lambda}_{c}^{\frac{1}{2}})\bm{z})}{\left(\exp((X_{i}\mathbf{\Lambda}_{c}^{\frac{1}{2}})\bm{z})+r\right)^{2}} - \sum_{j=0}^{p} \frac{\Lambda_{jk}^{\frac{1}{2}} \Lambda_{jl}^{\frac{1}{2}}}{C \sigma_{\beta_{j}}^{2}}.
\end{align}

To find bounds for $r_{c}$, we must now try to find bounds on the second derivatives given above and compute the matrix norm of the matrix made up of these bounds (which ultimately bounds $\nabla^{2} \log f_{c}^{(z)}(\bm{z})$ element-wise). For this example, we can find global and lower bounds of the second derivatives. Note however, we typically will expect better performance with the local bounds on $P^{\mathbf{\Lambda}_{c}}$ (as this will typically lead to the expected number of points we need to evaluate while performing Poisson thinning, $\kappa_{c}$, to be lower) despite these bounds being slightly more expensive to compute in practice.

\subsubsection{Global bounds of \texorpdfstring{$P^{\mathbf{\Lambda}_{c}}$}{the matrix norm}}

Note that $\frac{e^{ax}}{(e^{ax}+r)^{2}} \leq \frac{1}{4r}$ for all $x$ (where $a$ is some constant), so we can use this to obtain global bounds on the matrix norm in the transformed space. Note that this maximum occurs at $x=\frac{1}{a}\log(r)$. To find global bounds, we can use
\begin{equation}
    \sup\left[\left| \frac{\partial^{2} \log f_{c}^{(z)}(\bm{z})}{\partial z_{k} \partial z_{l}} \right| \right] = \sum_{i=1}^{n} \frac{(y_{i}+r)r|X\mathbf{\Lambda}_{c}^{\frac{1}{2}}|_{ik}\cdot|X\mathbf{\Lambda}_{c}^{\frac{1}{2}}|_{il}}{4r} + \sum_{j=0}^{p} \frac{\Lambda_{jk}^{\frac{1}{2}} \Lambda_{jl}^{\frac{1}{2}}}{C \sigma_{\beta_{j}}^{2}}.
\end{equation}

\subsubsection{Local bounds of \texorpdfstring{$P^{\mathbf{\Lambda}_{c}}$}{the matrix norm}} \label{sec:NB_phi_local_bounds}

Local bounds can be obtained if we can find local bounds for 
\begin{equation}
    \label{eq:G_r}
    G_{r}(\bm{z}) := \frac{\exp((X_{i}\mathbf{\Lambda}_{c}^{\frac{1}{2}})\bm{z})}{\left(\exp((X_{i}\mathbf{\Lambda}_{c}^{\frac{1}{2}})\bm{z})+r\right)^{2}}
\end{equation}
for $\bm{z}\in R^{(z)}$ and $i=1,\dots,n$. In that case, we have
\begin{align}
    \sup_{\bm{z} \in R^{(z)}}\left[\left| \frac{\partial^{2} \log f_{c}^{(z)}(\bm{z})}{\partial z_{k} \partial z_{l}} \right| \right]
    & = \sum_{i=1}^{n} \left[(y_{i}+r)r|X\mathbf{\Lambda}_{c}^{\frac{1}{2}}|_{ik}\cdot|X\mathbf{\Lambda}_{c}^{\frac{1}{2}}|_{il}\cdot\max_{\bm{z} \in R^{(z)}}\left\{G_{r}(\bm{z})\right\}\right] + \sum_{j=0}^{p} \frac{\Lambda_{jk}^{\frac{1}{2}} \Lambda_{jl}^{\frac{1}{2}}}{C \sigma_{\beta_{j}}^{2}}.
\end{align}

We can obtain bounds for $G_{r}(\bm{z})$ by noting that $\frac{\exp(x)}{(r+\exp(x))^{2}}\leq\frac{1}{4r}$ for all $x$ and this maximum is attained at $x=\log(r)$. Further note that $\frac{\exp(x)}{(r+\exp(x))^{2}}\leq\frac{1}{4r}$ is a uni-modal function (with mode at $x=\log(r)$ as noted). Now let
\begin{equation}
    F_{i}(\bm{z}):=(X_{i}\mathbf{\Lambda}_{c}^{\frac{1}{2}})\bm{z}=\sum_{j=1}^{d}(X_{i}\mathbf{\Lambda}_{c}^{\frac{1}{2}})_{j}\bm{z}_{j},
\end{equation}
then let $F_{i}^{\downarrow}:=\min_{\bm{z} \in R^{(z)}} F_{i}(\bm{z})$ and $F_{i}^{\uparrow}:=\max_{\bm{z} \in R^{(z)}} F_{i}(\bm{z})$ denote the minimum and maximum of $F_{i}(\bm{z})$ for $\bm{z} \in R^{(z)}$ respectively. Then we note that this can simply be computed in with a linear cost with $d$. Now, noting that $F_{i}(\bm{z})$ is linear and $\frac{\exp(x)}{(r+\exp(x))^{2}}\leq\frac{1}{4r}$ is uni-modal, after computing $F_{i}^{\downarrow}$ and $F_{i}^{\uparrow}$, there are two cases:
\begin{enumerate}
    \item If we have $\log(r) \in [F_{i}^{\downarrow}, F_{i}^{\uparrow}]$, then we know that for this hypercube $R^{(z)}$, we will attain the maximum $\frac{1}{4r}$.
    \item If $\log(r)\notin[F_{i}^{\downarrow}, F_{i}^{\uparrow}]$, then the maximum of $G_{r}(x)$ occurs at which ever point is the closest to $\log(r)$.
\end{enumerate}

Therefore local bounds can be obtained by minimising and maximising $F_{i}(\bm{z})$ for $\bm{z} \in R^{(z)}$. If this interval includes $\log(r)$, then the local maximum attains the global maximum, otherwise, the local maximum occurs at either of these intervals (whichever is closer to $\log(r)$).

This method for finding local bounds requires two optimisations of $F_{i}(\bm{z})$, but we note that we can actually obtain the bounds by only performing one optimisation. In particular, we can evaluate $F_{i}(\bm{z})$ at any arbitrary value $\hat{\bm{z}} \in R^{(z)}$ (we simply take this to be the centre of the hypercube). If we have $F_{i}(\hat{\bm{z}})>\log(r)$, then we just need only need minimise the function $F_{i}(\bm{z})$, since if we have $F_{i}^{\downarrow} < \log(r)$, then we know that $\log(r) \in [F_{i}^{\downarrow}, F_{i}^{\uparrow}]$, so the global maximum is attained. If $F_{i}^{\downarrow} > \log(r)$, then the maximum of $G(x)$ just occurs at $F_{i}^{\downarrow}$ and we can avoid the need to maximise the function $F_{i}(\bm{z})$. However, if conversely, we evaluate $F_{i}(\bm{z})$ at $\bm{z}=\hat{\bm{z}}$ and we have $F_{i}(\hat{\bm{z}})<\log(r)$, then we just need to maximise $F_{i}(\bm{z})$ for $\bm{z} \in R^{(z)}$ and apply the inverse of the same trick. To summarise, in order to find $\max_{\bm{z} \in R^{(z)}}\left\{G_{r}(\bm{z})\right\}$, we can apply \algoref{alg:G_max}.

\begin{algorithm}
    \caption{Computing the local bounds of $G_{r}(\bm{z})$ given in \eqref{eq:G_r} for $\bm{z} \in R^{(z)}$.}
    \label{alg:G_max}
    \begin{enumerate}
        \item Compute $F_{i}(\hat{\bm{z}})$ at some arbitrary value $\hat{\bm{z}} \in R^{(z)}$.
        \item If $F_{i}(\hat{\bm{z}})>\log(r)$:
        \begin{enumerate}
            \item Compute $F_{i}^{\downarrow}:=\min_{\bm{z} \in R^{(z)}} F_{i}(\bm{z})$.
            \item $\max_{\bm{z} \in R^{(z)}}\left\{\frac{\exp((X_{i}\mathbf{\Lambda}_{c}^{\frac{1}{2}})\bm{z})}{\left(\exp((X_{i}\mathbf{\Lambda}_{c}^{\frac{1}{2}})\bm{z})+r\right)^{2}}\right\}=
            \begin{cases}
            \frac{1}{4r} & \text{ if } F_{i}^{\downarrow} < \log(r), \\
            G(F_{i}^{\downarrow})=\frac{\exp(F_{i}^{\downarrow})}{\left(\exp(F_{i}^{\downarrow})+r\right)^{2}} & \text{ otherwise.}
            \end{cases}$
        \end{enumerate}
        \item Else (if $F_{i}(\hat{\bm{z}})<\log(r)$):
        \begin{enumerate}
            \item Compute $\max_{\bm{z} \in R^{(z)}} F_{i}(\bm{z})$.
            \item $\max_{\bm{z} \in R^{(z)}}\left\{\frac{\exp((X_{i}\mathbf{\Lambda}_{c}^{\frac{1}{2}})\bm{z})}{\left(\exp((X_{i}\mathbf{\Lambda}_{c}^{\frac{1}{2}})\bm{z})+r\right)^{2}}\right\}=
            \begin{cases}
            \frac{1}{4r} & \text{ if } F_{i}^{\uparrow} > \log(r), \\
            G(F_{i}^{\uparrow})=\frac{\exp(F_{i}^{\uparrow})}{\left(\exp(F_{i}^{\uparrow})+r\right)^{2}} & \text{ otherwise.}
            \end{cases}$
        \end{enumerate}
    \end{enumerate}
\end{algorithm}

\end{appendix}

\bibliography{biblio}

\end{document}